\documentclass[a4paper,12pt,numbered,print,index,oneside]{Classes/PhDThesisPSnPDF}

\usepackage[utf8]{inputenc}

\usepackage{algorithm}
\usepackage{algpseudocode}
\usepackage{braket}
\usepackage{bm}
\usepackage{qcircuit}
\usepackage{mathtools}
\usepackage{amsmath}
\usepackage[capitalize]{cleveref}
\usepackage{epigraph}
\usepackage{simpler-wick}
\usepackage{multicol}
\usepackage{hyperref}
\usepackage{tikz}
\usepackage{tikz-3dplot}
\usepackage{asymptote}
\usepackage{multicol}
\usepackage[english,spanish]{babel}
\usepackage{enumitem}
\usepackage{pdfpages}
\usepackage{blkarray}

\newtheorem{theorem}{Theorem}

\newtheorem{definition}{Definition}

\newtheorem{problem}{Problem}
\newtheorem{lemma}{Lemma}
\newcommand{\Prep}{\text{Prep}}
\newcommand{\Sel}{\text{Sel}}
\newcommand{\sgn}{\text{sgn}}
\newcommand{\polylog}{\text{poly}\log}
\newcommand{\poly}{\text{poly}}
\newcommand{\grad}{\nabla}
\newcommand{\Tr}{\text{Tr}}
\newcommand{\col}{\text{color}}
\newcommand{\Z}{\mathbb{Z}}
\newcommand{\xMapsto}[2][]{\ext@arrow 0599{\Mapstofill@}{#1}{#2}}
\def\Mapstofill@{\arrowfill@{\Mapstochar\Relbar}\Relbar\Rightarrow}

\DeclareMathOperator{\Ima}{Im}

\usepackage{comment}
\input{Preamble/preamble}

\title{Fault tolerant quantum algorithms}

\subtitle{Algoritmos cu\'anticos tolerantes a fallos}

\author{Pablo Antonio Moreno Casares}

\dept{Departamento de Física Teórica\\
Facultad de Ciencias Físicas}

\university{Universidad Complutense de Madrid}
\crest{\includegraphics[width=0.35\textwidth]{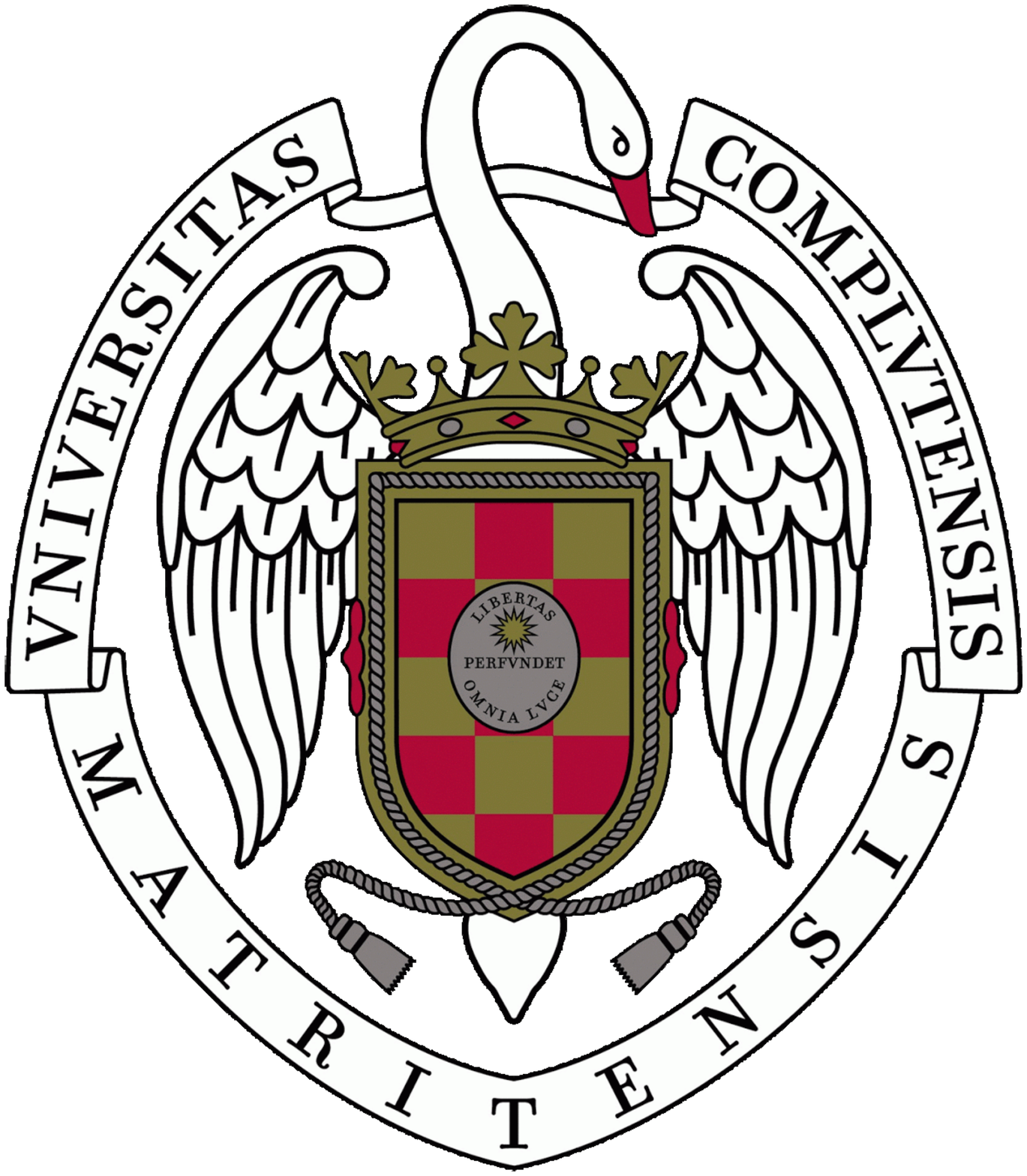}}

\supervisor{Miguel Ángel Mart\'in-Delgado Alcántara}

 \supervisorrole{Director: }

\supervisorlinewidth{0.67\textwidth}

\renewcommand{\submissiontext}{Memoria para optar al título de}

\degreetitle{Doctor en Física}

\degreedate{January 2023} 

\ifdefineAbstract
\fi

\ifdefineChapter
\fi

\begin{document}

\frontmatter

\includepdf[pages=-]{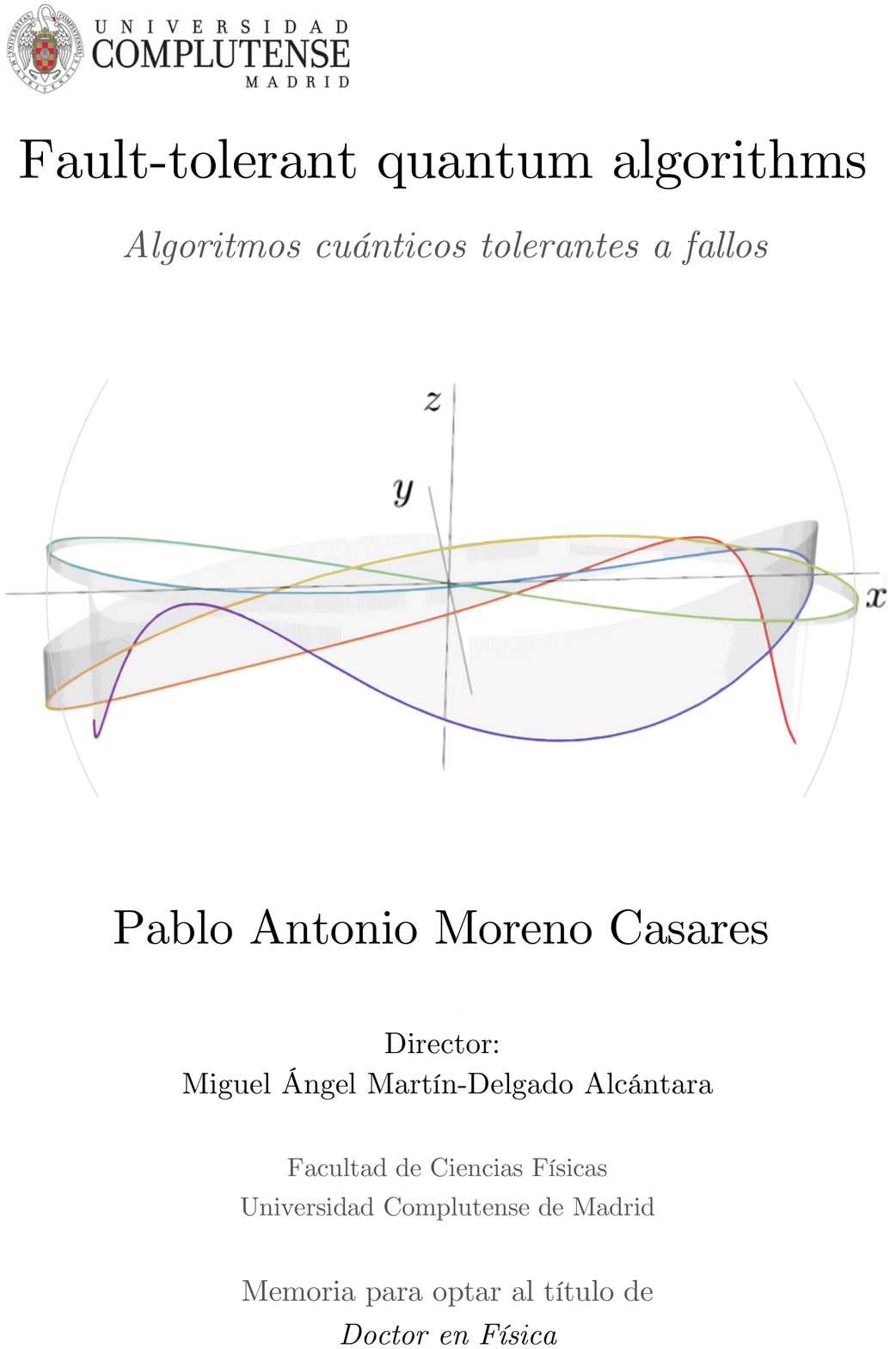}
\clearpage\thispagestyle{empty}\mbox{}\clearpage

\maketitle

\includepdf[pages=-]{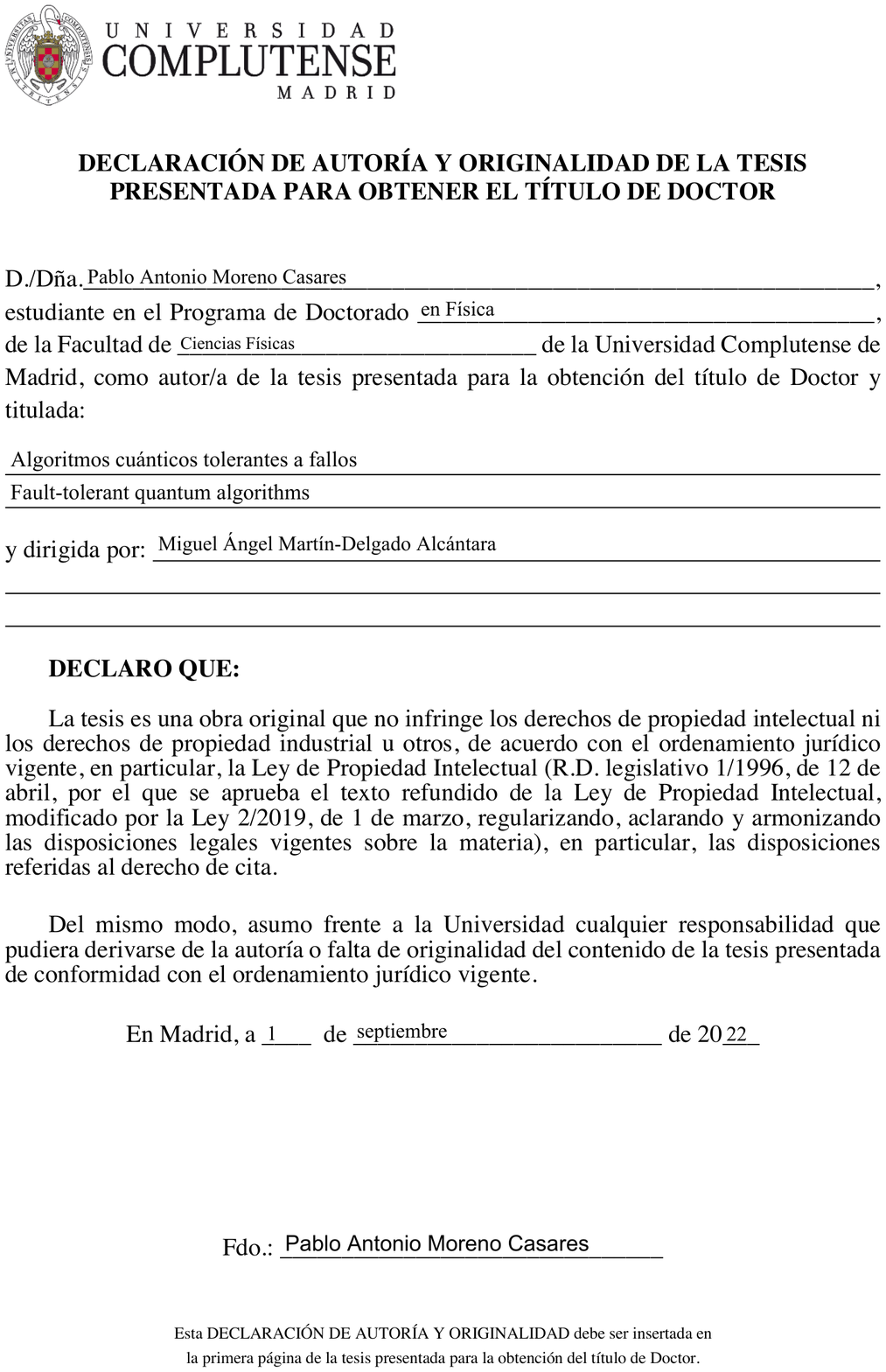}


\begin{dedication} 

Me gustaría dedicar esta tesis a mis abuelos que siempre han sido una fuente de inspiración, y mis padres y Pili, por su apoyo constante.

\end{dedication}

\begin{acknowledgements}

En primer lugar me gustaría agradecer al director de tesis, Miguel Ángel, por su dirección a lo largo de esta tesis, y muy en particular por la confianza depositada a la hora de elegir proyectos de investigación. Creo que esta tesis representa un éxito de supervisión por su parte. Así mismo, me gustaría agradecer a los miembros de nuestro grupo de investigación, como Roberto Campos o Santiago Varona, y con otros estudiantes del departamento de física, como Roberto Ruiz, que en ocasiones me han sugerido ideas sobre cómo abordar problemas técnicos. Además, quiero agradecer especialmente a Santiago Varona su disponibilidad para leer esta tesis con antelación y sugerir correcciones y cambios, lo que se ha reflejado en una mejora de la redacción y coherencia del texto.

También quiero dar las gracias a las personas con las que he colaborado en diversos artículos. Por un lado, el grupo de Juan Miguel (Alain, Modji, Roberto...) en \href{https://xanadu.ai/}{xanadu.ai} por su confianza y tratarme como parte del equipo. Además, agradezco mucho a Juan Miguel, Sam Pallister y Peter Jonhson la valoración positiva de mi trabajo. Por otro lado, estoy muy agradecido con el grupo de Jose Orallo, de Valencia, por una estancia espectacular. Jose es una de las personas más trabajadoras que conozco y es estupendo poder investigar con él. Espero que en el futuro podamos seguir trabajando juntos. Además, le agradezco su interés por el área de AI Safety, y creo que su experiencia en evaluación de modelos de inteligencia artificial puede ser muy útil.

En la misma línea, doy las gracias a muchas de las personas relacionadas con \href{https://www.effectivealtruism.org/}{Altruismo Eficaz} que me han apoyado durante estos años, y en especial a Ryan Carey y Jaime Sevilla
por su apoyo continuado, y a Victor Veitch por su predisposición a acogerme como postdoc. Me enorgullece formar parte de una comunidad a la que le preocupa mejorar el mundo cuanto sea posible. También me alegra poder contar con personas como Juan García, Pablo Melchor y Pablo Villalobos como amigos.

Finalmente, me gustaría agradecer a mi familia y a mis padres su apoyo incondicional a lo largo de estos años, así como a Pili porque es una persona maravillosa de la que estoy muy orgulloso. Sabes lo mucho que esta tesis se debe a tu ayuda constante.

\end{acknowledgements}

\selectlanguage{english}

\tableofcontents
\newpage
\begin{center}
\section*{Abstract}
\end{center}
\addcontentsline{toc}{section}{Abstract}

The framework of this thesis is fault-tolerant quantum algorithms, which can roughly be divided into the following non-disjoint families: a) Grover's algorithm and quantum walks, b) Shor's algorithm and hidden subgroup problems, c) quantum simulation algorithms, d) quantum linear algebra, and e) variational quantum algorithms. All of them are covered, to some extent, in this thesis.

Grover's algorithm and quantum walks are described in \cref{ch:Search}. We start by highlighting the central role that rotations play in quantum algorithms, explaining Grover's, why it is optimal, and how it may be extended. Key subroutines explained in this area are amplitude amplification and quantum walks, which will constitute useful parts of other algorithms. In this chapter, we present our Ref.~\cite{casares2022qfold}, where we explore the heuristic use of quantum Metropolis and quantum walk algorithms for solving an NP-hard problem. This method has been suggested as an avenue to digitally simulate quantum annealing and preparing ground states of many-body Hamiltonians.

In the third chapter, in contrast, we turn to the exponential advantages promised by the Fourier transform in the context of the hidden subgroup problem. However, since this application is restricted to cryptography, we later explore its use in quantum linear algebra problems. Here we explain the development of the original quantum linear solver algorithm, its improvements, and finally the dequantization techniques that would often restrict the quantum advantage to polynomial. In this chapter, we present two publications, Refs.~\cite{casares2020IP,casares2020active}, the former aimed at solving linear programming problems, and the latter at machine learning problems. As we shall see, both of them are restricted but in a different way: the first requires reading out the entire solution quantum state, while the second can be effectively dequantized.

\cref{ch:Chemistry} is concerned with quantum simulation. We will review classical quantum chemistry techniques, and then focus on Hamiltonian simulation and ground state preparation as the key problems to be solved. Hamiltonian simulation, in particular, will enable the use of quantum phase estimation which computes the eigenvalues or energies of a given quantum state. Our contribution~\cite{casares2021tfermion} constitutes a detailed analysis of the cost of many Hamiltonian simulation techniques present in the literature. Variational algorithms, while many times used as a Noisy Intermediate-Scale Quantum (NISQ) alternative to quantum phase estimation, will be presented here as a viable approach to ground state preparation, the other main problem. In contrast, in Ref.~\cite{delgado2022simulate}, we detail how to prepare the Hartree-Fock state in first quantization and plane waves, ideally suited for materials, and which does not require using those techniques. Combined with a state-of-the-art Hamiltonian simulation technique called qubitization, we can estimate the non-Clifford gate cost of running such an algorithm for realistic battery material properties.

Given the tradition of our group with error correction, we could not end this thesis without dedicating a final chapter to this topic. Here we explain the most important quantum error correction codes, the surface and color codes, and one extension of the latter, gauge color codes. They will show the complexity of implementing non-Clifford quantum gates, therefore validating their consideration as the bottleneck metric.

\selectlanguage{spanish}
\newpage
\begin{center}
\section*{Resumen}
\end{center}
\addcontentsline{toc}{section}{Resumen}

El marco conceptual de esta tesis son los algoritmos cuánticos tolerantes a fallos, que pueden dividirse aproximadamente en las siguientes clases no mutuamente excluyentes: a) algoritmo de Grover y paseos cuánticos, b) algoritmo de Shor y problemas de subgrupos ocultos, c) algoritmos de simulación cuántica, d) álgebra lineal cuántica, y e) algoritmos cuánticos variacionales. Todos ellos se tratan, en cierta medida, en esta tesis.

El algoritmo de Grover y los paseos cuánticos se explican en el capítulo \ref{ch:Search}. Comenzamos destacando el papel central que juegan las rotaciones en los algoritmos cuánticos, explicando el de Grover, por qué es óptimo, y cómo puede ser extendido. Las subrutinas clave explicadas en esta área son la amplificación de la amplitud y los paseos cuánticos, que serán partes importantes de otros algoritmos. En este capítulo presentamos nuestra Ref.~\cite{casares2022qfold}, donde exploramos el uso heurístico de los algoritmos de Metrópolis y paseos cuánticos para resolver problemas NP-difíciles. De hecho, este método ha sido sugerido como una vía para simular digitalmente el método conocido como `quantum annealing', y la preparación de estados fundamentales de Hamiltonianos `many-body'.

En el tercer capítulo, en cambio, nos centramos en las ventajas exponenciales que promete la transformada de Fourier en el contexto del problema de los subgrupos ocultos. Sin embargo, dado que esta aplicación está restringida a la criptografía, más adelante exploramos su uso en problemas de álgebra lineal cuántica. Aquí explicamos el desarrollo del algoritmo cuántico original para la resolución de sistemas lineales de ecuaciones, sus mejoras, y finalmente las técnicas de `descuantización' que a menudo restringen la ventaja cuántica a polinómica. En este capítulo se presentan dos publicaciones, Refs.~\cite{casares2020IP,casares2020active}, la primera orientada a la resolución de problemas de programación lineal, mientras que la segunda al aprendizaje automático. Como veremos, ambas están restringidas, pero de forma diferente: la primera requiere la lectura de estados cuánticos, mientras que la segunda puede ser efectivamente descuantizada.

El capítulo \ref{ch:Chemistry} se ocupa de la simulación cuántica. Revisaremos las técnicas clásicas de la química cuántica, y luego nos centraremos en la simulación Hamiltoniana y en la preparación del estado fundamental como problemas clave a resolver. La simulación Hamiltoniana, en particular, permitirá el uso de la estimación de fase cuántica, que calcula los valores propios o las energías de un estado cuántico dado. De hecho, nuestra contribución~\cite{casares2021tfermion} constituye una comparativa detallada del coste de muchas técnicas de simulación Hamiltoniana presentes en la literatura. Los algoritmos variacionales, aunque muchas veces utilizados como alternativa a la estimación cuántica de fase en sistemas ruidosos (NISQ), se presentarán aquí como un enfoque viable a la preparación del estado fundamental, el segundo problema principal. 
Por el contrario, en la Ref.~\cite{delgado2022simulate} detallamos cómo preparar el estado Hartree-Fock en primera cuantización y ondas planas, adecuado para materiales, que no requiere de dichas técnicas. Combinado con una técnica de simulación Hamiltoniana de última generación llamada qubitización, somos capaces de estimar el coste en puertas lógicas no-Clifford de ejecutar dicho algoritmo. Dicho algoritmo puede ser usado para predecir propiedades realistas de materiales de baterías eléctricas.

Dada la tradición de nuestro grupo en la corrección de errores, no podíamos terminar esta tesis sin dedicar un capítulo final a este tema. Aquí explicamos los códigos de corrección de errores cuánticos más importantes, los códigos de superficie y de color, y una extensión de estos últimos, los códigos de gauge de color. Así mostraremos la complejidad de implementar puertas cuánticas no-Clifford, validando su consideración como métrica de referencia.

\selectlanguage{english}
\newpage
\begin{center}
\section*{List of publications}
\end{center}
\addcontentsline{toc}{section}{List of publications}

\vspace{1em}
\begin{itemize}
\item Casares, Pablo A. M., Roberto Campos, and Miguel Angel Martin-Delgado. ``QFold: quantum walks and deep learning to solve protein folding.'' Quantum Science and Technology, 7.2 (2022): 025013.
\item Casares, Pablo A. M., and M. A. Martin-Delgado. ``A quantum active learning algorithm for sampling against adversarial attacks.'' New Journal of Physics, 22.7 (2022). 073026.
\item Casares, Pablo A. M., and Miguel Angel Martin-Delgado. ``A quantum interior-point predictor-corrector algorithm for linear programming.'' Journal of Physics A: Mathematical and Theoretical 53.44 (2020): 445305.
\item Casares, Pablo A. M., Roberto Campos, and Miguel Angel Martin-Delgado. ``TFermion: A non-Clifford gate cost assessment library of quantum phase estimation algorithms for quantum chemistry.'' Quantum (2022), 6:768.
\item Delgado, Alain, et al. ``Simulating key properties of lithium-ion batteries
with a fault-tolerant quantum computer.'' Physical Reviews A (2022), 106:032428 -- main co-author.
\item Campos, R., Casares, Pablo A. M., and Martin-Delgado, M. A. ``Quantum Metropolis solver: A quantum walks approach to optimization problems.'' (2022) arXiv preprint arXiv:2207.06462.
\item Escrig, G., Campos, R., Casares, Pablo A. M., and Martin-Delgado, M. A. ``Parameter estimation of gravitational waves with a quantum metropolis algorithm.'' (2022) arXiv preprint arXiv:2208.05506.
\end{itemize}

\noindent Additional articles not covered in this thesis, carried out during the internship, on the topic of classical Machine Learning:
\begin{itemize}
    \item Casares, Pablo A. M., et al. ``How General-Purpose Is a Language Model? Usefulness and Safety with Human Prompters in the Wild'' (2022), Proceedings of the 36th AAAI Conference on Artificial Intelligence (AAAI-22).
\end{itemize}


\newpage
\begin{center}
\section*{Conference contributions and internship}
\end{center}
\addcontentsline{toc}{section}{Conference contributions and internship}

\noindent During this thesis, the author carried out a 3-month internship in the research group of José Hernández Orallo in the field of classical Machine Learning, as a result of which Ref.~\cite{casares2022how} was published and presented as a poster at the prestigious \href{https://aaai.org/Conferences/AAAI-22/}{AAAI-22} conference.

\vspace{1pt}
\noindent Additionally, the author has carried out the following activities during his PhD:
\begin{itemize}
    \item Oral presentation of Ref.~\cite{casares2021tfermion} at the \href{https://meetings.aps.org/Meeting/MAR22/Content/4178}{APS March meeting 2022}.
    \item Oral presentation of Ref.~\cite{casares2022qfold} at the \href{https://www.quantumconf.eu/2022/}{Quantum conference in Bilbao in 2022}.

    \item Tutorial on the works of Ref.~\cite{casares2022qfold,campos2022quantum} during the \href{https://qce.quantum.ieee.org/2021/}{IEEE 2021 Quantum week}.
    \item Poster presentation of the work in Ref.~\cite{casares2022qfold} at the \href{https://ice-6.hbar.es/index.html}{ICE-6 
    Quantum Information in Spain conference in 2021}.
    \item Poster presentation of the work in Ref.~\cite{casares2022qfold} during the \href{https://www.mcqst.de/news-and-events/events/conference-2021.html}{Munich conference on Quantum Science and Technology 2021}.
    \item Poster presentation of the work in Ref.~\cite{casares2022qfold} in the \href{https://www.quantummachinelearning.org/qtml2021.html}{Quantum Techniques in Machine Learning 2021 conference}.
    
    \item Poster presentation of the work in Ref.~\cite{casares2020active} at the conference \href{https://premc.org/conferences/qtech-quantum-technology/}{QTech 2020}.
    \item Poster presentation of the work in Ref.~\cite{casares2020IP} during the \href{https://www.mcqst.de/mcqst2020/home/}{Munich conference on Quantum Science and Technology 2020}.
    \item Poster presentation of the work in Ref.~\cite{casares2020active} in the Quantum Techniques in Machine Learning 2020 conference.
    \item Workshop presentation in the \href{https://humanaligned.ai/}{Human-aligned AI Summer School} in 2022 on the topic of `How to produce high-quality research'.
    \item Attendance and organization of the \href{https://aisrp.org/}{AI Safety Research Program in 2020}.
    \item Attendance to the \href{https://humanaligned.ai/}{Human-aligned AI Summer School} in 2019.
    \item Attendance to the \href{https://aisafety.camp/}{AI Safety Camp} in 2019.
    \item Attendance to the \href{http://dalimeeting.org/dali2019b/workshop-05-04.html}{ELLIS Quantum Machine Learning workshop in San Sebastian in 2019.}
\end{itemize}




\printnomenclature

\mainmatter

\chapter{Introduction}  

\ifpdf
    \graphicspath{{Introduction/Figs/Raster/}{Introduction/Figs/PDF/}{Chapter1/Figs/}}
\else
    \graphicspath{{Introduction/Figs/Vector/}{Chapter1/Figs/}}
\fi



We are living an extraordinary time for quantum computing. Not even a year after the start of this thesis, in 2019, a team at Google released an article highlighting that they had been able to execute a quantum supremacy experiment, one experiment that is not feasible for classical computers~\cite{arute2019quantum}. This first experiment used random quantum circuits, but during these years other supremacy experiments based on boson sampling have also been carried out~\cite{zhong2020quantum,madsen2022quantum}. On the other hand, this last year we have started witnessing the first experiments that aim to fault-tolerantly implement a set of quantum gates~\cite{krinner2022realizing,ryan2021realization,postler2022demonstration,zhao2021realizing}, which open the door to fault-tolerant quantum algorithms, the main topic of this thesis, and the kind of algorithms that are likely to be most useful in the long term.

However, they are not the only kind of algorithms. Noisy intermediate-scale quantum algorithms (NISQ) have been proposed as a useful alternative while we are not able to achieve fault tolerance~\cite{bharti2021noisy}. The two most famous quantum algorithms in this category are the Variational Quantum Eigensolver~\cite{peruzzo2014variational} and the Quantum Approximate Optimization Algorithm~\cite{farhi2014quantum}. The most important limitation of these algorithms, when used in a non-fault-tolerant setup, is that it is difficult to surpass the capabilities of classical computers able to handily implement billions of logical gates with just a thousand quantum logical gates~\cite{coudron2020computations}. This intuition is the reason I have focussed on fault-tolerant quantum algorithms in this thesis.

The first and most famous quantum algorithms are Grover's~\cite{grover1998quantum} and Shor's~\cite{shor1999polynomial}, which will be the basis for the second and third chapters of this thesis. The second chapter will deal with quantum search, starting from the Grover algorithm and exploring quantum walks and amplitude amplification as important tools not only for searching itself but in other quantum algorithms. In this chapter, we present QFold~\cite{casares2022qfold}, an article and simulation experiment where we use a heuristic quantum walk to find the folded configuration of proteins. Unfortunately, the quantum advantage we measure is fairly small and unlikely to be directly useful, given the overhead of quantum error correction in the speed of quantum gates. On the other hand, heuristic quantum walks can digitally simulate quantum annealing processes, which might be used to prepare the ground state of fermionic Hamiltonians~\cite{lemieux2021resource}. In general, the quantum speedups one may find with quantum walks and amplitude amplification are polynomial, often quadratic, in nature. Moreover, this advantage is often diluted when parallelism is possible. In particular, to solve an unstructured search in some fixed amount of time, one only needs linearly more classical than quantum processors~\cite{gingrich2000generalized}.

The third chapter, in contrast, takes a look at the kind of problems where one may hope to find exponential speedups, mostly relying on the quantum Fourier transform. Shor's algorithm represents the paradigmatic example of these techniques, later generalized to the (Abelian) hidden subgroup problem and mostly used in cryptography applications. The price to pay for this large quantum advantage is the need for structure in the problem that is being addressed~\cite{aaronson2009need}. However, in 2008 there was a breakthrough that promised an exponential speedup in the ubiquitous problem of solving a linear system of equations in a very quantum way: with quantum input and output, and under sparsity and well-conditioning assumptions~\cite{harrow2009quantum}. This would become a line of research called quantum linear algebra, and it is used in two more articles during this thesis~\cite{casares2020active,casares2020IP}. The latter uses those techniques in combination with classical methods in the literature, to propose one of the first quantum interior point methods, aimed at solving linear programming problems. However, the limitations of quantum methods and the necessity to prepare and read quantum states again reduce the quantum advantage to polynomial. In the second paper, we found one machine learning application (adversarial examples) where a readout of the quantum state is not needed. However, in a second breakthrough, an undergraduate student called Ewin Tang showed that many of these quantum linear algebra algorithms could be `dequantized'~\cite{tang2019quantum}. Dequantization substitutes quantum computing for randomized classical computing with $\ell_2$-sampling~\cite{tang2018pca}, showing again that many algorithms, and in particular ours, will only ever achieve a polynomial speedup.

But if there is a research topic where one may expect quantum computing to be particularly useful, that is in the area of simulation of other quantum systems. Thus,~\cref{ch:Chemistry} will discuss its applications to chemistry. It has three main parts. In the first one, we review three of the most popular classical algorithms: Hartree-Fock, Density Functional Theory, and Coupled-Cluster theory. Then, we move on to Hamiltonian simulation, a core quantum technique that computes how a system evolves. From this section, I would like to highlight one technique, called \textit{qubitization}~\cite{low2019qubitization}, which allows performing in an optimal number of queries a large variety of functions of the Hamiltonian. This fact will be reflected in previous chapters too~\cite{martyn2021grand}, but its usefulness to implement Hamiltonian simulation will be particularly reflected here. Finally, in the third section, we discuss ground state preparation, perhaps the most challenging problem in quantum chemistry. This is perhaps the area where variational quantum algorithms, also known as quantum machine learning, will have the largest impact~\cite{huang2021provably,mcclean2021foundations}. In the interest of length, however, we will only review the basics. 

Two articles from my thesis belong to this chapter, Refs.~\cite{casares2021tfermion,delgado2022simulate}. In the first one, we present what is perhaps the first software library allowing us to compare a variety of quantum phase estimation algorithms: TFermion. This library analyzes the cost of several quantum phase estimation algorithms proposed in the literature and enables comparisons between them that were previously not possible. TFermion also led to a collaboration with quantum computing startup Xanadu, and ultimately to Ref.~\cite{delgado2022simulate}. In this reference, we provide a thorough review of how we may use one qubitization-based first-quantization quantum algorithm to analyze the properties of battery materials. Additionally, we perform two minor but important technical advances in the area of Hartree Fock state preparation in plane waves and first quantization, and on the extension of the original algorithm to non-cubic cells. Finally, we provide detailed estimates on the number of non-Clifford gates and the time a quantum computer would require to simulate a particular battery material.

The fifth and final chapter of the thesis will answer the question of why non-Clifford gates are often the most expensive ones to be implemented in a quantum computer. Its focus will be on quantum error correction, an area of research with a long tradition in our research group. Unfortunately, the area is wide enough to only provide a review of the main topics. We start by describing the basics of Calderbank–Shor–Steane (CSS) and stabilizer codes. Then, we explain the two most popular families of error correction codes: surface codes~\cite{kitaev1995quantum} and color codes~\cite{bombin2006topological,bombin2007topological}, the latter of which was a key breakthrough by my thesis director. Unfortunately, a theorem by Eastin and Knill proves that it is impossible to construct a stabilizer code that is capable of transversally implementing a universal set of gates~\cite{eastin2009restrictions}. Therefore, while surface and color codes are very attractive, we are forced to find a way to perform those gates fault-tolerantly. The two leading approaches are distilling one kind of gates (often non-Clifford gates) in another code, what is known as magic state distillation, or using subsystem codes. The first one is popular due to its low cost at high error regimes~\cite{beverland2021cost}, and also the reason non-Clifford gates are expensive. In this thesis, we have instead chosen to describe gauge color codes~\cite{bombin2015gauge}, as a beautiful subsystem code framework that generalize color codes.


\chapter{\label{ch:Search}Quantum Search}

\ifpdf
    \graphicspath{{Chapter1/Figs/Raster/}{Chapter1
    /Figs/PDF/}{Chapter1/Figs/}}
\else
    \graphicspath{{Chapter1/Figs/Vector/}{Chapter1/Figs/}}
\fi

\epigraph{One thing that should be learned from the bitter lesson is the great power of general-purpose methods, of methods that continue to scale with increased computation even as the available computation becomes very great. The two methods that seem to scale arbitrarily in this way are search and learning.}{Richard S. Sutton, {\it The Bitter Lesson}}

\section{Objectives}

\begin{itemize}
    \item Understand the mathematical techniques behind quantum walks and their properties.
    \item Understand the strengths and limitations of quantum walks as a search or Markov-chain mixing technique.
    \item Explore the kind of problems where quantum walks might be useful, with a special focus on the quantum Metropolis algorithm as a privileged application to combinatorial optimization.
    \item Find out the quantum advantage associated with a heuristic quantum Metropolis algorithm, the problems where it might be applied, and its usefulness.
\end{itemize}

\section{\label{sec:Grover}A tale of two rotations: Grover's algorithm}

Search is one of the few basic subroutines across many algorithms in computer science. Optimal planning, simulated annealing, and reinforcement learning, widely used in many and an increasing number of applications, can all be seen as ways of searching under different conditions. 
Perhaps the simplest case is when one has to look for a marked item in an unordered list. Lacking any structure to guide it, a classical computer cannot but iterate over all elements of the set, perhaps allowing some parallelization, in search for the target item. Consequently, if there are $N$ elements in the list, this implies computational complexity scaling as $O(N)$.

Can we do better? Surprisingly, yes: one may leverage quantum mechanics to improve how quickly one can search, which results in $O(\sqrt{N})$ steps.
Let us assume we have an oracle such that given item $x\in \{0,..., N-1\}$, checks whether it fulfills a given condition $f(x) = 1$.
The algorithm implementing this search is called after his discoverer, Lov Grover, and implements two rotations iteratively~\cite{grover1998quantum}, one of which uses this oracle. In fact, for almost any quantum algorithm, it is often intuitive to think in terms of rotations in the Hilbert space. The reason is that quantum operators, except for measurements, are elements of a Special Unitary group $SU(d)$, with $d$ representing the dimension of the Hilbert space. For the Grover algorithm, starting from the uniform superposition $H^{\otimes n}\ket{0}^{\otimes n} = \frac{1}{\sqrt{N}}\sum_{k=0}^{N-1}\ket{k}$, the two rotations are the following:
\begin{itemize}
    \item A phase rotation implemented by any procedure (oracle) that identifies the marked item:
    \begin{equation}
        O:\ket{x} \mapsto (-1)^{f(x)}\ket{x}.
    \end{equation}
    This can be done for example by outputting a bit $\ket{f(x)}$ and implementing a $Z$ gate over it, before uncomputing the bit
    \begin{equation}
        \ket{x}\ket{0} \mapsto \ket{x}\ket{f(x)}\mapsto (-1)^{f(x)}\ket{x}\ket{f(x)}\mapsto (-1)^{f(x)}\ket{x}\ket{0}.
    \end{equation}
    \item A second rotation called diffusion operator $U_s$, that implements a rotation over the initial state, taken to be a uniform superposition
    \begin{equation}
        U_s = H^{\otimes n}(\bm{1} - 2\ket{0}\bra{0} )H^{\otimes n}.
    \end{equation}
\end{itemize}

\begin{figure}
\[
\begin{array}{c}
\Qcircuit @C=0.9em @R=0.75em { 
  & \ket{0} & & \qw & \gate{H} & \qw & \multigate{4}{O}& \qw \gategroup{1}{9}{5}{11}{.7em}{--} & \gate{H} & \ctrlo{1} & \gate{H}& \qw & \multigate{4}{O} &\qw &\cdots & &  \multigate{4}{O}& \qw \gategroup{1}{19}{5}{21}{.7em}{--} & \gate{H} & \ctrlo{1} & \gate{H} & \qw \\
  & \ket{0} & & \qw & \gate{H} & \qw & \ghost{O}& \qw & \gate{H} & \ctrlo{0} & \gate{H}& \qw & \ghost{O} &\qw &\cdots & &  \ghost{O}& \qw & \gate{H} & \ctrlo{0} & \gate{H} & \qw \\
   & \vdots & & & & & & & & \vdots  & & & & & & & & & & \vdots  & &\\
  \\
  & \ket{0} & & \qw & \gate{H} & \qw & \ghost{O}& \qw & \gate{H} & \ctrlo{0} & \gate{H}& \qw & \ghost{O} &\qw &\cdots & &  \ghost{O}& \qw & \gate{H} & \ctrlo{0} & \gate{H} & \qw \\
}
\end{array}
\]
\caption{\textbf{Grover's algorithm.} The dashed box represents the diffusion operator $U_s$, while the oracle operator $O$ is explicitly indicated. After the initialization, operators $O$ and $U_s$ must be applied $O(\sqrt{N})$ times.}
\label{fig:Grover}
\end{figure}
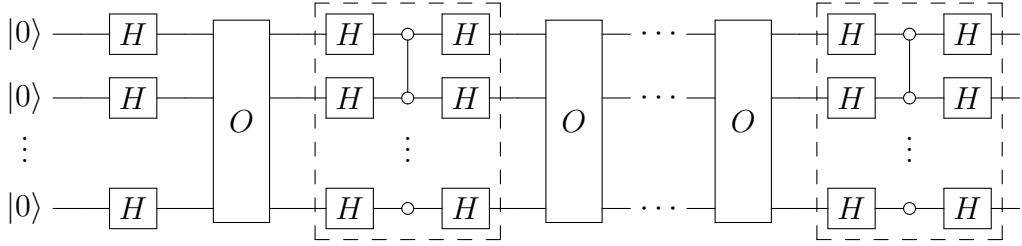

\begin{figure}
    \centering
    \includegraphics[width = \textwidth]{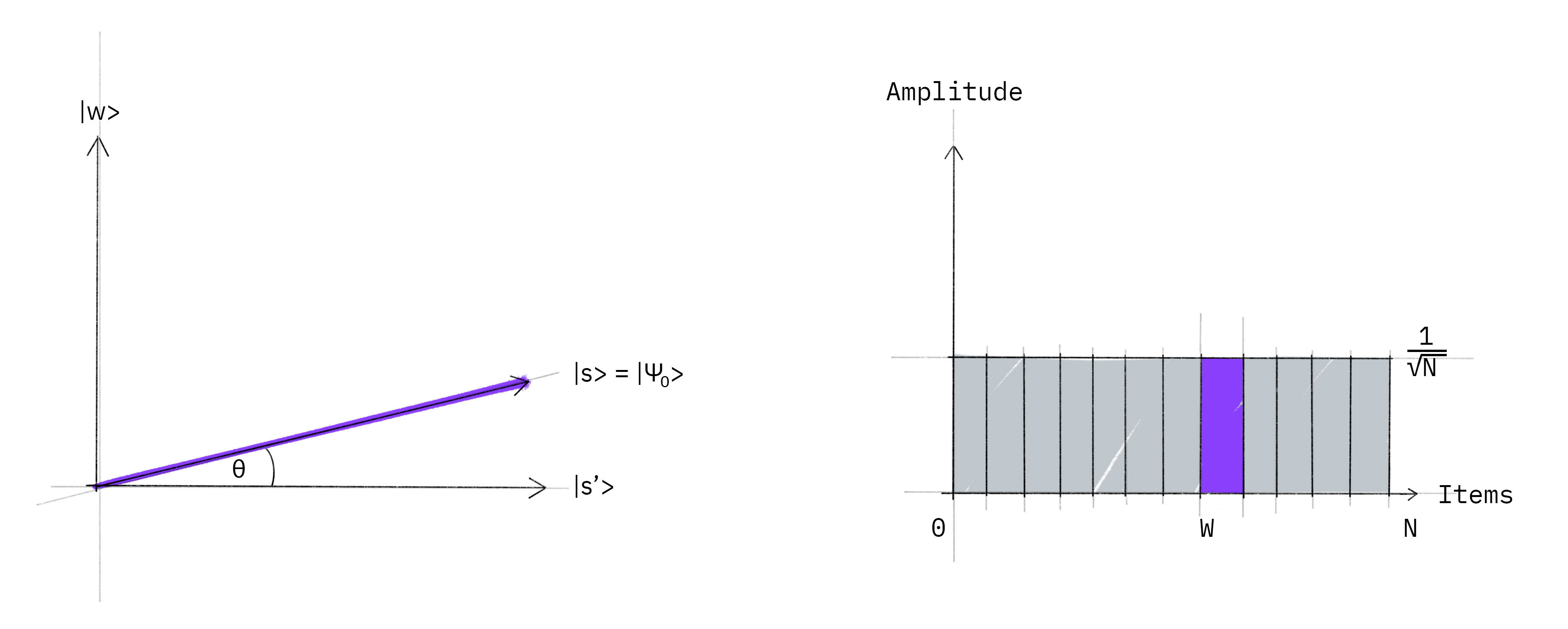}
    \includegraphics[width = \textwidth]{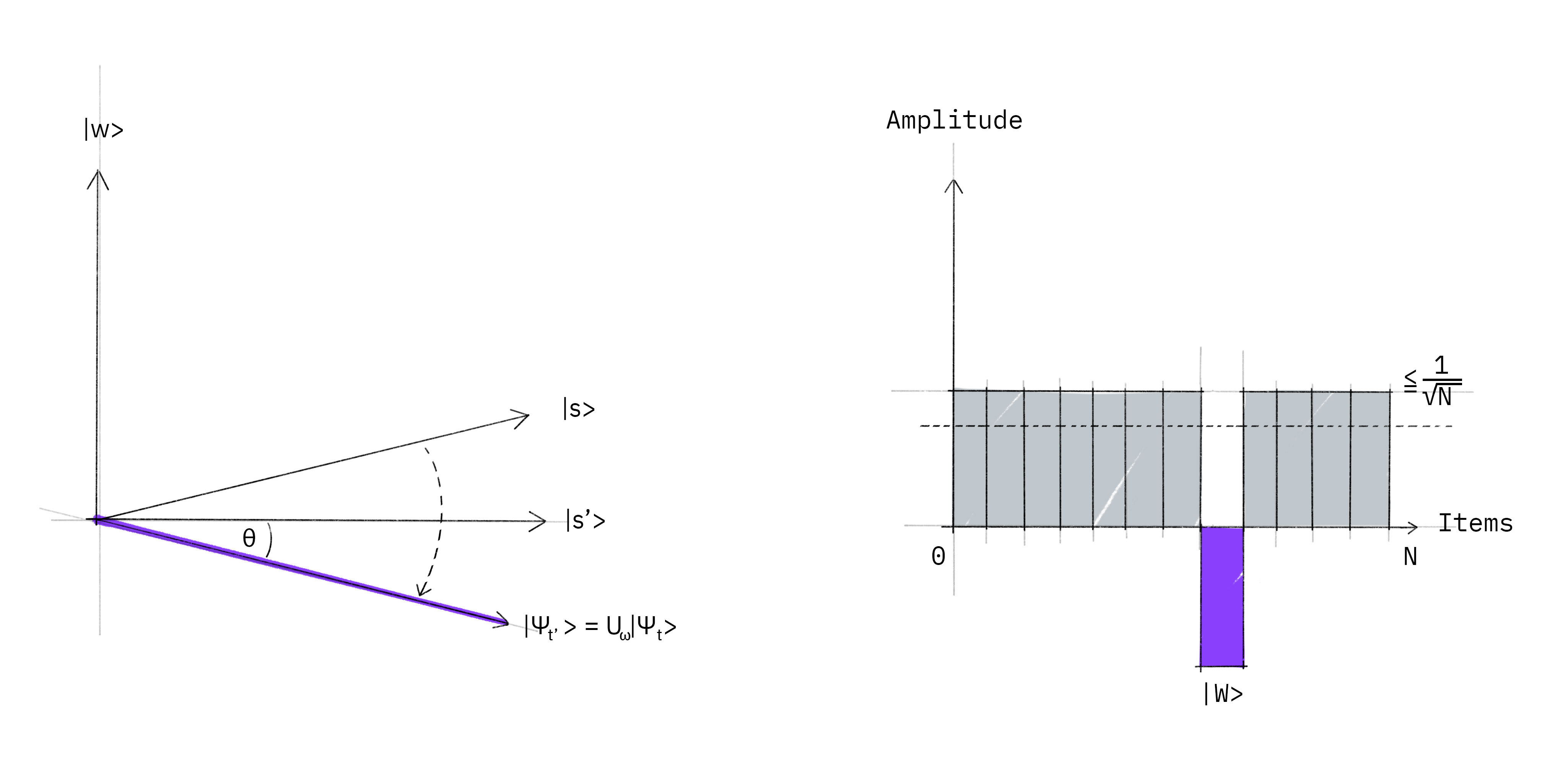}
    \includegraphics[width = \textwidth]{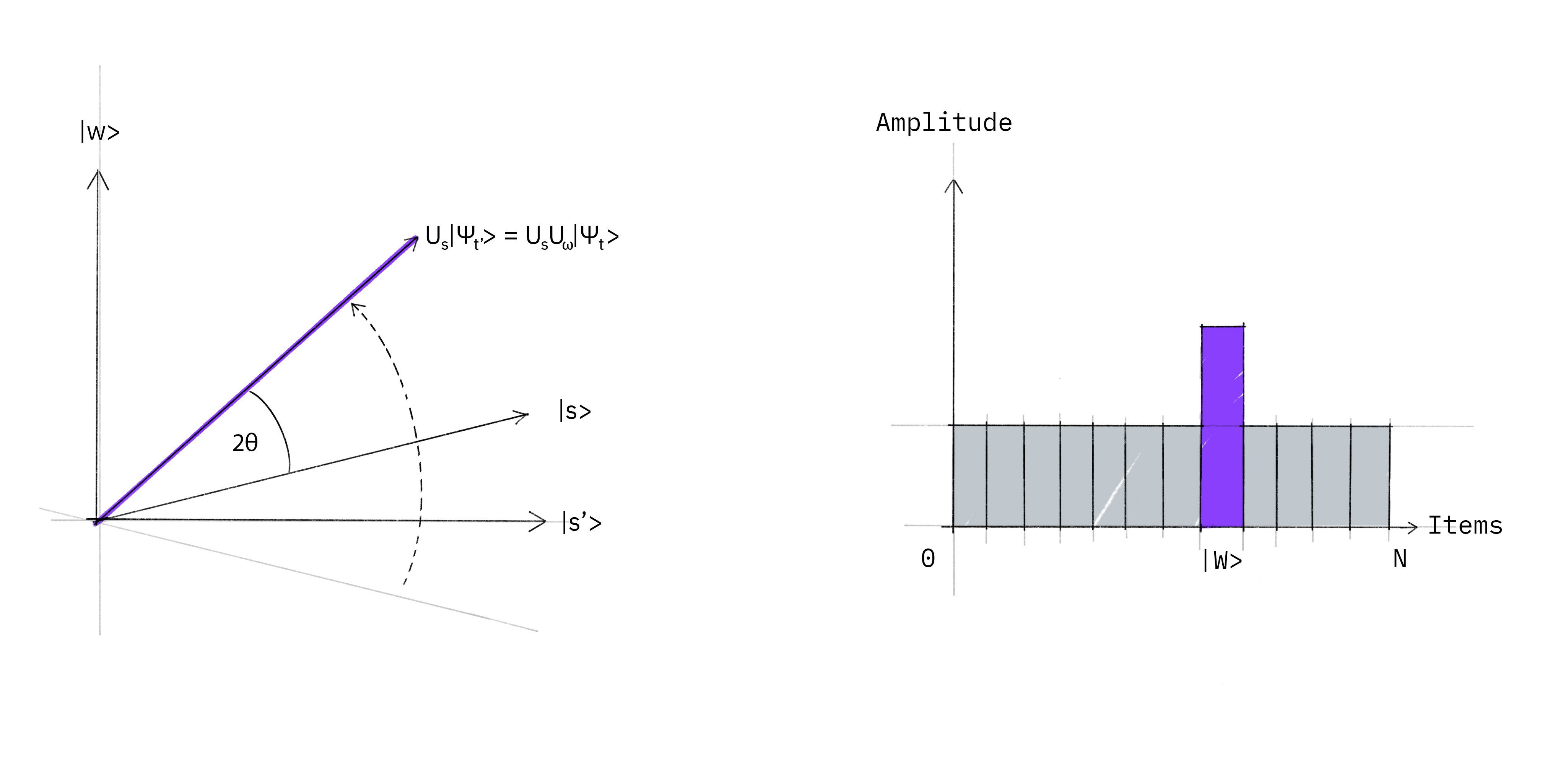}
    \caption{\textbf{Grover's algorithm rotations.} Top: initial state, a uniform superposition of marked and not marked states. Medium: Oracle rotation, which changes the sign of the marked states. Bottom: diffusion operator, that performs a rotation around the average. The angle $\sin\theta = \sqrt{M/N}$, for $M$ the number of marked items in a list of $N$ elements. Images from the \href{https://github.com/qiskit-community/qiskit-textbook/blob/463cf2f13529c7ac7f6dc09d7808f2731edcb2ef/content/ch-algorithms/grover.ipynb}{Qiskit textbook} under Apache 2.0 license.}
    \label{fig:Grover_steps}
\end{figure}
In~\cref{fig:Grover_steps}, we define $\theta := \arcsin \sqrt{\frac{M}{N}}$, where $M\ll N$ is the number of marked items. In each rotation, the angle grows by $2\theta$, so after $t$ time steps the angle will be $(2t+1)\theta$. Since the target angle is $\approx \frac{\pi}{2}$, the number of steps to take grows as $t\approx \frac{\pi}{4}\sqrt{\frac{N}{M}}$. In conclusion, by alternating these two rotations we can achieve a quadratic speedup over what is possible classically.

The next question one may ask is: can we do better? Unfortunately, the answer is no, as Bennett, Bernstein, Brassard, and Vazirani proved even before Grover discovered his algorithm~\cite{bennett1997strengths}. For simplicity assume a single marked item $\ket{\omega}$. Their procedure compares two circuits. The first one contains $T$ calls to the oracle $O = U_{\omega} = \bm{1}-2\ket{\omega}\bra{\omega}$, and other arbitrary but fixed rotations $U_1,\ldots,U_T$,~\cite{preskill1999lecture}
\begin{equation}
    \ket{\psi_\omega(t)} =  U_\omega U_T U_\omega \cdots U_\omega U_1 \ket{\psi(0)}.
\end{equation}
The second is the same circuit without any calls to the oracle
\begin{equation}
    \ket{\varphi(t)} = U_T U_{T-1} \cdots U_1 \ket{\psi(0)}.
\end{equation}
To compare them, we can check how much the operator $U_\omega$ changes the state at time step $t$
\begin{equation}
    || (U_\omega -\bm{1})\ket{\varphi(t)} || = 2|\braket{\omega|\varphi(t)}|.
\end{equation}
This can be interpreted as an error vector between the two preparations. In fact,
\begin{equation}
    ||\ket{\psi_\omega(T)} - \ket{\varphi(T)}|| \leq 2 \sum_{t=1}^T |\braket{\omega|\varphi(t)}|.
\end{equation}
Since simple arithmetic shows that~\cite[Eq. 6.166]{preskill1999lecture}
\begin{equation}
    \left(\sum_{t=1}^T c_t\right)^2 \leq T \sum_{t=1}^T c_t^2
\end{equation}
we have
\begin{equation}
    ||\ket{\psi_\omega(T)} - \ket{\varphi(T)}||^2 \leq  4 \left(\sum_{t=1}^T |\braket{\omega|\varphi(t)}|\right)^2 \leq 4T \sum_{t=1}^T |\braket{\omega|\varphi(t)}|^2\leq 4 T^2.
\end{equation}
To distinguish between states with certainty, we need states to be orthogonal. Consequently, we take $\ket{\omega}$ to be part of an orthonormal basis $\omega'$, and summing over it,
\begin{equation}
    \sum_{\omega'}||\ket{\psi_{\omega'}(T)} - \ket{\varphi(T)}||^2 \leq 4T \sum_{t=1}^T \underbrace{\sum_{\omega'}|\braket{\omega'|\varphi(t)}|^2}_{ =\braket{\varphi(T)|\varphi(T)} } = 4 T^2.
\end{equation}
The left-hand side of this expression can be lower-bounded by $2N-2\sqrt{N}$ \cite[Eq. 6.174]{preskill1999lecture}, so we have that $4 T^2 \geq 2N-2\sqrt{N}$, and finally $T \geq \sqrt{\frac{N-\sqrt{N}}{2}}$. In other words, if we aim to distinguish all $N$ orthogonal possible values of $\ket{\omega}$, we need to call $T = O(\sqrt{N})$ times the oracle $U_\omega$. Further, if we aim for a success probability of $1-\delta$, then one can show that we need $T\geq \sqrt{\frac{N}{2}}\sqrt{1-\sqrt{\delta}}$ \cite[Chapter 6]{preskill1999lecture}.


\subsection{\label{sec:AA}Amplitude Amplification}

Grover's algorithm is, in its original form, just intended for unordered search. We would now like to study some of its main use cases and extensions.
The first of such applications is known as amplitude amplification, and it is widely used in many settings~\cite{brassard2002quantum,grover1998quantum}. Let us suppose we have a quantum algorithm $A$ that probabilistically prepares a state $\ket{\psi}$:
\begin{equation}\label{eq:A_for_AA}
    A:\ket{1}_f\ket{0}_s \mapsto \alpha \ket{0}_f\ket{\psi}_s + \beta \ket{1}_f\ket{\psi^{\perp}}_s,
\end{equation}
where register $f$ indicates failure in preparing the target state, and register $s$ contains the state.
By identifying $\alpha = \sin \theta$ and $\beta = \cos \theta$, we can uncover a two-dimensional rotation amenable to a Grover-like search algorithm. We define the Amplitude Amplification operator~\cite{brassard2002quantum}
\begin{equation}\label{eq:Amplitude_Amplification}
    Q = - A R_s A^{\dagger} R_{t},
\end{equation}
where $R_s := (2\ket{10}\bra{10}_{fs}- \bm{1})$ and $R_{t} := (2\ket{0}\bra{0}_f - \bm{1}_f)$, as the rotation around the start and target states. To recover Grover's algorithm, $A$ would have to be equal to the Hadamard gate. Overall, this operator acts as
\begin{equation}
\begin{split}
    Q\ket{0}_f\ket{\psi}_s&=\cos(2\theta)\ket{0}_f\ket{\psi}_s-\sin(2\theta)\ket{1}_f\ket{\psi^{\perp}}_s,\\
    Q\ket{1}_f\ket{\psi^{\perp}}_s&=\sin(2\theta)\ket{0}_f\ket{\psi}_s + \cos(2\theta)\ket{1}_f\ket{\psi^{\perp}}_s.
\end{split}
\end{equation}
We can also diagonalize this operator, obtaining eigenvectors
\begin{equation}
    \ket{\psi_{\pm}} = \frac{1}{\sqrt{2}}\left(\ket{1}_f\ket{\psi^{\perp}}_s \pm i \ket{0}_f\ket{\psi}_s \right),
\end{equation}
with eigenvalues $e^{\pm 2i \theta}$ respectively. In other words,
\begin{equation} \label{eq:diag_Amplitude_Amplification}
    Q = e^{2i\theta}\ket{\psi_{+}}\bra{\psi_{+}} + e^{-2i\theta}\ket{\psi_{-}}\bra{\psi_{-}}.
\end{equation}
As we shall see, $Q$ is very helpful when we have to implement an operator $A$ multiple times, but do not want the success probability to vanish exponentially in the number of steps.

\subsection{\label{sec:Fixed_point_AA}Fixed point Amplitude Amplification and the first glimpse of Quantum Signal Processing}

A different modification we analyze in this section is the possibility of an algorithm with a fixed point. Note that Grover's algorithm has an inconvenient feature: if we do not know how many marked items there are, it is unclear how to choose the number of amplification steps. In fact, given the oscillatory nature of Grover's algorithm, it is possible to overshoot and amplify `too much'.

To remedy this, we present a second algorithm, also by Grover, which monotonically increases the probability of measuring the target~\cite{grover2005fixed}. Let $\ket{s}$ and $\ket{t}$ be the starting and target states, and $U$ a rotation between the subspaces spanned by them, such that $|\braket{t|U|s}|^2 = 1-\varepsilon$. By defining 
\begin{equation}
    R_s = \bm{1} - \left(1 - \exp\left(i\frac{\pi}{3}\right)\right)\ket{s}\bra{s}, \qquad
    R_t = \bm{1} - \left(1 - \exp\left(i\frac{\pi}{3}\right)\right)\ket{t}\bra{t},
\end{equation}
the operator $U R_s U^\dagger R_t U$ will fulfill $|\braket{t|U R_s U^\dagger R_t U|s}|^2 = 1-\varepsilon^3$~\cite{grover2005fixed}. If instead of $\pi/3$ we had used $\pi$, we would recover the original Grover algorithm.

We can now use this idea to monotonically improve the success probability in an Amplitude Amplification setting. We can use the recursion
\begin{equation}
    U_{m+1} = U_{m} R_s U_m^\dagger R_t U_m, \qquad U_0 = A,
\end{equation}
where $A$ is defined in~\eqref{eq:A_for_AA}.
The success probability scales as $|\braket{t|U_m|s}|^2 = 1 - \varepsilon^{3m}$. Expressed in the number of calls to the oracle $q_m$, $|\braket{t|U_m|s}|^2 = 1 - \varepsilon^{2q_m +1}$~\cite{grover2005fixed}. In contrast, the classical success probability is $1-\varepsilon^{c+1}$, for $c$ oracle calls.

This algorithm, while being optimal under the monotonic condition, loses the original quadratic speedup~\cite{yoder2014fixed}. To see it, imagine we want to find the marked item with failure probability $\delta$, such that the initial and target states have a small overlap probability $\lambda = 1-\varepsilon \rightarrow 0$. We want to find $q_m$ such that
\begin{equation}
    \delta \geq \varepsilon^{2q_m +1} \approx (1-\lambda)^{2q_m+1} \approx 1-(2q_m+1)\lambda,
\end{equation}
what implies that \cite[Eq. 16]{grover2006quantum}
\begin{equation}
    q_m \gtrsim \frac{1-\delta}{2\lambda} = O\left(\frac{1}{\lambda}\right),
\end{equation}
similar to the classical case. In contrast, Grover's original algorithm required $O(\lambda^{-1/2})$ calls to the oracle.

Can we find a way around it? We can, but it implies letting go of the monotonic assumption. As before, we assume we have an initial state $\ket{s} = A \ket{0}$, and want to prepare a target state $\ket{t} = e^{-i\xi}\ket{t'}$ such that $\braket{t'|s} = e^{i\xi}\sqrt{\lambda}$. In other words, $\ket{s} = \sqrt{\lambda}\ket{t} + \sqrt{1-\lambda^2}\ket{t^\perp}$. We are also provided with an oracle $O$ which marks the target state $\ket{t'}$. The objective is finding a quantum circuit $Q$ such that the success probability $P_L = |\braket{t'|Q|s}|^2$ is $1-\delta^2$, where $\delta\in [0,1]$, after $L$ calls to the oracle.

The key idea to tackle this problem~\cite{yoder2014fixed}, is to generate a sequence of generalized Grover rotations $G(\alpha,\beta)$
\begin{equation}
    Q = G(\alpha_L, \beta_L)\cdots G(\alpha_1, \beta_1), \qquad G(\alpha,\beta) = - S_s(\alpha) S_t(\beta),
\end{equation}
where 
\begin{equation}
    S_x(\alpha) := \bm{1} - (1-e^{i\alpha})\ket{x}\bra{x}.
\end{equation}
These reflection operators $S_s(\alpha)$ and $S_t(\beta)$ can also be understood as rotations in a 2-dimensional Block sphere-like space:
\begin{equation}
    R_{\varphi}(\theta) = \exp\left(-\frac{i}{2}\theta (Z\cos (\varphi)  + X\sin (\varphi) )\right).
\end{equation}
Using this rotation and taking $\phi = 2\arcsin(\sqrt{\lambda})$, we can rewrite~\cite{yoder2014fixed}
\begin{equation}
    S_t(\beta) = e^{i\beta/2}R_0(\beta), \qquad S_s(\alpha) = e^{-i\alpha/2}R_\phi(\beta).
\end{equation}
This formulation allows computing what angles $\alpha_i, \beta_i$ can be used to increase the success probability $P_L$ to $1-\delta^2$. For $l = (L-1)/2$ and $j\in [1, l]$ we should take~\cite{yoder2014fixed}
\begin{equation} \label{eq:Grover_generalized_angles}
    \alpha_j= -\beta_{l-j+1} = 2 \cot^{-1}\left(\tan(2\pi j/L ) \sqrt{1-\gamma^2}\right),
\end{equation}
with $\gamma^{-1} := T_{1/L}(1/\delta)$, and $T_L(\cos \theta):= \cos(L \theta)$ the $L^{th}$ first-order Chebyshev polynomial. Using these angles, one obtains that
\begin{equation}
    P_L = |\braket{t|Q|s}|^2 = 1-\delta^2 T_L^2(T_{1/L}(1/\delta)\sqrt{1-\lambda}).
\end{equation}
Chebyshev polynomials fulfill that if $|x|\leq 1$, then $|T_L(x)|\leq 1$. From this it is clear that if $|T_{1/L}(1/\delta)|\sqrt{1-\lambda}\leq 1$, then we can write $P_L\geq 1-\delta^2$ as wanted. This happens whenever $\lambda\geq 1- T_{1/L}^{-2}(1/\delta) = : w$. We will call $w$ the width. For large $L$ and small $\delta>0$ \cite[Eq. 2]{yoder2014fixed}
\begin{equation}
    w\approx \left(\frac{\log (2/\delta)}{L}\right)^2,
\end{equation}
what implies that we need to choose
\begin{equation}\label{eq:yoder_complexity}
    L \geq \frac{\log (2/\delta)}{\sqrt{\lambda}},
\end{equation}
recovering the quadratic speedup that characterizes Grover's original algorithm.

Furthermore, we can concatenate this procedure to improve the target $\delta$. Calling $Q_L(B)$ the rotation that uses $BA$ in place of $A$ for the definition of the initial state, and $\chi_1$, $\chi_2$ some phases, then if
\begin{equation}
    Q_{L_1}\ket{s}:= \sqrt{1-P_{L_1}(\lambda)}\ket{t^\perp} + \sqrt{P_{L_1}(\lambda)}e^{-i\chi_1}\ket{t},
\end{equation}
we also have that~\cite{yoder2014fixed}
\begin{equation}
    Q_{L_2}(Q_{L_1}) Q_{L_1}\ket{s}:= \sqrt{1-P_{L_2}(P_{L_1}(\lambda))}\ket{t^\perp} + \sqrt{P_{L_2}(P_{L_1}(\lambda))}e^{-i(\chi_1+\chi_2)}\ket{t}.
\end{equation}
Since $Q_{L_1}$ acts as a prefix, we can implement $Q_{L_1}$ and then decide whether to improve the result further.
Moreover, if we choose $\delta = 1$, from \eqref{eq:Grover_generalized_angles} we obtain $\alpha_j = \beta_j = \pm \pi$, thereby generalizing Grover's original algorithm. And if instead we chose $\delta = 0$, we obtain $-\alpha_1 = \beta_1 = \pi/3$~\cite{yoder2014fixed}, and using recursion as explained before we recover the $\pi/3$ monotonic algorithm, but as expected without the quadratic speedup from \eqref{eq:yoder_complexity}.

\section{\label{sec:Walks}Quantum walks}

While the extensions of Grover's algorithm that we have discussed so far deal with unordered databases or probabilistic algorithms, there are situations where we have more information about the structure of the search space. In particular, let us consider a graph $G(X, E)$, where $X$ denotes the vertices (states) and $E$ the edges (state transitions). We now define a Markov chain over this graph, that for each pair of vertices $x,y$ connected by an edge assigns a probability of transition, forming a matrix $P = (p_{xy})_{(x,y)\in E}$. We will analyze the particular case of discrete ergodic Markov chains.

\begin{definition}[Discrete ergodic Markov chain]
A Markov chain $\mathcal{P} = (p_{xy})_{(x,y)\in E}$ defined over a graph $G(X,E)$ is called ergodic, if $\exists t_0$ such that $\forall x, y \in X$, and starting from $x$ at time $t=0$ and following probability transitions dictated by $\mathcal{P}$, then $\forall t > t_0$, the probability of finding an item $y$ at time $t$ is greater than $0$. In mathematical notation,
\begin{equation}
    \exists t_0 \quad| \quad \forall x, y \in X,\quad  p(x, t=0) = 1 \Rightarrow p(y, t>t_0)>0.
\end{equation}
\end{definition}

Ergodic Markov chains have a unique stationary distribution $\pi$, that fulfills $ \pi^T P = \pi^T$, or in other words, the eigenvalue 1 has multiplicity 1 too \cite[Theorem 5.9]{koralov2007theory}. In this case, we define an eigenvalue gap $\delta = 1-\lambda$ where $\lambda$ is the second-largest absolute value of an eigenvalue of $\mathcal{P}$, the first being 1. This eigenvalue gap $\delta$ will determine the time required for the Markov chain to mix, or in other words, to approximate $\pi$.
Finally, Markov chains are called \textit{reversible} if they obey the detailed \textit{balance property}
\begin{equation} \label{eq:detailed_balance}
   p_{yx} \pi_x =  p_{xy}\pi_y,
\end{equation}
and are called \textit{symmetric} if $P = P^T$. Symmetric Markov chains display the uniform superposition as the stationary state~\cite{magniez2011search}.

\subsection{\label{sec:hitting}Hitting time and search algorithms}

There are two main tasks one can perform in a Markov chain when there is a set of marked items $M$,
\begin{itemize}
    \item \texttt{Detect}$^{(=k)}$: Check if $M = \emptyset$ under the promise that either $M=\emptyset$ or $|M| = k$.
    \item \texttt{Find}$^{(=k)}$: Find $m\in M$ under the promise that $|M| = k$.
\end{itemize}
If instead of superscript $(=k)$ we use $(\geq k)$, the same promise applies with greater or equal than. If no promise is given, then $k\geq 1$ is assumed.

Classically, these two tasks are the same: one cannot detect the presence of a marked node without finding it. A simple classical algorithm for finding a marked state $m\in M\subset X$ is the following,~\cite{magniez2011search}: 

\begin{algorithm}[h!]
\begin{algorithmic}[1]
\State \textbf{Input}: Ergodic Markov chain $\mathcal{P}$, graph $G(X,E)$, time $t$.
\State \textbf{Output}: Marked state $m$ or signal that not marked item exists.
\State Initialize $x$ sampled from the stationary distribution $\pi$.
\For{$t$ steps}:
\State If state reached $x$ is marked, output $x$.
\State Else, simulate 1 step of Markov chain $\mathcal{P}$ from $x$.
\EndFor
\State Output `no marked element exists'.
\end{algorithmic}
\caption{A basic classical search algorithm}\label{alg:Classical_search}
\end{algorithm}

Running \cref{alg:Classical_search} has three cost sources: $\texttt{S}$ for the set-up cost of initializing the data structure and sampling the initial $x$, often from the stationary distribution $\pi$ of the corresponding Markov chain, $\texttt{U}$ for updating the item according to $\mathcal{P}$, and $\texttt{C}$ for checking if it is a marked item. 

\begin{definition}[Hitting time]
We define the hitting time $HT(P, M)$ of a Markov chain $\mathcal{P}$ as the expected number of evaluations of $\texttt{U}$ required to find a marked item $m\in M$ with \cref{alg:Classical_search} starting from the projection of the stationary distribution $\pi$ into the set of unmarked states $\ket{\pi_U}= \Pi_U \ket{\pi}$.
\end{definition}

The Hitting time $HT$ from \cref{alg:Classical_search} in \textit{symmetric} Markov chains is $HT = t = O(1/(\delta \epsilon))$ \cite[Proposition 1]{magniez2011search}.
In such a case, its overall cost is $O\left(\texttt{S}+\frac{1}{\delta \epsilon}\left(\texttt{U}+\texttt{C}\right)\right)$.
Another slightly different algorithm is the following~\cite{magniez2011search}: 

\begin{algorithm}[h!]
\begin{algorithmic}[1]
\State \textbf{Input}: Ergodic Markov chain $\mathcal{P}$, graph $G(X,E)$, times $t_1$, $t_2$.
\State \textbf{Output}: Marked state $m$ or signal that not marked item exists.
\State Initialize $x$ sampled from the stationary distribution $\pi$.
\For{$t_1$ steps}:
\State If state reached $x$ is marked, output $x$.
\State Else, simulate $t_2$ steps of Markov chain $\mathcal{P}$ from $x$.
\EndFor
\State Output `no marked element exists'.
\end{algorithmic}
\caption{A more efficient classical search algorithm}\label{alg:Classical_search_efficient}
\end{algorithm}

Taking $t_2$ steps of the Markov Chain aims to mix the state again so that it is close to the stationary distribution $\pi$. In a symmetric Markov chain we can take $t_1 = O(1/\epsilon)$ and $t_2= O(1/\delta)$, resulting in \cref{alg:Classical_search_efficient} having complexity $O\left(\texttt{S}+\frac{1}{\epsilon }\left(\frac{1}{\delta}\texttt{U}+\texttt{C}\right)\right)$,  \cite[Proposition 1]{magniez2011search}.

\begin{figure}[ht!]
    \centering
    \includegraphics[width = .8\textwidth]{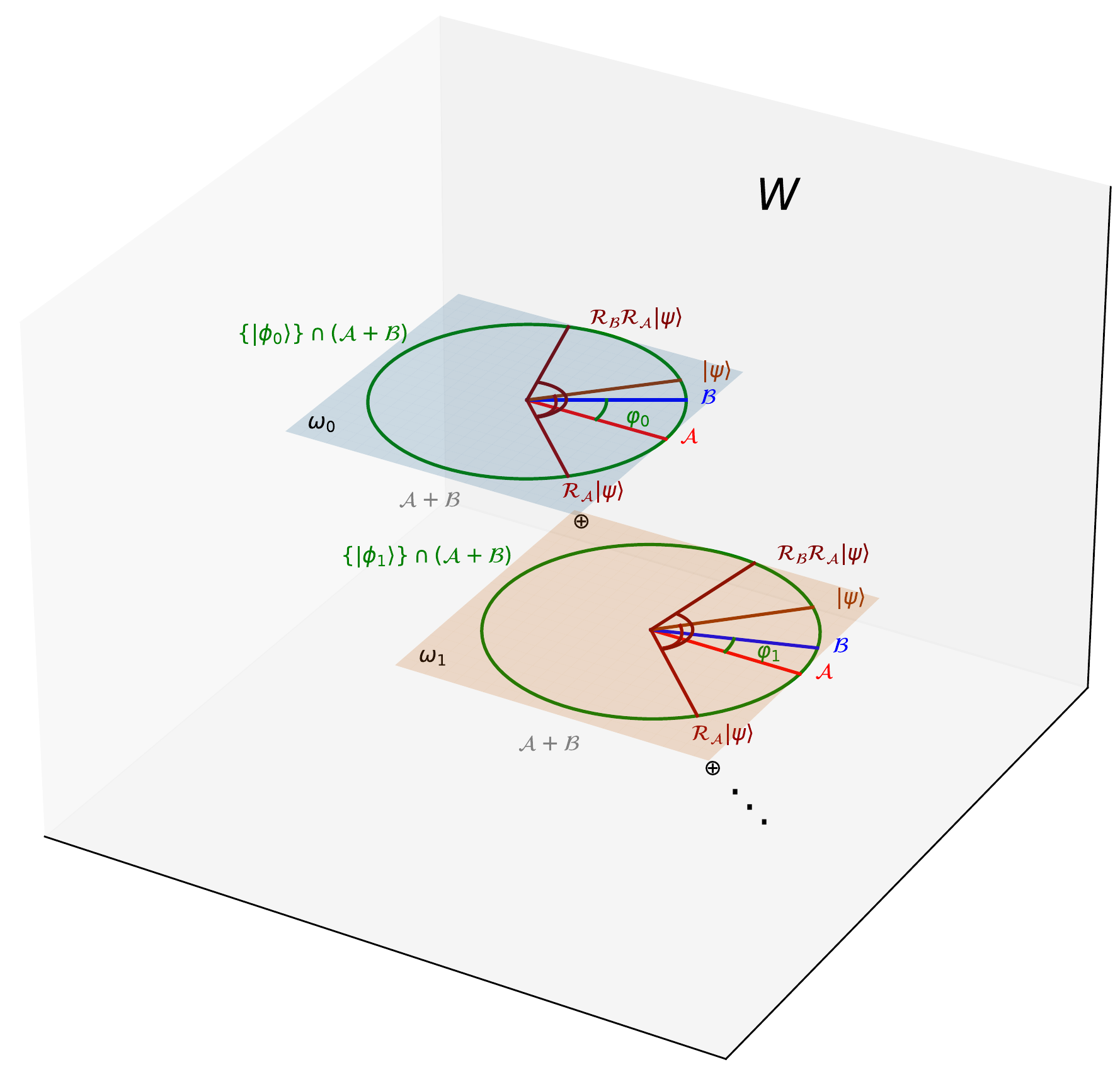}
    \caption{\textbf{Szegedy quantum walk.} Geometrical visualization of the action of a Szegedy quantum walk operator $W$, defined in \eqref{eq:W_definition}. $W$ performs a series of rotations that in the subspace $\mathcal{A} + \mathcal{B}$ may be written as a block diagonal  matrix, where each block is a 2-dimensional rotation $\omega_j = R(2\varphi_j)$ given by \eqref{eq:block rotation matrix in A+B}. This figure represents the direct sum of Grover-like rotations in the subspace spanned by $\mathcal{A}+\mathcal{B}$, and therefore $W$. Note that $\varphi_0=0$ because it corresponds to eigenvalue $\lambda_0=1$, but is represented non-zero for clarity.}
    \label{fig:quantum_walk}
\end{figure}

Our objective is to find quantum equivalent algorithms to \texttt{Detect} and \texttt{Find} which scale quadratically better in the hitting time. These algorithms will be called \textit{quantum walks}\footnote{In this thesis, we will focus on discrete-time quantum walks. However, it is also possible to define a continuous-time quantum walk by defining a Hamiltonian that dictates the evolution of the system. Such Hamiltonian depends on the graph, for example by dictating the allowed transitions between nodes. Continuous quantum walks are therefore implemented via Hamiltonian simulation techniques that we will see in \cref{ch:Chemistry}.}, the first of which was introduced by Ambainis for Johnson graphs in Ref.~\cite{ambainis2004quantum}. Later on, Szegedy proposed a more general walk that extends to ergodic reversible Markov chains~\cite{szegedy2004quantum}. To describe it, we start by defining a bipartite Hilbert space $\mathcal{H}\otimes \mathcal{H}$, where $\mathcal{H}$ represents the space of possible solutions $X$.
The update operator $U$ will now map
\begin{equation}
    U: \ket{x}\ket{0}\mapsto \ket{\alpha_x}:= \ket{x}\otimes \sum_{y\in X}\sqrt{p_{xy}}\ket{y},
\end{equation}
and similarly, we can define
\begin{equation}
    V: \ket{0}\ket{y}\mapsto \ket{\beta_y}:= \sum_{y\in X}\sqrt{p_{yx}}\ket{x}\otimes \ket{y}.
\end{equation}
The relation between these operators is $SU = VS $, where $S$ is the Swap operation between both Hilbert subspaces. Using them, we can define subspaces
\begin{equation}
    \mathcal{A}:= \text{span} \{\ket{x}\ket{0}: x\in X\}, \qquad
    \mathcal{B}:= U^\dagger V S \mathcal{A}= U^\dagger S U \mathcal{A}  .
    \label{Subspaces A and B}
\end{equation}
The projector on these subspaces
\begin{equation}
    \Pi_{\mathcal{A}}:= (\mathbf{1}\otimes \ket{0}\bra{0}),
    \qquad
    \Pi_{\mathcal{B}}:= U^\dagger V S (\mathbf{1}\otimes \ket{0}\bra{0}) S V^\dagger U = U^\dagger S U (\mathbf{1}\otimes \ket{0}\bra{0}) U^\dagger S U
\end{equation}
define rotations $R_\mathcal{A}$ and $R_\mathcal{B}$, which we can use to introduce the quantum walk operator
\begin{equation}
 W= R_\mathcal{B}R_\mathcal{A}= U^\dagger S U R_\mathcal{A} U^\dagger S U R_\mathcal{A}. \label{eq:W_definition}
\end{equation}

To analyze it, we can compare it with 
\begin{equation}
 \mathcal{M}:= U^\dagger V S = U^\dagger S U.   
 \label{eq:M_qw}
\end{equation}
Since we assumed that the Markov chain is reversible, using the detailed balance equation, we can find its matrix element on the subspace $\mathcal{A}$, $\braket{x,0|U^\dagger V S|y,0}= \sqrt{p_{yx}}\sqrt{p_{xy}} = \sqrt{\pi_x/\pi_y}p_{xy}$~\cite[Pag.~756]{yung2012quantum}. In matrix form, this is called the \textit{discriminant matrix}
\begin{equation}
    \mathcal{D} = D_\pi^{-1/2}\mathcal{P} D_\pi^{1/2}
    \label{eq:M_qw_reversible}
\end{equation}
where $D_\pi$ is the diagonal matrix containing the entries of $\pi$, the equilibrium distribution. Since the matrix $D_\pi$ is positive definite, 
$\mathcal{P}$, $\mathcal{D}$ and $\mathcal{M}$ have the same spectrum in the subspace defined by projector $\Pi_{\mathcal{A}}$, see Ref.~\cite[Pag. 413]{ambainis2020quadratic} and Ref.~\cite[Pag. 5]{apers2019quantum}. These eigenvalues $1 = \lambda_0>\lambda_1\geq \ldots \geq \lambda_{d-1}$ can be rewritten as phases $\lambda_j = \cos \varphi_j $, and we will denote the corresponding eigenstates as $\ket{\phi_j}\ket{0}$. 
Since~\cite[Eq. S21]{yung2012quantum}
\begin{equation}
    \braket{\phi_j,0 |\Pi_\mathcal{A} U^\dagger V S \Pi_\mathcal{A}|\phi_j,0} = \lambda_j\\
    = \lambda_j^\dagger =\braket{\phi_j,0 |\Pi_\mathcal{A} SV^\dagger U \Pi_\mathcal{A}|\phi_j,0},
\end{equation}
we have,
\begin{equation}
    \Pi_\mathcal{A} U^\dagger V S \Pi_\mathcal{A} = \Pi_\mathcal{A} SV^\dagger U \Pi_\mathcal{A}.
\end{equation}
We can use this, and the definition of the spectrum of $\mathcal{M}$, to compute~\cite[Eq. S19]{yung2012quantum}
\begin{subequations}
    \begin{equation}\label{eq:Project_M}
        \Pi_\mathcal{A}U^\dagger VS \ket{\phi_j}\ket{0} = \cos \varphi_j \ket{\phi_j}\ket{0},
    \end{equation}
    and
    \begin{equation}
        \Pi_\mathcal{B} \ket{\phi_j}\ket{0} =  U^\dagger V S \cos \varphi_j \ket{\phi_j}\ket{0}.
    \label{effect of B proyector}
    \end{equation}
\end{subequations}
Consequently, the subspace spanned by $\{\ket{\phi_j}\ket{0}, U^\dagger V S \ket{\phi_j}\ket{0} \}$ is preserved under $R_\mathcal{A}$ and $R_\mathcal{B}$. In such subspaces, the operation of $W$ is a rotation
\begin{equation}
w_j =
    \begin{pmatrix}
    \cos(2\varphi_j) & -\sin(2\varphi_j)\\
    \sin(2\varphi_j) &
    \cos(2\varphi_j)
    \end{pmatrix}.
    \label{eq:block rotation matrix in A+B}
\end{equation}

In contrast to the classical walk, where the eigenvalue gap was $\delta = 1-\lambda$, the quantum phase gap is defined as $\Delta := 2\varphi = 2\arccos \lambda$. The relation between both is $\Delta \geq 2\sqrt{1-|\lambda|^2}\geq 2\sqrt{\delta}$~\cite[Theorem 7]{magniez2011search}, which will ultimately originate the familiar quadratic speedup that also characterizes Grover algorithm~\cite{szegedy2004quantum}.

However, so far we have not explained how to implement either of the two tasks using Szegedy's quantum walks. Let us first consider a quantum version of \cref{alg:Classical_search} for \texttt{Detect}, presented in \cref{alg:Szegedy_quantum_detect}. What Szegedy did in his original article~\cite{szegedy2004quantum} is a so-called \textit{swap test}, between the stationary distribution and the same state evolved under $W'$ defined as in \eqref{eq:W_definition} for an absorbent Markov chain $\mathcal{P}'$. $\mathcal{P}'$ is defined as $\mathcal{P}$ with the transitions from marked to unmarked states deleted.
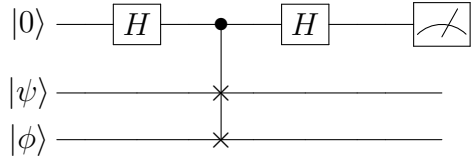
\begin{figure}[t!]
\centering
\mbox{
\Qcircuit @C=0.9em @R=0.75em { 
  & \ket{0} & & \qw & \gate{H} & \qw & \ctrl{4} & \qw & \gate{H} & \qw & \qw & \meter \\
  \\
  & \ket{\psi} & & \qw & \qw & \qw & \qswap & \qw & \qw & \qw & \qw & \qw \\
  \\
  & \ket{\phi} & & \qw & \qw & \qw & \qswap & \qw & \qw & \qw & \qw & \qw \\
}}
\caption{\textbf{Swap test.} Originally described in Re.~\cite{buhrman2001quantum}, the probability of measurement of $0$ of this algorithm is $\frac{1}{2}(1+\|\braket{\phi|\psi}\|^2)$. This circuit might be consequently used to estimate $\|\braket{\phi|\psi}\|^2$, to precision $\epsilon^{-1}$ with $O(\epsilon^{-2})$ measurements, or alternatively using quantum amplitude estimation with complexity $O(\epsilon^{-1})$. The complexity in the failure probability is logarithmic in both cases.}
\label{fig:Swap_test}
\end{figure}

\begin{algorithm}[h!]
\begin{algorithmic}[1]
\State \textbf{Input}: Ergodic symmetric Markov chain $\mathcal{P}$, graph $G(X,E)$, overlap $\epsilon$, eigenvalue gap $\delta$, target failure probability $p_f$.
\State \textbf{Output}: Whether a marked item exists.
\State Define $\mathcal{P}'$ as the absorbent Markov chain corresponding to $\mathcal{P}$, and $W'$ its corresponding quantum walk.
\For{$O(\log p_f^{-1})$ steps}:
\State Pick $t$ uniformly at random in $\left[1,O\left(\frac{1}{\sqrt{\delta\epsilon}}\right)\right]$.
\State Prepare the state $\frac{1}{\sqrt{2}}(\ket{0}+\ket{1})\ket{\pi}$, for $\ket{\pi}$ the stationary distribution.
\For{$t$ steps}:
\State Controlled on the first qubit, evolve the state using $W'$.
\EndFor
\State The resulting state is
\begin{equation}
 \frac{1}{\sqrt{2}}\left(\ket{0}\ket{\pi} +\ket{1}(W')^t\ket{\pi}\right).
\end{equation}
\State Perform a Hadamard on the first qubit, obtaining
\begin{equation} \label{eq:Szegedy_swap_test}
 \frac{1}{2}\left(\ket{0}(\bm{1} + (W')^t)\ket{\pi})+\ket{1}(\bm{1}  -  (W')^t)\ket{\pi})\right).
\end{equation}
\State Measure the first qubit.
\State If measurement is $\ket{1}$, output `Marked item detected'.
\EndFor
\State Output `No marked item exists' and stop.
\caption{Szegedy quantum detection \cite[Lemma 7]{szegedy2004quantum}}\label{alg:Szegedy_quantum_detect}
\end{algorithmic}
\end{algorithm}
To see why \cref{alg:Szegedy_quantum_detect} works, remember that $\pi^T \mathcal{P} = \pi$, and therefore $W\pi = \pi$. If no marked item exists, then $W' = W$, and consequently with probability $1$ we will measure $\ket{0}$ after \eqref{eq:Szegedy_swap_test}. $t\in[1, O(1/\sqrt{\delta\epsilon})]$ is chosen so that if there are $\epsilon|X|$ marked item, then $\|(1-(W')^t)\pi\|$ does not cancel out. Overall, the complexity of this algorithm is $\tilde{O}\left(S+\frac{1}{\sqrt{\delta \epsilon}}\left(\texttt{U}+\texttt{C}\right)\right)$.

We now focus on the problem of \texttt{Find}. We can remove the requirement of symmetry of the Markov chain, and only require that it is reversible (e.g., it fulfills \eqref{eq:detailed_balance}), while at the same time obtaining a quantum equivalent of \cref{alg:Classical_search_efficient}~\cite{magniez2011search}.
For that, we will use a subroutine called \textit{phase estimation}, which we will introduce in \cref{ch:Phase} and denoted here by $P$. Given a Hermitian operator $H$ with eigenstates $\ket{\phi_j}$ and eigenvalues $\lambda_j:=\cos \varphi_j$,
\begin{equation}
    P(H): \ket{\phi_j}\ket{\bm{0}}\rightarrow \ket{\phi_j}\ket{\tilde{\varphi}_j},
\end{equation}
where $\tilde{\varphi}_j$ is a binary approximation to $\varphi_j$.

The objective is to perform a Grover-like algorithm to amplify the marked states. Grover's algorithm would consist of rotations
\begin{equation}
    G = \left(2\sum_{m\in M}\ket{m}\bra{m} - \bm{1}\right)(2\ket{\pi}\bra{\pi}-\bm{1}).
\end{equation}
The first rotation requires the use of $\texttt{C}$, while the second uses $\texttt{S}$ that prepares the stationary state $\ket{\pi}$. However, this does not use the graph structure.

Our objective is to modify Grover's algorithm and avoid the use of $\texttt{S}$: instead of decomposing $(2\ket{\pi}\bra{\pi}-\bm{1}) = \texttt{S}(2\ket{\bm{0}}\bra{\bm{0}}-\bm{1})\texttt{S}^\dagger $, we can use a combination of step operator $U$ and $P(H)$ to perform a rotation over $\ket{\pi}$. This relies on the fact that we can use $P(H)$ to detect and apply a phase to the state with eigenvalue $1$. Thus, this procedure will be more appropriate when phase estimation $P(H)$ is cheaper than the quantum state preparation subroutine $\texttt{S}$. The proposed algorithm is the following.

\begin{algorithm}[h!]
\begin{algorithmic}[1]
\State \textbf{Input}: Ergodic reversible Markov chain $\mathcal{P}$, graph $G(X,E)$, overlap $\epsilon$, eigenvalue gap $\delta$, target failure probability $p_f$, $O(\log (\sqrt{\epsilon}\Delta p_f)^{-1} \Delta^{-1})$ auxiliary qubits $\ket{\cdot}_a$.
\State \textbf{Output}: A marked item if it exists, or `no marked element exists' message.
\State Choose $t$ uniformly at random from $[1, O(1/\epsilon)]$.
\State Prepare $\ket{\pi}\ket{0}\ket{0}_a^{\otimes \log p_f^{-1}\log \Delta^{-1}}$.
\For{$t$ steps}:
\State For any possible state $\ket{x}\ket{y}\ket{z}$, use check operator $\texttt{C}$ so that
\begin{equation}
    \ket{x}\ket{y}\ket{z}_a \mapsto (-1)^{x\in M} \ket{x}\ket{y}\ket{z}_a.
\end{equation}
\State Apply $P(W)$ $O(\log ( \sqrt{\epsilon} p_f)^{-1})$ times in new sets of auxiliary qubits, with precision $O(\Delta^{-1})$.
\State Compute the median of the estimated $\ket{\tilde{\lambda}_j}$s.
\State If \texttt{Median}($\{\ket{\tilde{\lambda}_j}\}$) = 1, apply a $(-1)$ phase.
\State Uncompute the median.
\State Uncompute the $P(W)$ operators.
\EndFor
\State Measure the first register
\If{$x\in M$}
\State Output $x$
\Else
\State Output `No element exists'.
\EndIf
\caption{Magniez quantum search \cite[Theorem 3]{magniez2011search}}\label{alg:Magniez_quantum_search}
\end{algorithmic}
\end{algorithm}

The reason we need to run $P(W)$ with precision $\Delta^{-1}$ is that this is the precision we need to distinguish between $\ket{\pi}$ and any other eigenstate of $W$. As we will see in \cref{ch:Phase}, achieving precision $\Delta^{-1}$ means a number of logical gates scaling as $O(\Delta^{-1})$. Furthermore, the median is used to ensure that the probability of failure decreases exponentially fast, using the Median Lemma \cite[Lemma 1]{nagaj2009fast}. The implementation of such \texttt{Median} protocol can be done with a reversible sorting protocol such as a sorting network. Overall, the complexity is $\tilde{O}\left(S + \frac{1}{\sqrt{\epsilon}}\left(C+\frac{1}{\Delta}U\right)\right)$, where $\Delta = O(\sqrt{\delta})$. Moreover, instead of using Grover's method, we can instead use the $\pi/3$ monotonically increasing algorithm~\cite{grover2005fixed}, or the quantum signal processing algorithm~\cite{yoder2014fixed} that we saw at the end of \cref{sec:Fixed_point_AA}.

\subsubsection{Quantum Fast Forwarding and an optimal algorithm}

So far, however, we have not given a closed expression for the Hitting Time. The Hitting Time can be expressed as \cite[Proposition 9]{krovi2016quantum}
\begin{equation}
    HT(P,M) = \sum_{k=1}^{d-|M|}\frac{|\braket{\phi'_k|\pi_U}|^2}{1-\lambda'_k},
\end{equation}
where $d$ is the number of eigenvalues $\lambda'_k\neq1$ of the absorbent $\mathcal{M}'$ defined similarly as in \eqref{eq:M_qw_reversible}, corresponding to eigenstates $\ket{\phi'_k}$ (see Ref.~\cite[Proposition 34]{krovi2016quantum}). The sum is explicitly designed to avoid the $|M|$ eigenstates with eigenvalue 1, the marked states, where the fraction would not be well-defined. If there are no transitions between marked states, it is easy to recognize the eigenvalue 1 states. In such a case, remember that if $\ket{m}\in M$, then $\mathcal{P}'\ket{m} = \ket{m}$ as the Markov chain $\mathcal{P}'$ is absorbent, so $\ket{m}$ is an eigenstate with eigenvalue 1.

The Hitting Time can sometimes be $O(1/\sqrt{\delta\epsilon})$ as we saw was the case for symmetric Markov Chains, but it may also be smaller~\cite{ambainis2020quadratic}. For that reason, the algorithms we have explained so far are not optimal. However, there exists one algorithm that approximately is, achieving a quadratic speedup in the Hitting Time up to polylogarithmic complexity~\cite{ambainis2020quadratic}. It is based on two main techniques.

First, it makes use of `extrapolated' quantum walks, defined as the quantum walk corresponding to
\begin{equation}
    \mathcal{P}(s) = (1-s)\mathcal{P} + s\mathcal{P}', \quad s\in [0,1),
\end{equation}
which is still a reversible Markov chain with a unique stationary distribution that we will call $\pi(s)$~\cite{krovi2016quantum}.

The second technique, called \textit{Quantum Fast Forwarding}, is more involved~\cite{apers2019quantum}. Its objective is to simulate an approximation to $\mathcal{D}^t$ in $O(\sqrt{t})$ applications of the quantum walk operator, even if that entails some failure probability. To do so, one first proves that $\Pi_{\mathcal{A}}\mathcal{M} = \mathcal{D}$ \cite[Lemma 1]{apers2019quantum}, as we already saw in \eqref{eq:Project_M}. Using this,
\begin{equation}
    \mathcal{D}\ket{\phi_j}\ket{0} = \cos\varphi_j \ket{\phi_j}\ket{0} \Rightarrow \mathcal{D}^t\ket{\phi_j}\ket{0} = \cos^t(\varphi_j) \ket{\phi_j}\ket{0}.
\end{equation}
We will further modify the definition of the quantum walk such that in this algorithm it will be 
\begin{equation}\label{eq:W_step_Ambainis}
    W = (2\Pi_{\mathcal{A}} - \bm{1})\mathcal{M},
\end{equation}
with $\mathcal{M}$ defined in \eqref{eq:M_qw}.
Since $W$ can be understood as 2 rotations, we have \cite[Proposition 1]{apers2019quantum} and~\cite{szegedy2004quantum}
\begin{equation} \label{eq:QFF_Chebyshev}
    \Pi_{\mathcal{A}} W^t \ket{\phi_j}\ket{0}= T_t(\mathcal{D}) \ket{\phi_j}\ket{0}= T_t(\cos\varphi_j) \ket{\phi_j}\ket{0} = \cos(t\varphi_j) \ket{\phi_j}\ket{0},
\end{equation}
where $T_t(\cos\varphi)$ is a Chebyshev polynomial of the first kind. In fact, in \cref{ch:Phase} and in particular, in equation \eqref{eq:W^n_quantum_walk}, we will see that powers of quantum walk implement Chebyshev polynomials in general.
Now, we compare
\begin{equation}
    \cos^{t}(\theta) = 1-\frac{t\theta^2}{2} + O(t^2\theta^4),\quad\text{and}\quad \cos(t'\theta) = 1 - \frac{t'^2\theta^2}{2} + O(t'^4\theta^4).
\end{equation}
From this we check that setting $t' = \sqrt{t}$, both expressions coincide up to second order. Notice the fact that $W^{l}$ implements $T_l(\cos\varphi_j)$ on the eigenstates $\ket{\phi_j}\ket{0}$, and we can use that to approximate $\cos^t(\varphi_j)$, the action of $\mathcal{D}^t$.
In particular, we will implement a Chebyshev series such that \cite[Lemma 3]{apers2019quantum}
\begin{equation}
    \left|\cos^t(\theta) - \sum_{l=0}^{\lceil \sqrt{C t} \rceil } p_l \cos(l t\theta)\right|\leq \epsilon,
\end{equation}
for some coefficients $p_l$ defined in their equation 6, and $C = 2\ln(2/\epsilon)$ the precision of the approximation. 
More specifically, we implement \cite[Equation 8]{apers2019quantum}
\begin{equation}\label{eq:QFF_series}
    \sum_{l=0}^\tau p_l \Pi_{\mathcal{A}}W^l\ket{\phi_j}\ket{0}
    =  \sum_{l=0}^\tau p_l T_l(\mathcal{D})\ket{\phi_j}\ket{0}
\end{equation}
for $\tau = O(\sqrt{t}\log\epsilon^{-1})$ to approximate $\mathcal{D}^t$, our objective.

We now have to explain how to implement the sum of Chebyshev terms. For that, we resort to a very standard technique that will be used in later chapters too. It is called \textit{Linear Combination of Unitaries} (LCU) decomposition. The idea is that if we want to implement \eqref{eq:QFF_series}, we use the following two operators:
\begin{subequations}\label{eq:Prep&Sel_HT}
\begin{equation}
    \Prep: \ket{0}_a\ket{\phi} \mapsto \frac{1}{\sqrt{\sum_j p_j^2}}\sum_l \sqrt{p_l}\ket{l}_a \ket{\phi},
\end{equation}
and
\begin{equation}
    \Sel:
    \ket{l}_a \ket{\phi} \mapsto \ket{l}_a T_l(\mathcal{D})\ket{\phi}.
\end{equation}
Note that $\Sel$ is a controlled application of \eqref{eq:QFF_Chebyshev}.
\end{subequations}
Combining them, we have
\begin{equation}
    (\ket{0}\bra{0}_a\otimes \bm{1})\Prep^\dagger \cdot \Sel \cdot \Prep \ket{0}_a \ket{\phi} = \ket{0}_a  \sum_{l} p_l T_l(\mathcal{D})\ket{\phi}.
\end{equation}
Using $Q = \Prep^\dagger \cdot \Sel \cdot \Prep$, we can implement the Chebyshev series with some failure probability.
Using these tools, Ref.~\cite{ambainis2020quadratic} proposes \cref{alg:QFF_quantum_search}.
\begin{algorithm}[h!]
\begin{algorithmic}[1]
\State \textbf{Input}: Ergodic reversible Markov chain $\mathcal{P}$, graph $G(X,E)$, upper bound to Hitting Time $HT$.
\State \textbf{Output}: A marked item if it exists, or `no marked element exists' message.
\State Define $T = O(HT)$ and $S =\left\{1-\frac{1}{r}: r\in 1,2,4,\ldots,2 ^{\log T+ O(1)}\right\}$
\For{$O(\log T)$ steps} Amplitude Amplification over
\State Use $\texttt{S}$ to prepare
\begin{equation}
    \sum_{t=1}^T \frac{1}{\sqrt{T}}\ket{t}\sum_{s\in S}\frac{1}{\sqrt{|S|}}\ket{s}\ket{\pi}.
\end{equation}
\State Perform measurement $\{\Pi_M, \bm{1}-\Pi_M\}$. If the output is marked, output it.
\State Quantum Fast Forward the state
\begin{equation} \label{eq:QFF_ambainis_algorithm_step}
    \ket{s}\ket{t}\ket{\pi_U}\mapsto \ket{1}\ket{t}\ket{s}\mathcal{D}^t(s)\ket{\pi_U} + \ket{0}\ket{\ldots}.
\end{equation}
with precision $\epsilon = O(\log^{-1}T)$, where $\ket{0}\ket{\ldots}$ encodes a failure state.
\EndFor
\State Measure the first register.
\If{$x\in M$}
\State Output $x$,
\Else
\State Output `No element exists'.
\EndIf
\caption{Fast-forward-based quantum search \cite[Theorem 3]{ambainis2020quadratic}}\label{alg:QFF_quantum_search}
\end{algorithmic}
\end{algorithm}

The Quantum Fast Forward subroutine has complexity $O(\sqrt{t\log \epsilon^{-1}})$ times the complexity of a single quantum walk step \eqref{eq:W_step_Ambainis}. The key result from Ref.~\cite{ambainis2020quadratic} is that we can lower bound the success amplitude as
\begin{equation}
    ||\Pi_M\mathcal{D}^t(s)\ket{\pi_U}||\geq \Omega(\log^{-1}T).
\end{equation}
in \eqref{eq:QFF_ambainis_algorithm_step}. Consequently, \cref{alg:QFF_quantum_search} has complexity \begin{equation}
    O(\texttt{S}\sqrt{\log \text{HT}}+\sqrt{\text{HT}}(U+\texttt{C})\sqrt{\log \text{HT}\log \log \text{HT}}),
\end{equation}
whenever an upper bound HT over the classical hitting time is known~\cite[Theorem~3]{ambainis2020quadratic}. If it is unknown, then one may repeat the search with an exponentially increasing number of steps until an element is found, or we flag that no element exists~\cite{boyer1998tight}. This increases the complexity by $O(\log \text{HT})$. Alternatively, we could use fixed-point amplitude amplification~\cite{yoder2014fixed}. In this case, we cannot use phase estimation to reflect over the stationary state, needing instead to use $\texttt{S}$ to prepare it from a computational basis state.

\subsection{\label{sec:Metropolis}Mixing time and Monte Carlo algorithms}

In the previous subsection, we analyzed the capability of quantum walks to find a marked item in a Markov chain. In this one, we would instead like to study the convergence to the limiting distribution. For example, if $\pi$ is the stationary distribution, we might define the mixing time as follows.
\begin{definition}[Mixing time]
The Mixing time MT is defined as
\begin{equation}
    MT(\epsilon) = \min\{t| \forall t'>t, \forall s_0, D(\mathcal{P}^{t'} s_0, \pi)\leq \epsilon\},
\end{equation}
where $D$ indicates the distance between the distributions
\begin{equation}
    D(p,q) = \frac{1}{2}\sum_{v=1}^N |p_v - q_v|,
\end{equation}
and $v$ indicates the vertices of the Markov chain.
\end{definition}
The validity of this definition rests on the fact that, for classical random walks,
\begin{equation}
    \lim_{t\rightarrow\infty} \mathcal{P}^t s_0 = \pi,
\end{equation}
independently of the initial state $s_0$~\cite{portugal2013quantum}.

If we try to do this equivalently for quantum walks, however, we fail: due to the walk being unitary, there is no limiting distribution. One can see this by noticing that \cite[Chapter~7]{portugal2013quantum}
\begin{equation}
    \frac{1}{2}\|\ket{\psi(t+1)}-\ket{\psi(t)}\|^2 = \frac{1}{2}\|W^t(W-\bm{1})\ket{\psi(0)}\|^2 = 1 - \mathfrak{R}(\braket{\psi(0)|W|\psi(0)}),
\end{equation}  
where $\mathfrak{R}$ indicates the real part. Since the norm above does not need to approach $0$ and is in fact independent of $t$, the quantum walk will not converge to a stationary state. For this reason, we will have to look for alternative approaches to obtaining some sense of limiting distribution.

One possible option to define a quantum equivalent is to take a time-averaged sampling, e.g., given some possibly entangled auxiliary register $\ket{a}$ and a vertex register $\ket{v}$
\begin{equation}\label{eq:average_prob}
    \lim_{T\rightarrow\infty}\bar{p}_v(T) = \lim_{T\rightarrow\infty} \frac{1}{T} \sum_{t=0}^{T-1} \sum_{a}|\braket{v,a|\psi(t)}|^2.
\end{equation}
Let us analyze whether $\bar{p}_v(T)$ converges. Imagine that we start from a quantum state
\begin{equation}
    \ket{\psi(0)} = \sum_{k,a} c_{a,k}\ket{\phi_k,a}.
\end{equation}
Operator $W$ can be similarly decomposed
\begin{equation}
    W = \sum_{k,a} e^{2 i \varphi_{k,a}}\ket{\phi_k,a}\bra{\phi_k,a}.
\end{equation}
The eigenvalues $\cos\varphi_{k,0}$ and the eigenvectors $\ket{\phi_{k},0}$ extend to the whole bipartite space $\cos\varphi_{k}$ and $\ket{\phi_k}\ket{0}$ that we previously defined in $\mathcal{A}$.
Consequently, the evolved state is
\begin{equation}
\ket{\psi(t)} = \sum_{k,a} c_{a,k}e^{2 i \varphi_{k,a}}\ket{\phi_k,a}. 
\end{equation}
Substituting into \eqref{eq:average_prob},
\begin{equation}
\begin{split}
    \bar{p}_v(T) &= \frac{1}{T}\sum_{t=0}^{T-1}\sum_{b}|\braket{v,b|\psi(t)}|^2\\
    &= \sum_{a,a',b}\sum_{k,k'} c_{a,k}c_{a',k'}^*\braket{\phi_{k},a|v,b}\braket{v,b|\phi_{k'},a'}\times \frac{1}{T}\sum_{t=0}^{T-1} e^{2 i(\varphi_{k,a}-\varphi_{k',a'})t}.
\end{split}
\end{equation}
On the other hand, \cite[Chapter 7]{portugal2013quantum}
\begin{equation}
    \lim_{T\rightarrow \infty}\frac{1}{T}\sum_{t=0}^{T-1} e^{2 i(\varphi_{k,a}-\varphi_{k',a'})t} =
    \begin{cases}
    1 & \varphi_{k,a}= \varphi_{k',a'}\\
    0 & \varphi_{k,a}\neq  \varphi_{k',a'}\\
    \end{cases},
\end{equation}
because if the eigenvalues are different, we have an average over the approximate roots of the unit that cancels out asymptotically. Consequently, since
\begin{equation}
    \lim_{T\rightarrow \infty}\bar{p}_v(T)= \sum_{a,b}\sum_{k} |c_{a,k}|^2|\braket{\phi_{k},a|v,b}|^2,
\end{equation}
we have proven the limiting distribution exits. However, this distribution depends on the initial state coefficients $c_{k, a}$, and not only on the first eigenstate but on all of them, in contrast to the classical random walk~\cite{portugal2013quantum}.

Using this limiting distribution, we can study a possible definition of the quantum convergence time. In contrast to the hitting time, the quantum advantage in the mixing time will depend on the specific graph, and while in some systems the quantum mixing time displays a quadratic advantage \cite[Table~7.1]{portugal2013quantum}, in others the quantum walks mix more slowly than their classical counterparts~\cite{chakraborty2020fast}.

\subsubsection{Monte Carlo algorithms}

The fact that the limiting probability distribution $ \lim_{t\rightarrow\infty}\bar{p}_v(t)\neq \pi$ limits the usefulness of the previous definition. Instead, one may wish to prepare a coherent version of the stationary distribution, $\ket{\pi}$. This task is precisely the definition of operator \texttt{S}, which was not previously explained but was used in the search algorithms above. Indeed, one possible way to solve the mixing problem would be running the finding problem in reverse: since all search algorithms start from the coherent stationary state and end up in a marked item if we had a marked item we could reverse the algorithm to find the stationary state. There are some problems, though, since a constant overlap with such marked states is not sufficient to achieve similar running time~\cite{chakraborty2020analog}.

There are some special cases where it is possible to prepare coherent versions of the stationary state using phase estimation and measurements: since we know that the stationary state has eigenvalue 1, using phase estimation and measurements may probabilistically project the system into the stationary state. This opens the door to performing a digital simulation of an adiabatic algorithm.

In adiabatic quantum simulated annealing~\cite{somma2007quantum, somma2008quantum,wocjan2008speedup}, we start with the stationary state $\ket{\pi_0}$ of a simple Markov chain $\mathcal{P}_0$. For example, any symmetric Markov chain has a uniform superposition as stationary state, $\ket{\pi_0} = H^{\otimes n}\ket{0}^{\otimes n}$~\cite{magniez2011search}. Then, we look for a list of Markov chains $\mathcal{P}_0,\ldots \mathcal{P}_r$ that interpolates between the initial simple Markov chain $\mathcal{P}_0$ and the target $\mathcal{P} = \mathcal{P}_r$. At this point, one strategy is to use phase estimation and projective measurements to make the system evolve from $\ket{\pi_i}$ to $\ket{\pi_{i+1}}$.

\begin{algorithm}[h!]
\begin{algorithmic}[1]
\State \textbf{Input}: List of ergodic Markov chains $\{\mathcal{P}_i\}_{i\in\{1,\ldots,r\}}$ with corresponding quantum walks $\{W_i\}_i$, such that their stationary states fulfill $|\braket{\pi_{i-1}|\pi_i}|^2 \geq p$.
\State \textbf{Output}: Stationary state $\ket{\pi}$ of the Markov chain $\mathcal{P}$.
\State Initialize the stationary distribution $\ket{\pi_0}$, often a uniform distribution.
\For{$i$ in ${1,\ldots,r}$}:
\State Implement phase estimation $P(W_i)$ with precision $O(1/\Delta)$.
\State Measure the eigenvalue.
\If{Eigenvalue is not 1}
\State Restart algorithm.
\EndIf
\EndFor
\State Output $\ket{\pi}$.
\end{algorithmic}
\caption{Quantum Zeno effect annealing}\label{alg:Zeno}
\end{algorithm}

This algorithm has complexity $O(1/\Delta)$, offering a quadratic speedup in the eigenvalue gap, as the mixing time of the classical algorithm scales as $O(1/\delta)$~\cite{wocjan2008speedup}. Two possible modifications of \cref{alg:Zeno} are the following. First, phase estimation might be substituted by randomized evolution~\cite{somma2008quantum}. In other words, using a random number of applications of the quantum walk $W_i$ at each step. Such a random number should be taken at uniform from $[0, O(1/\Delta)]$. A second one is the so-called Zeno effect with rewind~\cite{lemieux2021resource}. In this variant, instead of discarding the state whenever we measure an eigenvalue different from 1, we can rewind the process. Using the projectors $\Pi_j=\ket{\pi_j}\bra{\pi_j}$, $\Pi_j^\perp = \bm{1} - \Pi_j$ and $p_j = |\braket{\pi_{j}|\pi_{j+1}}|^2$, we find that transitions between $\Pi_j$ and $\Pi_{j+1}$, or $\Pi_j^\perp$ and $\Pi_{j+1}^{\perp}$ occur with probability $p_{j}$. Similarly, the transitions $\Pi_j^\perp\leftrightarrow \Pi_{j+1}$ and $\Pi_j\leftrightarrow \Pi_{j+1}^\perp$ occur with probabilities $1-p_j$. Consequently, even if at a given step we measure $\Pi_j^\perp$ we can later on recover one of the $\ket{\pi_i}$ states.

A third alternative to the use of projective measurements is to amplify the overlap between consecutive $\ket{\pi_j}$~\cite{wocjan2008speedup}. Since we are only given a lower bound for the overlap between two states, it is useful to use Grover's $\pi/3$ algorithm instead of Grover's original one. We define
\begin{equation}
    U_{i;0} = \bm{1};\qquad R_i = e^{i(\pi/3)}\Pi_i + \Pi_i^\perp;\qquad U_{i;m+1} = U_{i;m}\cdot R_i\cdot U_{i;m}^\dagger\cdot R_{i+1}\cdot U_{i;m},
\end{equation}
where $m$ controls the number of amplification steps and $i$ the Markov chain. Using these definitions, we can obtain
\begin{equation}
    \ket{\pi} = \prod_{j=0}^r U_{j;m}\ket{\pi_0},
\end{equation}
with error probability $\epsilon$, if we choose
\begin{equation}
    m \geq L = \frac{12r\log (2r/\epsilon)}{\log(1/(1-p))},
\end{equation}
and assume overlap $p_j \geq p$ for all $j$~\cite{wocjan2008speedup}. As before, the rotations around $\ket{\pi_j}$ are performed with the help of the phase estimation operator.

The quantum simulated annealing techniques explained above are particularly useful for preparing ground states of Hamiltonians. To achieve this, we can aim to prepare Gibbs states of the form
\begin{equation}\label{eq:Gibbs}
    \ket{\pi^\beta} = Z^{-1/2}\sum_\lambda e^{-\beta H/2}\ket{\lambda} = Z^{-1/2}\sum_\lambda e^{-\beta \lambda/2}\ket{\lambda},
\end{equation}
where $\ket{\lambda}$ is an eigenstate of the Hamiltonian with eigenvalue $\lambda = E_\lambda$, and $Z$ is the partition function that normalizes the state. When $\beta$, the inverse temperature, goes to infinity, only the eigenstate with the lowest energy survives.

These techniques can be especially helpful in the context of the Metropolis-Hastings algorithm. The key idea of the Metropolis-Hastings approach is to engineer a series of `rapidly' mixing Markov chains~\cite{metropolis1953equation,hastings1970monte}. To do so, one starts from the detailed balanced equation of a reversible Markov chain,~\eqref{eq:detailed_balance}, $p_{yx}\pi_x = p_{xy}\pi_y$. Consequently, 
\begin{equation}
    \frac{p_{yx}}{p_{xy}} = \frac{\pi_y}{\pi_x}.
\end{equation}
The next step is to divide each transition into a proposal probability $T_{y,x}$ and an acceptance probability $A_{y,x}$, such that
\begin{equation}
    p_{yx} \propto T_{y,x}A_{y,x}.
\end{equation}
Consequently, we need to choose $A_{y,x}$ such that
\begin{equation}
    \frac{A_{y,x}}{A_{x,y}} = \frac{\pi_y}{\pi_x}\frac{T_{x,y}}{T_{y,x}}.
\end{equation}
A common choice that fulfills the previous equation is to take
\begin{equation}
    A_{y,x} = \min\left(1, \frac{\pi_y}{\pi_x}\frac{T_{x,y}}{T_{y,x}} \right).
\end{equation}
In particular, we can take a parameterized family of Markov chains
\begin{equation}\label{eq:Metropolis_general}
    \mathcal{P}_{\beta}(y,x) = \begin{cases}
    T_{y,x} A_{\beta;y,x} & y\neq x\\
    1-\sum_{z} T_{z,x} A_{\beta;z,x} & y = x,
    \end{cases}
\end{equation}
where 
\begin{equation}\label{eq:Metropolis_main}
    A_{\beta;y,x} = \min \{1, e^{-\beta(E_y-E_x)}\},
\end{equation}
$E_x$ indicates the `energy' of a state $x$,
and $T_{y,x}=1/n_e = T_{x,y}$ for $n_e$ the number of edges connected to $x$ or $y$. This means that the Markov chain has stationary state $\pi_x = Z^{-1} e^{-\beta E_x}$, and in the limit $\beta\to\infty$, $\pi$ concentrates all its probability mass on the state $x$ with lowest energy $E_x$. Similarly, if $\beta = 0$, then $\pi$ is the uniform distribution over all states.

Classical simulated annealing has complexity~\cite{somma2008quantum}
\begin{equation}
    O\left(\frac{\|H\|}{\gamma}\frac{\log (d/\epsilon^2)}{\delta}\right),
\end{equation}
where $\delta$ is the eigenvalue gap of the Markov chains $\mathcal{P}_{\beta}$, $\gamma$ the energy gap of the target Hamiltonian, and $d = |\Omega|$ the size of the search space.
In contrast, \cref{alg:Zeno} has complexity \cite[Eq.~5]{somma2008quantum}
\begin{equation}
    O\left(\left(\frac{\|H\|}{\gamma}\right)^2\frac{\log^2 (d/\epsilon)\log d\log(1/\epsilon)}{\Delta}\right),
\end{equation}
and the one using Grover's $\pi/3$ amplitude amplification \cite[Eq.~3]{wocjan2008speedup},
\begin{equation}
    O\left(\frac{\|H\|}{\gamma}\frac{\log d}{\Delta}\log \left(\frac{\|H\|}{\gamma}\log d\right)\right).
\end{equation}
Finally, the Zeno algorithm has been generalized to quantum Hamiltonians, where the stationary state \cref{eq:Gibbs} contains a superposition of eigenvalues that no longer represent computational basis (classical) states. In this case, we can also obtain a quadratic speedup in the eigenvalue gap $\Delta$, see Eq.~S50 in Ref.~\cite{yung2012quantum}.

\section{\label{sec:QFold}QFold}

Proteins are one of the most fundamental molecules in biochemistry. In contrast to many others, such as lipids, they exhibit very large variability in composition, which makes them a good candidate for numerous biochemical tasks. Proteins are made of chains of amino acids, which in turn are composed of a backbone chain of an amino group -NH, a central carbon labeled $\alpha$, a carboxyl group -COOH, and a side chain of other atoms, also called radical, attached to the $\alpha$ Carbon. The different composition possibilities of the side chain define 20 amino acids.

As proteins play a central role in biochemistry, understanding their specific functions or even designing proteins with specific roles is a very important problem. However, this function is not only determined by their amino acid decomposition or by the order those amino acids appear in the protein chain, but also by how such a chain is folded into its natural configuration, often much more complex to determine. This folding is largely determined by two angles per amino acid, commonly called $\psi$ and $\phi$, see~\cref{fig:glycylglycine}.

\begin{figure}[t]
    \centering
    \includegraphics[width = .5\textwidth]{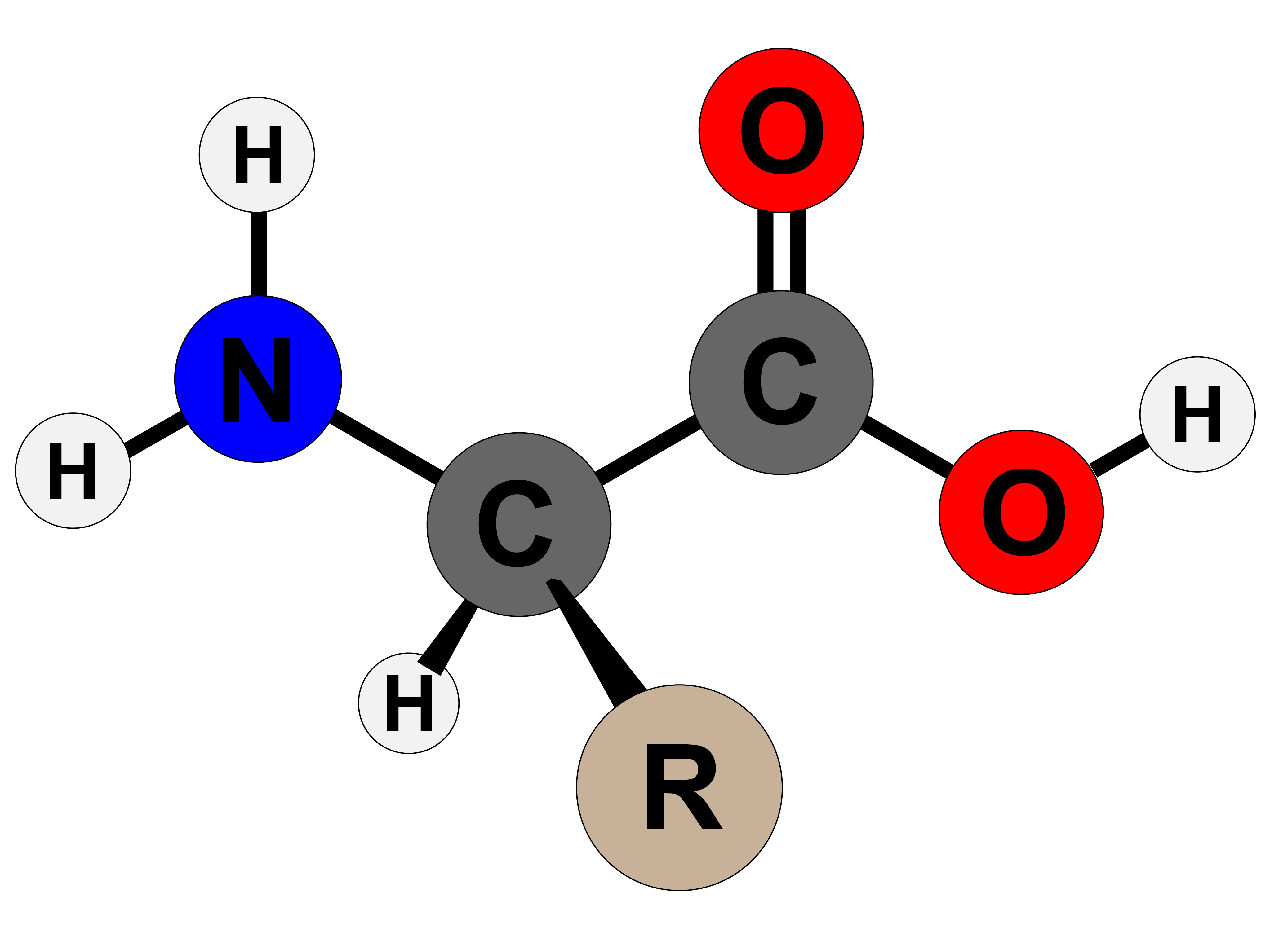}
    \caption{\textbf{Structure of an aminoacid.} The radical, indicated by the letter $R$, is a chain between 20 different options that determine the specific amino acid. Taken from \href{https://commons.wikimedia.org/wiki/File:Amino_Acid_Structure.png}{Wikipedia}, under CC-BY-SA 3.0 license.}
    \label{fig:aminoacid}
\end{figure}

\begin{figure}[t]
\centering
\includegraphics[width=\textwidth]{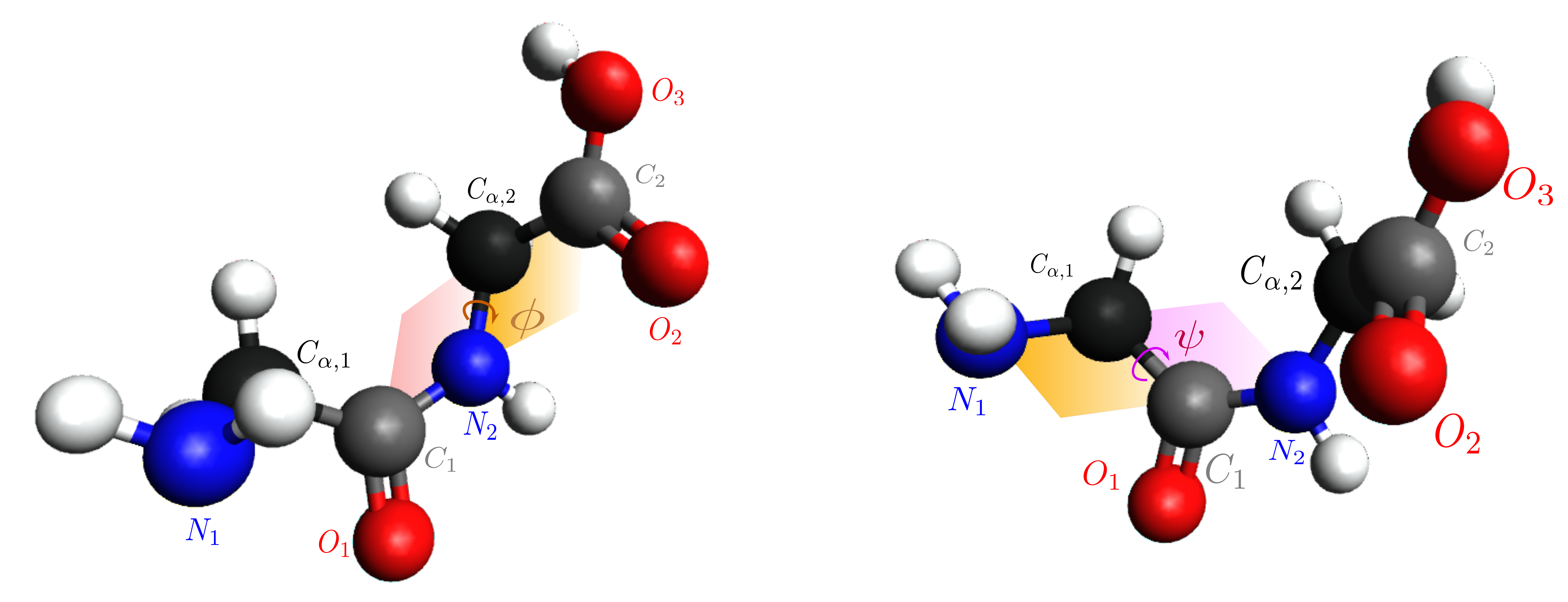}
\caption{\textbf{Example of the smallest dipeptide: the glycylglycine}. Each amino acid has the chain (Nitrogen-$C_{\alpha}$-Carboxy). Different amino acids would have a different side chain attached to the $C_\alpha$ instead of Hydrogen as is the case for Glycine. In each figure, we depict either angle $\phi$ or $\psi$. Angle $\psi$ is defined as the torsion angle between two planes: the first one defined by the three atoms in the backbone of the amino acid ($N_1$, $C_{\alpha,1}$, $C_1$), and the second by the same atoms except substituting the Nitrogen in that amino acid by the Nitrogen of the subsequent one: ($C_{\alpha,1}$, $C_1$, $N_2$). For the $\phi$ angle the first plane is made out of the three atoms in the amino acid ($N_2$, $C_{\alpha,2}$, $C_2$) whereas the second plane is defined by substituting the Carboxy atom in the amino acid by the Carboxy from the preceding amino acid: ($C_1$, $N_2$, $C_{\alpha,2}$).}
\label{fig:glycylglycine}\end{figure}

A breakthrough in the understanding of how proteins fold happened in 1961 when the scientist Christian Anfinsen showed that a denaturalized ribonuclease peptide (small protein) would always recover its shape and function if put back into its usual environment~\cite{anfinsen1961kinetics}. This and similar experiments led Anfinsen to postulate that proteins tend to fold in their thermodynamic ground state as determined by the aminoacid chain~\cite{anfinsen1973principles}, and would ultimately win him the Chemistry Nobel Prize in 1972.

While not universally true, as there are proteins with various stable configurations or others that have none, his `thermodynamic hypothesis' has guided the research in this area since then. Unfortunately, it is known that the protein structure prediction problem is NP-hard~\cite{hart1997robust}, even for the simplest toy models~\cite{berger1998protein}. This seems to be at odds with the fact that most proteins can fold in timescales of seconds or less. However, while baffling, scientists have so far not been able to find a computationally fast and inexpensive method to understand how proteins fold.

It is for this reason, and its NP-hardness, that this problem makes a good candidate for simulated annealing approaches, and for their quantum equivalent algorithms too. In fact, until very recently, simulated annealing has been the default approach for one of the most popular protein structure prediction computer libraries, Rosetta,~\cite{Rosetta}; and in distributed computational efforts such as Rosetta@Home~\cite{Rosetta@home,das2007rosetta@home}.
With the development of quantum walks over the last decade and a half, and the promise of quadratic speedups in a computationally hard problem, some effort has been put into developing tailored quantum algorithms~\cite{babbush2012construction,robert2021resource, perdomo2012finding, fingerhuth2018quantum, babej2018coarse, perdomo2008construction,outeiral2020investigating, wong2021quantum} for this problem and related ones~\cite{mulligan2020designing,banchi2020molecular}. Perhaps because the computational complexity of the problem does not decrease for toy models, almost all of these manuscripts have assumed simplified lattice models.

In contrast, in recent years powerful deep learning techniques have also been applied to this problem~\cite{AlphaFold,AlphaFoldv2}, producing extremely promising results in the Critical Assessment of Techniques for Protein Structure Prediction (CASP) competition~\cite{CASP}. In our article, we propose implementing a quantum annealing search starting from the state that the deep learning model has suggested~\cite{casares2022qfold}. The deep learning that we use as an initial guess will be \textit{Minifold}~\cite{ericalcaide2019minifold}, a small AlphaFold-inspired deep learning module with enough precision for our purposes. 

\subsection{\label{par:QFold_setup}The QFold algorithm}

\begin{figure}[h!]
\centering
\includegraphics[width=\textwidth]{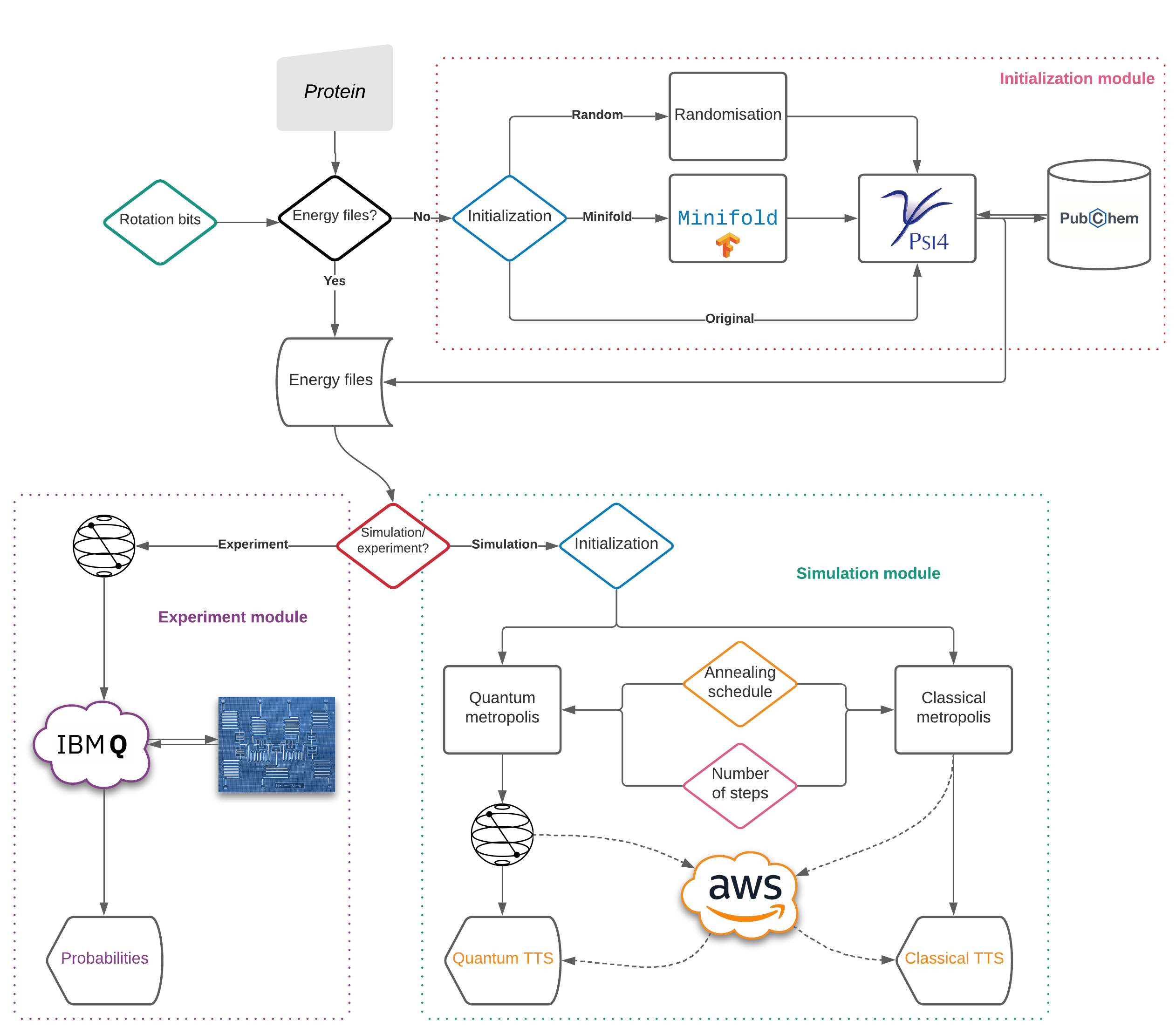}
\caption{\textbf{Flow chart of the QFold algorithm}~\cite{casares2022qfold}. QFold has several functionalities integrated, that could be summarized in an initialization module, a simulation module, and an experiment module. We denote by diamonds each of the decisions one has to make. The top part constitutes the initialization module, where \textit{Minifold}, a deep learning module, can be used to get a guess of the correct folding, and \textit{Psi4} and the original geometry from \textit{PubChem} are used to calculate the energies of rotations. The bottom half represents the \texttt{experiment} or \texttt{simulation} algorithms, which output either probabilities or quantum/classical TTS of the corresponding metropolis algorithms, making use of \textit{Qiskit}.
}
\label{fig:flow_chart}\end{figure}

Relying on Anfinsen's hypothesis, our work aims to use a quantum Metropolis algorithm to find the thermodynamic ground state of the system. This is a search problem, but since there is no oracle to mark the state, the objective is instead to find the stationary state of a rapidly mixing Markov chain that favors the states with the lowest energy. Consequently, our procedure will fit into the scheme of Monte Carlo algorithms discussed in the previous subsection, and in particular, the Metropolis-Hastings algorithm defined in \eqref{eq:Metropolis_general} and \eqref{eq:Metropolis_main}.

As we reviewed in that section, there are a few quantum Metropolis proposals that could work on this problem. However, most of them use the expensive quantum phase estimation subroutine and come with theoretical guarantees of performance, which is not the case in the commonly used classical Metropolis algorithm. As a result, Lemieux et al.~\cite{lemieux2019efficient} proposed to use a heuristic algorithm more in the spirit of the classical Metropolis algorithm. It simply applies quantum walk steps to make the state evolve over time
\begin{equation}
    \ket{\psi(T)} := \tilde{W}_T ... \tilde{W}_1 \ket{\pi_0}.
    \label{eq:L quantum walk steps}
\end{equation}
The quantum walk step $\tilde{W}$ is a modification of Sgedy's original algorithm that instead of duplicating $\mathcal{H}$ uses a coin. It uses 3 quantum registers: $\ket{\cdot}_S$ indicating the current state of the system, $\ket{\cdot}_M$ that indexes the possible moves one may take according to $T_{y,x}$ in \eqref{eq:Metropolis_general}, and $\ket{\cdot}_C$ the Boltzmann coin register. We may also have auxiliary registers $\ket{\cdot}_A$.
Then the quantum walk operator is defined as
\begin{equation}
    \tilde{W} = R V^\dagger B^\dagger F B V,
    \label{quantum walk operator tilde W}
\end{equation}
where $V$ prepares in register $\ket{\cdot}_M$ a superposition over all possible steps one may take; $B$ rotates the coin qubit $\ket{\cdot}_C$ to have amplitude of $\ket{1}_C$ corresponding to the acceptance probability indicated by \eqref{eq:Metropolis_main}; $F$ changes the $\ket{\cdot}_S$ register to the new configuration conditioned on the value of $\ket{\cdot}_M$ and $\ket{\cdot}_C = \ket{1}_C$; and $R = \bm{1}-2 \ket{0}\bra{0}_{MCA}$.

The metric chosen to compare the classical and quantum Metropolis algorithms is called Total Time to Solution~\cite{lemieux2019efficient}. Its role is to measure the expected number of quantum walk steps it would take to find a solution if we allow for restarts of the algorithm. Specifically,
\begin{equation}
    TTS(t):= t \frac{\log (1-\delta)}{\log ( 1-p(t))},
    \label{TTS}
\end{equation}
where $t$ is the number of steps we take in the algorithm, $\delta$ is some target probability threshold, and $p(t)$ is the success probability after $t$ steps. For each problem, we are interested in the minimum TTS, either classical or quantum, as this is the expected TTS we would recover if we were to take the optimal number of walk steps before measuring.

In our article, we explore different annealing schedules~\cite{van1987simulated} including
\begin{itemize}
\begin{subequations} \label{Annealing schedules}
    \item \texttt{Boltzmann} or \texttt{logarithmic} implements the famous logarithmic schedule~\cite{geman1984stochastic}
    \begin{equation}
        \beta(t) = \beta(1) \log(t e) = \beta(1) \log(t) + \beta(1).
    \end{equation}
    Notice that the multiplication of $t$ times $e$ is necessary to make a fair comparison with the rest of the schedules so that they all start in $\beta(1)$.
    \item \texttt{Cauchy} or \texttt{linear} implements the simple schedule given by
    \begin{equation}
        \beta(t) = \beta(1) t.
    \end{equation}
    \item \texttt{geometric} defines~\cite{kirkpatrick1983optimization}
    \begin{equation}
        \beta(t) = \beta(1) \alpha^{-t+1},
    \end{equation}
    where $\alpha<1$ is a parameter heuristically set to $0.9$. 
    \item And finally \texttt{exponential} uses
    \begin{equation}
        \beta(t) = \beta(1) \exp(\alpha (t-1)^{1/N}),
    \end{equation}
    where $\alpha$ is again set to $0.9$ and $N$ is the space dimension, which in this case is equal to the number of torsion angles. 
\end{subequations}
\end{itemize}

\begin{figure}[t]
\includegraphics[width=1.35\textwidth/2]{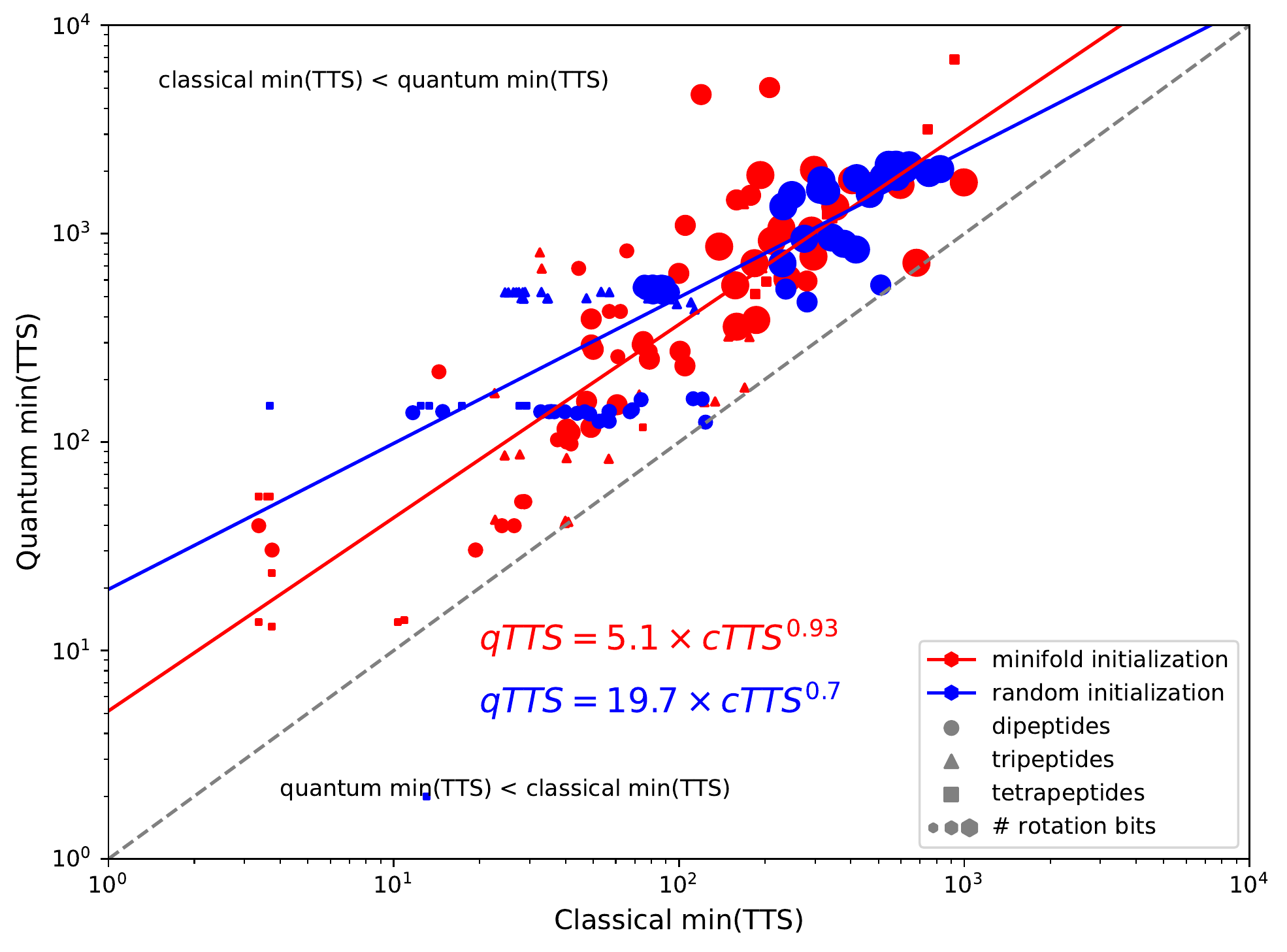}
\centering
\caption{\textbf{Comparison of the classical and quantum minimum TTS achieved for the simulation of the quantum Metropolis algorithm with $\beta = 10^3$}, for 10 dipeptides (with $b = 3, 4, 5$ rotation bits of precision in the angles), 10 tripeptides ($b = 2$) and 4 tetrapeptides ($b = 1$), also showing the different initialization options (random or minifold), and the best fit lines~\cite{casares2022qfold}. The dashed grey line separates the space where the quantum TTS is smaller than the classical TTS.
The key aspect to notice in this graph is that although for smaller instances the quantum algorithm does not seem to match or beat the times achieved by the classical Metropolis, due to the exponent being smaller than one (either $0.89$  or $0.53$ for minifold or random respectively) for average size proteins we can expect the quantum advantage to dominate. In the main text, we explain why the random initialization exponent seems more favorable than the minifold exponent and discuss further details respectively.
}
\label{fig:fixed_beta_TTS_slope}\end{figure}

\subsection{\label{ssec:QFold_results}Results}

\begin{figure}[t!]
\centering
\makebox[\textwidth][c]{\includegraphics[width=1.\textwidth]{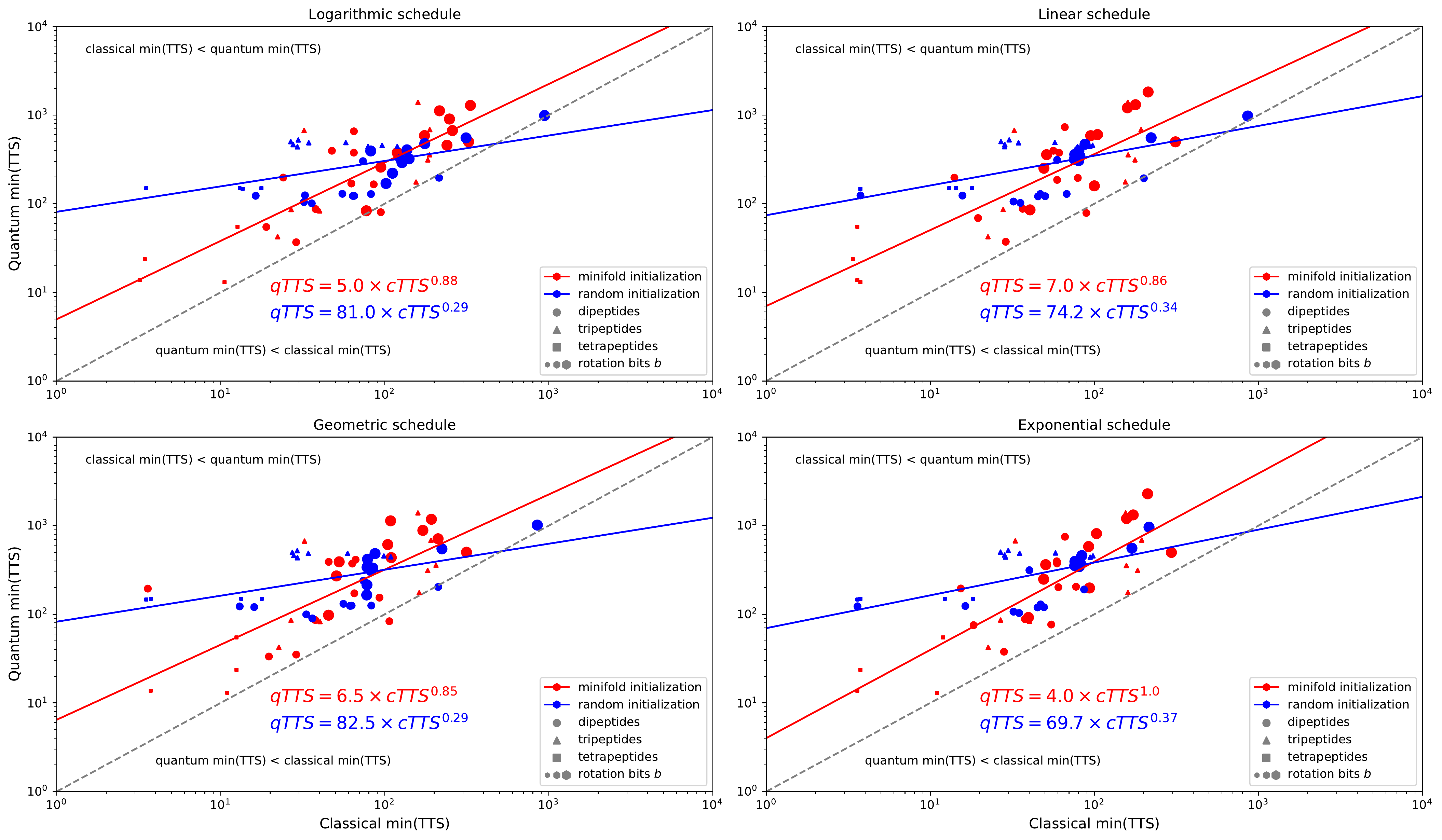}}
\caption{\textbf{Comparison of the classical and quantum minimum TTS achieved for a number of small peptides with several annealing schedules}~\cite{casares2022qfold}. This figure shows the different initialization options (random or minifold) and annealing schedules (\texttt{Boltzmann/logarithmic}, \texttt{Cauchy/linear}, \texttt{geometric} and \texttt{exponential}), and the best fit lines. The dashed grey line depicts the diagonal. The corresponding fit exponents are given in table \ref{tab:exponents}, where we observe a small polynomial quantum advantage. On the other hand, using an exponential schedule does not seem to give any advantage when used with a minifold initialization.}
\label{fig:var_beta_TTS_slope}\end{figure}

In our article, we analyze two kinds of experimental results: those coming from simulations and others from actual experiments in quantum computers. As mentioned, we test whether the quantum or classical Metropolis algorithms achieve a better Total Time to Solution. In the simulation results, we are particularly interested in understanding how this metric evolves for larger system sizes, and we depict them in~\cref{fig:fixed_beta_TTS_slope} and~\cref{fig:var_beta_TTS_slope}. In this chapter, we have thoroughly explained that the expected quantum advantage is polynomial, but since the algorithm is heuristic, we are interested in finding the actual exponent that encapsulates it. A quantum advantage will be revealed whenever the quantum minimum TTS grows as a power of the classical minimum TTS with an exponent lower than 1. For instance, we would measure a quadratic quantum advantage if the exponent were 0.5. We compute the exponent using the standard technique of linear least square fitting in the logarithmic scale for the classical and quantum minimum TTS achieved.

An important remark is in order. This figure seems to suggest that the quantum advantage when we use minifold as an initialization module is smaller than whenever we use a random initialization. While this may seem to indicate that our proposed deep learning initialization is actively harmful to the predictive power of the quantum Metropolis these results should be interpreted differently: for the smallest instances, and in particular, when using a random initialization, the minimum quantum TTS value is achieved for $t=2$. This implies that in such cases the algorithm is avoiding the use of the Metropolis algorithm, and instead repeatedly choosing and outputting random points from the search space. The reason is that in the smallest instances, the multiplicative prefactors of the quantum walk have a larger effect than the eigenvalue gap that determines the quantum advantage. Consequently, the quantum advantage found using the minifold initialization is likely to be more representative of the actual large-size behavior.

\begin{table}[b!]
\centering
\begin{tabular}{ |c||c|c|  }
 \hline
 \multicolumn{3}{|c|}{Fit exponents} \\
 \hline
 Schedule & Random initial. & Minifold initial.\\
 \hline
 Fixed $\beta$   &$0.70\pm 0.08$  & $0.93\pm 0.06$\\
 Logarithmic &   $0.29\pm 0.07$  & $0.88\pm 0.09$\\ 
 Linear &$0.34\pm0.07$  & $0.86\pm 0.11$\\
 Exponential    &$0.37\pm 0.07$  & $1.00\pm 0.12$\\
 Geometric &   $0.29\pm 0.07$  & $0.85\pm 0.18$\\
 \hline
\end{tabular}\caption{\textbf{Table of scaling exponents for different annealing schedules and initialization options}~\cite{casares2022qfold}. The peptides are the same, except that for fixed $\beta$ we have also included dipeptides with 5 bits of precision, which is costly for the rest of the schedules. For fixed $\beta$, the value heuristically chosen was $\beta = 1000$, while the initial $\beta$ value in each of the schedules, defined in \eqref{Annealing schedules}, is $\beta(1) = 50$. The uncertainty is expressed via the standard deviation in the expected exponent, calculated with the bootstrapping method \cite{efron1992bootstrap}.}
\label{tab:exponents}
\end{table}

The estimated quantum advantages indicated in~\cref{tab:exponents}, and shown in~\cref{fig:fixed_beta_TTS_slope} and~\cref{fig:var_beta_TTS_slope}, largely match the value 0.7549 computed in the original reference by Lemieux et al.~\cite{lemieux2019efficient} for a different problem, although are slightly worse. Unfortunately, I believe the most important conclusion that can be extracted from these simulations is rather a negative result: as argued in~\cite{babbush2021focus} and mentioned in our article, it is very unlikely that the $\simeq0.9$ polynomial quantum advantage that we observe might be of any use if one takes into account the error correction overhead required to operate a quantum computer. On the other hand, our results are exploratory, due to the incapability to simulate large enough systems without an actual quantum computer, so different modifications might achieve larger and more useful quantum advantages. Additionally, we assumed that the optimal choice of parameters is the same for both quantum and classical metropolis algorithms, but this choice is not necessarily true.

\begin{figure}[t!]
    \centering
    \includegraphics[width = 1.35\textwidth/2]{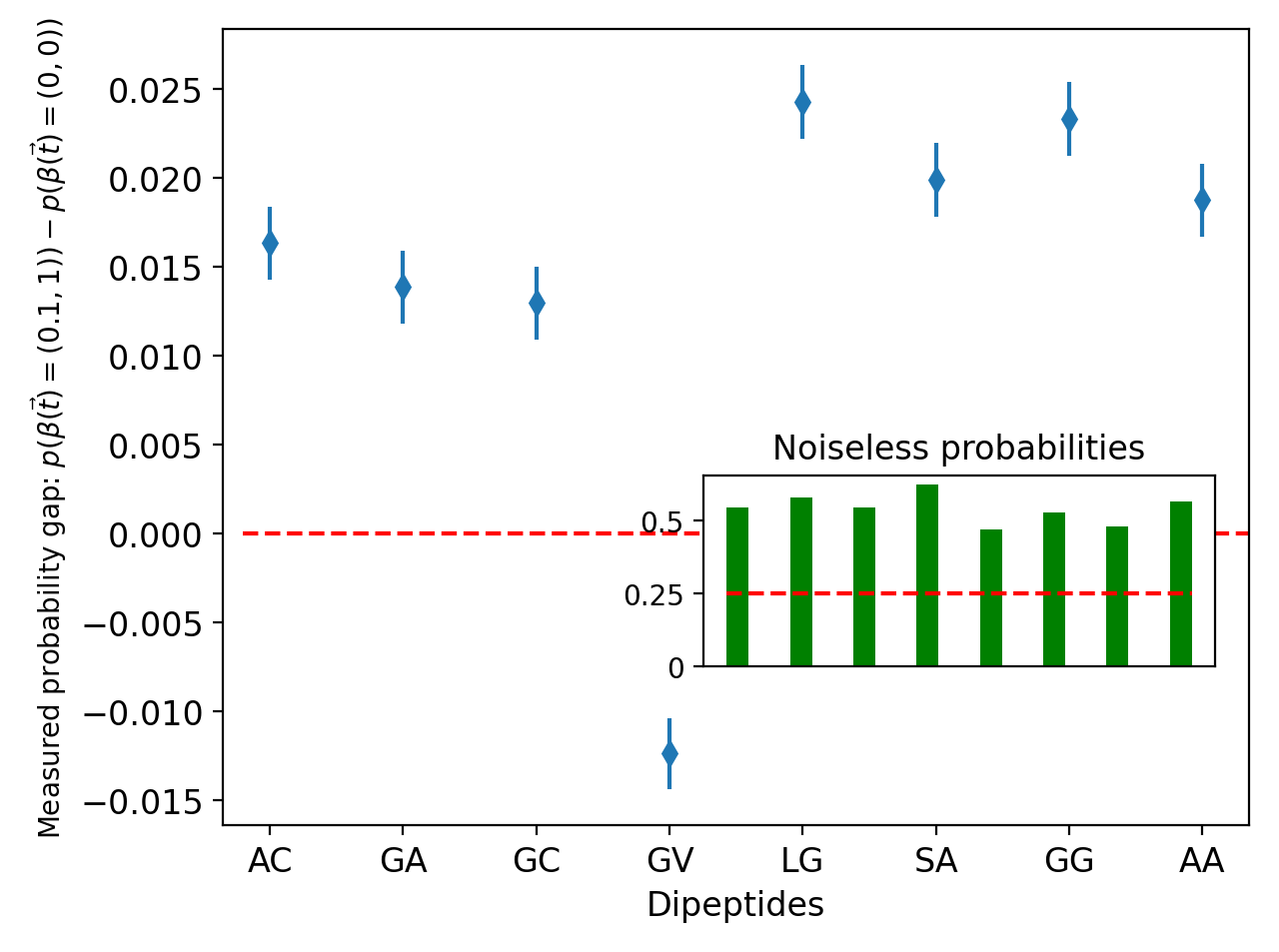}
    \caption{\textbf{Results from hardware measurements} corresponding to the experimental realization of the quantum Metropolis~\cite{casares2022qfold}. For each dipeptide we perform a student t-test to check whether the average success probabilities for $\beta(\bm{t}) = (0,0)$ and $\beta(\bm{t}) = (0.1,1)$ are actually different \cite{student1908ttest}. The largest p-value measured in all 8 cases is $3.94\cdot 10^{-18}$, indicating that in all cases the difference is significant. For each dipeptide, we run 163840 times the circuit, and for the baseline 204800 times.}
    \label{fig:Hardware}
\end{figure}

In addition to simulation results, we also execute a small proof-of-concept experiment in an actual IBM Quantum processor. Since NISQ devices such as this one are very limited in the depth of the circuits they can execute, we are restricted to very small search spaces, with only four positions. Consequently, our objective here is just to understand whether we can overcome the error in the circuit and measure probability differences between setting $\beta = 0$, and $\beta \neq 0$. The results are depicted in~\cref{fig:Hardware} and indicate that in 7 out of the 8 dipeptides tested, the quantum algorithm correctly increases the probability of measuring the correct state. Unfortunately, we do not have a good model for why the remaining peptide achieves a statistically significant negative result, although it is reasonable to assume this is due to the physical imperfections of the superconducting chip.

Beyond our article, recent years have seen tremendous advances in the capabilities of deep learning models to predict the structure of proteins. It is encouraging to see that the latest AlphaFold model can often achieve precision similar to costly laboratory methods~\cite{AlphaFoldv2}. Moreover, these models have been open-sourced, followed by the processing of large databases of proteins that were previously unavailable~\cite{AlphaFold_database}. On the quantum walk side, the heuristic quantum Metropolis algorithm was later on used for quantum ground state preparation, see~\cite{lemieux2021resource}. In our group, its application to other problems in Machine Learning~\cite{campos2022quantum} and Bayesian inference~\cite{escrig2022parameter} has been explored too.

\section{Results}

\begin{itemize}
    \item We have explained Grover's algorithm, as well as several of their generalizations such as amplitude amplification, and fixed-point amplitude amplification. The latter has the advantage that probability does not decrease significantly if we apply more amplification steps than necessary. 
    \item In~\cref{sec:Walks} we have analyzed the quantum walk technique, which displays a similar mathematical structure as Grover's algorithm and may also be understood as its generalization. 
    \item In particular, we have explained how quantum walks can often achieve a quadratic speedup in search problems due to the quadratically larger quantum eigenvalue gap, $\Delta = O(\sqrt{\delta})$. We have indicated a series of techniques that extend the range of search problems where quadratic advantages might be found, culminating in the Quantum Fast Forwarding algorithm that applies to ergodic Markov chains.
    \item Such quadratic advantage is however not always possible in mixing problems, where the objective is to prepare the stationary state of the Markov chain.
    \item One exception to the previous rule is the Metropolis algorithm: We have studied how the Metropolis algorithm may benefit from the quantum walk technique, providing a quadratic advantage to prepare stationary distributions of Markov chains, with applicability to hard combinatorial optimization problems.
    \item We have computationally analyzed a heuristic quantum Metropolis algorithm which provides a modest quantum advantage in the exponent in protein folding~\cite{casares2022qfold}, machine learning~\cite{campos2022quantum} and Bayesian inference problems~\cite{escrig2022parameter}. This advantage should be especially helpful in cases where the problem is NP-hard, and therefore it is not possible to exploit the problem structure. A clear example of this is precisely protein folding.
    \item We have further analyzed several annealing schedules for the heuristic quantum metropolis algorithm and found that most give a small improvement over constant-$\beta$ walks. These results confirm and strengthen similar results found in one previous article by Lemieux et al.,~\cite{lemieux2019efficient}, for simpler Ising chain models.
\end{itemize}

\chapter{\label{ch:Phase}Quantum Linear Algebra}  

\ifpdf
    \graphicspath{{Chapter2/Figs/Raster/}{Chapter2/Figs/PDF/}{Chapter1/Figs/}}
\else
    \graphicspath{{Chapter2/Figs/Vector/}{Chapter2/Figs/}}
\fi

\epigraph{Not only do these new algorithms promise exponential speedups over classical algorithms, but they do so for eminently-practical problems, involving machine learning, clustering, classification, and finding patterns in huge amounts of data. So, do these algorithms live up to the claims? That’s a simple question with a complicated answer.}{Scott J. Aaronson, {\it Quantum Machine Learning Algorithms: Read the Fine Print}}


\section{Objectives}

\begin{itemize}
    \item Understand the mathematical techniques used in quantum linear algebra.
    \item Understand the limitations of quantum linear algebra techniques, in particular, those related to data loading and readout.
    \item Use quantum linear algebra techniques to improve state-of-the-art linear programming methods.
    \item Understand dequantization and how it limits many of the applications of quantum linear algebra to Machine Learning.
\end{itemize}

\section{\label{sec:QFT}The Fourier transform \& phase estimation algorithms} 

In the previous chapter, we explored applications related to quantum search, starting with Grover's algorithm. In this one, we want to understand how we can deal with linear algebra problems, which are present in many other computer science applications. To do so, we first have to explain the \textit{quantum Fourier transform}, which will enable the implementation of the \textit{quantum phase estimation} algorithm mentioned in the previous chapter. Additionally, phase estimation will be the key subroutine in Shor's algorithm, arguably the most famous quantum algorithm together with Grover's. 

As a motivation for the latter, let us introduce a related one. Let $\oplus$ denote bitwise binary addition such that for example $011 \oplus 101 = 110$.

\begin{problem}[Simon]\label{prob:Simon}
 Let $f: 2^{n}\mapsto 2^{n}$ such that $f(i) = f(j)$ if and only if $i\oplus s = j$ for some secret $s$. Find $s$.
\end{problem}

To solve this problem, Simon proposed the following quantum algorithm~\cite{simon1997power}:

\begin{algorithm}[h!]
\begin{algorithmic}[1]
\State \textbf{Input}: Oracle $f: \ket{i}\ket{0}\mapsto \ket{i}\ket{f(i)}$ fulfilling the promise of \cref{prob:Simon}.
\State \textbf{Output}: String $s$ solving the corresponding Simon \cref{prob:Simon}.
\State Use Hadamard gates to create a uniform superposition, 
\begin{equation}
    \frac{1}{\sqrt{2^n}}\sum_{i}\ket{i}\ket{0}.
\end{equation}
\State Query the oracle $f$.
\State (Optionally) measure the second qubit register on the computational basis, obtaining
\begin{equation}
    \frac{1}{2}(\ket{i}+\ket{i\oplus s})\ket{f(i)}.
\end{equation}
\State Apply a second Hadamard to the first register, obtaining
\begin{equation}
    \frac{1}{\sqrt{2^n}}\sum_j\left((-1)^{i\cdot j}\ket{j}+ (-1)^{(i\oplus s) \cdot j}\ket{j}\right) = \frac{1}{\sqrt{2^n}}\left(\sum_j (-1)^{i\cdot j} (1+(-1)^{s \cdot j})\ket{j}\right).
\end{equation}
\State Measure $j$. Each $\ket{j}$ will have non-zero amplitude only if $s\cdot j = 0$.
\State Repeat until one has obtained sufficient (linearly independent) samples from $j$, and solve the associated linear system of equations.
\end{algorithmic}
\caption{Simon's algorithm}\label{alg:Simon}
\end{algorithm}

Once we have obtained $k$ linearly independent samples, the probability of subsequently measuring another independent sample is $(2^{n-1}-2^k)/2^{n-1}\geq 1/2$. Since the linear system of equations can be solved in time $O(n^3)$, the overall complexity is $O(n^3)$. In contrast, Simon proved that any classical algorithm requires $\Omega(\sqrt{2^n})$ calls to the oracle to find collisions with high probability~\cite{simon1997power}.

Shor's algorithm solves a similar periodicity-related problem.
\begin{problem}[Period finding]\label{prob:Shor}
 Let $f: \mathbb{Z}\mapsto \{0,...,N-1\}$ such that $f(i) = f(j)$ if and only if $i = j \mod r$. Find $r$.
\end{problem}
There is an important difference, though. In Simon's problem, the addition was bit-wise, while in this case, the periodicity is a global property. It is for this reason that Hadamard gates will not be enough. Instead, we will introduce the quantum Fourier transform, which is related to the Hadamard transform by
\begin{equation}
    H = \begin{pmatrix}
    1 & 1\\
    1 & -1
    \end{pmatrix} = F_2
\end{equation}
The Fourier transform is a unitary transformation defined by the matrix
\begin{equation}
    F_N = \frac{1}{\sqrt{N}}
    \begin{pmatrix}
    1 & 1      & 1        & 1        &\ldots   & 1\\
    1 & \omega_N & \omega_N^2 & \omega_N^3 &\ldots & \omega_N^{N-1}\\
    1 & \omega_N^2 & \omega_N^4 & \omega_N^6 & \ldots & \omega_N^{2(N-1)}\\
    \vdots &\vdots &\vdots &\vdots &\ddots &\vdots\\
    1 & \omega_N^{N-1} &  \omega_N^{2(N-1)} & \omega_N^{3(N-1)} & \ldots & \omega_N^{(N-1)^2}
    \end{pmatrix}
    = \frac{1}{\sqrt{N}}\left(e^{\frac{2\pi i }{N}(j\cdot k)}\right)_{jk},
\end{equation}
where $\omega_N = e^{2\pi i/N}$.

\begin{figure}
\centering
\Qcircuit @C=1em @R=0.65em { 
  & \qw & \gate{\hat{F}_2}\qwx[4] & \qw & \qw & \qw  & \qw & \qw & \gate{\hat{F}_2}\qwx[2]& \qw& \qw & \qw& \qw & \gate{\hat{F}_2}\qwx[1] & \qw & \qw & \qw & \qw \\
  & \qw & \qw & \gate{\hat{F}_2}\qwx[4] & \qw & \qw & \qw  & \qw &  \qw&\gate{\hat{F}_2}\qwx[2]& \qw & \qw& \qw& \gate{\hat{F}_2} & \qw & \qswap\qwx[3] & \qw & \qw\\
  & \qw & \qw & \qw & \gate{\hat{F}_2}\qwx[4]  & \qw  & \qw  & \qw & \gate{\hat{F}_2}& \qw& \qw & \gate{e^{0\pi i/ 4}}& \qw& \gate{\hat{F}_2}\qwx[1] & \qw & \qw& \qw & \qw\\
  & \qw & \qw & \qw & \qw  & \gate{\hat{F}_2}\qwx[4] & \qw  & \qw& \qw & \gate{\hat{F}_2}& \qw& \gate{e^{2\pi i/ 4}}& \qw& \gate{\hat{F}_2} & \qw & \qw & \qswap\qwx[3] & \qw\\
  & \qw & \gate{\hat{F}_2} & \qw & \qw & \qw  & \qw & \gate{e^{0\pi i/ 8}}& \gate{\hat{F}_2}\qwx[2]& \qw& \qw& \qw& \qw& \gate{\hat{F}_2}\qwx[1] & \qw & \qswap & \qw & \qw\\
  & \qw & \qw & \gate{\hat{F}_2} & \qw & \qw  & \qw & \gate{e^{2\pi i/ 8}}& \qw& \gate{\hat{F}_2}\qwx[2]& \qw& \qw& \qw& \gate{\hat{F}_2} & \qw& \qw& \qw & \qw\\
  & \qw & \qw & \qw& \gate{\hat{F}_2} & \qw & \qw  & \gate{e^{4\pi i/ 8}}& \gate{\hat{F}_2}& \qw& \qw& \gate{e^{0\pi i/ 4}}& \qw & \gate{\hat{F}_2}\qwx[1] & \qw& \qw& \qswap\qwx& \qw\\
  & \qw & \qw & \qw & \qw & \gate{\hat{F}_2}  & \qw & \gate{e^{6\pi i/ 8}}& \qw & \gate{\hat{F}_2}& \qw& \gate{e^{2\pi i/ 4}}& \qw& \gate{\hat{F}_2} & \qw& \qw& \qw& \qw\\
}
\caption{\textbf{Quantum Fermionic Fast Fourier Transform for 8 modes.} The $\hat{F}_2$ gate is defined in \eqref{eq:F_2}, while the single-qubit gates are $Z$-like rotations. This circuit is the same as depicted in Fig. 1~\cite{ferris2014fourier}.}
\label{fig:FFT}
\end{figure}
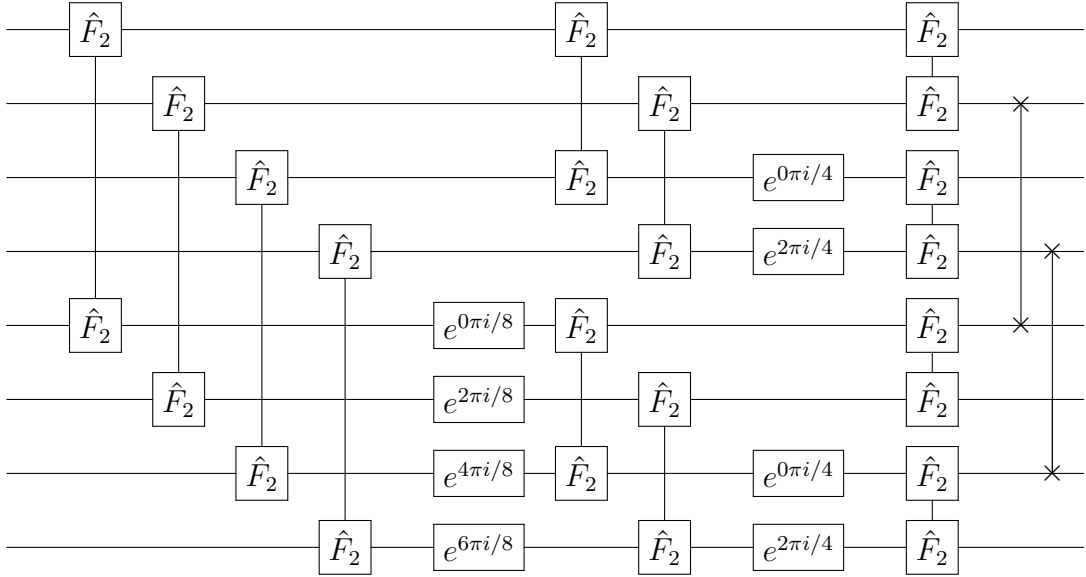

Applying one such matrix classically naively takes $O(N^2)$ time, but can be reduced to $O(N\log N)$ by noting that vector coordinates are transformed as
\begin{equation}
\begin{split}
    \bar{\bm{v}}_j = \sum_{k} \omega_N^{j\cdot k}  \bm{v}_k &= \frac{1}{\sqrt{N}}\left(\sum_{\text{even }k}\omega_N^{jk}\bm{v}_k + \omega_N \sum_{\text{odd }k}\omega_N^{j(k-1)}\bm{v}_k \right)\\
    &= \frac{1}{\sqrt{2}}\left(\frac{1}{\sqrt{N/2}}\sum_{\text{even }k}\omega_{N/2}^{jk/2}\bm{v}_k + \omega_N^j \sum_{\text{odd }k}\omega_{N/2}^{j(k-1)/2}\bm{v}_k \right),
\end{split}
\end{equation}
where the bar indicates the Fourier transform. The key takeaway is that we can compute the $n$-Fourier transform with 2 $(n-1)$-Fourier transforms. Moreover, we can decompose the $(n-1)$-Fourier transforms, initiating a recursion process, which spans a tree of depth $O(\log N)$ and $N/2$ terms at the deepest level. This results in complexity $O(N\log N)$, and is commonly known as the \textit{fast Fourier transform} (FFT)~\cite{cooley1965algorithm}.

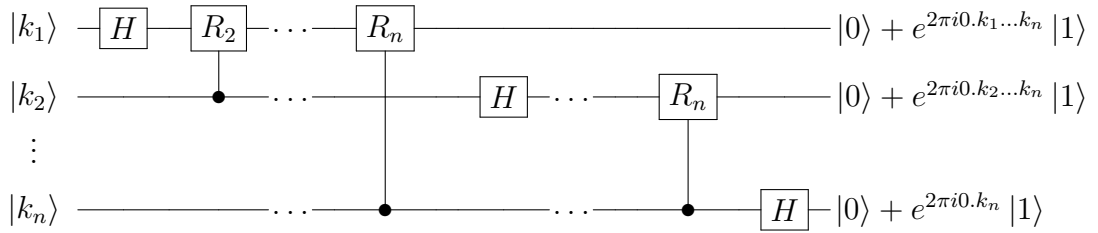
\begin{figure}
\[
\begin{array}{c}
\Qcircuit @C=0.7em @R=0.75em { 
  & \ket{k_1} & & & \gate{H} & \qw & \gate{R_2} & \qw & \ldots & & \qw & \gate{R_{n}} &\qw  &\qw   &\qw & \qw &\qw  &\qw   &\qw & \qw &\qw  &\qw   &\qw & \qw & & & & & &\ket{0} + e^{2\pi i 0.k_{1}\ldots k_{n} }\ket{1}\\
  & \ket{k_2} & & & \qw & \qw & \ctrl{-1} & \qw & \ldots & &  \qw& \qw & \qw& \qw & \gate{H}& \qw & \ldots & & \qw  & \qw & \gate{R_{n}} &\qw   &\qw & \qw & & & & & &\ket{0} + e^{2\pi i 0.k_{2}\ldots k_{n} }\ket{1}\\
   & \vdots \\
  \\
  & \ket{k_n} & & & \qw & \qw & \qw & \qw & \ldots & &  \qw & \ctrl{-4}& \qw & \qw &\qw & \qw & \ldots &  & \qw  & \qw & \ctrl{-3} &\qw & \gate{H} & \qw & & & & &\ket{0} + e^{2\pi i 0.k_n }\ket{1}  \\
}
\end{array}
\]
\caption{\textbf{Quantum Fourier transform.} Notice that the qubit representation of the quantum Fourier transform comes out reversed, and can be reordered using swaps if desired. Each $R_l$ gate is defined by \eqref{eq:R_l}. The QFT circuit requires $O(n^2)$ gates, for $n$ qubits.}
\label{fig:QFT}
\end{figure}
In fact, this technique has also been used in quantum computing, especially for fermions~\cite{ferris2014fourier}, see \cref{fig:FFT}. In this case, the fermionic operator $\hat{F}_2$ is implemented over two qubits
\begin{equation}\label{eq:F_2}
    \hat{F}_2 = 
    \begin{pmatrix}
    1 & 0 & 0 & 0\\
    0 & 2^{-1/2} & 2^{-1/2} & 0\\
    0 & 2^{-1/2} & -2^{-1/2} & 0\\
    0 & 0 & 0 & -1
    \end{pmatrix},
\end{equation}
where the last $-1$ is due to the anticommutation relation of fermions~\cite{ferris2014fourier}. This matrix operator can be decomposed into two C-NOT gates sandwiching a controlled Hadamard gate in the opposite direction, followed by a C-$Z$ to implement the final phase.

However, since the Fourier transform is unitary, there is also a direct implementation over $n = \log N$ qubits. To understand how to generate it, we rewrite~\cite{de2019quantum}
\begin{equation}\label{eq:QFT}
\begin{split}
    F_N\ket{k} = \frac{1}{\sqrt{N}}\sum_{j = 0}^{N-1}e^{\frac{2\pi i}{N}jk}\ket{j} = \frac{1}{\sqrt{N}}\sum_{j = 0}^{N-1}e^{\frac{2\pi i}{N}(\sum_l j_l 2^{-l})k}\ket{j_1}\ldots\ket{j_n}\\
    = \frac{1}{\sqrt{N}}\sum_{j = 0}^{N-1}\left(\prod_{l=1}^n e^{\frac{2\pi i}{N}( j_l 2^{-l})k}\right)\ket{j_1}\ldots\ket{j_n} = \bigotimes_{l = 1}^n \frac{1}{\sqrt{2}}\left(\ket{0}+e^{2\pi i k / 2^l}\ket{1}\right).
\end{split}
\end{equation}
We note that the exponent in $e^{2\pi i k / 2^l}$ would have equivalent effect as the fractional part of its binary expression, $e^{2\pi i 0.k_{n-l+1}\ldots k_n}$~\cite{de2019quantum}. This suggests the quantum Fourier circuit depicted in \cref{fig:QFT}, with rotation $R_l$ defined by 
\begin{equation}\label{eq:R_l}
    R_l = \begin{pmatrix}
    1 & 0\\
    0 & e^{2\pi i/ 2^l}
    \end{pmatrix}.
\end{equation}

The (inverse) quantum Fourier transform is an important step in what might be considered one of the most useful quantum subroutines: the quantum phase estimation algorithm~\cite{kitaev1995quantum}. Imagine we have a unitary operator $U$ with eigenstates $\ket{\psi_i}$ and eigenvalues $\lambda_i = e^{2\pi i\varphi_i}$.
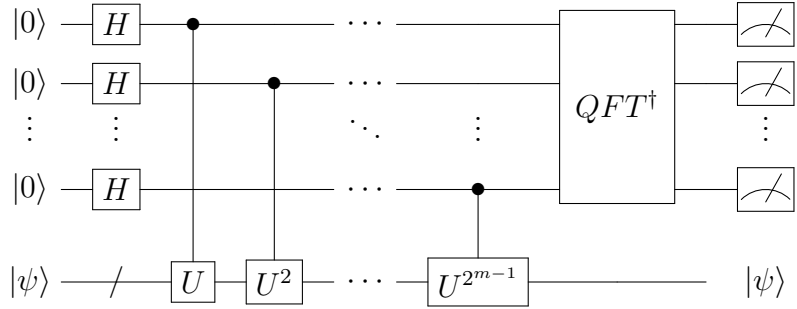
\begin{figure}\centering
\mbox{
\Qcircuit @C=1em @R=0.5em {
&\ket{0} & &\gate{H} & \ctrl{8} & \qw & \qw & \cdots & & \qw & \multigate{5}{QFT^{\dagger}} & \qw & \meter\\
&\ket{0} & & \gate{H} & \qw & \ctrl{7} & \qw & \cdots & & \qw & \ghost{QFT^{\dagger}} & \qw & \meter\\
& \vdots & & \vdots & & & & \ddots & &  \vdots & & & \vdots\\
& & & & & & & & & & \\
& & & & & & & & & & \\
&\ket{0} & & \gate{H} & \qw & \qw & \qw & \cdots & & \ctrl{3} & \ghost{QFT^{\dagger}} & \qw & \meter
 \\
& & & & & & & & & & & \\
& & & & & & & & & & & \\
&\ket{\psi} & & \qw{/} & \gate{U} & \gate{U^2} & \qw & \cdots & & \gate{U^{2^{m-1}}} & \qw & \qw & \ket{\psi}  \\
}}
\caption{\textbf{Quantum phase estimation algorithm}. The first part of the circuit implements a phase-kickback, whose phase is then estimated using an inverse quantum Fourier transform~\cite{nielsen2002quantum}. Note that obtaining $t$ bits of accuracy with a probability of failure smaller than $p_f$ requires $2^{m}-1$ calls to $U$ for $m = t+\left\lceil \log_2\left(\frac{1}{2}+\frac{1}{2p_f}\right) \right\rceil$~\cite{cleve1998quantum}, so the complexity of the algorithm grows as $O(\epsilon^{-1})= O(2^{t})$.}
\label{fig:qpe}
\end{figure}
Quantum phase estimation has two main steps. First, we implement controlled versions of the phase kickback using operator $U$, for each eigenstate $\ket{\psi}$ and corresponding eigenvalue $\varphi$, synthesizing state
\begin{equation}
\begin{split}
    \frac{1}{2^{t/2}}\left(\ket{0} + e^{2\pi i 2^{t-1}\varphi}\ket{1}\right)\otimes \left(\ket{0} + e^{2\pi i 2^{t-2}\varphi}\ket{1}\right)\otimes\ldots \otimes \left(\ket{0} + e^{2\pi i 2^{0}\varphi}\ket{1}\right)\otimes \ket{\psi} = \\
    \frac{1}{2^{t/2}}\left(\ket{0} + e^{2\pi i 0.\varphi_t}\ket{1}\right)\otimes \left(\ket{0} + e^{2\pi i 0.\varphi_{t-1}\varphi_t}\ket{1}\right)\otimes\ldots \otimes \left(\ket{0} + e^{2\pi i 0.\varphi_1\ldots\varphi_t}\ket{1}\right)\otimes \ket{\psi}.
\end{split}
\end{equation}
Then, we identify this state with the output of \eqref{eq:QFT}. As such, an inverse QFT can recover the binary expression of the phase implemented by $U$. The corresponding circuit can be seen in \cref{fig:qpe}. 
Interestingly, the inverse quantum Fourier transform is not necessary. We already know that a single-qubit Hadamard gate is the same as a single-qubit QFT. We can use this fact to iteratively refine an estimate of a single qubit quantum phase estimation~\cite{kitaev1995quantum, wiebe2016efficient}.

As a very simple use case, we can use phase estimation together with amplitude amplification to obtain the \textit{amplitude estimation} algorithm, by phase estimating the eigenvalues from \eqref{eq:diag_Amplitude_Amplification},~\cite{brassard2002quantum}.
Now let us turn to the \cref{prob:Shor}. The following \cref{alg:Shor} solves it \cite[Sec. 5.4.1]{nielsen2002quantum} using quantum phase estimation, and can be employed for Shor's original purpose of finding prime factors. 

\begin{algorithm}[h!]
\begin{algorithmic}[1]
\State \textbf{Input}: Oracle $f: \ket{j}\ket{0}\mapsto \ket{j}\ket{f(j)}$ fulfilling the promise of \cref{prob:Shor}.
\State \textbf{Output}: Value $r$ solving \cref{prob:Shor}.
\State Use Hadamard gates to create a uniform superposition 
\begin{equation}
    \frac{1}{\sqrt{2^n}}\sum_{j=0}^n\ket{j}\ket{0}.
\end{equation}
\State Query the oracle $f$, obtaining
\begin{equation}
    \frac{1}{\sqrt{2^n}}\sum_{j}\ket{j}\ket{f(j)} \approx \frac{1}{\sqrt{2^n}}\frac{1}{\sqrt{r}}\sum_{j=0}^n\sum_{\ell = 0}^r e^{2\pi i j\ell/r}\ket{j}\ket{\hat{f}(\ell)},
\end{equation}
where $r$ is the period, because one defines
\begin{equation}
    \ket{\hat{f}(\ell)} = \frac{1}{\sqrt{r}}\sum_{j=0}^{r-1}e^{-2\pi i \ell/r}\ket{f(j)} \Rightarrow \ket{f(j)} = \frac{1}{\sqrt{r}} \sum_{j=0}^{r-1}e^{2\pi i \ell/r}\ket{\hat{f}(\ell)}.
\end{equation}
The approximate equality is because $2^n$ is not necessarily a multiple of $r$.
\State Apply an inverse quantum Fourier transform, obtaining a state close to
\begin{equation}
    \frac{1}{\sqrt{r}}\sum_{\ell = 0}^{r-1}\ket{\widetilde{\ell/r}}\ket{\hat{f}(\ell)}.
\end{equation}
\State Measure the first register, obtaining with high probability an approximation to $\ell_1/r$ for a random $\ell_1\in\{0,\ldots,r-1\}$.
\State Repeat the procedure above, and obtain another approximation of the phase, for a randomly chosen $\ell_2\in\{0,\ldots,r-1\}$.
\State Obtain fractions $\ell'_1/r_1$ and  $\ell'_2/r_2$ that approximate the phases measured in the previous steps, using the continued fraction algorithm (see \cite[Box 5.3]{nielsen2002quantum} and \cite[Section 5.4]{de2019quantum}). 
\State If $\ell'_1$ and $\ell'_2$ contain no common factors (what happens with probability at least $1/4$ \cite[Eq. 5.58]{nielsen2002quantum}), then $r$ is the smallest common multiple of $r_1$ and $r_2$.
\end{algorithmic}
\caption{Period finding algorithm}\label{alg:Shor}
\end{algorithm}

\subsection{\label{ssec:HSP}The hidden subgroup problem}

This problem that we just explained is an instance of a more general one called the hidden subgroup problem (HSP). To explain it, we need some basic notions of group theory.
\begin{definition}[Group]
Let $G$ be a set of elements with an operation $\times: G\times G\mapsto G$ such that
\begin{itemize}
    \item $g_1\times(g_2\times g_3) = (g_1\times g_2)\times g_3$ for any $g_1,g_2,g_3\in G$.
    \item There is an identity term $1\in G$ such that $1\times g = g = g\times 1$ for any $g\in G$.
    \item For any $g\in G$ there exists $g^{-1}\in G$ such that $g\times g^{-1} = 1= g^{-1}\times g$.
\end{itemize}
\end{definition}

A group is called Abelian if $g\times h = h\times g$, and a set $H\subseteq  G$ is a subgroup if it is itself a group. A set of elements $T$, denoted as $\braket{T}$, is called generating set of $G$ if using elements in $T$ and $\times$ we can generate any $g\in G$. Finally, given a subgroup $H$, a (left) \textit{coset} of $H$ is a set $gH := \{gh|h\in H\}$ for $g\in G$. Cosets of $H$ are either identical or disjoint~\cite{de2019quantum}.
Using these definitions, we can pose the following general problem.

\begin{problem}[Hidden subgroup problem]\label{prob:HSP}
Let $G$ be a group and $f:G\mapsto S$  a function, where $S$ is a finite set. Let $f$ have the property that there exists a hidden subgroup $H$ such that $f(g) = f(g') \Leftrightarrow gH = g'H$. The goal is to find $H$ or its generators.
\end{problem}

Simon's and period finding problems described above are particular instances of the hidden subgroup problem. For Simon's problem, the hidden subgroup is $\{0,s\}$ for $s\in \{0,1\}^n$. In the period-finding problem, on the other hand, we have $H = \{0, r, 2r,\ldots, |\text{Im} f|-r\} = \braket{0,r}$, where the sum is performed module $|Im f|$, for the oracle function $f$. 

The hidden subgroup problem (HSP) is very common in cryptography. For example, the RSA cryptographic scheme can be attacked with Shor's algorithm, which finds the decomposition in primes of large composite numbers. More generally, there is a quantum algorithm that allows us to solve it whenever the group is finite and Abelian. To explain it, we first introduce some additional definitions.

\begin{definition}[Representation]
A representation $\rho$ of a group $G$ is a map $\rho:G\mapsto GL_\mathbb{C}(V)$ to the general linear group of complex invertible matrices (automorphisms), such that $\rho(g_1)\cdot\rho(g_2) = \rho(g_1\times g_2)$ for any $g_1$, $g_2$ in $G$. Further, a representation is called irreducible if it has no trivially invariant subspaces, where the dimension of the representation is the dimension of the vector space $V$. In finite groups, this means that such representation cannot be decomposed in the direct sum of lower-dimensional representations.
\end{definition}

Every column of the Fourier transform is a map $\chi: \mathbb{Z}_N\mapsto \mathbb{C}$ defined as $\chi_k(j) = e^{\frac{2\pi i}{N}jk}= \omega_N^{jk}$. Since $\chi_k(j+j')= \chi_k(j)\chi_k(j')$, $\chi_k$ is a 1-dimensional representation (also called \textit{character}) of $\mathbb{Z}_N$. The characters also form a $N$-dimensional basis
\begin{equation}
    \ket{\chi_k} = \frac{1}{\sqrt{N}}\sum_{j=0}^{N-1}\chi_k(j)\ket{j} = \frac{1}{\sqrt{N}}\sum_{j=0}^{N-1}\omega_N^{jk}\ket{j},
\end{equation}
and the Fourier transform maps between this and the computational basis: $F:\ket{k}\mapsto \ket{\chi_k}$.

An important feature of finite Abelian groups is that they are isomorphic to a direct product $\mathbb{Z}_{N_1}\times \ldots \times\mathbb{Z}_{N_l}$. In this product of groups, the Fourier transform is defined as the tensor product of Fourier transforms for each of the $\mathbb{Z}_N$.

Finally, to explain a general finite Abelian hidden subgroup problem we define the \textit{dual subgroup} $\hat{G}$ whose elements are the characters $\chi_k$, together with pointwise multiplication. Then, for any subgroup $H\subseteq G$, let
\begin{equation}
    H^{\perp} = \{\chi_k|\chi_k(h)= 1,\quad \forall h\in H\},
\end{equation}
of dimension $|G|/|H|$. These definitions are sufficient to propose a general finite Abelian hidden subgroup problem algorithm~\cite{kitaev1995quantum, mosca1998hidden}, which we describe in \cref{alg:Abelian_HSP}.
\begin{algorithm}[h!]
\begin{algorithmic}[1]
\State \textbf{Input}: Oracle $f: G\rightarrow S$ for some set $S$ fulfilling the promise of \cref{prob:HSP}, that is, $f$ is constant on cosets of hidden subgroup $H$.
\State \textbf{Output}: Generators of hidden subgroup $H$.
\For{$O(\log |G|)$ steps}
\State Initialize $\ket{0}^{\otimes \log |G|}\ket{0}^{\otimes \log |S|}$.
\State Create a superposition over $G$:
\begin{equation}
\frac{1}{\sqrt{|G|}}\sum_{g\in G}\ket{g}\ket{0}.
\end{equation}
\State Compute $f$:
\begin{equation}
\frac{1}{\sqrt{|G|}}\sum_{g\in G}\ket{g}\ket{f(g)}.
\end{equation}
\State Measure the second register obtaining $f(s)$ for some unknown $s\in G$:
\begin{equation}
\frac{1}{\sqrt{|H|}}\sum_{h\in H}\ket{s+h}.
\end{equation}
\State Apply the quantum Fourier transform of $G\approx\mathbb{Z}_{N_1}\times \ldots \times\mathbb{Z}_{N_l}$ to the state obtaining
\begin{equation}
\begin{split}
    \frac{1}{\sqrt{|H|}}\sum_{h\in H}\ket{\chi_{s+h}} &= \frac{1}{\sqrt{|G||H|}}\sum_{h\in H}\sum_{g\in G} \chi_{s+h}(g)\ket{g}\\
    &= \frac{1}{\sqrt{|G||H|}}\sum_{g\in G}\chi_{s}(g) \sum_{h\in H}\chi_{h}(g)\ket{g} = \sqrt{\frac{|H|}{|G|}}\sum_{g: \chi_g\in H^\perp}\chi_{s}(g)\ket{g},
\end{split}
\end{equation}
where the last equal holds because
\begin{equation}
    \sum_{h\in H}\chi_{h}(g) = \sum_{h\in H}\chi_{g}(h)=
    \begin{cases}
    |H| & \text{if }\chi_g\in H^\perp,\\
    0 & \text{if }\chi_g\notin H^\perp.
    \end{cases}
\end{equation}
\State Measure the first register. Since all $|\chi_g(s)|^2=1$, obtain one $\ket{g}:\chi_g\in H^\perp$ uniformly at random.
\EndFor
\State Solve the constrain satisfaction problem $\chi_g(h) = 1$ $\forall h\in H$ to find generators of $H$.
\end{algorithmic}
\caption{Finite Abelian hidden subgroup problem algorithm}\label{alg:Abelian_HSP}
\end{algorithm}

On the other hand, we can also define a non-Abelian quantum Fourier transform~\cite{de2019quantum}:
\begin{equation}
    F:\ket{g}\mapsto \sum_{\rho\in \hat{G}}\sqrt{\frac{\dim G}{|G|}}\ket{\rho}\sum_{i,j}\rho(g)_{i,j}\ket{i,j}.
\end{equation}
This quantum Fourier transform may still sometimes be computed efficiently, for example for the symmetric group. However, it does not necessarily lead to an efficient quantum algorithm because there is not a single measurement basis that provides enough information to reconstruct $H$~\cite{moore2008symmetric}. Furthermore, while it is possible to solve the non-Abelian problem with only logarithmically many \textit{positive operator-valued measurements} (POVMs)~\cite{de2019quantum}, constructing those measurements is not efficient in general~\cite{ettinger2004quantum}.

In specific cases, it is sometimes possible to solve the hidden subgroup problem even if the group is not Abelian. Some examples are the normal~\cite{hallgren2003hidden}, nil-2~\cite{ivanyos2008efficient}, and solvable~\cite{watrous2001quantum} groups. In contrast, other problems such as the graph isomorphism problem still lack an efficient solution, and the most efficient available algorithm is Babai's classical quasi-polynomial algorithm, which takes time $O(\exp(\log^{O(1)} |G|))$~\cite{babai2016graph}.

\section{\label{sec:Algebra}Linear Algebra} 

In the previous section, we introduced the quantum Fourier transform and phase estimation algorithms, and we have shown that there are important problems that can be solved exponentially faster if we use them. Perhaps the next question is whether there are problems with less structure where we can also find this kind of exponential advantage. As indicated in the previous chapter, exponential advantages are much more useful than quadratic ones because they ensure that we can overcome the hardware slowdown derived from the need to perform error correction~\cite{babbush2021focus}.   

\subsection{\label{ssec:HHL}Solving linear systems of equations}

One important problem that can be solved using quantum phase estimation is the following:
\begin{problem}[Quantum linear system of equations]\label{prob:HHL}
Let $\ket{b}\in\mathbb{C}^n$ be a quantum state, and $A\in\mathbb{C}^{n\times n}$. Prepare the quantum state $\ket{x} = A^{-1}\ket{b}$.
\end{problem}
Since solving linear systems of equations are such a common subroutine in many algorithms, certainly solving them quickly opens the door to large efficiency improvements in many algorithms. As a result, the so-called HHL algorithm~\cite{harrow2009quantum}, named after their discoverers, was met with a lot of excitement in the community. \cref{alg:HHL} assumes $A$ is a Hermitian matrix, but it is easy to convert an arbitrary problem $A x = b$ to Hermitian writing instead
\begin{equation}
    \begin{pmatrix}
    0 & A\\
    A^\dagger & 0
    \end{pmatrix}
    \begin{pmatrix}
    \bm{0}\\
    \bm{x}
    \end{pmatrix}
    =
    \begin{pmatrix}
    \bm{b}\\
    \bm{0}
    \end{pmatrix}.
\end{equation}
If the eigenvalues of $A$ are $\{\lambda_j\}_j$, the new eigenvalues would be $\{\pm \lambda_j\}_j$. This procedure also works for non-square matrices $A$.
\begin{algorithm}[h!]
\begin{algorithmic}[1]
\State \textbf{Input}: State $\ket{b}\in\mathbb{C}^n$, sparse-access oracle access to the entries of the (Hermitian) matrix $A\in\mathbb{C}^{n\times n}$.
\State \textbf{Output}: State $\ket{x}= A^{-1}\ket{b}\in\mathbb{C}^n$.
\State Initialize state $\ket{b}\ket{0}\ket{0}= \sum_{j} \beta_j \ket{u_j}\ket{0}\ket{0}$, the formal decomposition of $\ket{b}$ as a superposition of eigenvectors $\ket{u_j}$ of $A$.
\State Implement phase estimation on $U = e^{-iAt} = \sum_j e^{-i \lambda_j t}\ket{u_j}\bra{u_j}$, obtaining state $\sum_{j} \beta_j \ket{u_j}\ket{\Tilde{\lambda}_j}\ket{0}$, where $\ket{\Tilde{\lambda_j}}$ is an approximation to the binary expression of eigenvalue $\ket{\lambda_j}$.
\State Apply the rotation
\begin{equation}\label{eq:Rotation_HHL}
    \sum_{j} \beta_j \ket{u_j}\ket{\Tilde{\lambda}_j}\ket{0} \mapsto \sum_{j} \beta_j \ket{u_j}\ket{\Tilde{\lambda}_j}\left(\sqrt{1-\frac{C^2}{\Tilde{\lambda}_j}}\ket{0} + \frac{C}{\tilde{\lambda}_j} \ket{1} \right)
\end{equation}
for some $C \geq \lambda_j$. Thus $C=O(1/\kappa)$, for $\kappa$ the condition number.
\State Uncompute the phase estimation to erase the second register.
\State Amplitude-amplify the component of the state with the third register at $\ket{1}$, and obtain
\begin{equation}
    \ket{x}\propto \sum_{j} \frac{\beta_j}{\tilde{\lambda}_j} \ket{u_j} \approx A^{-1}\ket{b}.
\end{equation}
\end{algorithmic}
\caption{Quantum linear system of equations algorithm (HHL)~\cite{harrow2009quantum}.}\label{alg:HHL}
\end{algorithm}

The rotation \eqref{eq:Rotation_HHL} can be applied efficiently because we can perform a series of rotations controlled by the binary expression of $\ket{\tilde{\lambda_j}}$.
To analyze the complexity of \cref{alg:HHL}, note that the cost comes from two main sources: the phase estimation subroutine, and the required amplitude amplification. If we assume that $A$ has no more than $s$ non-zero entries per row and column, then the Hamiltonian simulation of $e^{-iA t}$, which will be discussed later on in more detail, might be implemented at cost $\tilde{O}(t s^2)$~\cite{berry2007efficient}. Since we want to estimate each eigenvalue to precision $\epsilon$ and each eigenvalue could be as small as $1/\kappa$, the overall cost of the quantum phase estimation step is $\tilde{O}(ts^2)=\tilde{O} (s^2 \kappa \epsilon^{-1})$. Remember that the condition number $\kappa$ is defined as
\begin{equation}
    \kappa = \left|\frac{\max_i \sigma_i}{\min_i \sigma_i}\right|,
\end{equation}
where $\sigma_i$ are the singular values which, for square matrices, match the absolute value of eigenvalues $|\lambda_i|$.

Second, for amplitude amplification, the largest complexity
happens when almost all eigenvalues are large, so that coefficients become $\frac{1}{\kappa \lambda_i}\approx \frac{1}{\kappa}$. Consequently, the overall algorithm has complexity $\tilde{O} (s^2 \kappa^2 \epsilon^{-1})$. As we can see, if the size of the matrix is $N$, the algorithm depends only polylogarithmically on that factor. 

We have seen how the basic quantum linear system algorithm can be implemented. In contrast, if one wishes to solve this problem classically, the fastest algorithm is the conjugate gradient method~\cite{shewchuk1994introduction}, which has complexity $O(N s \sqrt{\kappa} \log \epsilon^{-1})$ for positive definite matrices $A$ and $O(N s \kappa \log \epsilon^{-1})$ otherwise. The contrast between these complexities is an exponential speedup in $N$. However, notice also that the classical algorithm provides a classical readout of the entire solution, which is not possible from a single copy of the quantum state $\ket{x}$. Furthermore, the complexity of parameters other than $N$ is worse. Fortunately, there are situations where these limitations do not represent a problem and a useful application of the quantum linear system algorithm might be carried out, see for example Ref.~\cite{clader2013preconditioned}.

\subsection{\label{ssec:Extensions}Improving the QLSA performance}

\paragraph{\label{par:Condition_number}Condition number.}

Let us now see how to improve the complexity of those other parameters, starting with the dependence on $\kappa$. In the previous discussion, we highlighted how one $O(\kappa)$ contribution comes from small $\lambda_i$ eigenvalues, which forces us to use a high precision in the phase estimation protocol. The second $O(\kappa)$ contribution, in contrast, happens when eigenvalues are large, decreasing the acceptance probability. This suggests performing amplitude amplification more times over those states or branches of the algorithm that display less expensive amplitude amplification. While perhaps too technical to explain here, the \textit{variable time amplitude amplification} algorithm of Ambainis~\cite{ambainis2010variable} does precisely this. It uses a version of Amplitude Estimation~\cite{brassard2002quantum} and the Median Lemma \cite[Lemma 1]{nagaj2009fast} to estimate eigenvalues from less to more precision in consecutive steps. At any step, if the eigenvalue is sufficiently large, the success part of the step after the rotation is amplified. In such a way, the success part of larger eigenvalues is amplified for more steps, but the initial amplifications are less costly because phase estimation is implemented with smaller precision.

Note though, that reducing the complexity of the condition number $\kappa$ to linear might not be sufficient to justify using \cref{alg:HHL} because we need $\kappa$ to be polylogarithmic in $N$ if we want to maintain the exponential speedup. For example, most linear systems that appear when trying to solve the finite element methods often display $\kappa = O(\text{poly} N)$~\cite{brenner2008mathematical,bank1989conditioning}. For that reason, it is worth considering preconditioning as a way to reduce $\kappa$. Preconditioning are methods whose objective is to find a matrix $M$ such that the system $M A x = M b$ has a better condition number than the original system of equations, at a relatively low cost. The best possible preconditioner would be $M = A^{-1}$, but this is equivalent to solving the system. 

Ref.~\cite{clader2013preconditioned} suggests a method to perform preconditioning on the fly on the quantum computer. This requires two main features from matrices $A$ and $M$: a) that only local information of $A$ is known, and b) that the resulting matrix $MA$ is equally sparse as $A$. 
To meet these two criteria, they use a quantum version of Sparse Preconditioners with Approximate Inverses (SPAI)~\cite{grote1997parallel,chow2000priori}. These algorithms minimize \cite[Eq. 10]{clader2013preconditioned}
\begin{equation}
    \|MA-\bm{1}\|_F^2 = \sum_{k=0}^N \|(MA-\bm{1})e_k\|_F^2,
\end{equation}
where $e_k = (0,\ldots,0,1,0\ldots,0)^T$, e.g., the $k^{th}$ column of the identity matrix; and $\|A\|_F:= \sqrt{\sum_{i,j}A_{ij}^2}$ is the Frobenius norm. To carry out this task, we choose a sparsity pattern of $M$, often taken to be the same as that of $A$. This results in $N$ $s\times s$ small systems of equations, each of which can be solved in superposition in time $O(s^3)$. Similarly, one also has to multiply $M\ket{b}$, but in the end one can successfully reduce $\kappa$ with only a multiplicative overhead polynomial in $s = O(\text{poly}\log N)$. Ref.~\cite{clader2013preconditioned} also shows that this technique can be used to calculate the electromagnetic scattering cross-section of an arbitrary target using the finite element methods~\cite{jin2015finite} with exponential speedup over known classical methods.

\paragraph{\label{par:Sparsity}Sparsity.}

A second important generalization is to go beyond the sparse-access oracle system and allow for dense matrices. To do so, we will assume that using quantum random access memory (qRAM)~\cite{giovannetti2008quantum} we can implement the following unitaries with precision $\epsilon^{-1}$ in time $O(\text{poly}\log(N/\epsilon))$ for $A\in \mathbb{R}^{N\times N}$~\cite{kerenidis2016quantum}
\begin{equation}\label{eq:quantum_accessible_data_structure}
    U_\mathcal{M}:\ket{i}\ket{0}\rightarrow \frac{1}{||A_{i\cdot}||}\sum_{j}A_{ij}\ket{ij},\qquad
    U_\mathcal{N}:\ket{0}\ket{j}\rightarrow \frac{1}{||A||_F}\sum_i ||A_{i \cdot}|| \ket{ij},
\end{equation}
where $||A_{i \cdot}||$ stands for the $\ell_2$-norm of the $i^{th}$ row of $A$. This data access is sometimes called the \textit{quantum accessible data structure}. Note that
\begin{equation}
    \braket{i,0|U_\mathcal{M}^\dagger U_\mathcal{N}|j,0} = \frac{||A_{i \cdot}||}{||A||_F}\frac{A_{i,j}}{||A_{i \cdot}||} = \frac{A_{i,j}}{||A||_F}.
\end{equation}
Using these operators implemented via qRAMs, we can efficiently perform one step of the quantum walk~\cite[Theorem 5.1]{kerenidis2016quantum}
\begin{equation}
    W = (2U_\mathcal{M} U_\mathcal{M}^\dagger  - \bm{1}) (2U_\mathcal{N} U_\mathcal{N}^\dagger - \bm{1}) = R_\mathcal{M}R_\mathcal{N}.
\end{equation}
We analyze this quantum walk as we did for Grover: the rotation implemented by $W$ will be twice the angle $\theta_i/2$ between $U_\mathcal{M}\ket{u_i}$ and $U_\mathcal{N}\ket{v_i}$, for the singular value decomposition $A = \sum_i \sigma_i \ket{u_i}\bra{v_i}$. Indeed, $\theta_i$ depends on $\sigma_i$ as $\cos (\theta_i/2) = \frac{\sigma_i}{\|A\|_F}$ \cite[Lemma 5.3]{kerenidis2016quantum}. This matches the relation between the rotation angles $\varphi_i$ and the eigenvalues $\lambda_i$ that occurred in quantum walks, except for a factor of 2 in the definition of $\theta_i$, to respect the original paper convention. 
\begin{algorithm}[h!]
\begin{algorithmic}[1]
\State \textbf{Input}: Matrix $A$, operators $U_\mathcal{M}$ and $U_\mathcal{N}$, initial state $\ket{x}=\sum_i x_i \ket{v_i}$, precision $\epsilon$.
\State \textbf{Output}: State $\sum_i x_i \ket{v_i}\ket{\bar{\sigma}_i}$, for $\bar{\sigma_i}\approx \sigma_i$.
\State Initialize $\ket{x}=\sum_i x_i \ket{v_i}$.
\State Apply $U_\mathcal{N}$ to the state $\ket{x}$, obtaining $U_\mathcal{N}\ket{x} = \sum_i x_i U_\mathcal{N}\ket{v_i}$. $U_\mathcal{N}\ket{v_i}$ are eigenvectors of $W$ with eigenvalues $e^{i\theta_i}$, \cite[Lemma 5.3]{kerenidis2016quantum}.
\State Perform phase estimation of $W$ with precision $\epsilon$ on state $U_\mathcal{N}\ket{x}$, obtaining $\sum_i x_i (U_\mathcal{N}\ket{v_i})\ket{\bar{\theta}_i}$ for $|\bar{\theta}_i-\theta_i|\leq \epsilon$.
\State Compute $\bar{\sigma}_i = \cos{\bar{\theta}_i}$ in binary reversible arithmetic, so the state is $\sum_i x_i (U_\mathcal{N}\ket{v_i})\ket{\bar{\sigma}_i}$.
\State Uncompute $U_\mathcal{N}$ on the first register, outputting $\sum_i x_i \ket{v_i}\ket{\bar{\sigma}_i}$.
\end{algorithmic}
\caption{Quantum singular value estimation~\cite{kerenidis2016quantum}.}\label{alg:QSVE}
\end{algorithm}

Using \cref{alg:QSVE}, we can estimate the singular values of $A$, which are related to the eigenvalues through $|\lambda_i| = \sigma_i$. To ascertain the sign, we can implement the same algorithm with $A+\epsilon\bm{1}$ and compare the results. Finally, we can use this technique to substitute the quantum phase estimation algorithm in \cref{alg:HHL}, to solve the linear system of equations with a dense matrix A~\cite{wossnig2018quantum}. The complexity, instead of linear in the sparsity parameter $s$, will now depend on the Frobenius norm of the matrix $\|A\|_F$.

\paragraph{\label{par:Precision}Precision.}
Finally, the last parameter that can be improved is the precision complexity, which in the original quantum linear system of equations was $O(\epsilon^{-1})$. Notice that the complexity of $O(\epsilon^{-1})$ is due to the phase estimation, so to improve the complexity of the overall algorithm we need to bypass its use. For this, we will use the Linear Combination of Unitaries (LCU) approach that we have already mentioned previously, with operators $\Prep$ and $\Sel$ defined as in \eqref{eq:Prep&Sel_HT}. Imagine that we want to implement operator $U = \sum_i \alpha_i U_i$. Then, using
\begin{equation}\label{eq:Prep&Sel}
    \Prep \ket{\bm{0}} \mapsto \frac{1}{\sqrt{\sum_i \alpha_i}}\sum_i \sqrt{\alpha_i} \ket{i};\qquad 
    \Sel = \sum_i \ket{i}\bra{i}\otimes U_i
\end{equation}
we can define $W = \Prep^\dagger \cdot \Sel \cdot \Prep$, which maps
\begin{equation}
    W\ket{\bm{0}}\ket{\psi} = \frac{1}{\sum_i \alpha_i}\ket{\bm{0}}\sum_i \alpha_i U_i \ket{\psi} + \ket{(\bm{0}\psi)^{\perp}} 
\end{equation}
Then, one can use (variable time) amplitude amplification to amplify the part of the state that we are interested in.

Ref.~\cite{childs2017quantum} proposes two possible LCU compositions that can be implemented with gate complexity $O(\log \epsilon^{-1})$. The first is to approximate the application of matrix $A^{-1}$ based on its corresponding Fourier series, where unitaries $U_j = e^{-i A t_j}$. This amounts to implementing the Hamiltonian simulation of $A$, as done for phase estimation. Since both $A^{-1}$ and each of the $e^{-i A t_j}$ operators will be diagonal on the same basis of eigenvalues, one may propose a series decomposition for scalar eigenvalues, which generalizes to the operator. The authors of Ref.~\cite{childs2017quantum} showed that
\begin{equation}\label{eq:1/x_Childs}
    \frac{1}{x} = \frac{i}{\sqrt{2\pi}}\int_0^\infty dy \int_{-\infty}^\infty dz ze^{-z^2/2}e^{-ixyz}
\end{equation}
can be approximated by the series \cite[Lemma 11]{childs2017quantum}
\begin{equation}
    h(x) = \frac{i}{\sqrt{2\pi}} \sum_{j=0}^{J-1} \Delta_y\sum_{k=-K}^K \Delta_z z_k e^{-z_k^2/2}e^{-ix y_j z_k},
\end{equation}
for $z_k = k\Delta_z$, $y_j = j\Delta_y$, $J = \Theta(\frac{\kappa}{\epsilon}\log (\kappa/\epsilon))$, $K = \Theta(\kappa\log (\kappa/\epsilon))$, $\Delta_y = \Theta(\epsilon/\sqrt{\log (\kappa/\epsilon)})$, $\Delta_z = \Theta(1/\kappa\sqrt{\log (\kappa/\epsilon)})$; to precision $\epsilon^{-1}$ in the domain $D_\kappa = [-1,-1/\kappa]\cup[1/\kappa,1]$. The key to avoid the $O(\epsilon)$ complexity from $J$ is to use Hadamard gates to prepare a uniform superposition over $\ket{j}$. Then, the $\Sel$ operator will be defined as \cite[Eq. 52]{childs2017quantum}
\begin{equation}
    \Sel = i \sum_{j=0}^J \sum_{k=-K}^K \ket{j,k}\bra{j,k}\otimes \sgn(k) e^{-i A y_j z_k}, 
\end{equation}
which can be efficiently implemented based on the binary representation of $j$ and $k$, avoiding the complexity on $\kappa$.

The second option is to implement a Chebyshev series based on quantum walks. We already saw a hint of this in \eqref{eq:QFF_Chebyshev}. We first decompose $A^{-1}$ as a Chebyshev series $\sum_i \alpha_i \mathcal{T}_i(A)$. Then, we explain how to perform the $\mathcal{T}_i(A)$ Chebyshev polynomial operators of the first kind using Szegedy-like quantum walks. Defining $T$ as
\begin{equation}
    T: \ket{j}\ket{0}\mapsto \ket{j}\ket{\psi_j} =  \ket{j}\otimes \frac{1}{\sqrt{s}}\sum_{k\in [N]: A_{jk}\neq 0} \left(\sqrt{A^*_{jk}}\ket{k} + \sqrt{1-|A^*_{jk}|}\ket{k+N}\right)
\end{equation}
for $s$ the sparsity and $S$ a swap between the two registers, one can take the quantum walk to be $W = S(2T T^\dagger - 1)$. Similarly to Szegedy quantum walks, for each eigenvalue $\lambda$ of $H = A/s$, $W$ has block-diagonal form in the invariant subspace span$\{T\ket{\lambda},ST\ket{\lambda}\}$~\cite[Lemma 15]{childs2017quantum}
 \begin{equation}
    W = \bigoplus_\lambda \begin{pmatrix}
     \lambda & -\sqrt{1-\lambda^2}\\
     \sqrt{1-\lambda^2} & \lambda
     \end{pmatrix}_\lambda.
 \end{equation}
Then, one can prove
\begin{equation}\label{eq:W^n_quantum_walk}
    W^n = \bigoplus_\lambda
    \begin{pmatrix}
    \mathcal{T}_n(\lambda) & -\sqrt{1-\lambda^2}\mathcal{U}_{n-1}(\lambda)\\
    \sqrt{1-\lambda^2}\mathcal{U}_{n-1}(\lambda) & \mathcal{T}_n(\lambda)
    \end{pmatrix}_\lambda
\end{equation}
by induction, where $\mathcal{U}_n$ represent Chebyshev polynomials of the second kind. The case $n=1$ is clearly true because $\mathcal{T}_1(\lambda) = \lambda$ and $\mathcal{U}_0(\lambda)=0$. Then
\begin{equation}
\begin{split}
    W^nW &= \bigoplus_\lambda
\begin{pmatrix}
\lambda\mathcal{T}_n(\lambda) -(1-\lambda^2)\mathcal{U}_{n-1}(\lambda) & -\sqrt{1-\lambda^2}(\mathcal{T}_n(\lambda) +\lambda\mathcal{U}_{n-1}(\lambda))\\
\sqrt{1-\lambda^2}(\mathcal{T}_n(\lambda) +\lambda\mathcal{U}_{n-1}(\lambda)) & \lambda\mathcal{T}_n(\lambda) -(1-\lambda^2)\mathcal{U}_{n-1}(\lambda)
\end{pmatrix}_\lambda\\
&= \bigoplus_\lambda    
\begin{pmatrix}
    \mathcal{T}_{n+1}(\lambda) & -\sqrt{1-\lambda^2}\mathcal{U}_{n}(\lambda)\\
    \sqrt{1-\lambda^2}\mathcal{U}_{n}(\lambda) & \mathcal{T}_{n+1}(\lambda)
\end{pmatrix}_\lambda
\end{split}
\end{equation}
because $\mathcal{T}_{n+1}(\lambda) = \lambda\mathcal{T}_n(\lambda) -(1-\lambda^2)\mathcal{U}_{n-1}(\lambda)$ and $\mathcal{U}_{n}(\lambda) = \mathcal{T}_n(\lambda) +\lambda\mathcal{U}_{n-1}(\lambda)$~\cite{bateman1953higher}. Additionally, Lemma 14 in Ref.~\cite{childs2017quantum} shows that
\begin{equation}
 g(\lambda) = 4\sum_{j = 0}^{j_0}(-1)^{j}\left(\frac{\sum_{i=j+1}^{b}{2b\choose b+u}}{2^{2b}}\right)\mathcal{T}_{2j+1}(\lambda)   
\end{equation}
approximates $\lambda^{-1}$ in the domain $D_\kappa$. As a result, we can use the LCU decomposition $\Prep^\dagger \cdot \Sel\cdot \Prep$ to synthesize this function with cost polylogarithmic in precision $\epsilon^{-1}$. This Chebyshev approach is marginally more efficient than the Fourier approach, even if in both cases the complexity is polylogarithmic. Both techniques also allow for variable time-amplitude amplification to keep the complexity in the condition number linear. To do so, they only require a low precision application of quantum phase estimation. The discussion of variable time amplitude amplification applied to these two methods can be found in Section 5 in Ref.~\cite{childs2017quantum}.

\subsection{\label{ssec:Qubitization}Qubitization}

The obvious next step is to generalize from the preparation of $A^{-1}$ to an arbitrary function $f[H]= \sum_\lambda f(\lambda) \ket{\lambda}\bra{\lambda}$ for a Hamiltonian $H$. Since $f[H]$ does not need to be unitary, in practice this will mean finding a unitary operator $U$ such that
\begin{equation}
    U\ket{G}_a\ket{\psi}_s = \ket{G}_af[H]\ket{\psi}_s + \sqrt{1-\|f[H]\ket{\psi}\|^2}\ket{G_\psi^\perp}_{as},
\end{equation}
for some states $\ket{G}$ and $\ket{G_\psi^\perp}$ such that $\bra{G_\psi^\perp}_{as}(\ket{G}_a\otimes\bm{1}_s) = 0$. Usually $\ket{G}$ is taken to be just $\ket{G} = \ket{0}$. Notice how $U$ operates on the state register and an additional auxiliary one. This can be visually expressed as
\begin{equation}
    U = \begin{pmatrix}
        f[H] & \cdot\\
        \cdot & \cdot\\
    \end{pmatrix},
\end{equation}
which, for this reason, is called \textit{block encoding}. The main problem with $U$, an example of which is a linear combination of units, is that it has some failure probability, as we already saw, and therefore requires amplitude amplification. 

Instead, Ref.~\cite{low2019qubitization} proposes to find a different operator $W$, also encoding $H = \braket{G|W|G}$, but performing a $SU(2)$ rotation. In other words, we are looking for a quantum walk operator that can be expressed as
\begin{equation}\label{eq:qubitization_quantum_walk}
    W = \bigoplus_\lambda \begin{pmatrix}
        \lambda & -\sqrt{1-\lambda^2}\\
        \sqrt{1-\lambda^2} & \lambda
    \end{pmatrix}_\lambda = \bigoplus_\lambda e^{-Y_\lambda \theta_\lambda}
\end{equation}
for $\theta_\lambda = \cos^{-1}\lambda$.
To build this operator, they propose the construction
\begin{equation}
    W = ((2\ket{G}\bra{G}-\bm{1})_a\otimes \bm{1}_s) S U,
\end{equation}
for an operator $S$ to be determined. They also observe that if $U^2= \bm{1}$, as is the case for $U = \Prep^\dagger \cdot \Sel\cdot \Prep$, then one can take $S = \bm{1}$, and automatically fulfill the conditions required for $W$ to implement \eqref{eq:qubitization_quantum_walk} \cite[Lemma 8]{low2019qubitization}:
\begin{equation}
    \bra{G}_a SU \ket{G}_a = H; \qquad \bra{G}_a SUSU \ket{G}_a = \bm{1}.
\end{equation}
Else, one builds $U' = \ket{0}\bra{0}\otimes U + \ket{1}\bra{1}\otimes U^\dagger$ and $S' = (\ket{1}\bra{0}+\ket{0}\bra{1})\otimes \bm{1}_{as}$, and taking $\ket{G'}= \frac{1}{\sqrt{2}}(\ket{0}+\ket{1})\ket{G}$, one can verify
\begin{equation}
\begin{split}
    \bra{G'}_a S'U' \ket{G'}_a = \bra{G'}_a U' \ket{G'}_a = \frac{1}{2}(H+H^\dagger) = H;\\
    \bra{G'}_a S'U'S'U' \ket{G'}_a = \bra{G'}_a (U')^\dagger U' \ket{G'}_a =\bm{1}.
\end{split}
\end{equation}
The procedure for generating and implementing this operator $W$ from a block encoding $U$ is called \textit{qubitization}. Quantum walks, as we have seen, have the advantage of implementing Chebyshev polynomials given by \eqref{eq:W^n_quantum_walk}. To implement arbitrary functions $f[H]$ we need to add additional degrees of freedom to rotation $W$. This can be done with the 2-dimensional operator
\begin{equation}
    Z_\phi = ((1+e^{-i\phi})\ket{G}\bra{G}-\bm{1}) = \bigoplus_\lambda \begin{pmatrix}
        e^{-i\phi} & 0\\
        0 & 1
    \end{pmatrix}_\lambda,
\end{equation}
and conjugating $W$ by $Z_\phi$,
\begin{equation}
\begin{split}
    W_\phi &= Z_{\phi-\pi/2} W Z_{-\phi+\pi/2}  \\
    &= \bigoplus_\lambda \begin{pmatrix}
        \lambda & -ie^{-i\phi} \sqrt{1-|\lambda|^2}\\
        -ie^{+i\phi} \sqrt{1-|\lambda|^2}
    \end{pmatrix}_\lambda = \bigoplus_\lambda e^{-i\theta_\lambda (\cos(\phi) X_\lambda + \sin(\phi) Y_\lambda)}.
\end{split}
\end{equation}
The qubitization circuit will be composed of $N$ $W_\phi$ operators, for different $\phi$ angles $W_{\vec{\phi}} = W_{\phi_N}\ldots W_{\phi_1}$.
Using $W_{\vec{\phi}}$, we can implement any function $f[H] = A[H] + iC[H]$, where $A[H]$ and $C[H]$ are real polynomials of degree $ N$ (or $N/2$, depending on the specific technique used), of equal parity (respectively opposite parity). The determination of the corresponding angles is called \textit{quantum signal processing}  (see~\cite{low2017optimal} and Theorems 3 and 4 in~\cite{low2019qubitization}), and has been used specifically for the Hamiltonian simulation of $e^{-i H t}$, which in turn can be used for phase estimation in Chemistry applications that we will study in \cref{ch:Chemistry}. Crucially, the determination of these angles is computationally efficient and has been studied to have complexity scaling as $O(N^3 \polylog(N/\epsilon))$~\cite{chao2020finding,haah2019product}.

\section{\label{sec:Interior}Interior point methods}

The first problem we addressed in this thesis was the study of Linear Programming problems using the quantum linear algebra tools described above~\cite{casares2020IP}. Linear programming is a field where one has to optimize a linear function of multiple variables, subject to multiple constraints. For example, given $A\in \mathbb{R}^{m\times n}$, $\bm{c}\in \mathbb{R}^n$ and $\bm{b}\in \mathbb{R}^m$, the problem is to find $\bm{x}\in \mathbb{R}^n$ such that:
\begin{subequations}
    \begin{equation}
      \text{minimizes } \bm{c}^T \bm{x}  
    \end{equation}
    \begin{equation}
      \text{subject to } A \bm{x} \geq \bm{b}, \qquad \bm{x}\geq \bm{0}.  
    \end{equation}\label{LP}
\end{subequations}
This is called the primal problem (LP). The dual problem (DP) is closely related: finding $\bm{y}\in\mathbb{R}^m$ such that
\begin{subequations}
    \begin{equation}
        \text{maximizes } \bm{b}^T \bm{y}
    \end{equation}
    \begin{equation}
        \text{subject to } A^T \bm{y} \leq \bm{c}.
    \end{equation}\label{LD}
\end{subequations}
These two problems are connected by the primal-dual gap, which for linear programming problems is $0$~\cite{matousek2006understanding}:
\begin{equation}
    \bm{b}^T \bm{y} - \bm{c}^T \bm{x} = 0.
    \label{dual gap}
\end{equation}
\begin{figure}
    \centering
    \includegraphics[width = 0.5\textwidth]{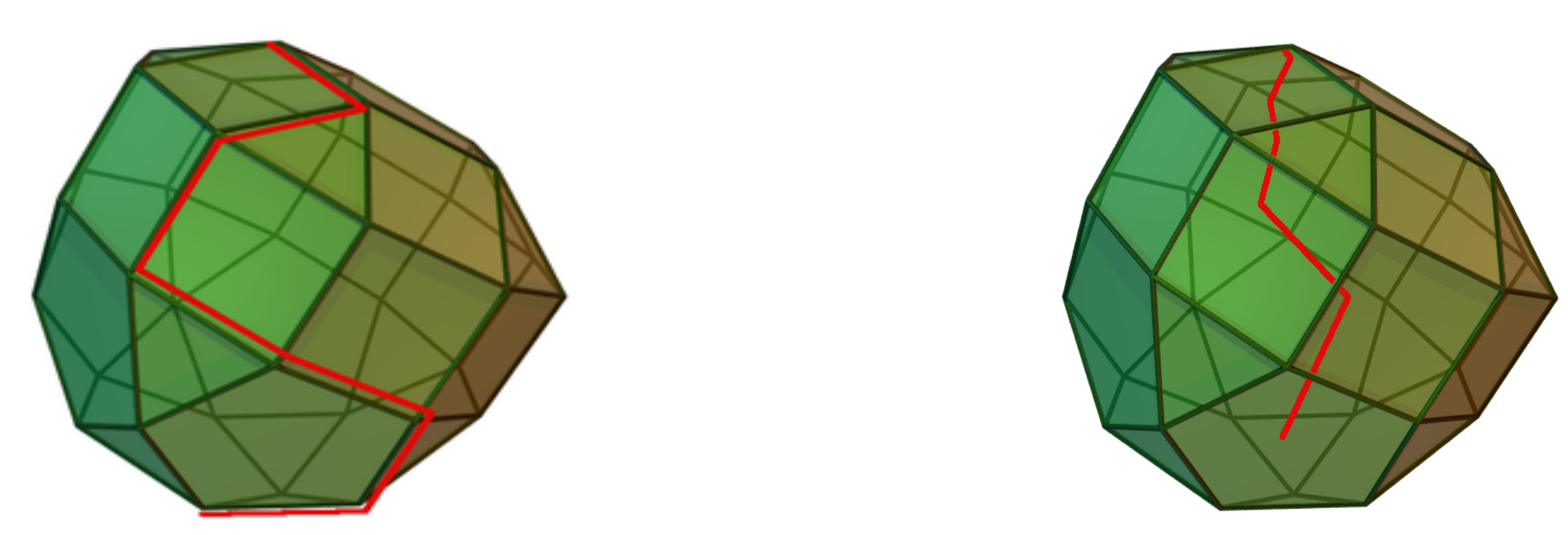}
    \caption{\textbf{Linear Programming problem, and different algorithms: simplex method (left) and interior point method (right).} The polyhedron represents the space of feasible solutions, e.g. those that fulfill the constraints. Such constraints are represented by planes, the faces of the polyhedra. The difference between simplex methods and interior point methods is how they transverse the space, from the border or the interior of the polyhedra. Images taken from Wikipedia for the \href{https://ca.m.wikipedia.org/wiki/Fitxer:Simplex-method-3-dimensions.png}{simplex method} and the \href{https://commons.wikimedia.org/wiki/File:Interior-point-method-three-dimensions.png}{interior point method}, under CC-BY-SA 3.0 license.}
    \label{fig:Interior_point}
\end{figure}

This kind of problem is abundant in operation research, yet simple enough to have been studied in detail. To solve these problems, two families of methods stand out in the literature. The first is the simplex method~\cite{dantzig1951application}, proposed by George Dantzig in 1947, which is simple and very fast in practice and consequently widely used~\cite{murty1983linear}. The second important family of algorithms is the family of interior point methods, which in contrast to the simplex method are (weakly) polynomial in the worst case. The first algorithm of this kind was proposed in 1979 by Khachiyan, and had complexity $O(n^6 L)$ for $L = \sum_{i,j} \lceil \log (A_{ij}+1)\rceil+1$~\cite{khachiyan1979polynomial}. The idea was subsequently improved by Karmarkar~\cite{karmarkar1984new} making it easily implementable. A review of how this family of techniques can be found in Ref.~\cite{potra2000interior}.

For our quantum algorithm, we take one important and efficient classical interior point method by Ye et al.,~\cite{ye1994nl}, and quantize some of the linear algebra that it uses. The method can be classified as a homogeneous self-dual algorithm because it solves both the dual and primal problems simultaneously by tackling a homogeneous problem whose solution can be mapped to those of the original primal and dual problems. Such a homogeneous problem (HLP) is
\begin{equation}
   \min \theta
\end{equation}  
such that ($\bm{x}\geq \bm{0},\tau\geq 0, \tau\in\mathbb{R}$):
\begin{align}
\begin{array}{r@{\,}lr@{\,}lr@{\,}lr@{\,}lcl}
           & &+A&\bm{x} &-\bm{b}&\tau &+\bar{\bm{b}}&\theta & =  & \bm{0}\\
       -A^T&\bm{y}&  &  &+\bm{c}&\tau &-\bar{\bm{c}}&\theta &\geq& \bm{0}\\
       +\bm{b}^T&\bm{y}&-\bm{c}^T&\bm{x}& & & +\bar{z}&\theta &\geq &0\\
        -\bar{\bm{b}}^T& \bm{y}&+\bar{\bm{c}}^T&\bm{x}&-\bar{z}&\tau& & & = & -(\bm{x}^0)^T \bm{s}^0-1
\label{eq:Homogenous_equations}
\end{array}
\end{align}
with 
\begin{equation}
\begin{split}
    &\bar{\bm{b}}:=\bm{b}-A\bm{x}^0, \qquad \bar{\bm{c}}:=\bm{c}-A^T\bm{y}^0-\bm{s}^0,\\ &\bar{z}:=\bm{c}^T\bm{x}^0+1-\bm{b}^T\bm{y}^0.
    \label{b and c bar}
\end{split}
\end{equation}
It is called homogeneous because it has a single non-zero constraint. Ye's algorithm~\cite{ye1994nl} will then follow the so-called central path through the inner space of the polyhedra corresponding to the self-dual (HLP) problem
\begin{equation}
    \mathcal{C}=\left\{(\bm{y},\bm{x},\tau,\theta, \bm{s},k)\in \mathcal{F}_h^0:
    \begin{pmatrix}
        X\bm{s}\\
        \tau k
        \end{pmatrix}
        =\frac{\bm{x}^T \bm{s}+\tau k}{n+1} \bm{1} \right\},
\end{equation}
until it reaches the solution.
The volume around the central path is called its neighborhood
\begin{equation}\
    \mathcal{N}(\beta)=\left\{(\bm{y},\bm{x},\tau,\theta, \bm{s},k)\in \mathcal{F}_h^0:\left|\left|\begin{pmatrix}
        X \bm{s}\\
        \tau k
        \end{pmatrix}-\mu \bm{1}\right|\right|
        \leq\beta \mu
        \text{ where }\mu=\frac{\bm{x}^T \bm{s}+\tau k}{n+1} \right\}.
        \label{eq:central path neighbourhood}
\end{equation}

Equations \eqref{eq:Homogenous_equations} and \eqref{eq:central path neighbourhood} will give rise to a dense linear system of equations that will ultimately need solving at each step of the algorithm,
\begin{equation}
\begin{blockarray}{ccccccc}
& m &  n &  1 &  1 & n & 1 \\
\begin{block}{c(cccccc)}
     m& 0 & A & -\bm{b} & \bar{\bm{b}} & 0  &  0  \\
     n & -A^T & 0 & \bm{c} & -\bar{\bm{c}} & -1 & 0 \\
     1 &\bm{b}^T & -\bm{c}^T & 0 &\bar{z} & 0 & -1 \\
     1& -\bar{\bm{b}}^T & \bar{\bm{c}}^T & -\bar{z} &0 & 0 & 0 \\
     n& 0 & S^t & 0 &  0 &  X^t & 0\\
     1& 0 &  0 & k^t & 0  &  0 & \tau^t\\
    \end{block}
    \end{blockarray}
    \qquad
    \begin{blockarray}{c}
    \\
    \begin{block}{(c)}
    \bm{d_y}\\
    \bm{d_x}\\
    d_\tau\\
    d_\theta\\
    \bm{d_s}\\
    d_k\\
    \end{block}
    \end{blockarray}\quad=\quad
    \begin{blockarray}{c}
\\
    \begin{block}{(c)}
    \bm{0}\\
    \bm{0}\\
    0\\
    0\\
    \gamma^t\mu^t 1_{n\times 1}-X^t \bm{s}^t\\
    \gamma^t\mu^t -\tau^tk^t\\
    \end{block}
    \end{blockarray}.
    \label{Matrix system}
\end{equation}
Since the system of equations does not need to be sparse, we rely on the dense linear system of equations algorithm of Ref.~\cite{wossnig2018quantum} via block encoding and assume access to the quantum-accessible data structure. 

Ye's algorithm~\cite{ye1994nl} is also a predictor-corrector method because it takes two kinds of steps: a first one to predict which direction should we modify the current state, and a corrector, which refines the solution. 
\begin{enumerate}
    \item \textit{Predictor step:} Solve \eqref{Matrix system} with $\gamma^t=0$ for $\bm{d}_{\bm{v}^t}$ where $\bm{v}^t=(\bm{y}^t,\bm{x}^t,\tau^t,\theta^t, \bm{s}^t,k^t)\in\mathcal{N}(1/4)$. Then find the biggest step length $\delta$ such that 
\begin{equation}
\bm{v}^{t+1}=\bm{v}^{t}+\delta \bm{d}_{\bm{v}^t}
\label{predictor sum}
\end{equation}
is in $\mathcal{N}(1/2)$, and update the values accordingly. Then $t \leftarrow t+1$.

\item \textit{Corrector step:} Solve \eqref{Matrix system} with $\gamma^t=1$ and set 
\begin{equation}
\bm{v}^{t+1}=\bm{v}^{t}+ \bm{d}_{\bm{v}^t}, \label{corrector sum}
\end{equation}
which will be back in $\mathcal{N}(1/4)$. Update $t \leftarrow t+1$.
\end{enumerate}
The algorithm will continue until some stopping conditions are met, see section IIC in Ref.~\cite{casares2020IP}. When that happens, the solution of the primal and dual problem are given $\bm{x}^*/\tau^*$ and $(\bm{y}^*/\tau^*,\bm{s}^*/\tau^*)$, respectively. Note that if the primal and dual solutions exist, then $\tau^*>0$.

Most of the complexity of the algorithm is due to the necessity to read out the resulting quantum state. Not only because tomography is expensive, but because it is necessary to ensure that the result will not fall outside the neighborhood $\mathcal{N}(\beta)$ for specific $\beta$ values. There are also limitations to our paper. First, in our article, we make the explicit assumption of a target error for each component of the solution, rather than the vector $\ell_2$-distance with the exact solution. While this is perhaps not usual, it also makes sense as otherwise $\epsilon$ will have a hidden dependence on the number of variables or constraints. Second, the system of equations will become increasingly bad conditioned, making $\kappa$ very large. This large condition number may also affect the norm of the solution of each system of equations, and therefore the complexity of the precision required. Since the quantum tomography algorithm has quadratic scaling with precision, this would make our algorithm unsuitable for our purposes.

Overall, the $\kappa$ dependence will likely make the performance of our algorithm worse than that of the equivalent classical algorithms. Finally, it is also worth highlighting that instead of using the exact-feasible approach of~\cite{ye1994nl}, and given the inexact nature of the solution of the system of equations recovered by tomography, a more appropriate approach might be to `quantize' inexact interior point algorithms, as proposed in later work~\cite{augustino2021infeasible,mohammadisiahroudi2022efficient}.

\section{\label{sec:SVM}Support Vector Machines}

A second linear algebra problem we tackled in this thesis is related to machine learning~\cite{casares2020active}. We use quantum linear algebra techniques within an active learning framework, to make machine learning systems robust to adversarial examples. Adversarial examples are better understood in the context of classification systems. They are slightly modified versions of correctly classified examples that are, however, miss-identified by the classifier~\cite{szegedy2013intriguing}. Often, these changes are imperceptible to humans, yet they completely change the output of the machine learning model.

One of the reasons that have been argued for their existence is that modern machine learning models operate in very high dimensional spaces, while the actual dimension of the problem is much lower~\cite{khoury2018geometry}. For example, an image classifier will encode each image in a giant $3\times n\times n$-dimensional tensor, where $n$ is the number of pixels in each direction, and 3 the colors needed to specify the red, green, and blue components of the pixel. Meanwhile, the objects these images depict live in a much lower effective dimension. The model has to partition the high-dimensional space with decision boundaries, but since the space volume is exponential in the dimension of the system, it becomes hard to make sure there is not a direction in which the decision changes abruptly. Finding these adversarial examples is often as simple as perturbing the system in the direction of the largest gradient of the decision function~\cite{goodfellow2014explaining}.

Adversarial examples are dangerous if machine learning models are intended to make decisions in real life. For this reason, there have been efforts in preventing them. In our article, we provide a way to use theorem 11 from Ref.~\cite{khoury2018geometry}, which ensures that training the model on a $\delta$-cover of the subclasses makes the model robust against these examples for sufficiently small $\delta$. Subclasses must be understood in this context as manifolds of a lower dimension embedded in the high dimensional space. How large $\delta$ is, will depend on the distance between classes in the high dimensional space. Setting $\delta$ very low makes the coverage more fine-grained and more robust, but also more expensive, so it is important to reliably find the distance between classes.

That being the objective, we use Support Vector Machines, which naturally define the margin or distance of separation between classes. In particular, a linear Support Vector Machine might be defined as the decision boundary provided by
\begin{equation}
    \vec{w}\cdot \vec{x} - b = 0,
\end{equation}
where $2/|\vec{w}|$ is chosen to be the margin of the SVM, for example, the separation between the two classes.
\begin{figure}
    \centering
    \includegraphics[width = 0.5\textwidth]{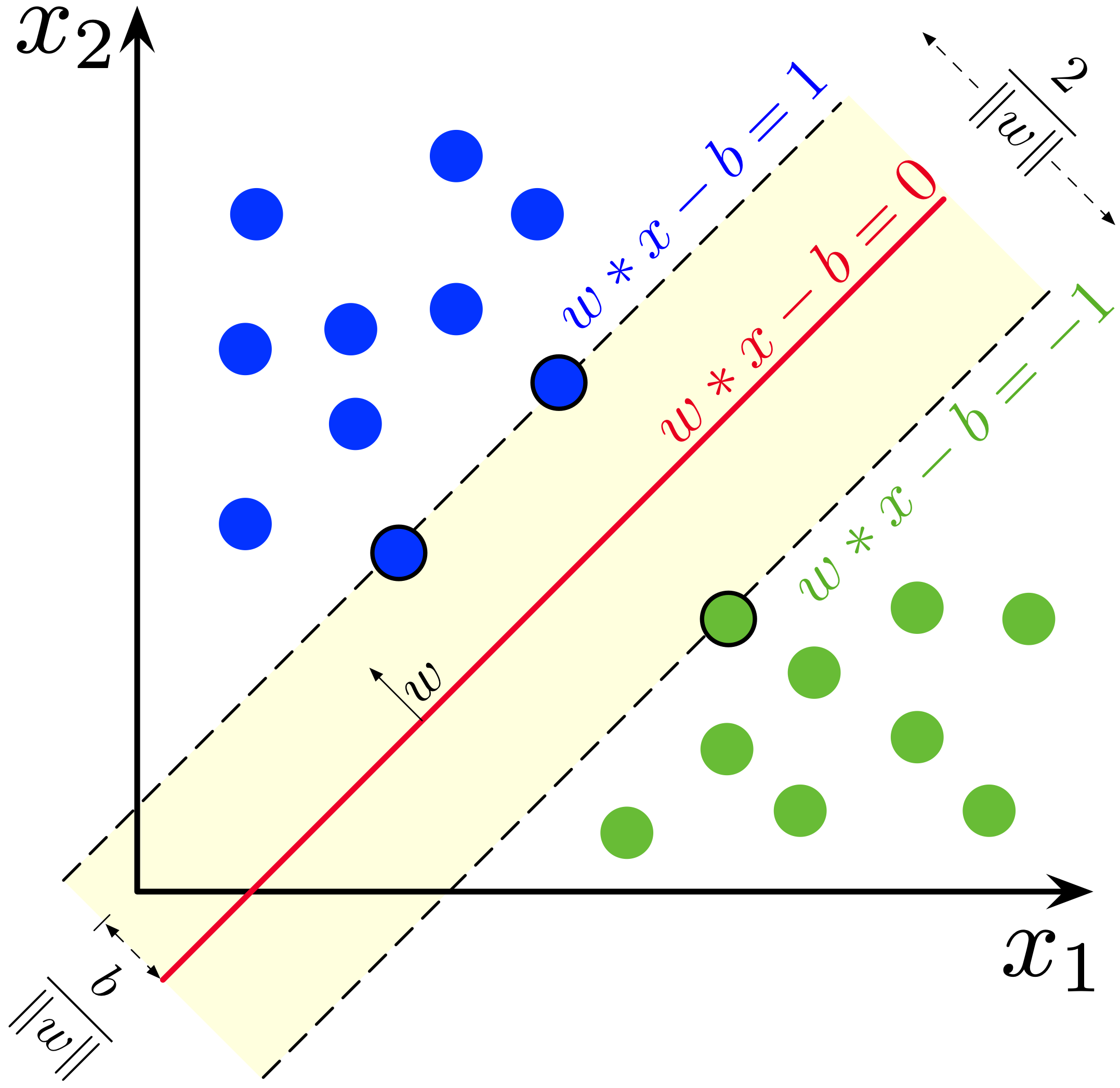}
    \caption{\textbf{Support Vector Machine.} The margin of the support vector machine is chosen so that one class lies in $\vec{w}\cdot \vec{x} -b \geq 1$ and the other $\vec{w}\cdot \vec{x} -b \leq -1$. This is equivalent to saying that the margin is $2/|\vec{w}|$. Image taken from \href{https://commons.wikimedia.org/wiki/File:SVM_margin.png}{Wikipedia} under license CC-BY-SA 4.0.}
    \label{fig:SVM}
\end{figure}
Since we want to make sure we have found the minimum distance between two classes, what we can do is find examples from both classes that minimize the margin of the trained Support Vector Machine, see~\cref{fig:SVM}. This way we will find an accurate approximation to the maximum allowed $\delta$. We conceptualize this objective in the `informativeness', defined as $P_{c}(\vec{x}_{n+1})\cdot |\vec{w}_{n+1}|$, where $\vec{x}_{n+1}$ is the candidate point to be added to the training set, $P_c(\vec{x}_{n+1})$ evaluates the probability that this new point is in class $c$, and $|\vec{w}_{n+1}|$ is proportional to the inverse size of the margin of the SVM if the point (or example) was added to such class. In conclusion, we want to find examples that maximize the `informativeness', so we can add them to the training set, and find an accurate approximation of the maximum allowed $\delta$.

The quantum algorithm we use in this case is a quantization of the Support Vector Machine proposed in Ref.~\cite{rebentrost2014quantum}. The basic idea is to solve the linear system of equations
\begin{equation}
    F
    \begin{pmatrix}
    b\\
    \vec{\alpha}
    \end{pmatrix}
=
\begin{pmatrix}
0 & \vec{1}^T\\
\vec{1} & K+\gamma^{-1}\mathbf{1} 
\end{pmatrix}
    \begin{pmatrix}
    b\\
    \vec{\alpha}
    \end{pmatrix}
    =    \begin{pmatrix}
    0\\
    \vec{y}
    \end{pmatrix}.
    \label{eq:SVM_equations}
\end{equation}
where $K_{ij} = \vec{x}_i\cdot \vec{x}_j$. Then, the solution vector $(b, \vec{\alpha})$ encoded as a quantum state can be used in a swap test (see \cref{fig:Swap_test}) to calculate the class to which a given point $\vec{x}_{n+1}$ could be added and with which probability.

Similarly, once we have prepared $\ket{b, \vec{\alpha}}$, we can compute $\vec{w}_{n+1}$ using the definition $\vec{w}= \sum_{i=1}^{n+1} \alpha_i \vec{x}_i$. To do so, we use linear algebra to multiply  
\begin{equation}
A\ket{b,\vec{\alpha}} := 
\begin{pmatrix}
0 & x_{1,1} & ... & x_{1,m}\\
\vdots &    &     & \vdots \\
0 & x_{n+1,1} & ... & x_{n+1,m}
\end{pmatrix}
\begin{pmatrix}
b\\
\alpha_1\\
\vdots \\
\alpha_{n+1}
\end{pmatrix} = 
\begin{pmatrix}
\sum_{i=1}^m  \alpha_j x_{1,j}\\
\vdots \\
\sum_{i=1}^m  \alpha_j x_{n+1,j}\\
\end{pmatrix}
= \vec{w}.
 \label{second matrix system}
\end{equation}
This requires a similar approach to the linear combination of units that we use to implement $A^{-1}$ as a Fourier or Chebyshev series, but for $A$ defined as above. Once we have done that, we can obtain the norm $|\vec{w}_{n+1}|$ as a function of the success probability in the application of $F^{-1}$ and $A$ to the solution. Overall, we can see that provided qRAM access to the data this procedure can be implemented in polylogarithmic complexity in the number of points and the dimension of each element of the dataset.

\section{\label{ssec:Dequantization}Dequantization}

In the previous two sections, we have exemplified how to apply quantum linear algebra techniques to solve optimization and machine learning problems. However, it is also the case that these algorithms rely on qRAM access to the data whose complexity is not counted in the final algorithm, only its execution time. Is there a way to compare them with classical algorithms in a fair setting? This is the question we aim to answer in this section.
The key to this comparison lies in the data access procedure. To compare them, Ewin Tang introduced the following concept that mimics the qRAM requirements in a classical setting~\cite{tang2019quantum}.
\begin{definition}[Sample and query access]
We say we have $O(T)$ sample and query access to a vector $x\in \mathbb{C}^N$, if we can in time $O(T)$, a) $\ell_2$-sample indices $i$ according to probabilities $\frac{|x_i|^2}{\|x\|^2}$; and b) query an index $i$ to obtain $x_i$, and query for $\|x\|$ too. For a matrix $A\in \mathbb{C}^{N\times M}$, we similarly say that we have sample and query access, if we have sample and query access to the vector of norms of rows of A, $(|A_{1,*}|,\ldots,|A_{N,*}|)$, and sample and query access to each row of $A$ taken as a vector. We will denote query access by $Q(x)$, sample and query access by $SQ(x)$, and if only an upper bound $(1+\nu)\|x\|$ to the norm is known, $SQ^\nu(x)$.
\end{definition}
This data access is similar to the one provided by qRAMs, since the qRAM allows one to query for any entry, and measuring the prepared state provides sample access. Similarly happens for the sparse-access oracle input model assumed by \cref{alg:HHL}.

Using this definition, we can implement three key subroutines in a classical computer that will help us mimic the quantum algorithms in a classical computer, as long as those algorithms operate in a low rank (low dimensional) subspace~\cite{tang2018pca}. The key is to substitute the state preparation assumptions from the qRAM, with a $\ell_2$-sampling assumption in classical computing. This is the reason it is called \textit{dequantization}. Specifically, we will say we can dequantize a quantum algorithm if the quantum algorithm runs in time $T$ and the classical algorithm in time $O(\poly (T))$.
The first dequantization subroutine is a dequantized swap test, which allows computing the inner product between two vectors.
\begin{algorithm}[h!]
\begin{algorithmic}[1]
\State \textbf{Input}: $SQ(x)\in \mathbb{C}^N$, $Q(y)\in \mathbb{C}^N$. Precision $\epsilon^{-1}$, failure probability $\delta$.
\State \textbf{Output}: An $\epsilon$-precise estimation of $\braket{x|y}$ with probability $1-\delta$.
\State Let $s = O(\frac{1}{\epsilon^2}\log \frac{1}{\delta})$
\State Measure $i_1,\ldots,i_s$ from $x$.
\State Compute $z_j = x_j^\dagger y_j \frac{\|x\|^2}{|x_j|^2}$ for $j\in\{i_1,\ldots,i_s\}$.
\State Distribute the measurements in $O(\log \frac{1}{\delta})$ groups and compute the average within each group.
\State Compute and output the median of the averages.
\end{algorithmic}
\caption{Inner product estimation}\label{alg:Dequantized_swap_test}
\end{algorithm}
This `dequantized' inner product algorithm achieves precision $\epsilon^{-1}$ with a probability of failure $\delta$.
\begin{lemma}[Inner product, proposition 4.2 in~\cite{tang2019quantum}].
Given $SQ^\nu(\vec{x})$ and $Q(\vec{y})$ we can estimate $\braket{\vec{x}|\vec{y}}$ to precision $\epsilon$ and probability $\geq 1-\delta$ in time $O\left(T\epsilon^{-2}\log \delta^{-1}\right)$.
\end{lemma}
The quantum advantage in this subroutine stems from the use of amplitude estimation, a quadratic advantage.

The second subroutine allows obtaining sample and query access from the result of a shallow matrix-vector product. Shallow, because the length of the vector $k$ will be relatively small compared to the number $N$ of rows in the matrix. The technique used is based on rejection sampling: if we can sample from one distribution $Q(s)$, and want to sample from $P(s)$ while being able to compute probabilities from both, we can:
\begin{enumerate}
    \item Sample $i$ from $Q$.
    \item With probability $\frac{P(i)}{MQ(i)}$ output $i$, else restart, where $M = \max_i \frac{P(i)}{Q(i)}$.
\end{enumerate}
In our case, the target distribution $P$ will be $Ww$ from which we can compute the entries in complexity $O(k)$, and the input probability $Q$, sampling from $\|W_{*,j}\|$ with probability $\|W_{*,i}\|^2|w_i|^2$.
\begin{algorithm}[h!]
\begin{algorithmic}[1]
\State \textbf{Input}:  $SQ(V^\dagger)\in \mathbb{C}^{k\times N}$, $Q(w)\in\mathbb{C}^k$.
\State \textbf{Output}:  $SQ^{\nu}(Vw)$.
\State \texttt{Query}: On input $s$, compute and output $(Vw)_s$.
\State \texttt{Sample}: Run \texttt{RejectionSampling} until success, or $kC(V,w)\log\frac{1}{\delta}$ failures.
\State \texttt{Norm}: Run \texttt{RejectionSampling} $\frac{k}{\nu^2}C(V,w)\log\frac{1}{\delta}$ times, and let $p$ be the success rate. Output $pk \sum_i |w_i|^2 \|V_{*,i}\|^2$.
\Function{RejectionSampling}{$SQ(V^\dagger)$, $Q(w)$}
\State Sample $i\in[k]$ proportional to $\|V_{*,i}\|^2|w_i|^2$.
\State Sample $s\in [N]$ from $\|V_{*,i}\|$.
\State Compute $p_s = \frac{(Vw)_s^2 }{k\sum_j (V_{s,j}w_j)^2}=\frac{(\sum_j V_{s,j}w_j)^2}{k\sum_j (V_{s,j}w_j)^2}$.
\State With probability $p_s$, output $s$; else failure $\emptyset$.
\EndFunction
\end{algorithmic}
\caption{Shallow matrix-vector product sample and query access}\label{alg:Dequantized_matrix_vector}
\end{algorithm}

Using algorithm~\cref{alg:Dequantized_matrix_vector}, one can prove the following result.
\begin{lemma}
[Thin matrix-vector product, proposition 4.3 in~\cite{tang2019quantum}].
Given a matrix $V\in \mathbb{C}^{n\times k}$, $w\in \mathbb{C}^k$, and given $SQ(V^\dagger)$ and $Q(w)$, we can obtain $SQ^\nu(Vw)$ with success probability $\geq 1-\delta$ and complexities
\begin{enumerate}
    \item query in time $O(Tk)$,
    \item sample in time $O(Tk^2 C(V,w) \log \delta^{-1})$,
    \item query the norm in time $O(Tk^2 C(V,w) \nu^{-2}\log \delta^{-1})$.
\end{enumerate}
where $C(V,w) = \sum_i \|w_i V_{*,i}\|^2/|Ww|^2$, and $V_{*,i}$ is the $i^{th}$ column of $V$.
\end{lemma}
Notice how the product of any row of $V$ and $w$ is computed directly, given that $k\ll N$. Quantum algorithms will obtain a quantum advantage in this case when $k$ is large, but this technique is often used in combination with a low-rank approximation of a matrix $S$ resulting from the third subroutine, which we explain next.
\begin{algorithm}[h!]
\begin{algorithmic}[1]
\State \textbf{Input}:  $SQ(A)\in \mathbb{C}^{M\times N}$, singular value cutoff $\sigma$, error $\epsilon$, failure probability $\delta$.
\State \textbf{Output}:  $SQ(S)\in \mathbb{C}^{N\times q}$, $Q(U)\in \mathbb{C}^{q\times \ell}$, $Q(\Sigma)\in\mathbb{C}^{\ell \times \ell}$.
\State $K = \frac{\|A\|_F^2}{\sigma^2}$ and $q = \Theta(\frac{K^2}{\epsilon^2}\log(\frac{1}{\delta})$.
\State $\ell_2$-sample $q$ row indices, from distribution $\frac{\|A_{i,*}\|_F^2}{\|A\|_F^2}$, rescaling $S_{i_r,*} = \frac{\|A\|_F }{\sqrt{q}\|A_{i_r,*}\|}A_{i_r,*}$.
\State Sample $q$ column indices from distribution $\mathcal{F}$, obtained from first uniformly sampling $r\sim [q]$, and then $\ell_2$-sampling $c$ from $S_{r,*}$.
\State Let $W$ be the resulting matrix from rescaling $W_{*,c} = \frac{\|A\|_F}{\sqrt{q}\|S_{*,q_c}\|}S_{*,q_c}$.
\State Compute singular vectors $u^{(1)},\ldots, u^{(\ell)}$ of $W$ corresponding to singular values $\sigma^{(1)},\ldots,\sigma^{(\ell)}$ larger than $\sigma$.
\State Output $SQ(S)$, the matrix $U\in\mathbb{R}^{q\times\ell}$ composed of $u^{(i)}$ taken as columns, and $\Sigma$ the diagonal matrix with entries $\sigma^{(i)}$.
\end{algorithmic}
\caption{Low-rank singular vector matrix decomposition}\label{alg:Dequantized_svd}
\end{algorithm}

This algorithm allows us to perform a low-rank approximation of $A$ in the subspace spanned by the sampled left singular vectors $D = A V V^\dagger$ for $V = S^\dagger U \Sigma^{-1}$.
\begin{theorem}[Low-rank singular vector approximation, theorem 4.4 in~\cite{tang2019quantum}]\label{theorem:low_rank_decomposition}
Suppose $O(T)$-time $SQ(A)$, $A\in \mathbb{C}^{n\times d}$; a singular value threshold $\sigma$ and an error parameter $\epsilon \in (0, \sqrt{\sigma/||A||_F}/4]$. Let $K = ||A||_F^2/\sigma^2$. Then, in time
\begin{equation}
    O\left(\frac{K^{12}}{\epsilon^6}\log^3\delta^{-1} + T \frac{K^8}{\epsilon^4}\log^2\delta\right)
\end{equation}
we output $SQ(S)$, $S\in \mathbb{C}^{q\times n}$, $U\in \mathbb{C}^{q\times l}$, $\Sigma\in \mathbb{R}^{l\times l}$, with $l = \Theta(K^4\epsilon^{-2}\log^2 \delta^{-1})$. These matrices implicitly describe the low-rank approximation of $A$, $D = A V V^\dagger$, with $V = S^\dagger U \Sigma^{-1}$. Additionally, with probability $\geq 1-\delta$, $||A-D||^2_F\leq||A-A_l||^2_F+\epsilon||A||_F^2$.
\end{theorem}
These three subroutines are very powerful and allow to dequantize several linear algebra problems where the quantum algorithm operates in low dimensional subspaces. 

\subsection{\label{ssec:Examples_dequantization}Examples}

\paragraph{Recommendation systems.} The first example of a dequantization algorithm was applied to the quantum recommendation system algorithm by Kerenidis and Prakash~\cite{kerenidis2016quantum}. In the recommendation problem, we are given a large matrix of $A\in \mathbb{R}^{M\times N}$ (users and products) and one has to decide what product has a greater fit with a given user. Mathematically, the objective is to project row $A_i$ corresponding to the $i^{th}$ user, into the space spanned by the singular vectors with the largest singular values, representing the products he or she is more likely to enjoy. In other words, sample from $A_i V V^{\dagger} = D_i$.
Kerenidis and Prakash used a simple strategy, shown in \cref{alg:Quantum_recommendations_algorithm}.
\begin{algorithm}[h!]
\begin{algorithmic}[1]
\State \textbf{Input}: Quantum accessible data structure of preferences matrix $A$ (see \eqref{eq:quantum_accessible_data_structure}), threshold $\sigma$, error $\epsilon$, user $i$.
\State \textbf{Output}: Sample access to any row of the low rank projection of $A$, $\hat{A}$.
\State Use the quantum accessible data structure to initialize state $\ket{A_i}$ corresponding to user $i$, formally represented as a sum over singular vectors $\ket{A_i} = \sum_j \alpha_j \ket{v_j}$.
\State Perform singular value estimation, $\sum_j \alpha_j \ket{v_j}\ket{\bar{\sigma}_j}$, using \cref{alg:QSVE}.
\State Flag those singular values that are above threshold $\sigma$, $\sum_j \alpha_j \ket{v_j}\ket{\bar{\sigma}_j}\ket{\bar{\sigma}_j\geq \sigma}$,
\State Uncompute the singular values, $\sum_j \alpha_j \ket{v_j}\ket{0}\ket{\bar{\sigma}_j\geq \sigma}$.
\State Measure the last register and postselect on $\ket{\bar{\sigma}_j\geq \sigma}$.
\State Measure and output the first register on the computational basis: the result is a sample from the projection of $\ket{A_i}$, into the space spanned by the singular vectors with the largest singular values.
\end{algorithmic}
\caption{Quantum recommendation system algorithm}\label{alg:Quantum_recommendations_algorithm}
\end{algorithm}

How can we dequantize this algorithm? Recall that we said that $D = A V V^\dagger$ represents a low-rank projection of the $A$ matrix. In Ref.~\cite{tang2019quantum}, showed how to obtain sample and query access to the singular value decomposition of $V$, proving \cref{theorem:low_rank_decomposition}. In this case, we are interested in the projection of a single row of $A$, so we want to sample from $A_i (S^\dagger U \Sigma) (S^\dagger U \Sigma)^\dagger = (A_i S^\dagger) (U \Sigma \Sigma^\dagger U^\dagger) S$. The first parentheses can be understood as a `small' number of inner products (\cref{alg:Dequantized_swap_test}), where small indicates the target rank of the decomposition. The resulting vector can then be multiplied by the second parentheses, which is a product of small matrices. Finally, we can also perform a shallow matrix-vector product with $S$ (\cref{alg:Dequantized_matrix_vector}) to obtain sample and query access to the resulting distribution.

\paragraph{Linear system of equations.} Another subroutine we can `dequantize' is the HHL algorithm for low-rank matrices. This is a very important condition, as otherwise \cref{alg:Dequantized_svd} fails. The quantum linear system \cref{alg:HHL} aims to prepare $\ket{x} = A^{-1} \ket{b}$. The dequantized algorithm will in contrast provide sample and query access to $A^+ b$, where $A^+$ is the Moore-Penrose pseudo inverse that fulfills that if $A$ is invertible then $A^{-1}= A^+$~\cite{moore1920reciprocal}. The objective is to sample from 
\begin{equation}
    A^+ b = (A^T A)^+ A^T b \approx \sum_{i=1}^k\frac{1}{\Sigma_{ii}}v_i v_i^T A^T b.
\end{equation}
For this decomposition, we will need the Singular Value Decomposition of \cref{alg:Dequantized_svd}. Then, notice that $v_i^T A^T b = \Tr(A^T v_i^T b)$ can be computed via a swap test between two-order tensors $A^T$ and $b v_i^T$ (see for example Ref.~\cite{tang2018pca}). This provides query access to $Q(v_i^T A b/\Sigma_{ii})$. Then, since we have $SQ(V)$, we also have sample and query access to an approximation of the objective, $SQ(A^+ b)$, via the shallow matrix-vector product in \cref{alg:Dequantized_matrix_vector}.

\paragraph{Support Vector Machine and adversarial examples.} Finally, we can also dequantize our active learning algorithm to sample against adversarial examples. The idea is simple: first, we use the method just explained above to gain sample and query access to $SQ(F^+ \vec{y})$, see \eqref{eq:SVM_equations}. Once we have that, we aim to find the norm 
\begin{equation}
    ||A\ket{b,\vec{\alpha}}||^2 = \braket{b,\vec{\alpha}|A^\dagger A|b,\vec{\alpha}},
\end{equation}
which can be computed as an inner product $\braket{a|b}$ with
\begin{equation}
    a = \sum_{i}\sum_j \sum_k A_{ji}||A_{k,*}||\ket{i}\ket{j}\ket{k},
\qquad 
 b = \sum_{i}\sum_j \sum_k \frac{w_j w_k  A_{ki}}{||A_{k,*}||}\ket{i}\ket{j}\ket{k},
\end{equation}
similarly to the protocol used for Supervised Clustering dequantized in Ref.~\cite{tang2018pca}. Key to the success of this dequantization technique is the fact that the examples from both classes live in a low-dimensional subspace, and therefore we can perform low-rank operations, which we leverage in the quantum algorithm too. The resulting quantum advantage of using quantum linear algebra instead of its dequantized counterpart will consequently be polynomial.

\section{Results}
\begin{itemize}
    \item We have reviewed Shor's algorithm and its generalization to the Hidden Subgroup Problem over arbitrary Abelian groups.
    \item We have explained the basic quantum linear algorithm solver and described the complexity improvements carried out over time to its dependence over most parameters, culminating in qubitization as a flexible technique. Qubitization is intimately related to quantum signal processing, which allows to synthesize arbitrary polynomial functions of the encoded Hamiltonian.
    \item In Ref.~\cite{casares2020IP} we have described a quantum interior point method that improved over previously available results, establishing at the time the state-of-the-art in quantum algorithms for linear programming.
    \item An important technical achievement in that paper was to ensure that the polynomial complexity in the precision would not hamper the convergence of the classical algorithm in which it is inspired.
    \item However, due to the dependence on the preparation and readout of the data often relying on qRAMs, the quantum advantage is limited. Another generic limitation to this class of quantum and classical algorithms is the polynomial dependence on the condition number, which in turn becomes increasingly larger as we proceed along the algorithm, thus why interior point methods are called weakly polynomial.
    \item In Ref.~\cite{casares2020active} we also explored how to apply quantum linear algebra to Machine Learning problems. In particular, we have tackled the problem of adversarial examples using quantum Support Vector Machines. Finally, we have explained that quantum active learning techniques might help provide guarantees on the robustness of a classical training set.
    \item The usefulness of the technique in this area relies heavily on the problem having a low-rank nature, which can also be exploited by classical algorithms. Consequently, we have also indicated how our algorithm may be dequantized, and thus implemented in a classical computer, with a polynomial slowdown.
    \item These points illustrate that the problems we aim to target with quantum linear algebra techniques should be chosen carefully, as generic speedups are often polynomial and not exponential. Consequently, their usefulness will depend more strongly on the error correction overhead.
\end{itemize}
\chapter{\label{ch:Chemistry}Quantum Chemistry}

\ifpdf
    \graphicspath{{Chapter3/Figs/Raster/}{Chapter3/Figs/PDF/}{Chapter3/Figs/}}
\else
    \graphicspath{{Chapter3/Figs/Vector/}{Chapter3/Figs/}}
\fi

\epigraph{The program that Fredkin is always pushing, about trying to find a computer simulation of physics, seems to me to be an excellent program to follow out...  And I'm not happy with all the analyses that go with just the classical theory, because nature isn't classical, dammit, and if you want to make a simulation of nature, you'd
better make it quantum mechanical, and by golly it's a wonderful problem, because it doesn't look so easy.}{Richard P. Feynman, {\it Simulating Physics with Computers}}

\section{Objectives}

\begin{itemize}
    \item Understand the most common classical ab-initio chemistry techniques.
    \item Building on the quantum walk and linear algebra techniques explained in previous chapters describe the available Hamiltonian simulation techniques, as well as the tradeoffs present in this and other algorithmic choices.
    \item Benchmark the different Hamiltonian simulation techniques.
    \item Describe a problem where quantum phase estimation might be applied, and provide rigorous cost estimates.
\end{itemize}

\section{Introduction}

As indicated in the quote, one of the most promising problems to which one can apply quantum computing is the study of quantum systems, interesting for basic science purposes but also for the chemical and material science industries. The reason quantum computing is so well suited for chemical problems is that quantum states can be represented exactly on a quantum computer, and the evolution of (closed) systems is unitary and relatively efficient to implement.

In this chapter, we will explore algorithms tailored for this research area. Quantum phase estimation will in particular stand a key role in these algorithms and will make use of Hamiltonian simulation as a key subroutine. While quantum phase estimation allows to efficiently compute the energy corresponding to an eigenstate, preparing the corresponding ground state is often hard. Consequently, this chapter is divided into three main sections: the first section reviews basic classical algorithms for quantum algorithms; the second explains Hamiltonian simulation and its application to quantum phase estimation; and finally, the third introduces different quantum state preparation techniques.

\section{\label{sec:Classical_Chemistry}Classical quantum chemistry}

\subsection{\label{sec:HF}Hartree-Fock}

The Hartree-Fock procedure is considered a basic step in the calculation of an approximation to the ground state, and a stepping stone for more complex procedures afterward. The quantum chemistry problems we are interested in assume access to the molecular Hamiltonian. Let $\eta$ and $L$ represent the number of electrons and nuclei, respectively, indexed by $i$ and $\ell$, and at positions $r_i$ and $R_\ell$. Let also $Z_\ell$ and $m_\ell$ be the charge and mass of nuclei $\ell$. Knowing that each particle has kinetic energy and that there is a Coulomb potential between charged particles, we can write the molecular Hamiltonian as
\begin{equation} \label{eq:molecular_hamiltonian}
    H = \underbrace{-\sum_{i=1}^\eta \frac{\nabla_i^2}{2m_e}}_{T}  \underbrace{-\sum_{i=1}^\eta \sum_{\ell = 1}^L \frac{Z_\ell}{\|R_\ell-r_i\|}}_{U}+\underbrace{\sum_{\substack{i\neq j\\ i,j=1}}^\eta \frac{1}{2}\frac{1}{\|r_i-r_j\|}}_{V}  \underbrace{-\sum_{\ell=1}^L\frac{\nabla_\ell^2}{2m_\ell}}_{T_{\text{nuclei}}}+\underbrace{\sum_{\substack{\ell\neq \kappa\\ \ell,\kappa=1}}^L \frac{1}{2}\frac{Z_\ell Z_\kappa}{\|R_\ell-R_\kappa\|}}_{V_{\text{nuclei}}}.
\end{equation}
We want to solve the Schrödinger equation, $H\ket{\psi} = E\ket{\psi}$, and find both $\ket{\psi}$ and $E$. However, this problem is very complicated, so we will make a series of approximations. The first such approximation is the so-called Born-Oppenheimer approximation: since the mass of the nuclei is much larger than the mass of the electrons $m_\ell\gg m_e$, we can decouple the degrees of freedom of the nuclei from the rest of the wave function, and analyze them classically~\cite{born1927quantentheorie}. With this approximation, the molecular Hamiltonian \eqref{eq:molecular_hamiltonian} is reduced to the electronic Hamiltonian:
\begin{equation}\label{eq:Electronic_hamiltonian}
    H = T + U + V,
\end{equation}
with $T$, $V$, and $U$ defined above as the kinetic, potential, and external potential operators. We also define the one-body term $h = T+U$, leaving $V$ as the two-body component.
Additionally, the Hartree-Fock procedure approximates the wave function of the electrons as a product of disentangled wave functions for each electron, with the characteristic antisymmetrization of fermionic wavefunctions
\begin{equation} \label{eq:Slater_determinant}
    \ket{\psi} = \frac{1}{\sqrt{\eta!}}\sum_{\sigma\in S_\eta}\sgn(\sigma) \ket{\phi_{\sigma(1)}}\otimes\ldots\otimes \ket{\phi_{\sigma(\eta)}} = \frac{1}{\sqrt{\eta!}}
    \begin{vmatrix}
    \phi_1(\bm{x}_1) & \phi_1(\bm{x}_2) & \ldots & \phi_1(\bm{x}_\eta)\\
    \phi_2(\bm{x}_1) & \phi_2(\bm{x}_2) & \ldots & \phi_2(\bm{x}_\eta)\\
    \vdots & \vdots & \ddots & \vdots\\
    \phi_\eta(\bm{x}_1) & \phi_\eta(\bm{x}_2) & \ldots & \phi_\eta(\bm{x}_\eta)\\
    \end{vmatrix}.
\end{equation}
The latter form gives such wavefunction the name of \textit{Slater determinant}, and since there is no entanglement, the Hartree-Fock procedure is referred to as a mean-field approach. To simplify the notation, we will use $\ket{i}:= \ket{\phi_i}$, and will also variationally enforce this basis functions to be orthogonal $\braket{i|j} = \delta_{i,j}$.

We want to compute the energy of this wavefunction,
\begin{equation}
    E = \braket{\psi|H|\psi} = \int \psi^\dagger(\bm{r}) H \psi(\bm{r}) d\bm{r}.
\end{equation}
Since the wavefunctions of individual electrons are disentangled, we can compute the one-electron one-body Hamiltonian matrix elements on individual orbitals,
\begin{equation}
\braket{p|h(1)|q}
= \int d\bm{x} \phi^{\dagger}_p(\bm{x}) \left(\frac{\grad^2}{2} + \sum_{\ell=1}^L \frac{Z_\ell}{\|R_\ell-r\|}\right)\phi_q(\bm{x}),
\end{equation}
where $\bm{x} = (r,\sigma)$ for $\sigma$ the spin and $r_i$ the spatial coordinates.
Similarly, the two-electron two-body term matrix elements might be written as
\begin{equation}
\braket{pq|V(2)|rs}
= \frac{1}{2} \int d\bm{x}_1 d\bm{x}_2 \phi^{\dagger}_p(\bm{x}_1) \phi^{\dagger}_q(\bm{x}_2) \left( \frac{1}{\|r_1-r_2\|}\right)\phi_r(\bm{x}_1)\phi_s(\bm{x}_2).
\end{equation}
This last integral form is written by chemists in notation $[pr|qs]$, in contrast to physicists' notation of $\braket{pq|rs}$. 

Overall, the Hartree-Fock energy is obtained by minimizing $\braket{\psi|H|\psi}$ subject to the orthogonality constraints we mentioned. From this procedure, we obtain~\cite[Eq. 2.110]{szabo2012modern},
\begin{equation}
    E_{HF} = \sum_{p=1}^\eta \braket{p|h|p} + \sum_{p>q}^\eta [pp|qq] - \sum_{p>q}^\eta [pq|qp].
\end{equation}
Since $ \phi^\dagger(\bm{x}) \phi(\bm{x}) d\bm{r}$ represents a probability density, the first term makes sense as the weighted average of the one-body Hamiltonian term when the electron is in orbital $\phi_i$. Similarly, the term called Coulomb term
\begin{equation}\label{eq:HF_Coulomb_term}
    [pp|qq] = \frac{1}{2}\int d\bm{x}_1 d\bm{x}_2 \phi^{\dagger}_p(\bm{x}_1) \phi_p(\bm{x}_1) \left( \frac{1}{\|r_1-r_2\|}\right)\phi^{\dagger}_q(\bm{x}_2)\phi_q(\bm{x}_2)
\end{equation}
can also be interpreted as a two-variable expected value, in integral form.
Finally, the `exchange' term 
\begin{equation}\label{eq:HF_exchange_term}
    [pq|qp] = \frac{1}{2}\int d\bm{x}_1 d\bm{x}_2 \phi^{\dagger}_p(\bm{x}_1) \phi_q(\bm{x}_1) \left( \frac{1}{\|r_1-r_2\|}\right)\phi^{\dagger}_q(\bm{x}_2)\phi_p(\bm{x}_2)
\end{equation}
arises from the antisymmetry of the Hartree-Fock wave function, but has no classical meaning associated. Other terms are canceled out by orthonormality of the molecular wavefunctions $\ket{\phi}$.

Next, we separate the spin from the spatial part of the wavefunction. For example, $\phi_i(\bm{x}) d\bm{x} = \varphi_i(\bm{r})\sigma(\omega) d\bm{r}d\omega$. This will lead, for example, to decomposing
\begin{equation}
    \braket{p|h|q} = \sum_\omega \sigma_p(\omega)^\dagger\sigma_q(\omega) \int d\bm{r} \varphi_p^\dagger(\bm{r}) \varphi_q(\bm{r}) =: \left[\sum_{\omega} \sigma_p^\dagger(\omega)\sigma_q(\omega)\right] (p|h|q),
\end{equation}
where $(i|h|j)$ denotes the spatial integral. Similarly,
\begin{equation}
\begin{split}
    [pr|qs] := \frac{1}{2} \int d\bm{x}_1 d\bm{x}_2 \phi^{\dagger}_p(\bm{x}_1) \phi_r(\bm{x}_1)  \left( \frac{1}{\|r_1-r_2\|}\right) \phi^{\dagger}_q(\bm{x}_2) \phi_s(\bm{x}_2) =: \\
    \left[\sum_{\omega_1} \sigma_p^\dagger(\omega_1)\sigma_r(\omega_1)\right] \left[\sum_{\omega_2} \sigma_q^\dagger(\omega_2)\sigma_s(\omega_2)\right] (pr|qs),
\end{split}
\end{equation}
where rounded brackets indicate the spatial component,
\begin{equation}
    (pr|qs) := \frac{1}{2} \int d\bm{r}_1 d\bm{r}_2 \varphi^{\dagger}_p(\bm{r}_1) \varphi_r(\bm{r}_1)  \left( \frac{1}{\|r_1-r_2\|}\right) \varphi^{\dagger}_q(\bm{r}_2) \varphi_s(\bm{r}_2).
\end{equation}
This can be further simplified by taking into account that
\begin{equation}
    \sum_\omega \sigma_p^\dagger(\omega)\sigma_q(\omega) = \delta_{\sigma_p,\sigma_q}.
\end{equation}
The terms $\braket{p|h|p}$ and $[pp|qq]$ will consequently always survive, but the exchange term may not, depending on the spin of $\ket{p}$ and $\ket{q}$ involved.

Overall, the Hartree-Fock Hamiltonian acting on a molecular orbital $\ket{i}$ can be written as
\begin{equation}
    H\ket{\phi_i(\bm{x}_1)} = h(\bm{x}_1)\phi_i(\bm{x}_1) + \sum_{j\neq i}\underbrace{\int d\bm{x}_2 \frac{|\phi_j(\bm{x}_2)|^2}{|r_2-r_1|}}_{J_j(\bm{x}_1)}\phi_i(\bm{x}_1) - \sum_{j\neq i}\underbrace{\int d\bm{x}_2 \frac{\phi_j^\dagger(\bm{x}_2)\phi_i(\bm{x}_2)}{|r_2-r_1|}}_{K_j(\bm{x}_1)} \phi_j(\bm{x}_1).
\end{equation}
This is called the Fock operator~\cite[Eq. 3.16]{szabo2012modern}
\begin{equation}
    f(\bm{x}) = h(\bm{x}) + \sum_j(J_j(\bm{x})-K_j(\bm{x})),
\end{equation}
and leads to the eigenvalue equation that determines the Hartree Fock energy~\cite[Eq. 3.17]{szabo2012modern}
\begin{equation}\label{eq:Fock_energy}
    f(\bm{x})\phi_i(\bm{x}) = \epsilon_i\phi_i(\bm{x}).
\end{equation}
If we solve this equation, then $E_{HF} = \frac{1}{2}(\sum_i \epsilon_i  + \braket{i|h|i})\neq \sum_i \epsilon_i$. The reason is that, defining $J_{i,j} = \braket{i|J_j|i} = [jj|ii]$ and $K_{i,j}= \braket{i|K_j|i} = [ji|ij]$, 
\begin{equation}
    \sum_{i}\epsilon_i = \sum_i \braket{i|h|i} +  \sum_{i,j} (J_{i,j} - K_{i,j}) \neq \sum_i \braket{i|h|i} +  \sum_{i>j} (J_{i,j} - K_{i,j}) = E_{HF},
\end{equation}
so by directly adding the individual orbital energies $\epsilon_i$, we would be double-counting the interaction between electrons in those molecular orbitals.

Without adding more structure to the orbitals, \eqref{eq:Fock_energy} is a complicated integro-differential equation. For this reason, Roothaan proposed decomposing each molecular orbital $\ket{\phi_i}$ as a linear combination of atomic orbitals~\cite{roothaan1951new}:
\begin{equation}
    \phi_i = \sum_\mu C_{\mu,i} \tilde{\phi}_\mu,
\end{equation}
where $\{\tilde{\phi}_\mu\}$ are the basis of non-orthogonal atomic orbitals. Then, defining the Fock matrix in this new basis as
\begin{equation}
    F_{\mu,\nu} = \int d\bm{x} f(\bm{x}) \tilde{\phi}_\mu^\dagger(\bm{x})\tilde{\phi}_\nu(\bm{x})
\end{equation}
and the orbital overlap matrix as 
\begin{equation}
    S_{\mu,\nu} = \int d\bm{x} \tilde{\phi}_\mu^\dagger(\bm{x})\tilde{\phi}_\nu(\bm{x}),
\end{equation}
we rewrite \eqref{eq:Fock_energy} as~\cite[Eq. 3.35]{szabo2012modern}
\begin{equation}\label{eq:Roothaan}
    \sum_\nu F_{\mu,\nu} C_{\nu,i} = \epsilon_i \sum_\nu S_{\mu,\nu}C_{\nu,i}\Rightarrow F C = S C \epsilon.
\end{equation}
A key aspect to note is that the Fock matrix will depend on the molecular orbitals through $f(\bm{x})$, but finding these coefficients requires solving \eqref{eq:Roothaan}. For this reason, we resort to a self-consistent procedure to generate molecular orbitals and solve the Hartree-Fock equation.
\begin{algorithm}[t!]
\begin{algorithmic}[1]
\State \textbf{Input}: Access to the electronic Hamiltonian \eqref{eq:Electronic_hamiltonian}.
\State \textbf{Output}: Molecular orbitals $\{\phi_i\}$ and their energies $\epsilon_i$.
\State Initialize the overlap matrix of atomic orbitals $S_{\mu,\nu}$.
\State Guess initial Molecular Orbital coefficients $C_{\mu,i}$.
\While{$C_{\mu,i}$ has not converged}:
    \State Form the Fock matrix $F_{\mu,\nu}$. \Comment{Requires computing $f(\bm{x})$}
    \State Solve the generalized eigenvalue equation \eqref{eq:Roothaan} and find $C_{\mu,i}$.
\EndWhile
\end{algorithmic}
\caption{Hartree-Fock procedure}\label{alg:Hartree-Fock}
\end{algorithm}
From this algorithm, the costliest step is computing the Fock operator $f(\bm{x})$ which involves up to $O(N^4)$ molecular orbital integrals, due to the 4 indices in \eqref{eq:HF_Coulomb_term} and \eqref{eq:HF_exchange_term}. This can be reduced to $O(N^3)$ using density-fitting approximations~\cite{werner2003fast}, or if the size of the system is large enough, neglecting coefficients of far apart atomic orbital integrals leads to $O(N^2)$. Finally, a technique called `fast multipole' methods may bring that complexity down to $O(N)$ in the latter case~\cite{greengard1987fast}.

\subsection{\label{sec:DFT}Density Functional Theory}

In the previous section, we studied how a disentangled wave function allows us to approximate the ground state and energy. The problem with wave functions, however, is that since they have to model $\eta$ fermions, their dimension will scale as $\mathbb{C}^{3\eta}$, making it difficult to represent large systems accurately.

A possible alternative is to use the density functional
\begin{equation}\label{eq:density_functional}
\begin{split}
    n(\bm{r}) &= \int d\bm{r}_1\ldots d\bm{r}_\eta \psi^\dagger(\bm{r}_1,\ldots,\bm{r}_\eta)(\delta(\bm{r}-\bm{r}_1) + \ldots \delta(\bm{r}-\bm{r}_\eta))\psi(\bm{r}_1,\ldots,\bm{r}_\eta)\\
    &= \eta \int d\bm{r}_2\ldots d\bm{r}_\eta |\psi(\bm{r},\bm{r}_2,\ldots,\bm{r}_\eta)|^2
\end{split}
\end{equation}
as a basic function to model the system, since its dimension will only be $\mathbb{R}^3$. Note that this density functional is different from the probability density $\rho(r) = \psi^\dagger \psi$.
However, is the density functional a good `basic variable' to model chemical properties in which we might be interested? In a breakthrough article in 1964, Hohenberg and Kohn proved two theorems that answered positively this question~\cite{hohenberg1964inhomogeneous}. 

\begin{theorem}[Hohenberg and Kohn]
In any many-body system interacting under an electronic Hamiltonian, there is a one-to-one relation between the ground state electronic density $n_0(\bm{r})$ and the external potential $U$ up to a constant.
\end{theorem}
The proof is carried out by reductio ad absurdum. Let us suppose that we have two external potentials $U^{(1)}$ and $U^{(2)}$, which necessarily lead to different ground states $\psi^{(1)}$ and $\psi^{(2)}$ but the same ground-state electronic density $n_0(\bm{r})$. Since we have stated they are the ground state, we have\footnote{Following Richard Martin's book~\cite{martin2020electronic} and original proof by Hohenberg and Kohn, we assume that the Hamiltonians are not degenerate, but this condition can be relaxed, see~\cite{kohn1985highlights}.}
\begin{equation}
\begin{split}
    E^{(1)} = \braket{\psi^{(1)}|H^{(1)}|\psi^{(1)}} < \braket{\psi^{(2)}|H^{(1)}|\psi^{(2)}} &= \braket{\psi^{(2)}|H^{(2)}|\psi^{(2)}} - \braket{\psi^{(2)}|H^{(1)} - H^{(2)}|\psi^{(2)}}=\\
    &= E^{(2)} -  \int d\bm{r} (U^{(1)} - U^{(2)})n_0(\bm{r}).
\end{split}
\end{equation}
Similarly, we can find that 
\begin{equation}
    E^{(2)} < E^{(1)} -  \int d\bm{r} (U^{(2)} - U^{(1)})n_0(\bm{r}),
\end{equation}
and adding them together we reach the conclusion that
\begin{equation}
    E^{(2)} + E^{(1)} < E^{(2)} + E^{(1)},
\end{equation}
which is a contradiction. In summary, the definition of the density functional \eqref{eq:Fock_energy} entails that the same wave function implies the same density functional, and in this theorem, we have shown the opposite: if we have two ground states of different Hamiltonians, then they will result in different density functionals. Consequently, there is a one-to-one relationship between density functionals and external potentials.

The second theorem introduces the energy functional:
\begin{theorem}[Hohenberg and Kohn]
There exists a total energy functional $E_{HK}[n]$ valid for any external potential $U$, which can be minimized with respect to the energy, and find the ground state energy $E[n]_{HK}(\bm{r})$ and density $n_0(\bm{r})$.
\end{theorem}
The form of this energy functional is
\begin{equation}
    E_{HK}[n] = T[n] + V[n] + \int d\bm{r} U(\bm{r}) n(\bm{r}) + \sum_{\ell,\kappa = 1}^L E_{\ell,\kappa}
\end{equation}
where $\sum_{\ell,\kappa = 1}^L E_{\ell,\kappa}$ is the nuclei-nuclei interaction energy. Since this Hohenberg-Kohn functional corresponds to the expectation value of the Hamiltonian, the energy of the ground state and the energy density can be found by minimizing the expected energy $\braket{\psi|H|\psi}$ via the variational principle.
\begin{equation}
    \frac{\delta}{\delta n(\bm{r})} \left[E_{HK}[n] - \mu \int d\bm{r}' n(\bm{r}')\right] = 0,
\end{equation}
with $\mu$ a Lagrange multiplier.
The first two terms, additionally, are independent of the external potential, and therefore the same for any system. As such, it is called universal functional,
\begin{equation}
    F_{HK}[n]= T[n]+V[n].
\end{equation}
However, while we know how to perform the calculation of the external potential component of the energy functional, it is not clear how to describe the kinetic and potential functionals. 

To overcome this limitation, a year later Kohn and Sham proposed an ansatz that assumes that the ground state density functional $n_0(\bm{r})$ is equal to that of a non-interacting system, with effective (often local) potential $V_{KS}[n]$~\cite{kohn1965self}.
For such a non-interacting system, we can take the Hartree-Fock approximation
\begin{equation}
    n(\bm{r}) = \sum_{i=1}^\eta |\phi_i(\bm{r})|^2.
\end{equation}
The kinetic functional will be defined as
\begin{equation}
    T_s[n] = -\frac{1}{2}\sum_{i=1}^\eta \braket{\phi_i|\grad^2|\phi_i} = \frac{1}{2}\sum_{i=1}^\eta \int d\bm{r}|\grad \phi_i|^2.
\end{equation}
Similarly, the Hartree-Fock electron-electron energy functional can now be written as 
\begin{equation}
    E_{HF}[n] = \frac{1}{2}\int d\bm{r} d\bm{r}' \frac{n(\bm{r})n(\bm{r}')}{|\bm{r}-\bm{r}'|}.
\end{equation}
Using all of this, the energy of the Kohn-Sham model can be written as~\cite[Eq. 7.5]{martin2020electronic}
\begin{equation}
    E_{KS}[n] = T_s[n] +\int d\bm{r} U(\bm{r})n(\bm{r}) +  E_{HF}[n] + \sum_{\ell,\kappa = 1}^L E_{\ell,\kappa} + E_{xc}[n],
\end{equation}
where $E_{xc}[n]$ is called the exchange-correlation functional because it captures the remaining correlation and exchange interaction that the non-interacting kinetic and potential functionals do not. Formally,
\begin{equation}
    E_{xc}[n] = F_{HK}[n] - (T_s[n] + E_{HF}[n]) = \braket{T}-T_s[n] + \braket{V} - E_{HF}.
\end{equation}
In practice, we will use different parameterized functionals, depending on the computational budget, precision, and use case.
To minimize the overall energy we can use the variational principle
\begin{equation}
    \frac{\delta E_{KS}[n]}{\delta \phi_i^\dagger(\bm{r})} = \frac{\delta T_s}{\delta \phi_i^\dagger(\bm{r})} + \left[\frac{\delta \braket{U}}{\delta n(\bm{r})} + \frac{E_{HF}[n]}{\delta n(\bm{r})} + \frac{\delta E_{xc}[n]}{\delta n(\bm{r})} \right]\frac{\delta n(\bm{r})}{\delta \phi_i^\dagger(\bm{r})} = 0,
\end{equation}
subject to orthonormality of the molecular orbitals $\braket{\phi_i|\phi_j} = \delta_{i,j}$ as in the Hartree-Fock procedure. In the above equation, we also know that
\begin{equation}
    \frac{\delta T_s[n]}{\delta \phi_i^\dagger(\bm{r})} = -\frac{1}{2}\grad^2\phi_i(\bm{r}), \qquad
    \frac{\delta n(\bm{r})}{\delta \phi_i^\dagger(\bm{r})} = \phi_i(\bm{r}),\qquad \frac{\delta \braket{U}}{\delta n(\bm{r})} = U(\bm{r}).
\end{equation}
Using the Lagrange multiplier method, we can obtain the Kohn-Sham Schrödinger-like equations, similar to \eqref{eq:Fock_energy},~\cite[Eq. 7.11]{martin2020electronic}
\begin{equation}\label{eq:KS_Fock_energy}
    H_{KS}\phi_i(\bm{r}) = \epsilon_i\phi_i,
\end{equation}
with the Kohn-Sham Hamiltonian
\begin{equation}
\begin{split}
    H_{KS} = -\frac{1}{2}\grad^2  + V_{KS}(\bm{r}) = -\frac{1}{2}\grad^2 + U(\bm{r}) + \frac{\delta E_{HF}[n]}{\delta n(\bm{r})} + \frac{\delta E_{xc}[n]}{\delta n(\bm{r})}\\
    = -\frac{1}{2}\grad^2 + U(\bm{r}) + V_{HF}[n](\bm{r})  +  V_{xc}[n](\bm{r}).
\end{split}
\end{equation}
Equation \eqref{eq:KS_Fock_energy} is solved by a self-consistent procedure similar to the Hartree-Fock method \cref{alg:Hartree-Fock}, and therefore has similar complexity, $O(N^3)$ if we use density fitting techniques~\cite{werner2003fast}.
In this equation, the electron-electron potential functional is
\begin{equation}
    V_{HF}[n](\bm{r}) := \frac{\delta E_{HF}[n]}{\delta n(\bm{r})} =\frac{1}{2}\int d\bm{r}' \frac{n(\bm{r}')}{|\bm{r}-\bm{r}'|},
\end{equation}
and $V_{xc}[n]$ is a parametrized functional, taken of the form
\begin{equation}
    V_{xc}[n](\bm{r}) = \frac{\delta E_{xc}[n](\bm{r})}{\delta n(\bm{r})}.
\end{equation}
The exchange-correlation functional $E_{xc}[n](\bm{r})$ can be written in a similar way as
\begin{equation}
    E_{xc}[n] = \int d\bm{r} n(\bm{r})\epsilon_{xc}([n],\bm{r}),
\end{equation}
with $\epsilon_{xc}$ an energy density per electron at point $\bm{r}$, which depends only on the electronic density. 
It is common to break up the correlation and exchange functionals:
\begin{equation}
\begin{split}
    V_{xc}[n](\bm{r}) &= \epsilon_{xc}[n](\bm{r}) + n(\bm{r})\frac{\partial \epsilon_{xc}[n](\bm{r})}{\partial n(\bm{r})}\\
    &= \epsilon_{x}[n](\bm{r}) + n(\bm{r})\frac{\partial \epsilon_{x}[n](\bm{r})}{\partial n(\bm{r})} + \epsilon_{c}[n](\bm{r}) + n(\bm{r})\frac{\partial \epsilon_{c}[n](\bm{r})}{\partial n(\bm{r})}.
\end{split}
\end{equation}

There are many exchange-correlation functionals according to complexity or accuracy. The simplest is called local (spin) density approximation (LSDA), which depends uniquely on the (spin) density $n(\bm{r})$. For example, based on the uniform electron gas, Dirac proposed the Slater exchange functionals~\cite{dirac1930note}
\begin{equation}
    \epsilon^{\text{LDA}}_x[n] =  - c_x n^{1/3}(\bm{r}); \qquad \epsilon^{\text{LSDA}}_x[n] =  - 2^{-1/3}c_x (n^{1/3}_\uparrow(\bm{r})+ n^{1/3}_\downarrow(\bm{r})),
\end{equation}
for a constant $c_x$.
A further step in complexity is the Generalized Gradient Approximation (GGA) functionals that depend on $\grad n(\bm{r})$ too, such as~\cite{becke1988density}
\begin{equation}
    \epsilon_{x}^{\text{B88}} = \epsilon_{x}^{\text{LDA}} - \frac{\beta n^{1/3}(\bm{r})}{1+6\beta x \sinh^{-1}x}, \qquad x = \frac{|\grad n(\bm{r})|}{n^{4/3}(\bm{r})};
\end{equation}
or even Meta-GGA, with dependence on $\grad^2 n(\bm{r})$.

The most accurate functionals currently in use are hybrid functionals introduced by Becke in 1993,~\cite{becke1993new}. Inspired by the adiabatic theorem, his idea was to assume one can use annealing over the Hamiltonian
\begin{equation}
    H_\lambda = T + U(\lambda) + \lambda V,
\end{equation}
where $U(\lambda)$ is adjusted to ensure that $n(\bm{r})$ is constant. At $\lambda =0$ we get the non-interacting Hartree-Fock limit, while at $\lambda=1$ we recover the interacting Hamiltonian. Annealing, we take
\begin{equation}
    E_{xc}[n] = \int_0^1 d\lambda \braket{\psi|V_{xc}(\lambda)|\psi}.
\end{equation}
We now make two approximations: first, that $V_{xc}(\lambda)$ is linear in $\lambda$, so
\begin{equation}
    E_{xc}[n] \approx \frac{1}{2} \left(\braket{\psi|V_{xc}(0)|\psi} + \braket{\psi|V_{xc}(1)|\psi}\right).
\end{equation}
As mentioned, Becke assumed that $\braket{\psi|V_{xc}(0)|\psi} = E^{\text{HF}}_x[n]$ is the Hartree-Fock exchange functional, and in the non-interacting limit the correlation is $0$. His second approximation was to take $\braket{\psi|V_{xc}(1)|\psi} \approx E^{\text{LDA}}_{xc}[n]$. Overall, the hybrid functional mixes Hartree-Fock and LDA functionals
\begin{equation}
    E_{xc}^{\text{Becke}}[n] = \frac{1}{2}(H^{HF}_x[n] + E^{\text{LDA}}_{xc}[n]).
\end{equation}
One may also choose GGA or meta-GGA functionals instead of LDA. This led to the B3LYP, the most widely used functional nowadays~\cite{stephens1994ab}
\begin{equation}
    \epsilon_{xc}^{\text{B3LYP}} = (1-a)\epsilon_{x}^{\text{LSDA}} + aE^{\text{HF}}_{x} + b\Delta \epsilon_{x}^{\text{B88}} + (1-c)\epsilon_{c}^{\text{VWN}} + c \epsilon_{c}^{\text{LYP}},
\end{equation}
where $a$, $b$ and $c$ are constants, $\epsilon_{c}^{\text{VWN}}$ is the correlation functional of~\cite{vosko1980accurate} of type LSDA, $\epsilon_{c}^{\text{LYP}}$ the correlation functional from~\cite{lee1988development} of type GGA, and $\Delta \epsilon_{x}^{\text{B88}}$ the gradient correction of the B88 GGA exchange functional~\cite{becke1988density, becke1993density}. Finally, it is worth mentioning that `double hybrid' functionals exist, which include an MP2 (Møller-Plesset to second order~\cite{moller1934note}) term, or in other words, perturbation theory of the Hartree-Fock solution. There have also been attempts to use machine learning to model the exchange-correlation functional~\cite{kirkpatrick2021pushing} in especially challenging situations. More information on density functional theory can be found in chapters 6-8 of Martin's book~\cite{martin2020electronic}. 
 
\subsection{\label{sec:CC}Coupled-Cluster}

In the previous subsection, we have studied Density Functional Theory, which adds a parametrized exchange-correlation functional ansatz that should be able to capture energy corrections that the Hartree-Fock self-consistent procedure is not capable of obtaining. However, being an ansatz, even if in practice they can be very precise, they often come without much accuracy warranty. For this reason, we will come back to the wave function as a basic variable, and explain a post-Hartree-Fock method that is very precise: the Coupled-Cluster method. As we will see, this method is also the basis for the Unitary Coupled-Cluster ansatz used for the Variational Quantum Eigensolver.

The basic idea of the Coupled Cluster procedure is to expand the wave function as excitations of the Hartree-Fock state $\ket{\phi_0}$:
\begin{equation}
    \ket{\phi_{\text{CC}}} = e^{S} \ket{\phi_0},
\end{equation}
for a normal operator $S$. If $S$ consists of a sum $T$ of fermionic excitation operators of occupied orbitals and annihilation of virtual orbitals, then this flavor of Coupled Cluster is called Traditional Coupled-Cluster:
\begin{align}\label{eq:TCC}
    T = \sum_{k} T_k,\qquad T_k = \frac{1}{(k!)^2}\sum_{i_1<\ldots< i_k}^{\text{occ}}\sum_{a_1<\ldots< a_k}^{\text{vir}} t_{i_1\ldots i_{k}}^{a_1\ldots a_k} a_{a_1}^\dagger \ldots a_{a_k}^\dagger a_{i_1} \ldots a_{i_k},
\end{align}
with the $a_{i_j}$ and $a_{i_j}^\dagger$ fermionic annihilation and creation operators, and $t_{i_1\ldots i_{k}}^{a_1\ldots a_k}$ the amplitudes we have to fix. We will also use the notation
\begin{equation}
   \hat{\tau}_{i_1\ldots i_{k}}^{a_1\ldots a_k} = a_{a_1}^\dagger \ldots a_{a_k}^\dagger a_{i_1} \ldots a_{i_k}.
\end{equation}
If we consider up to order $\eta$ operators $T_k$, we will reconstruct the Full Configuration Interaction wave function, which takes into account all possible excitations from the Hartree-Fock Slater determinant. A clear advantage of Coupled-Cluster over other methods is its preservation of size consistency, meaning that if we have two disentangled subsystems, $\ket{\phi_A}$ and $\ket{\phi_B}$, the joint Coupled-Cluster exponential operator
\begin{equation}
    e^{S_A + S_B}\ket{\phi_A}\ket{\phi_B} = e^{S_A}\ket{\phi_A}\otimes e^{S_B}\ket{\phi_B},
\end{equation}
and the energy of the joint system will be the sum of the energy of the subsystems.

There are two main ways to use Coupled-Cluster in chemistry problems. The first is the projective approach, based on
\begin{equation}
    H e^{S}\ket{\phi_0} = E e^{S}\ket{\phi_0}.
\end{equation}
We can project the previous equation into
\begin{equation}
    \bra{\phi_0}\hat{\tau}_{i_1\ldots i_{k}}^{a_1\ldots a_k}= \bra{\phi^{a_1\ldots a_j}_{i_1\ldots i_j}},
\end{equation}
as well as into the Slater determinant $\bra{\phi_0}$.
From the projection, we get the following set of equations, called `unlinked'~\cite{anand2022quantum}
\begin{align}
    \braket{\phi_0|He^{T}|\phi_0} &= E \braket{\phi_0|e^{T}|\phi_0} = E,\\
    \braket{\phi^{a_1\ldots a_j}_{i_1\ldots i_j}|He^{T}|\phi_0} &=  E \braket{\phi^{a_1\ldots a_j}_{i_1\ldots i_j}|e^{T}|\phi_0}.
\end{align}
We can expand the first expression
\begin{equation}
\begin{split}
    \braket{\phi_0|H|\phi_{\text{CC}}} &= \underbrace{\braket{\phi_0|H|\phi_0}}_{E_{HF}} + \underbrace{\braket{\phi_0|HT_1|\phi_0}}_{0} + \underbrace{\braket{\phi_0|H(T_2 + \frac{1}{2}T_1^2)|\phi_0}}_{\neq 0}\\
    &+ \underbrace{\braket{\phi_0|H(T_3 + T_2T_1 + \frac{1}{6}T_1^3)|\phi_0} + \ldots}_{0}
\end{split}
\end{equation}
The fact that $\braket{\phi_0|HT_1|\phi_0} =0$ is due to the Brillouin theorem. By the definition of the molecular orbitals in the Hartree-Fock procedure,
\begin{equation}
    \braket{\phi_0|HT_1|\phi_0} = \sum_{i}\sum_a \braket{\phi_0|H|\phi^{a}_i} = \braket{a|h|i} + \sum_{b} \braket{a|J_j -K_j|i},
\end{equation}
which is an off-diagonal matrix element and, since the Hartree-Fock procedure precisely aims to diagonalize the Fock matrix on the basis of molecular orbitals, it vanishes. Operators with 3 or more also vanish due to Slater-Condon rules: the Hamiltonian operator can only de-excite up to 2 orbitals, so the molecular orbitals being orthonormal, this leads to a non-vanishing contribution from only the $0^{\text{th}}$ and $2^{\text{nd}}$ orders. While only $T_1$ and $T_2$ contribute to the energy, these operators might in turn depend on other excitation terms $T_i$ for $i>2$.\footnote{\label{foot:Thouless}A related and very useful result is known as Thouless theorem~\cite{thouless1960stability}, which states that $e^{T_1}$ maps a single Slater determinant state into another (non-orthogonal) single determinant. Thus $e^{T_1}$ might be understood as a basis change. A didactic explanation can be found in this nice post by Joshua Goings, \href{https://joshuagoings.com/2013/11/26/644/}{https://joshuagoings.com/2013/11/26/644/}.}

An alternative set of equations is the set of `linked' couple cluster equations:
\begin{align}
    \braket{\phi_0|e^{ -T}He^{T}|\phi_0} &= E,\\
    \braket{\phi^{a_1\ldots a_j}_{i_1\ldots i_j}|e^{ -T}He^{ T}|\phi_0} &= 0.
\end{align}
We can expand the similarity transformed Hamiltonian $\bar{H} = e^{-T}He^{T}$ as
\begin{equation}\label{eq:Hadamard_lemma}
    \bar{H} = e^{-T}He^{T} = H  +  [H,T] + \frac{1}{2!}[[H,T],T] + \frac{1}{3!}[[[H,T],T],T] + \ldots
\end{equation}
While for an arbitrary $S$ the expansion may not converge, if we take it as $T$ in equation \eqref{eq:TCC}, the series for the expected value is truncated at the fourth level:
\begin{equation}\label{eq:truncated_Hadamard_lemma}
\begin{split}
    \bar{H}  =  e^{T^\dagger}He^{T}  =  H   +   [H,T] 
    + \frac{1}{2!} [[H,T],T]
    + \frac{1}{3!} [[[H,T],T],T]  + \frac{1}{4!} [[[[H,T],T],T],T] .
\end{split}
\end{equation}
The reason for this is the generalized Wick theorem (see~\cite[Section 3.7 and Chapter 10]{shavitt2009many}), which states that for any two normal operators $A$ and $B$
\begin{equation}\label{eq:Wick}
\wick{
    [A,B] = AB- BA = (\colon AB\colon + \colon \c1 A \c1 B \colon) - (\colon BA \colon + \colon \c1 B \c1 A \colon),
}
\end{equation}
where $\wick{\colon \c1 A \c1 B \colon}$ represents the sum of normal products where there is at least one contraction, and $\colon  A B \colon$ the sum of normal ordered operators. In particular,
\begin{equation}
    \colon a^{\dagger}_i a_j \colon = a^{\dagger}_i a_j, \qquad \colon a_j a^{\dagger}_i \colon = - a^{\dagger}_i a_j, \qquad 
    \wick{
    \c a_i^\dagger \c a_j = \delta_{ij},
    }, \qquad \wick{
    \c a_i \c a_j} = 0 = \wick{ \c a_i^\dagger \c a_j^\dagger
    }.
\end{equation}
Since there is an even number of creation and annihilation operators in each $\tau$ operator~\cite{shavitt2009many}, $\colon AB\colon = \colon BA \colon$. Then, \eqref{eq:Wick} tells us that,
\begin{equation}
\colon AB\colon = \colon BA \colon\quad 
\Rightarrow \quad
\wick{
    [A,B] =  \colon \c1 A \c1 B \colon - \colon \c1 B \c1 A \colon .
}
\end{equation}
In other words, in our problem the commutator $[A, B]$ only depends on sums of products with at least one contraction.
We also know that the different terms in any operator $T_k$ only contain creation operators of virtual orbitals and annihilation of occupied ones. Consequently, all contractions of operators $\hat{\tau}_{i_1\ldots i_{k}}^{a_1\ldots a_k}$ are $0$, and thus they commute with one another as expected.
Finally, since each electronic Hamiltonian term contains up to 4 annihilation/creation operators, from the fifth order in \eqref{eq:Hadamard_lemma} there will be no annihilation/creation operators in the Hamiltonian to contract, making the corresponding $\wick{\colon \c1 A \c1 B \colon =0}$ and leading to \eqref{eq:truncated_Hadamard_lemma}.

Furthermore, given $T_k$ defined as in \eqref{eq:TCC},
\begin{equation}
    \bra{\phi_0}T_i = 0,
\end{equation}
because $T_i$ would apply creation operators to occupied orbitals, and annihilation operators to virtual ones. Taking into account that the Hamiltonian $H$ can only de-excite up to two orbitals, we use the same Slater-Condon rule and Brillouin theorem as above, to obtain
\begin{equation}
    E = E_{HF} + \braket{\phi_0|[H,T_2]|\phi_0}+ \frac{1}{2}\braket{\phi_0|[[H,T_1],T_1]|\phi_0}.
\end{equation}
Either the `linked' or `unlinked' set of equations can then be solved algebraically.

Finally, it is worth mentioning that projective methods are not the only flavor of Coupled-Cluster, we also have variational methods which involve minimizing
\begin{equation}
    E = \frac{\braket{\phi_0|e^{S^\dagger}He^{S}|\phi_0}}{\braket{\phi_0|e^{S^\dagger}e^{S}|\phi_0}}.
\end{equation}
One advantage of this variational approach over the projective approach is that the recovered energy is always necessarily an upper bound of the ground-state energy. 
However, it is not possible to apply the same truncation that we used for the similarity transformed Hamiltonian $\bar{H}$, notice the different sign in the $e^{S^\dagger}$ exponential. Additionally, if $S$ is not anti-Hermitian (e.g. $S= T$), the denominator might also be nontrivial. To ensure $S$ is indeed Hermitian, we might instead choose $S= T - T^\dagger$, which leads to Unitary Coupled-Cluster. We will see that this is a very natural ansatz to implement in a quantum computer.

\section{\label{sec:Chemistry_QPE}Hamiltonian simulation}

In the previous section, we analyzed three classical algorithms to solve problems in quantum chemistry. We now focus on quantum algorithms. As mentioned at the beginning of the chapter, quantum computing offers two features that make it very attractive for chemistry and material science applications. First, it is capable of representing quantum systems exactly, without approximations. Second, the unitary evolution of closed quantum systems can be naturally implemented in a quantum computer. In this section, we review the problem of how to implement this evolution, under the name of Hamiltonian simulation.

Hamiltonian simulation is a key technique to solve the two most important problems in quantum chemistry: preparing the ground state of a Hamiltonian and computing its energy. Other applications include preparing and analyzing thermal and excited states.
To see what we mean by Hamiltonian simulation, we start from the Schrödinger equation,
\begin{equation}
    H\psi = i\hbar \frac{d}{dt} \psi.
\end{equation}
As our electronic Hamiltonian \eqref{eq:Electronic_hamiltonian} is time-independent, the evolution of a quantum state $\psi$ can be modeled as
\begin{equation}
    \psi(x,0) = \sum_n a_n  \psi_{E_n}(x)\Rightarrow
    \psi(x,t) = \sum_n a_n  e^{-iE_n t/\hbar} \psi_{E_n}(x),
\end{equation}
where $a_n$ are the amplitudes, $\psi_{E_n}$ the eigenstates of the Hamiltonian $H$ with eigenvalues $E_n$. From this equation, it is clear that if $|a_0|\approx 1$ and we know how to implement the Hamiltonian simulation of $e^{-iHt/\hbar}$, we can use the phase estimation algorithm introduced in \cref{ch:Phase} to find $E_0$, the ground state energy. It is worth noticing however that using quantum phase estimation is not the only fault-tolerant method that can be used to estimate energies. Not treated in this thesis but very promising are the recent techniques by Lin and Tong and subsequent work, that assume access to small fault-tolerant quantum computers \cite{lin2020near,wang2022quantum,zhang2022computing}.

\subsection{\label{ssec:Fermionic_map}Quantization and fermion to qubit mapping}

\paragraph{First quantization.}
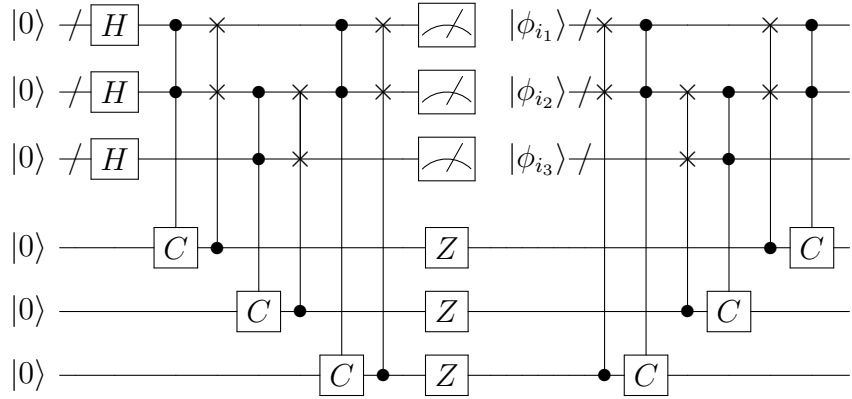
\begin{figure}[t!]
\[
\begin{array}{c}
 \Qcircuit @C=0.5em @R=0.75em { 
  & \ket{0} &  & & {/}\qw & \gate{H}  & \ctrl{1} & \qswap & \qw & \qw & \ctrl{1} & \qswap & \qw & \meter & & & &  \ket{\phi_{i_1}} & & & {/}\qw& \qswap & \ctrl{1}  & \qw & \qw & \qswap & \ctrl{1} & \qw \\
  & \ket{0} &  & & {/}\qw & \gate{H}  & \ctrl{3} & \qswap\qwx & \ctrl{1} & \qswap & \ctrl{5} & \qswap\qwx & \qw & \meter  & & & & \ket{\phi_{i_2}} & & & {/}\qw& \qswap\qwx & \ctrl{5}  & \qswap & \ctrl{1} & \qswap\qwx & \ctrl{3} & \qw \\
  & \ket{0} &  & & {/}\qw & \gate{H}  & \qw & \qw & \ctrl{3} & \qswap\qwx & \qw & \qw & \qw & \meter & & & & \ket{\phi_{i_3}} & & & {/}\qw& \qw & \qw  & \qswap\qwx & \ctrl{3} & \qw & \qw & \qw \\
  \\
  & \ket{0} &  & & \qw & \qw & \gate{C} & \ctrl{-3}  & \qw & \qw & \qw & \qw & \qw & \gate{Z} & \qw & \qw & \qw & \qw  & \qw & \qw  & \qw & \qw& \qw & \qw  & \qw  & \ctrl{-3} & \gate{C} & \qw \\
  & \ket{0} &  & & \qw & \qw  & \qw & \qw & \gate{C} & \ctrl{-4} & \qw & \qw  & \qw & \gate{Z} &  \qw & \qw & \qw  & \qw & \qw  & \qw & \qw & \qw & \qw & \ctrl{-3} & \gate{C} & \qw  & \qw & \qw \\
  & \ket{0} &  & & \qw & \qw& \qw & \qw & \qw & \qw & \gate{C} & \ctrl{-5} & \qw & \gate{Z} &  \qw & \qw & \qw  & \qw & \qw  & \qw & \qw  & \ctrl{-5} & \gate{C} & \qw & \qw & \qw  & \qw & \qw 
 }
\end{array}
\]
\caption{\label{fig:Antisymmetrization}\textbf{Antisymmetrization circuit.} Example of an antisymmetrization circuit for three electrons. The operation $C$ represents a comparison test controlled on the two registers that are being compared. The seed register (top left) is measured to post-select on the collision-free subspace. The $Z$ gates perform the phase flip when swapping two registers. At the end of the circuit, the auxiliary record qubits (bottom register) can be discarded as they are disentangled~\cite{berry2018improved}. This circuit can be extended to an arbitrary number of electrons $\eta$ by increasing the size of the sorting network and adding additional auxiliary qubits for each required comparison and swap.}
\end{figure}
Before getting into details of how to implement Hamiltonian simulation, however, we need to decide how to represent the quantum state and Hamiltonian. The first such choice is whether to represent the state in first or second quantization, in other words, whether the logical qubits of our system will represent the state of each electron, or the occupancy of each orbital, respectively.
For example, if we have $\eta$ disentangled electrons in our system, occupying orbitals $i_1,\ldots,i_\eta$, then the corresponding first quantized quantum state is
\begin{equation}\label{eq:antisymmetrization}
\sum_{\sigma\in S_\eta}\frac{(-1)^{\pi(\sigma)}}{\sqrt{\eta!}}\ket{\phi_{\sigma(i_1)}}\otimes \ldots \otimes\ket{\phi_{\sigma(i_\eta)})}.
\end{equation}
For $N$ orbitals, we need $\eta\log_2 N$ qubits to represent the state.
Note how this state is antisymmetric, what can be achieved from $\ket{\phi_{i_1}}\ket{\phi_{i_2}}\ldots\ket{\phi_{i_\eta}}$ by a procedure first described in~\cite{berry2018improved}, and depicted in~\cref{fig:Antisymmetrization}. This protocol can be carried out with minimal overhead, and fortunately only needs to be implemented once during the state preparation, because the Hamiltonian simulation operators will preserve antisymmetry throughout the algorithm.

In contrast, a second quantized state does not need to be explicitly antisymmetrized, because the creation and annihilation operators account for Fermi statistics automatically, but in contrast needs $N$ qubits for $N$ orbitals. If $N\gg \eta$, this implies that the second quantization requires more qubits to represent the state. Moreover, since we have to replicate Fermi statistics, the mapping from fermions to qubits is non-trivial: we have to keep track of occupation numbers as well as the parity. 

\paragraph{Jordan-Wigner mapping.} 
The Jordan-Wigner mapping represents~\cite{wigner1928paulische}
\begin{equation}
    a_j^\dagger \to Z_1\otimes\ldots \otimes Z_{j-1}\otimes (\sigma^+)_j\otimes \bm{1},
\end{equation}
for $X_j$, $Y_j$, $Z_j$ the Pauli operators acting on qubit $j$, and $\sigma^{+} = (X + iY)/2$.
This means that operator $a_j^\dagger$ acts non-trivially on $j$ qubits: the first $j-1$ qubit Pauli gates record the phase, and the last Pauli operator creates a particle in qubit $j$. Using this mapping, a Hermitian operator $a_p^\dagger a_q + a_q^\dagger a_p $ is implemented as~\cite{babbush2018encoding}
\begin{equation}\label{eq:Jordan-Wigner}
    a_p^\dagger a_q + a_q^\dagger a_p \mapsto 
    \begin{cases}
    \bm{1}-Z & p=q,\\
    \frac{X_p\Vec{Z}X_q + Y_p\Vec{Z}Y_q}{2}  & p\neq q,\\
    \end{cases}
\end{equation}
where $X_p\Vec{Z}X_q = X_p \otimes Z_{p+1}\otimes \ldots \otimes Z_{q-1}\otimes X_{q}$ with $p<q$, and similarly for $Y_p\Vec{Z}Y_q$. In summary, the Jordan-Wigner mapping implies acting on up to $O(N)$ qubits per fermionic Hamiltonian operator. Extending this mapping to dimensions higher than one has been discussed in Ref.~\cite{verstraete2005mapping}.

\paragraph{Parity mapping.} The dual of the Jordan-Wigner mapping is the parity mapping, where the parity is saved in a single qubit, but the occupation number in $O(N)$~\cite{cao2019quantum},
\begin{equation}
    a_j^\dagger \to \bm{1}\otimes  Z_{j-1}\otimes(\sigma^+)_j\otimes X_{j+1}\otimes\ldots \otimes X_{N}.
\end{equation}
The mapping from Jordan-Wigner
to parity mapping encoded states is
\begin{equation}
    \ket{i_1}\otimes\ket{i_2}\otimes\ldots \ket{i_N}\to \ket{i_1}\otimes\ket{i_1\oplus i_2}\otimes\ldots\otimes \ket{\bigoplus_{j=1}^N i_j},
\end{equation}
with $\oplus$ denoting binary sum.

\paragraph{Bravyi-Kitaev mapping.} Is there a way to avoid the $O(N)$ cost of the Jordan-Wigner and parity mappings? The Bravyi-Kitaev encoding ensures that both parity and occupation require no more than $O(\log_2 N)$ qubit operators~\cite{bravyi2002fermionic}. First proposed for $N = 2^n$ a power of two, it uses qubits to encode sums of occupation numbers. The mapping from the Jordan-Wigner mapping to the Bravyi-Kitaev matrix is carried out by the basis change~\cite{cao2019quantum}
\begin{equation}
    \ket{i_1}\otimes\ket{i_2}\otimes\ldots \ket{i_N}\to \ket{b_1}\otimes\ket{b_2}\otimes\ldots \ket{b_N}, \qquad
    b_k = \sum_{l=1}^k [\beta_n]^l_k i_l,
\end{equation}
where $\beta_0 = (1)$ and~\cite{seeley2012bravyi}
\begin{equation}
    \beta_n =
    \left(
    \begin{array}{c c c| c c c}
     & & & \multicolumn{3}{c}{\leftarrow 1 \rightarrow}\\
     & \beta_{n-1} & & & \bm{0} &\\
     &  & & &  & \\
     \hline
     &  & & &  & \\
     & \bm{0} & & & \beta_{n-1} &\\
     &  & & &  &
    \end{array}
    \right).
\end{equation}
For example,
\begin{equation}
    \beta_3 = 
    \begin{pmatrix}
    1 & 1 & 1 & 1 & 1 & 1 & 1 & 1\\
    0 & 1 & 0 & 0 & 0 & 0 & 0 & 0\\
    0 & 0 & 1 & 1 & 0 & 0 & 0 & 0\\
    0 & 0 & 0 & 1 & 0 & 0 & 0 & 0\\
    0 & 0 & 0 & 0 & 1 & 1 & 1 & 1\\
    0 & 0 & 0 & 0 & 0 & 1 & 0 & 0\\
    0 & 0 & 0 & 0 & 0 & 0 & 1 & 1\\
    0 & 0 & 0 & 0 & 0 & 0 & 0 & 1
    \end{pmatrix}.
\end{equation}
We can check that each column of these matrices contains up to $n = \log_2 N$ $1$s, which leads to $O(\log_2 N)$ weight creation and annihilation operators~\cite{seeley2012bravyi}. In fact, when $N$ is a power of $2$, operators $a_j^\dagger$ have weight no larger than $\log_2 N$~\cite{havlivcek2017operator}. In exchange for faster simulations, the detailed explanation of the form of creation and annihilation operators becomes complicated and would require a large explanation, which the interested reader may instead find in Ref.~\cite{seeley2012bravyi}.

For the sake of completeness, we mention that a closely related technique, called the Bravyi-Kitaev tree method, also achieves $O(\log_2 N)$ weight for the operators. While in practice this technique retrieves higher weight operators than the Bravyi-Kitaev method, it also allows reducing the number of qubits if $N$ is not a power of 2~\cite{havlivcek2017operator}.\footnote{An implementation of Jordan-Wigner and Bravyi-Kitaev operators can be found in \href{https://quantumai.google/openfermion/tutorials/jordan_wigner_and_bravyi_kitaev_transforms}{one tutorial on the topic in the Openfermion library}.} Other encodings include Ref.~\cite{babbush2017exponentially} based on the configuration-interaction matrix, Steudtner's method for which the number of gates to implement $a^\dagger$ and $a$ is independent on the number of basis functions~\cite{steudtner2018fermion,steudtner2019methods}, and Ref.~\cite{kirby2022second}, whose complexity on both the number of gates and qubits is polylogarithmic in the number of basis functions. Refs.~\cite{derby2021compact,derby2021acompact} suggest a particularly compact mapping in the number of fermionic modes (basis functions) per qubit, and finally the proposal of Ref.~\cite{whitfield2016local} proposes using auxiliary qubits to restrict the mapping to local operators.

\subsection{\label{ssec:Basis}Basis choice}
In addition to first or second quantization and the mapping to qubits, we have to choose a set of basis functions to generate the wave function. The two most common options are Gaussian functions and plane waves. 

\paragraph{\label{par:Gaussian}Gaussian functions.}  Gaussian functions aim to generalize the well-known hydrogen atom orbitals
\begin{equation}
    \phi_{n,l,m}(r,\theta,\phi)= R_{nl}(r)Y_{l,m}(\theta,\phi),
\end{equation}
where $Y_{l,m}(\theta,\phi)$ are spherical harmonics, and the radial component is a product of Laguerre polynomials and a negative exponential term in the radius~\cite{griffiths2018introduction}. The closest choice of orbitals to emulate such behavior are Slater-type orbitals (STO),
\begin{equation}
    R_n^{\text{STO}}(r) = N r^{n-1}e^{-\zeta r},
\end{equation}
for $N$ a normalizing term and $\zeta$ a diffuseness parameter~\cite{cao2019quantum}. While Slater-type orbitals can be used as a basis and have the advantage of correct asymptotic behavior, computing the corresponding Hamiltonian integrals we saw in \cref{sec:HF} must be done numerically because no analytical solutions are known. Instead, for simplicity Gaussian-type orbitals (GTO) are often used,
\begin{equation}
    R_n^{\text{GTO}}(r) = N r^{n-1}e^{-\zeta r^2},
\end{equation}
which exhibit Gaussian behavior and are easily analytically integrable as a result. Gaussian-type orbitals can be written in spherical coordinates, with spherical harmonics modeling the angular component, or in Cartesian coordinates, with a factor $x^iy^jz^k$ instead of the spherical harmonics, $\Vec{r} = (x,y,z)$ and $i+j+k$ the angular momentum.

Since Gaussian-type orbitals do not exhibit the correct asymptotic behavior in the exponent, often a linear combination of Gaussian orbitals (\textit{primitives}) is used to emulate a single Slater-type orbital (\textit{contraction}). Then, the Hartree-Fock or Density Functional Theory procedures will compute a linear combination of the contracted atomic orbitals to form molecular orbitals. A review of the different families of contractions of GTOs can be found in Ref.~\cite{jensen2013atomic}. On this basis, the Hamiltonian is usually written as
\begin{equation}\label{eq:Gaussian_Hamiltonian}
    H = \sum_{\alpha,\beta\in \{\uparrow,\downarrow\}}\sum_{p,q,r,s=1}^{N/2}h_{pqrs}a_{p,\alpha}^\dagger a_{q,\beta}^\dagger a_{r,\beta} a_{s,\alpha} + \sum_{\sigma \in\{\uparrow,\downarrow\}}\sum_{p,r=1}^{N/2}h_{p,r}a_{\sigma,p}^\dagger a_{\sigma,r}, 
\end{equation}
with one and two body coefficients,
\begin{equation}
    h_{pr} = \braket{p|T + U|r},\qquad h_{pqrs} = \braket{p,q|V|r,s}.
\end{equation}
Gaussian functions are well suited for isolated molecules, where the wave function exponentially vanishes at infinity.

\paragraph{\label{par:PWs}Plane waves.} The alternative to Gaussian basis functions are plane waves which, due to their periodicity, are better suited for periodic materials. If we consider a cell of volume $\Omega$, these functions are\footnote{More generally, the previous equation assumes a single $k$-point in the center of the Brillouin zone, the dual of the primitive unit cell, with unit vectors $\bm{b}_i$. For any other $k$-point, the wave functions take the form \begin{equation}
    \varphi_{p,k}(\bm{r}) = \sqrt{\frac{1}{\Omega}}e^{i(\bm{G}_p+\bm{k})\cdot \bm{r}},
\end{equation}
where 
\begin{equation}
    \bm{k} = \sum_{i=1}^3 \frac{n_i}{N_i} \bm{b}_i,
\end{equation}
$N_i$ indicates the number of plane waves in each dimension, and $n_i$ is an integer.
This is connected to Bloch's theorem~\cite{bloch1929quantenmechanik}, which states that $\psi(\bm{r}) = e^{i \bm{k}\cdot \bm{r}} u(\bm{r})$, where the potential $u(\bm{r})$ has the periodicity of the lattice. Adding more $k$-points is used to take into account correlations between electrons and nuclei of different unit cells, growing the system towards the thermodynamic limit.}
\begin{equation}
    \varphi_p(\bm{r}) = \sqrt{\frac{1}{\Omega}}e^{i\bm{G}_p\cdot \bm{r}},
\end{equation}
where $\bm{G}_p$ is a reciprocal lattice vector indexed by $\bm{p}\in\mathcal{G} = \left[-\frac{N^{1/3}}{2}+1, \frac{N^{1/3}}{2}-1\right]^3$.

On this basis, the Hamiltonian takes the form~\cite{babbush2018low}
\begin{align}
T&=\sum_{i=1}^\eta\sum_{p\in \mathcal{G}}\frac{\|\bm{G}_p\|^2}{2}\ket{\bm{p}}\bra{\bm{p}}_i, \label{eq:t_op}\\
U&=-\frac{4\pi}{\Omega}\sum_{i=1}^\eta\sum_{q\in \mathcal{G}}\sum_{\substack{ \nu\in \mathcal{G}_0  \\ (\bm{q}-\bm{\nu}) \in \mathcal{G} }}\frac{\sum_{I=1}^L Z_I e^{i\bm{G}_\nu \cdot \bm{R}_I}}{\|\bm{G}_\nu\|^2}\ket{\bm{q-\nu}}\bra{\bm{q}}_i \label{eq:u_op},\\
V&=\frac{2\pi}{\Omega}\sum_{i\neq j}^\eta\sum_{p,q\in \mathcal{G}}\sum_{\substack{\nu\in \mathcal{G}_0 \\ (\bm{p} + \bm{\nu}) \in \mathcal{G} \\ (\bm{q}-\bm{\nu}) \in \mathcal{G}  } } \frac{1}{\|\bm{G}_\nu\|^2}\ket{\bm{p+\nu}}\bra{\bm{p}}_i\otimes \ket{\bm{q}-\bm{\nu}}\bra{\bm{q}}_j \label{eq:v_op},
\end{align}
where $\mathcal{G}_0:= \mathcal{G}\backslash \{(0,0,0)\}$. Note the approximation, called aliasing or dualling, of making the momentum exchange be confined to lie within $\mathcal{G}_0$~\cite{remler1990molecular,mcclain2017gaussian}.

Similarly, on the plane-wave basis and second quantization, the Hamiltonian will take the form~\cite[Appendix B]{babbush2018low}
\begin{equation}
\begin{split}
    H =\underbrace{\frac{2\pi}{\Omega}\sum_{\substack{(p,\sigma)\neq (q,\sigma')\\
    \nu\neq 0}}\frac{c_{p,\sigma}^\dagger c_{q,\sigma'}^\dagger c_{q+\nu,\sigma'} c_{p-\nu,\sigma}}{\|\bm{G}_\nu\|^2}}_{V}
\underbrace{+\frac{1}{2}\sum_{p,\sigma}\|\bm{G}_p\|^2 c_{p,\sigma}^\dagger c_{p,\sigma}}_{T}
\underbrace{-\frac{4\pi}{\Omega}\sum_{\substack{p\neq q;\\ I,\sigma}}\left(Z_I \frac{e^{i\bm{G}_{q-p}\cdot \bm{R}_I}}{\|\bm{G}_{p-q}\|^2}\right)c_{p,\sigma}^\dagger c_{q,\sigma}}_{U}.
\label{eq:Plane wave Hamiltonian}
\end{split}
\end{equation}
One useful feature of plane waves is that they diagonalize the kinetic operator.
In contrast, we can Fourier transform the Hamiltonian, to obtain the dual basis Hamiltonian~\cite[Appendix C]{babbush2018low}
\begin{equation}
    \begin{split}
        H &=\underbrace{\frac{1}{2N}\sum_{p,q,\nu,\sigma}\|\bm{G}_\nu\|^2 \cos[\bm{G}_\nu\cdot \bm{r}_{q-p}] a_{p,\sigma}^\dagger a_{q,\sigma}}_{T}\\
        &\underbrace{-\frac{4\pi}{\Omega}\sum_{\substack{p, I,\sigma\\\nu\neq 0}}\left( \frac{Z_I\cos[\bm{G}_{\nu}\cdot (\bm{R}_I-\bm{r}_p)]}{\|\bm{G}_\nu\|^2}\right)n_{p,\sigma}}_{U}\underbrace{+\frac{2\pi}{\Omega}\sum_{\substack{(p,\sigma)\neq (q,\sigma')\\ \nu\neq 0}}\frac{\cos[\bm{G}_\nu \cdot \bm{r}_{p-q}]}{\|\bm{G}_\nu\|^2}n_{p,\sigma}n_{q,\sigma'}}_{V},
    \end{split}
    \label{eq:Dual plane wave Hamiltonian}
\end{equation}
in which the external potential and potential operators will be diagonal. The number of terms in the Hamiltonian in plane waves scales as $O(N^3)$ with the number of basis functions $N$, while in dual wave basis they scale as $O(N^2)$. This is more favorable than the $O(N^4)$ scaling in \eqref{eq:Gaussian_Hamiltonian}. However, this comparison is not fair since the basis used is different, and therefore $N$ is not the same either. To compare them, we should analyze the error behavior of both basis sets, which in both cases will asymptotically decrease as $O(1/N)$~\cite{kutzelnigg1992rates,shepherd2012convergence}. This, and a thorough literature review, leads Ref.~\cite[Appendix E]{babbush2018low} to state that one needs approximately 10 to 20 times more plane waves than Gaussian basis functions to model a system to the same accuracy, as long as the system is periodic. This comparison, however, must be taken with care.

Finally, it is also possible to simulate isolated molecules in plane waves, by taking an $8\Omega$ volume cell, and ensuring that at a distance larger than $D=\sqrt[3]{\Omega}$ the Coulomb interaction vanishes. This means that instead of a Fourier amplitude proportional to $\frac{4\pi}{\|\bm{G}_\nu\|^2}$, we get
\begin{equation}
    4\pi \frac{1-\cos[|\bm{G}_\nu|D]}{\|\bm{G}_\nu\|^2}
\end{equation}
in \eqref{eq:t_op}, \eqref{eq:u_op} and \eqref{eq:v_op} ~\cite[Appendix E]{babbush2018low}. In the dual basis, the approximation is even simpler, just dropping all terms $n_pn_q$ for which $|\bm{r}_p-\bm{r}_q|> D$.

\paragraph{\label{par:Bloch}Bloch and Wannier basis functions.}

Other basis functions have also been used throughout the literature including \textit{Bloch wave basis} (also known as \textit{band fermion basis}), whose defining feature is to diagonalize the one-body Hamiltonian in periodic materials~\cite{clinton2022towards}. Mathematically, if the one-body Hamiltonian is
\begin{equation}
   T+U = \sum_{\bm{k},p,q,\sigma} \underbrace{\left[\frac{|\hbar(\bm{k}+\bm{G}_p)|^2}{2m}\delta_{\bm{G}_p,\bm{G}_q}+ U_{\bm{G}_p-\bm{G}_q}\right]}_{h_{\bm{k},\bm{G}_p-\bm{G}_q}}c_{\bm{k}+\bm{G}_p,\sigma}^\dagger c_{\bm{k}+\bm{G}_q,\sigma}.
\end{equation}
Choosing, as we did on the plane wave basis explanation, a single crystal cell $\bm{k}$, we find can find the unitary transformation $S_{n,\bm{G}}(\bm{k})$ that diagonalizes the one-body Hamiltonian 
\begin{equation}
h_{\bm{k},\bm{G}_p-\bm{G}_q} = \sum_n S^\dagger_{\bm{G}_p,n}(\bm{k}) \epsilon_n(\bm{k}) S_{n,\bm{G}_q}(\bm{k}).
\end{equation}
Index $n$ will indicate an occupation band, and can take as many values as can reciprocal vector lattices $\bm{G}$. Then, the Bloch basis functions are defined as~\cite{clinton2022towards}
\begin{equation}
    \phi_{\bm{k},n}(\bm{r}) = \sqrt{\frac{1}{\Omega}} e^{i\bm{k}\cdot \bm{r}}\sum_{\bm{G}} e^{i\bm{G}\cdot \bm{r}}S_{n,\bm{G}}(\bm{k}).
\end{equation}

If we are interested in real-space basis functions, we can modify Bloch basis functions into \textit{Wannier basis functions}. In their simplest form, they can be written as~\cite{wannier1937structure}
\begin{equation}
    \textit{W}_{\bm{R}}(\bm{r}) = \textit{W}_0(\bm{r}-\bm{R}) = \sum_{\bm{k},n}e^{-i\bm{k}\cdot \bm{R}}\phi_{\bm{k},n}(\bm{r}) = \sum_{\bm{k},n}\phi_{\bm{k},n}(\bm{r}-\bm{R}).
\end{equation}
More complex options may include a gauge transformation of the Bloch basis, which allow localizing the orbitals thus reducing the size of the required basis.

\subsection{\label{ssec:Hamiltonian_simulation}Hamiltonian simulation techniques}

In this subsection, we review the main techniques of Hamiltonian simulation, that is, of implementing $e^{-iHt}$. This will be important as a subroutine in quantum phase estimation and state preparation, as we will see. An up-to-date reference of which technique is most efficient in each case can be found in~\cite{babbush2023quantum}. Since we are interested in Hamiltonian simulating the electronic Hamiltonian \eqref{eq:Electronic_hamiltonian}, we will assume the Hamiltonian is a Linear Combination of Unitaries from now on, $H = \sum_{l=1}^L a_l H_l$, with $a_l > 0$.

\paragraph{\label{par:Trotterization}Trotter.}

The Trotter-Suzuki decomposition was the first method proposed to implement Hamiltonian simulation~\cite{ortiz2001quantum} and does not require the $H_l$ above to be unitary but only Hermitian. The key idea is to decompose $e^{-i t \sum_{l} a_l H_l}$ as a product of terms of the form $e^{-i t H_l}$. Since this induces an error dependent on the time length of each time segment, we divide the total time simulation into many small segments. For example, the first-order Trotter formula is
\begin{equation}\label{eq:first_order_Trotter}
    e^{-iH t} = \underbrace{\left(\prod_l e^{-ia_l H_l t/r}\right)^r}_{\mathcal{S}_1(H; t/r)} + O\left(\sum|[H_{l_1},H_{l_2}]|t^2/r\right).
\end{equation}
The error will also multiplicatively depend on the commutator of the different $H_l$ terms, such that if for example they all commute, the Hamiltonian simulation can be fast-forwarded, e.g., implemented in a single time segment. In other words, in such a case, we can take $t$ as large as wished and $r=1$ still get an accurate result.

Similarly, we can propose a second-order formula to decrease the simulation error, 
\begin{equation}\label{eq:second_order_Trotter}
    e^{-iH t} = \underbrace{\left(\left(\prod_{l=1}^L e^{-ia_l H_l t/2r}\right)\left(\prod_{l=L}^1 e^{-ia_l H_l t/2r}\right)\right)^r}_{\mathcal{S}_2(H; t/r)}
    + O\left(\sum|[[H_{l_1},H_{l_2}],H_{l_3}]|t^3/r^2\right),
\end{equation}
and inductively~\cite{suzuki1991general,childs2019faster,su2021nearly},
\begin{equation}
    \mathcal{S}_{2k}(H;t/r) = \mathcal{S}^2_{2k-2}(H;p_k t/r)\mathcal{S}_{2k-2}(H;(1-4p_k) t/r)
    \mathcal{S}^2_{2k-2}(H;p_k t/r),
\end{equation}
with $p_k = 1/(4-4^{1/(2k-1)})$.
The Hamiltonian simulation error $\epsilon_{HS}$ will decrease as~\cite{mcardle2022exploiting}
\begin{align}
\left|\left|e^{-iH t/r} - \mathcal{S}_k(H; t/r)\right|\right|_2\leq W_k \left(\frac{ t}{r}\right)^{k+1}\leq\frac{\epsilon_{HS}}{r},\\
W_k = O\left(\max_{\bm{i}} [[\ldots[H_{l_{i_1}},H_{l_{i_2}}], H_{l_{i_3}}],\ldots],H_{l_{i_{k+1}}}] \right).
\end{align}
While the complexity of the method is in any case polynomial on the Hamiltonian simulation error $\epsilon_{HS}$, some techniques can help reduce it. The first is to introduce randomness either in the ordering of the $a_l H_l$ terms~\cite{childs2019faster}, or treat the $a_l/\lambda$ for $\lambda = \sum a_l$ as probabilities of applying $H_l$ for fixed amounts of time~\cite{campbell2019random}. Other randomization protocols have also been explored~\cite{wan2021randomized}.

The second line of research aims to bound to the norm of the commutators~\cite{kivlichan2020improved,campbell2021early,su2021nearly,mcardle2022exploiting}. For $N\gg \eta$, the so-called SHC-bound~\cite{su2021nearly,mcardle2022exploiting} scales as $O(N^3)$, while for $N$ closer to $\eta$ tighter bounds can be found in Ref.~\cite{mcardle2022exploiting}. Some of these use the fermionic semi-norm
\begin{equation}
    \|X\|_\eta := \max_{\phi,\psi}|\braket{\phi|X|\psi}|_{\eta},
\end{equation}
where $\ket{\phi}$ and $\ket{\psi}$ contain $\eta$ fermions. This semi-norm can be used to eliminate nonphysical terms, which lead to higher error than possible in the commutators, due to the particle-conserving nature of our system.

\paragraph{\label{par:Taylor}Taylor series.}

While Trotter series decomposition is popular and flexible, its error scales only inverse polynomially. Post-Trotter methods aim to improve this situation with a polylogarithmic error complexity. One possibility is Taylor series decomposition:
\begin{equation}
    U_r = e^{-iHt/r}\approx \sum_{k= 0}^K \frac{1}{k!}(-iHt/r)^k =
    \sum_{k=0}^K \sum_{l_1,...,l_k=1}^L \underbrace{\frac{(-it/r)^k}{k!}a_{l_1}...a_{l_k}}_{b_j}\underbrace{H_{l_1}...H_{l_k}}_{U_j}.
\end{equation}
To implement this LCU decomposition $\sum_{j}b_j U_j$, we use the same $\Prep$ and $\Sel$ operators that we have defined on other occasions:
\begin{equation}
    \Prep:\ket{0}\mapsto \sum_{j}\sqrt{b_j}\ket{j},\qquad
    \Sel:\ket{j}\ket{\psi}\mapsto \ket{j}U_{j}\ket{\psi},
\end{equation}
Then, we can perform $U_{LCU}^{Tay} = (\Prep^\dagger\otimes \bm{1}) \Sel (\Prep\otimes \bm{1})$, which exhibits some failure probability, as the action of $\Prep^\dagger$ does not fully return the value of the auxiliary register $\ket{j}$ to $\ket{0}$. Consequently, it is convenient to implement the algorithm in short time segments of order $\tau = \ln 2$, so that (oblivious) amplitude amplification can eliminate such error~\cite{berry2015simulating}. Overall, the value of $K$ can be chosen to be $\left\lceil-1+2\frac{\log (\epsilon_{HS}/r)}{\log \log (\epsilon_{HS}/r)+1}\right\rceil$~\cite[Lemma 5]{low2019qubitization}.

\paragraph{\label{par:Qubitization}Qubitization.}

Since the Hamiltonian is already written as a linear combination of unitaries, we can instead describe $U_{LCU} =  (\Prep^\dagger\otimes \bm{1}) \Sel (\Prep\otimes \bm{1})$ with different prepare and select operators
\begin{equation}
    \text{Prepare}:\ket{0}\mapsto \sum_{l}\sqrt{a_l}\ket{l},\qquad
    \text{Select}:\ket{l}\ket{\psi}\mapsto \ket{l}H_l\ket{\psi}.
\end{equation}
The matrix form of $U_{LCU}$ is
\begin{equation}
    U_{LCU} = \begin{pmatrix}
    H/\lambda & \cdot\\
    \cdot & \cdot \\
    \end{pmatrix},
\end{equation}
a \textit{block encoding} as we explained in~\cref{ssec:Qubitization}. This operator can also be studied by its action on state $\ket{0}\ket{\psi}$, 
\begin{equation}
U_{LCU}\ket{0}\ket{\psi} = \ket{0}\frac{H}{\lambda}\ket{\psi} + \sqrt{1-\frac{\|H\ket{\psi}\|}{\lambda}}\ket{1}\ket{\psi^\perp}.
\end{equation}
While this operator also has some probability of failure that requires amplitude amplification, an alternative is to use $\Prep$ and $\Sel$ as part of a quantum walk operator $Q$, acting in the same way as $U_{LCU}$ in the previous equation
\begin{equation}
\begin{split}
    Q\ket{0}\ket{\psi_k}&=\cos(\theta_k)\ket{0}\ket{\psi_k}-\sin(\theta_k)\ket{1}\ket{\psi_k^\perp},\\
    Q\ket{1}\ket{\psi_k^\perp}&=\cos(\theta_k)\ket{1}\ket{\psi_k^\perp}+\sin(\theta_k)\ket{0}\ket{\psi_k},
\end{split}
\end{equation}
for $\cos\theta_k= \frac{E_k}{\lambda}$ and $(\ket{\psi_k}, E_k)$ eigenvector and eigenvalue tuples. Notice that $\theta_k$ here plays the role of $2\theta$ in \eqref{eq:diag_Amplitude_Amplification} or $2\varphi_j$ in \eqref{eq:block rotation matrix in A+B}. In other words,
\begin{equation}
    Q = \bigoplus_k \begin{pmatrix}
    \frac{E_k}{\lambda} & -\sqrt{1-\frac{E_k^2}{\lambda^2}}\\
    \sqrt{1-\frac{E_k^2}{\lambda^2}} & \frac{E_k}{\lambda}
    \end{pmatrix}_k =  \bigoplus_k e^{-iY_k \theta_k}.
\end{equation}
To build operator $Q$, Ref.~\cite[Corollary 9]{low2019qubitization} suggests using
\begin{equation}
    Q = \underbrace{\Prep(2\ket{0}\bra{0}\otimes \bm{1}-\bm{1})\Prep^\dagger}_{\text{Rotation }1}\cdot \underbrace{\Sel}_{\text{Rotation }2}
\end{equation}
whenever $U_{LCU}^2=1$, as is the case here. Diagonalizing it, we get
\begin{equation}
    Q = \bigoplus_k\left( e^{i\theta_k}\ket{\theta_k}\bra{\theta_k}+e^{-i\theta_k}\ket{-\theta_k}\bra{-\theta_k}\right).
\end{equation}
Provided eigenstate $\ket{\psi_0}$, performing quantum phase estimation directly over $Q$ is sufficient to recover $\pm\theta_0$, which allows computing the energy as $E_0 = \lambda \cos(\pm \theta_0)$ with no Hamiltonian simulation error~\cite{berry2018improved}.

This protocol, called \textit{qubitization}, is even more powerful when combined with a technique called \textit{quantum signal processing}~\cite{low2017optimal}, which we have mentioned previously. Introducing
\begin{equation}
    Z_\phi := (1+e^{-i\phi})\Prep\ket{0}\bra{0}\Prep^\dagger -\bm{1} = \bigoplus_k \begin{pmatrix}
    e^{-i\phi} & 0\\
    0 & 1
    \end{pmatrix}_k,
\end{equation}
one can form operators 
\begin{equation}
W_\phi = Z_{\phi-\pi/2} Q Z_{-\phi+\pi/2} = 
\bigoplus_k \begin{pmatrix}
    -i e^{-i\phi} & 0\\
    0 & 1
    \end{pmatrix}_k
    \begin{pmatrix}
    \frac{E_k}{\lambda} & -\sqrt{1-\frac{E_k^2}{\lambda^2}}\\
    \sqrt{1-\frac{E_k^2}{\lambda^2}} & \frac{E_k}{\lambda}
    \end{pmatrix}_k
    \begin{pmatrix}
    i e^{-i\phi} & 0\\
    0 & 1
    \end{pmatrix}_k.
\end{equation}
A string of these, $W_{\vec{\phi}} = W_{\phi_Q}\ldots W_{\phi_1}$, can be used to synthesize $A[H] + iB[H]$, where $A$ and $B$ are polynomials of degree $Q$ (or $Q/2$) of $\cos(\theta_k/2)$ and $A$ and $B$ have equal~\cite[Theorem 3]{low2019qubitization} (or respectively opposite,~\cite[Theorem 4]{low2019qubitization}) parity. For example, this quantum signal processing technique can be used to simulate $e^{-iHt}$ with cost polylogarithmic in the precision parameter, by decomposing it into a Jacobi-Anger series~\cite[Theorem 1]{low2019qubitization}. Furthermore, quantum signal processing is Hamiltonian-query optimal, and is not restricted to Linear Combination of Unitaries but can also be used with other oracle access models such as sparse or density matrices.

\paragraph{\label{par:Interaction_picture}Interaction picture.}

While using qubitization we can perform errorless Hamiltonian simulation suitable for phase estimation, the algorithm still bears a linear dependence on the one-norm of the Hamiltonian $\lambda$. Implementing the simulation in the interaction picture aims to reduce such a complexity factor. In particular, if $H = A+B$, one can form the interaction picture Hamiltonian 
\begin{equation}
    H_I(t) = e^{iAt}B(t) e^{-iAt},
\end{equation}
and if $\|A\|\gg \|B\|$, this will reduce the norm of the Hamiltonian from $\|A+B\|$ to $\|B\|$. In this framework, the state will evolve as~\cite{low2018hamiltonian}
\begin{equation}
    \ket{\psi(t)} = e^{-iAt}\mathcal{T}\left[e^{-i\int_0^t H(s)ds}\right] \ket{\psi(0)}.
\end{equation}
In summary, we have to implement two parts, $e^{-iA t}$ and the time-ordered exponential~\cite{kieferova2019simulating,low2018hamiltonian}. The former might be easy to implement if all operators in $A$ commute, so this is a practical requirement for the interaction picture algorithm. The latter requires a Dyson series expansion
\begin{equation}
\begin{split}
    U(t) = \mathcal{T}\left[e^{-i\int_0^t H(s)ds}\right] = \sum_{k = 0}^\infty (-i)^k D_k, \qquad
    D_k = \frac{1}{k!}\int_{0}^t...\int_{0}^t\mathcal{T}[H(t_k)...H(t_1)]d^k t,
\end{split}
\end{equation}
which, similarly to the Taylor series, has logarithmic complexity in the Hamiltonian simulation precision $\epsilon_{HS}^{-1}$. The required block encoding of operator $B$ makes use of a Linear Combination of Unitaries, and operators $\Prep_B$ and $\Sel_B$, allowing to perform a block encoding of a time segment of $e^{-i(A+B)\tau}$ as~\cite{su2021fault}
\begin{equation}
\begin{split}
e^{-i(A+B)\tau}= e^{-iA \tau}\lim_{\substack{K\rightarrow \infty\\ M\rightarrow \infty}}\sum_{k=0}^K\frac{(-i\tau)^k}{M^k k!}\sum_{m_1=0}^{M-1} \ldots \sum_{m_k=0}^{M-1}\\
\Big(e^{-i\tau(-1/2-m'_k)A/M} B e^{-i\tau(m'_k-m'_{k-1})A/M} B\ldots 
B e^{-i\tau(m'_2-m'_{1})A/M} B e^{-i\tau(m'_1+1/2)A/M}\Big)\\
\approx \left(\bra{0}\Prep_B^\dagger\right)^{\otimes K}
\sum_{k=0}^K \frac{(-i\lambda_B\tau)^k}{M^k k!}\sum_{m_1,\ldots,m_k =0}^{M-1}
\Big(e^{-i\tau(M-1/2-m'_k)A/M} \Sel_B   \\
e^{-i\tau(m'_k-m'_{k-1})A/M}\Sel_B \ldots\Sel_B e^{-i\tau(m'_2-m'_{1})A/M} \Sel_B 
e^{-i\tau(m'_1+1/2)A/M}\Big)\Big(\Prep_B \ket{0}\Big)^{\otimes K},
\end{split} 
\end{equation}
with $m'_1,\ldots,m'_k$ the (time) ordered integers $m_1,\ldots,m_k$. Due to the use of block encodings, this approach similarly requires splitting the total time evolution into small time segments and using oblivious amplitude amplification. As a consequence, the cost scales polylogarithmically with the inverse Hamiltonian simulation error $\epsilon_{HS}$. Importantly, this technique may also be naturally used for time-dependent Hamiltonians, by taking $A$ as the time-independent component and $B$ as the time-dependent one.

\paragraph{\label{par:rank}Rank factorization.} 

While not a Hamiltonian simulation technique of its own, rank factorization approaches have been used in the literature to reduce the 1-norm of the Hamiltonian in the context of second quantization and qubitization or to simplify the implementation of $\Prep$ and $\Sel$. Note that the Hamiltonian 1-norm is a multiplicative factor in many of the state-of-the-art techniques above, and the reason why interaction picture and other methods were used in the first place~\cite{loaiza2022reducing}. Further, rank factorization methods have been found helpful when used in combination with QROM~\cite{babbush2018encoding}. We briefly describe the single and double-rank techniques, noting that the state of the art is currently at the tensor hypercontraction techniques described in Ref.~\cite{lee2020even,oumarou2022accelerating}. We follow the explanations in the appendices of this last reference.

We already discussed that the second-quantized Hamiltonian can be written as a one and two body term, see~\eqref{eq:Gaussian_Hamiltonian}. Specifically, 
\begin{equation}
    T = \frac{1}{2}\sum_{\sigma \in \{\uparrow, \downarrow\}} \sum_{p,q=1}^{N/2}T_{pq} (a_{p,\sigma}^\dagger a_{q,\sigma} + a_{q,\sigma}^\dagger a_{p,\sigma}),
\end{equation}
and 
\begin{equation}
    V = \frac{1}{8}\sum_{\alpha,\beta \in \{\uparrow, \downarrow\}} \sum_{p,q,r,s=1}^{N/2}V_{pqrs} (a_{p,\alpha}^\dagger a_{q,\alpha} + a_{q,\alpha}^\dagger a_{p,\alpha})(a_{r,\beta}^\dagger a_{s,\beta} + a_{s,\beta}^\dagger a_{r,\beta}).
\end{equation}
The next step is to Jordan-Wigner map these terms, using~\eqref{eq:Jordan-Wigner}. If we define
\begin{equation}
    Q_{pq\sigma} = \begin{cases}
        X_{p,\sigma}\vec{Z} X_{q,\sigma} & p< q,\\
        Y_{p,\sigma}\vec{Z} Y_{q,\sigma} & p> q,\\
        - Z_{p,\sigma} & p = q,
    \end{cases}
\end{equation}
we can rewrite
\begin{equation}
    T = \frac{1}{2}\sum_{\sigma \in \{\uparrow, \downarrow\}} \sum_{p,q=1}^{N/2} T_{pq}Q_{pq\sigma} + \sum_{p = 1}^{N/2} T_{pp}\bm{1},
\end{equation}
and 
\begin{equation}
    V = \frac{1}{8}\sum_{\alpha,\beta \in \{\uparrow, \downarrow\}} \left(\sum_{p,q,r,s=1}^{N/2}V_{pqrs}Q_{pq\alpha}Q_{rs\beta}+ \sum_{p,q,r=1}^{N/2}V_{pqrr}Q_{pq\alpha}+\sum_{p,r,s=1}^{N/2}V_{prrs}Q_{rs\beta} + \sum_{p,r = 1}^{N/2}V_{pprr}\bm{1}\right).
\end{equation}
This suggests rearranging
\begin{equation}
    T' = \frac{1}{2}\sum_{\sigma \in \{\uparrow, \downarrow\}} \sum_{p,q=1}^{N/2} (T_{pq} + \sum_{r=1}^{N/2} V_{pqrr}) Q_{pq\sigma},
\end{equation}
and
\begin{equation}
    V' = \frac{1}{8}\sum_{\alpha,\beta \in \{\uparrow, \downarrow\}} \sum_{p,q,r,s=1}^{N/2}V_{pqrs}Q_{pq\alpha}Q_{rs\beta},
\end{equation}
plus a term proportional to the identity which can be omitted in the Hamiltonian simulation. The single-rank factorization consists of performing a Cholesky decomposition such that we can approximate
\begin{equation}
    V_{pqrs}\approx \sum_{l=1}^L W_{pq}^{(l)} W_{rs}^{(l)}.
\end{equation}
Consequently, the two-body term might be approximated by~\cite{berry2019qubitization}
\begin{equation}
    W = \frac{1}{8}\sum_{l=1}^L\left(\sum_{\sigma \in \{\uparrow, \downarrow\}} \sum_{p,q}^{N/2} W_{pq}^{(l)} Q_{pq\sigma} \right)^2.
\end{equation}
The double-rank factorization goes a step further and diagonalizes each $W_{pq}^{(l)}$, approximating $V$ with~\cite{von2021quantum}
\begin{equation}
    F = \frac{1}{2}\sum_{l=1}^L U_l \left( \sum_{\sigma \in \{\uparrow, \downarrow\}} \sum_{p=1}^{\Xi^{(l)}} f_p^{(l)}n_{p,\sigma} \right)^2 U_l^\dagger.
\end{equation}
Discarding again the terms proportional to the identity, after the Jordan-Wigner mapping this operator looks
\begin{equation}
    F' = \frac{1}{8}\sum_{l=1}^L U_l \left( \sum_{\sigma \in \{\uparrow, \downarrow\}} \sum_{p=1}^{\Xi^{(l)}} f_p^{(l)}Z_{p,\sigma} \right)^2 U_l^\dagger.
\end{equation}

\section{\label{sec:State_prep}Quantum state preparation}

Before diving into the different techniques to prepare ground states, let us briefly discuss the complexity of this and previous tasks.
\begin{definition}[Bounded-error Quantum Polynomial-time (BQP)]
This complexity class is composed of all problems solvable by a quantum Turing machine in polynomial time and probability of error at most $1/3$.
\end{definition}
BQP is the class of problems that are `efficiently' solvable by a universal quantum computer. It is the quantum equivalent to the Bounded-error Probabilistic Polynomial-time (BPP), where the quantum Turing machine is substituted by a classical Turing machine. Running the algorithm multiple times, the $1/3$ failure probability might be exponentially reduced. A larger complexity class is the following.

\begin{definition}[Quantum Merlin-Arthur (QMA)]
Complexity class of binary decision problems where positive instances can be efficiently verified by a BQP solver with success probability 2/3, and negative instances rejected with similar probability.
\end{definition}
In more poetical words, Merlin, a powerful agent with unbounded computational resources, wants to convince Arthur (with access to a universal quantum computer) of the character, positive or negative, of one instance of the problem. Is there a quantum state and a polynomial-time single-interaction protocol by which Arthur can correctly verify or reject such a quantum state?
This definition is the quantum equivalent of the NP complexity class, which would apply to a classical BPP verifier instead.

We know that Hamiltonian simulation~\cite{kassal2008polynomial}, quantum phase estimation~\cite{wocjan2006several} and the HHL algorithm~\cite{harrow2009quantum} are BQP-complete problems, any other BQP problem can be reduced to them. Similarly, finding the non-interacting Kohn-Sham functional for a time-dependent electronic density is also BQP~\cite{whitfield2014computational}. In contrast, distinguishing the ground state of a two-body Hamiltonian provided the promise of an energy gap~\cite{kempe2006complexity}, and finding the universal functional in DFT~\cite{schuch2009computational}, are both QMA problems. However, while finding the ground state of general electronic Hamiltonians is computationally difficult, this does not preclude the possibility of finding average-case efficient algorithms for quantum systems found in nature. After all, those natural systems are often able to find the ground state, so we might as well.
The situation is similar to the one we encountered in \cref{sec:QFold} with protein folding, and we hope to similarly find an efficient algorithm in practice.

Similar to our approach with protein folding, one idea is to use digital quantum simulation of quantum annealing. Indeed, the techniques developed by Ref.~\cite{lemieux2019efficient} and that we use in~\cite{casares2022qfold} are well-tailored to finding the ground state of quantum Hamiltonians~\cite{lemieux2021resource, yung2012quantum}. We already saw that the complexity of quantum Metropolis algorithms for this purpose scales as $O(\Delta^{-1})$, where $\Delta$ is the eigenvalue gap. Overall, it should be noted that this problem is so hard, that not even quantum computers are expected to offer an exponential quantum advantage~\cite{lee2022there}.

\subsection{\label{ssec:UCC}Unitary coupled-cluster \& variational quantum eigensolver}

One alternative to quantum phase estimation is called the variational quantum eigensolver. It is based on the following idea: let the qubit-mapped Hamiltonian be
\begin{equation}
    H = \sum_{i,\alpha}h_{i,\alpha} \sigma_{\alpha}^i + \sum_{i,j,\alpha,\beta}h_{i,j,\alpha,\beta} \sigma_{\alpha}^i \sigma_{\beta}^j +\ldots,
\end{equation}
where $\sigma$ are Pauli operators and $\alpha,\beta,\ldots \in \{x,y,z\}$, and $i,j,\ldots$ indicate the qubit where such Pauli operator should be applied. Then, the energy of the system can be computed as the weighted average of the expectation values of each of the qubit Pauli string operators
\begin{equation}\label{eq:Hamiltonian_expected_value}
    E = \braket{H} = \sum_{i,\alpha}h_{i,\alpha} \braket{\sigma_{\alpha}^i} + \sum_{i,j,\alpha,\beta}h_{i,j,\alpha,\beta} \braket{\sigma_{\alpha}^i \sigma_{\beta}^j} +\ldots,
\end{equation}
The variational quantum eigensolver is based on the idea that we can prepare the ground state by variationally finding the state $\ket{\psi}$ that minimizes the ground state energy
\begin{equation}
    E = \min_{\psi}\frac{\braket{\psi|H|\psi}}{\braket{\psi|\psi}}.
\end{equation}
The question then is to find a good ansatz for the state $\ket{\psi}$. The most popular choice is to use the Unitary Coupled-Cluster applied to the Hartree-Fock state as a way to parametrize the ground state:
\begin{equation}
    \ket{\psi} = e^{T-T^\dagger} \ket{\phi_0}.
\end{equation}
The operator in the exponent is anti-Hermitian, so the exponential is a unitary operator that can be implemented in a quantum computer. To implement it, sometimes classical Møller-Plesset 2 (perturbation theory over Hartree-Fock) coefficients might be used as a possible educated guess for the initialization for the parameters~\cite{romero2018strategies}, and we will also need one of the Hamiltonian simulation techniques described in the previous section \cref{ssec:Hamiltonian_simulation}. For experiments in NISQ devices, this often entails implementing Trotter simulation of $U(\bm{t}) = e^{\frac{1}{2}\sum_{j=1}^J  t_j( \tau_j -  \tau_j^\dagger)}$. Then, to optimize the parameters $t_j$ of energy $E$ in \eqref{eq:Hamiltonian_expected_value}
it is customary to use any variation of gradient descend (Stochastic Gradient Descend, ADAM, ...), where gradients are computed via the parameter shift rule~\cite{bergholm2018pennylane,schuld2019evaluating,wierichs2022general}:
\begin{equation}
    \frac{\partial E}{\partial t_j} = \frac{1}{2\sin \alpha}\sum_{k} h_k\left(\Tr[P_kU^\dagger(\bm{t}_+)\rho_0U(\bm{t}_+)]- \Tr[P_kU^\dagger(\bm{t}_-)\rho_0U(\bm{t}_-)]\right),
\end{equation}
for $\rho_0 = \ket{\phi_0}\bra{\phi_0}$, $P_k$ the Pauli string operators in \eqref{eq:Hamiltonian_expected_value}, $h_k$ their weights, $\bm{t}_{\pm} = (t_1, \ldots, t_{j-1}, t_{j}\pm \alpha, t_{j+1}, \ldots, t_{J})$, and finally $\alpha$ an arbitrary parameter that should be chosen as $\pi/4$ for maximum accuracy.

The previous procedure, however, is very costly due to the sheer number of parameters $t_j$ to optimize in any molecule. For that reason, other ansätze have been proposed. For example, the Qubit Coupled-Cluster suggests using
\begin{equation}
    U(\bm{t}) = \prod_j e^{t_j P_j},
\end{equation}
where $P_j$ are Pauli string operators (not necessarily fermionic operators) that conserve the number of particles in the system~\cite{ryabinkin2018qubit}.

Another possibility is using a small ansatz and progressively adding more terms according to their importance. This is the central idea of the adaptive derivative assembled pseudo-Trotter VQE (ADAPT-VQE)~\cite{grimsley2019adaptive}. If we have a pool of $\{\tau_j\}$ operators, at any given step $i$ one chooses, with replacement, the operator $\tau_{j_i}$ with the largest gradient and adds it to the state preparation
\begin{equation}
    \ket{\psi} = e^{t_{j_i}(\tau_{j_i}-\tau_{j_i}^\dagger)}\ldots e^{t_{j_2}(\tau_{j_2}-\tau_{j_2}^\dagger)}
    e^{t_{j_1}(\tau_{j_1}-\tau_{j_1}^\dagger)}\ket{\phi_0}.
\end{equation}
After choosing the operator, one optimizes the value of the associated parameter $t_{j_i}$ and freezes it. This ansatz has the advantage that it can be stopped at arbitrary lengths.

While the previous quantum methods are based on the variational approach of the Unitary Coupled Cluster, we know from our discussion in \cref{sec:CC} that there is also a projective approach. This is the main idea of the projective quantum eigensolver (PQE), where one aims to minimize the residuals~\cite{stair2021simulating}
\begin{equation}
    r_m = \braket{\phi_m|U^\dagger(\bm{t}) H U(\bm{t})|\phi_0}\to 0.
\end{equation}
This forms a system of non-linear equations that might be solved with classical methods. The projective approach also allows for an adaptive version called selective, that progressively grows 
\begin{equation}
    H \to e^{t_{j_i}(\tau_{j_i}^\dagger-\tau_{j_i})}\ldots e^{t_{j_2}(\tau_{j_2}^\dagger-\tau_{j_2})}
    e^{t_{j_1}(\tau_{j_1}^\dagger-\tau_{j_1})} 
    H 
    e^{t_{j_1}(\tau_{j_1}-\tau_{j_1}^\dagger)} e^{t_{j_2}(\tau_{j_2}-\tau_{j_2}^\dagger)}\ldots
    e^{t_{j_i}(\tau_{j_i}-\tau_{j_i}^\dagger)}.
\end{equation}
Note that in this case the operator is built in reverse order, and the $i^{th}$ operator is chosen to zero out the largest residual at step $i$~\cite{stair2021simulating}. 

Finally, let us briefly mention that variational circuits such as these may find a problem of vanishing gradients, known as barren plateaus~\cite{mcclean2018barren}. The origin of this phenomenon is often the set of gates of the ansatz resembling a random set of unitaries. Fortunately, using the set of fermionic operators of the Unitary Coupled-Cluster as ansatz may avoid this problem~\cite{cerezo2021variational,farhi2014quantum,hadfield2019quantum}. A detailed discussion of the quantum variants of Unitary Coupled-Cluster can be found at Ref.~\cite{anand2022quantum}, and of variational algorithms and techniques in general in Ref.~\cite{cerezo2021variational}.

\subsection{\label{ssec:State_prep_project}Projection methods}

\paragraph{\label{par:Krylov}Imaginary time evolution and quantum Krylov subspace.} Besides implementing excitations of the Hartree-Fock as a way to prepare the ground state of the system, an alternative idea is to filter out the excited states. For example, in the imaginary time evolution, one seeks to implement an approximation to operator $e^{-\tau H}$~\cite{motta2020determining}. It is then prescribed to use tomography to select unitary operators $e^{-i\tau A}$ that approximate the action of $e^{-\tau H}$. 

This imaginary time evolution technique might also be used to generate excited states. In particular, we can generate a Krylov subspace by iteratively applying the $e^{-\tau H}$ operator, $\ket{\psi_l} = e^{-l \tau H}\ket{\psi_0}$, where $\ket{\psi_0}$ is often taken the Hartree-Fock state $\ket{\phi_0}$. The even-$l$ states $\ket{\psi_{0}}$, $\ket{\psi_{2}}$, $\ket{\psi_{4}}$, ... form a basis to describe the ground state~\cite{motta2020determining}. The Hamiltonian matrix elements of these states are then computed as $\braket{\psi_l|H|\psi_{l'}} = \braket{\psi_{(l+l')/2}|H|\psi_{(l+l')/2}}$, and diagonalizing it, one can get an approximation to the ground state and excited states. This method is known as QLanczos~\cite{motta2020determining}, and may even be used to prepare multi-reference states~\cite[Section 2.3]{stair2020multireference}: if the approximate ground state of the Hamiltonian contains similar amplitudes of various Slater determinants, we can find them by repeatedly measuring it. The generation of Krylov states and diagonalization may then be repeated with each of the Slater determinant reference states~\cite[Figure 1]{stair2020multireference}.
A similar approach of diagonalizing the Hamiltonian subspace spanned by the most relevant Slater determinants is also pursued in Ref.~\cite{tubman2018postponing} but in the Configuration Interaction framework.


\paragraph{\label{par:LCU_filtering}Linear Combination of Unitaries.}
Instead of using tomography to find operators that project out the higher energy states, we can also use Linear Combination of Unitaries to synthesize a function that probabilistically projects into the ground state. For example, taking inspiration from the linear system of equation solver, we can implement linear combinations of unitaries that simulate $H^{-1}$ as decomposed in \eqref{eq:1/x_Childs}. More generally, we can find a similar linear combination of unitaries to implement $H^{-K}$ for any $H$ with a positive spectrum~\cite{kyriienko2020quantum}, or $H^K$ if the spectrum is negative~\cite{bespalova2021hamiltonian}, either of which may require shifting the Hamiltonian by a constant.
Other alternatives include approximating $\cos^K H$~\cite{ge2019faster} or $e^{-t^2 H^2}$~\cite{keen2021quantum}. 

The key is that once a block encoding of these operators is applied, we have to amplify the success probability, via standard amplitude amplification, or the fixed point method from \cref{sec:Fixed_point_AA}. The cost of this projection algorithm will depend crucially on two main parameters: $\Delta$, the eigenvalue gap, and $\gamma$, the overlap of the initial state with the ground state. The former plays a role in the choice of parameter $K$ necessary to distinguish the ground state, while the latter will be determinant in the number of rounds of amplitude amplification required to have a success probability close to 1. Furthermore, $\Delta$ will appear divided by the Hamiltonian one-norm $\lambda$ due to the block encoding. Consequently, during amplitude amplification, the number of calls to the oracle implementing Hamiltonian simulation $U_H$ will scale as $\tilde{O}(\lambda/(\Delta \gamma))$, and $O(1/\gamma)$ calls to $U_I$, the oracle preparing the initial state, will be required. 
 

\begin{figure}[t!]
    \centering
    \includegraphics[width=1.5\textwidth/2]{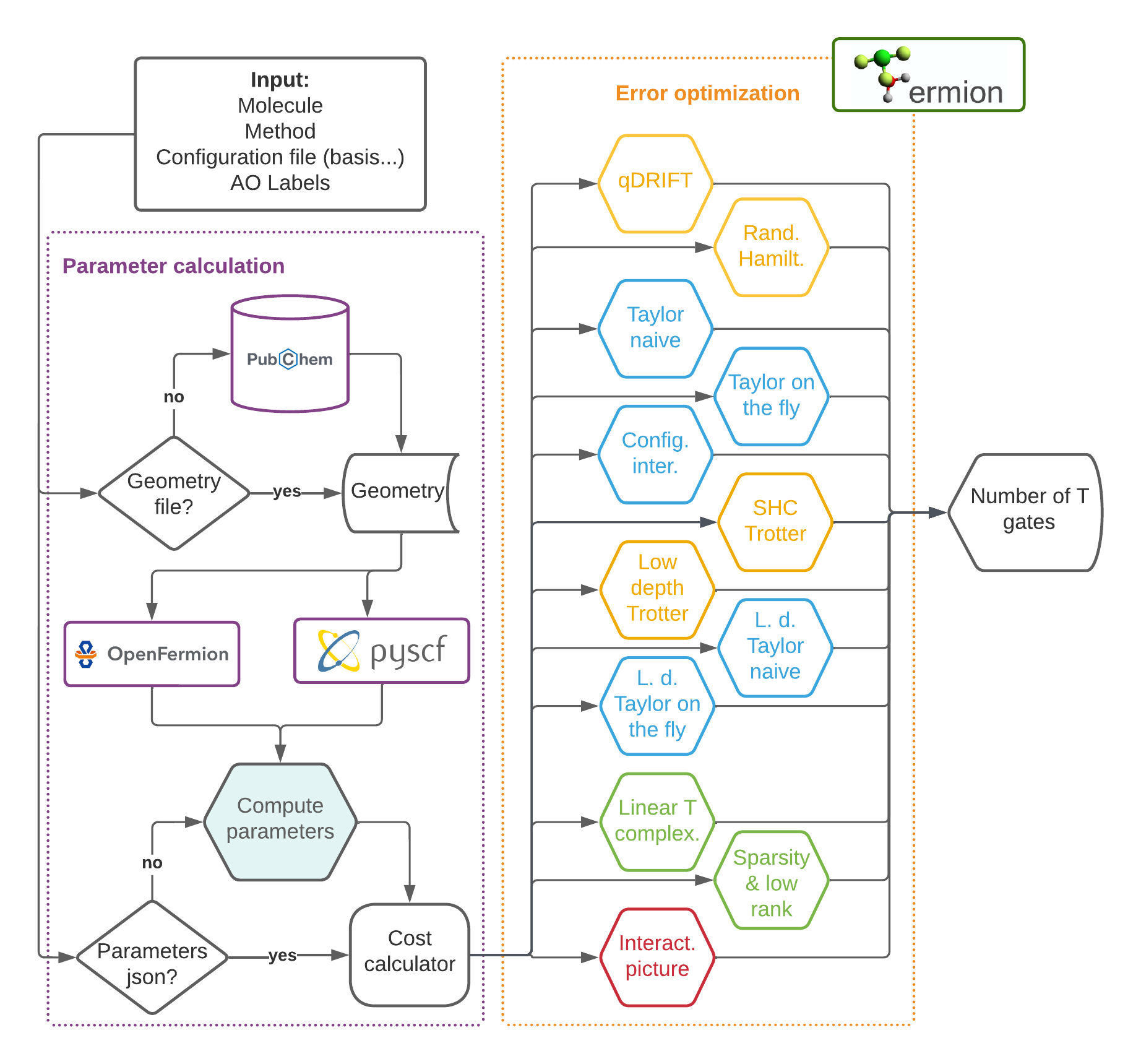}
    \caption{\textbf{Flowchart of the architecture of TFermion}~\cite{casares2021tfermion}, divided into two parts: the first one centered on the computation of the parameters needed for the cost estimate; and a second one on using such parameters to compute the number of T-gates. Methods are colored according to the Hamiltonian simulation technique used: yellow for Trotter, blue for Taylor series, green for qubitization, and red for interaction picture.}
    \label{fig:Architecture}
\end{figure}

\paragraph{\label{Qubitization_filtering}Qubitization.} Having discussed qubitization and signal processing as natural successors to the linear combination of unitaries approach, the reader might not be surprised to find out that this technique can also be used to filter out unwanted eigenstates. 
The objective, as in the previous cases, is to synthesize a Grover-like rotation that amplifies the ground state. The natural choice for such rotation is a sign function affecting only the lowest eigenstate, 
\begin{equation}
    R_{<\mu} = \sum_{k:\lambda_k < \mu}\ket{\phi_k}\bra{\phi_k}-\sum_{k:\lambda_k > \mu}\ket{\phi_k}\bra{\phi_k},
\end{equation}
and to implement one such block encoding we will need an approximation to the ground state energy and the eigenvalue gap $\Delta$, just as in the above case.\footnote{This procedure to implement reflections might also be used in \cref{sec:hitting} in general and in \cref{alg:Magniez_quantum_search} in particular, to implement quantum search algorithms without the need for quantum phase estimation, even if asymptotic complexity is similar.} Specifically, Ref.~\cite[Lemma 3]{lin2020near} shows that one needs a polynomial of degree $O\left(\frac{\lambda}{\Delta}\log \epsilon^{-1}\right)$ to approximate the sign function to error $\epsilon$, except in the segment $[\mu-\Delta/2, \mu+\Delta/2]$. This polynomial can be prepared with a similar number of queries to the Hamiltonian block encoding oracle $U_H$~\cite[Theorem 1]{lin2020near}. Furthermore, if desired, a block encoding of a projector can be prepared as
\begin{equation}
    P_{<\mu} = \frac{1}{2}(R_{<\mu}+\bm{1}),
\end{equation}
that will succeed to prepare the ground state with a probability of at least $\gamma^2$. Consequently, this also requires (fixed point) amplitude amplification. Overall, the cost is $O(\frac{\lambda}{\Delta \gamma}\log \epsilon^{-1})$ calls to the Hamiltonian block encoding $U_H$, and $O(1/\gamma)$ calls to the initial state preparation oracle $U_I$. This procedure also needs an upper bound estimate $\mu$ of the ground state energy, and if not available a binary search procedure based on limited-precision amplitude estimation must be used~\cite[Lemma 7]{lin2020near}.

\begin{figure}[t!]
    \includegraphics[width=\textwidth/2]{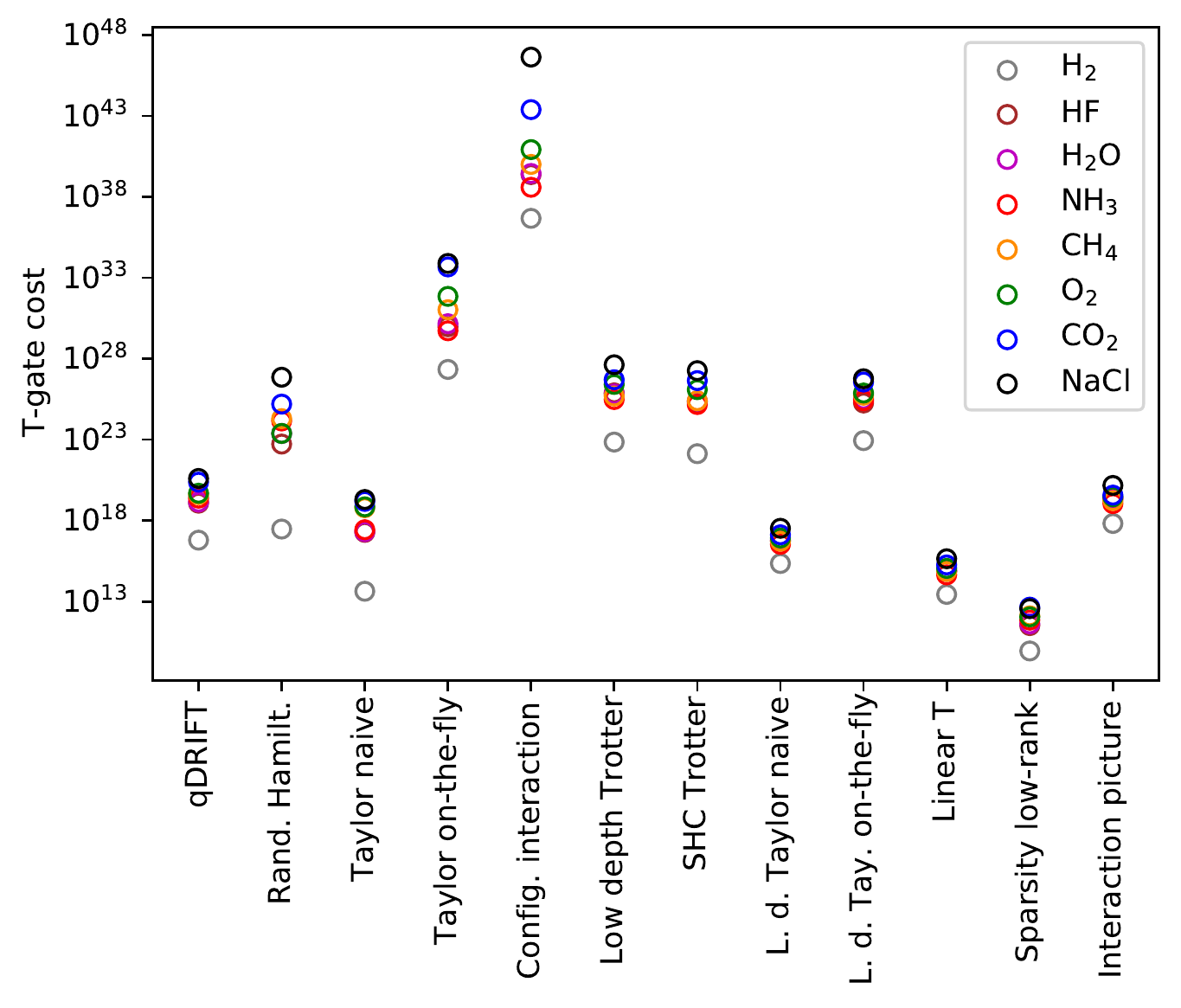}
    \includegraphics[width=\textwidth/2]{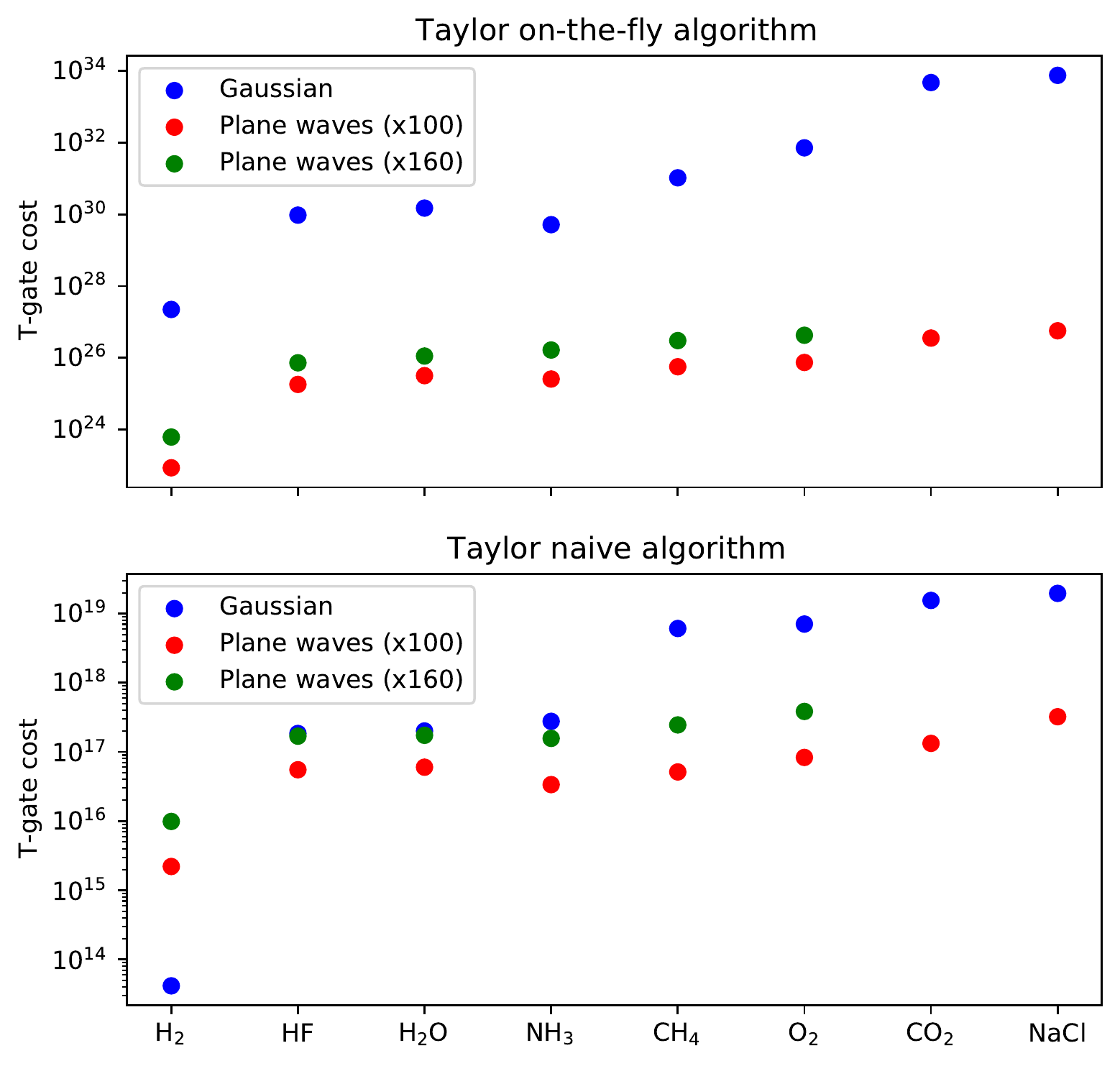}
    \caption{\textbf{Examples of the results that can be obtained with TFermion}~\cite{casares2021tfermion}. Left: the choice of basis and Hamiltonian simulation used will have a profound impact on the overall cost of the quantum phase estimation algorithm. Right: Our library allows us to compare the cost of the same algorithm with a different basis (plane waves vs Gaussians) provided a fair comparison between the number of functions on either basis can be established.}
    \label{fig:TFermion_examples}
\end{figure}

\section{\label{sec:TFermion}TFermion}

Since quantum chemistry and material science seem such good applications of quantum computing, there has been abundant research not only on developing new algorithms but on estimating the actual cost of implementing those. Starting with the seminal work of Ref.~\cite{reiher2017elucidating} on FeMoco, a molecule that is capable of fixing atmospheric nitrogen, there have been numerous articles trying to estimate the cost of using quantum phase estimation algorithms to compute chemical reaction rates~\cite{reiher2017elucidating,von2021quantum}, and analyze battery properties~\cite{kim2022fault,delgado2022simulate} or biological enzymes~\cite{goings2022reliably}. 

The cost is often measured in the number of non-Clifford gates, for example, T or Toffoli gates. T gates are $\pi/8$ single-qubit rotations, while Toffoli gates are the quantum equivalent of classical AND gates. These are the most costly gates because they cannot be implemented transversally in the surface~\cite{kitaev2003fault,fowler2012surface} or color codes~\cite{bombin2006topological,bombin2007topological}, and thus require costly magic state distillation~\cite{gidney2019efficient} or other more expensive alternatives~\cite{beverland2021cost}.

On the other hand, choosing the appropriate algorithm to tackle a given problem is not obvious and requires a deep understanding of each of the available algorithms. In this context, TFermion is a library our group has developed to facilitate the T gate cost estimation of performing phase estimation over a given molecule, with the different methods proposed in the literature~\cite{casares2021tfermion}.

In our article, we explain how several quantum phase estimation implementations can be compiled into a circuit, as well as give the first estimates of their T-gate cost. This library may find use cases such as comparing the cost of different algorithms. As an example, we analyzed the impact of the basis used for Taylor series-based quantum phase estimation, see~\cref{fig:TFermion_examples}. Note however that this comparison should be taken with care as we did not compute the amount of error induced in the algorithm due to the use of a finite number of basis functions in each case. Future extensions to the library should include newer algorithms, the number of qubits required, or the cost of state preparation.

\begin{figure}[bt!]
\includegraphics[width=.5\textwidth]{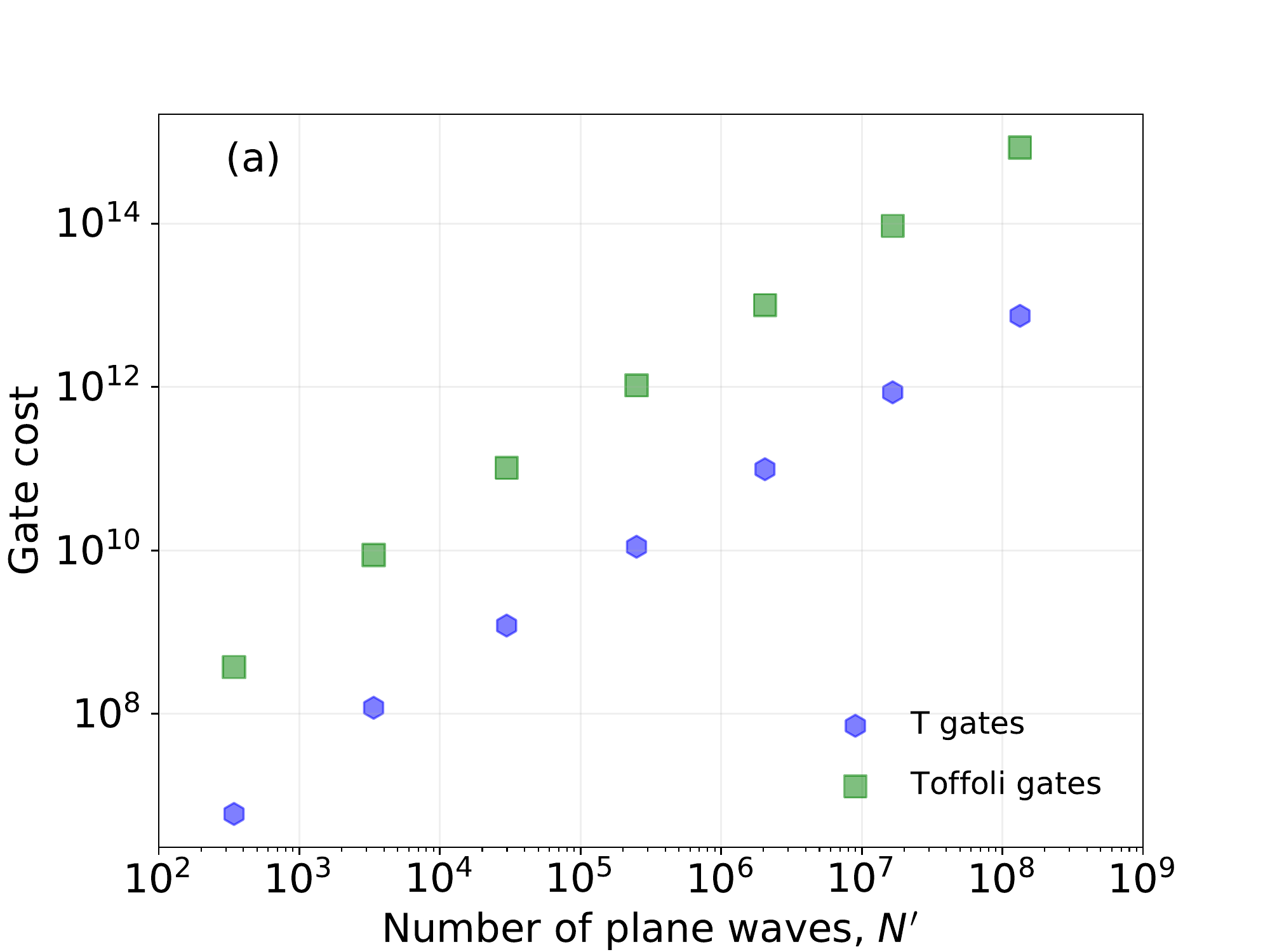}
\includegraphics[width=.5\textwidth]{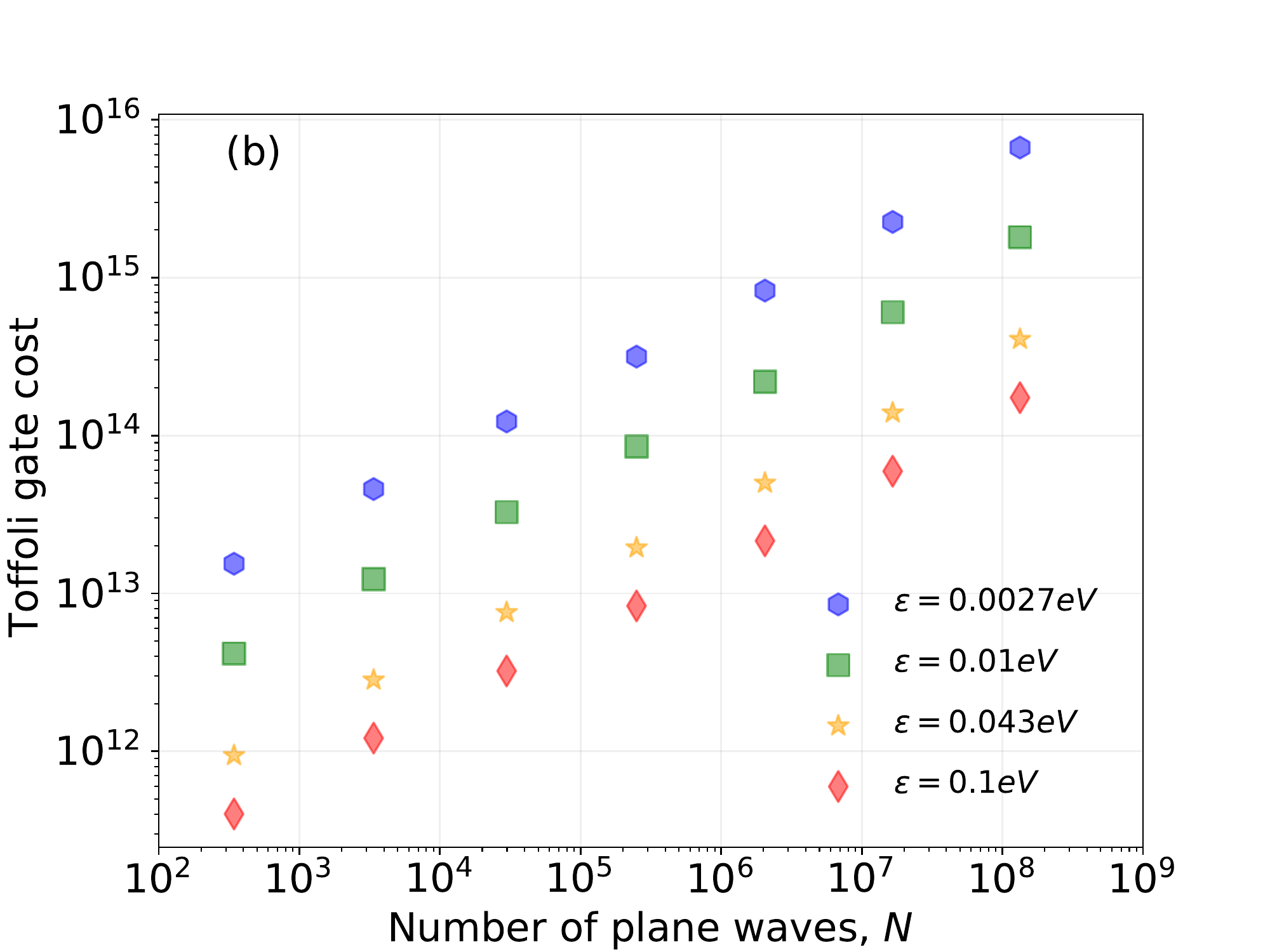}
\caption{\label{fig:Li_cost}\textbf{Non-Clifford gate cost for the execution of Quantum Phase Estimation} algorithm in a fault-tolerant quantum algorithm for $\mathrm{Li}_2\mathrm{FeSiO}_4$ cathode material~\cite{delgado2022simulate}. Left: Toffoli and T-gate cost of performing the Givens rotations explained below to perform the basis change from molecular orbital to plane wave basis, for different basis sizes. Right: Toffoli cost of the quantum phase estimation as a function of the number of basis functions, and target precision. Different error targets are required for the estimation of different properties.}
\end{figure}

\section{\label{sec:Batteries}Lithium batteries}

The second article on this topic presented in this thesis, Ref.~\cite{delgado2022simulate}, is a study on the advantages and cost of analyzing lithium battery properties using quantum phase estimation. In our objective toward decarbonization of energy generation, batteries still represent a notable technological limitation. Therefore, being able to better predict their properties computationally has substantial value. The cathode is one of the key components of the battery and will oftentimes determine some of its most important properties. For our work, we opted for $\mathrm{Li}_2\mathrm{FeSiO}_4$, a common candidate material for the cathode of batteries~\cite{manthiram2020reflection}.
During discharge, it undergoes the following chemical reaction,
\begin{equation}
\mathrm{Li}_x\mathrm{FeSiO}_4 + (2-x)\mathrm{Li} \rightarrow \mathrm{Li}_2\mathrm{FeSiO}_4.
\label{eq:cathode_reaction}
\end{equation}
In other words, $\mathrm{Li}$ atoms are inserted in the structure of the material. The voltage can then be computed as
\begin{equation}
V = -\frac{ \left[ E_{\mathrm{Li}_2\mathrm{FeSiO}_4} - E_{\mathrm{Li}_x\mathrm{FeSiO}_4} - (2-x)E_\mathrm{Li} \right]}{(2-x)e}.
\label{eq:voltage}
\end{equation}
Other properties such as the ionic mobility inside the cathode, and the thermal stability of the material, can also be predicted from the ground state energy of the different phases of the material.

In our article, we suggest using quantum phase estimation to compute these ground-state energies, which is an efficient approach as described throughout this chapter. The currently most promising implementation is based on a first quantization and plane waves representation of the quantum state, and qubitization as the Hamiltonian simulation~\cite{babbush2019quantum,su2021fault}. This choice of this combination stems from the use of plane waves to simulate a periodic material, first-quantization to reduce the one-norm of the Hamiltonian, and qubitization as a rapid Hamiltonian simulation technique.

\begin{figure}[ht!]
\centering
\includegraphics[width=1.35\textwidth/2]{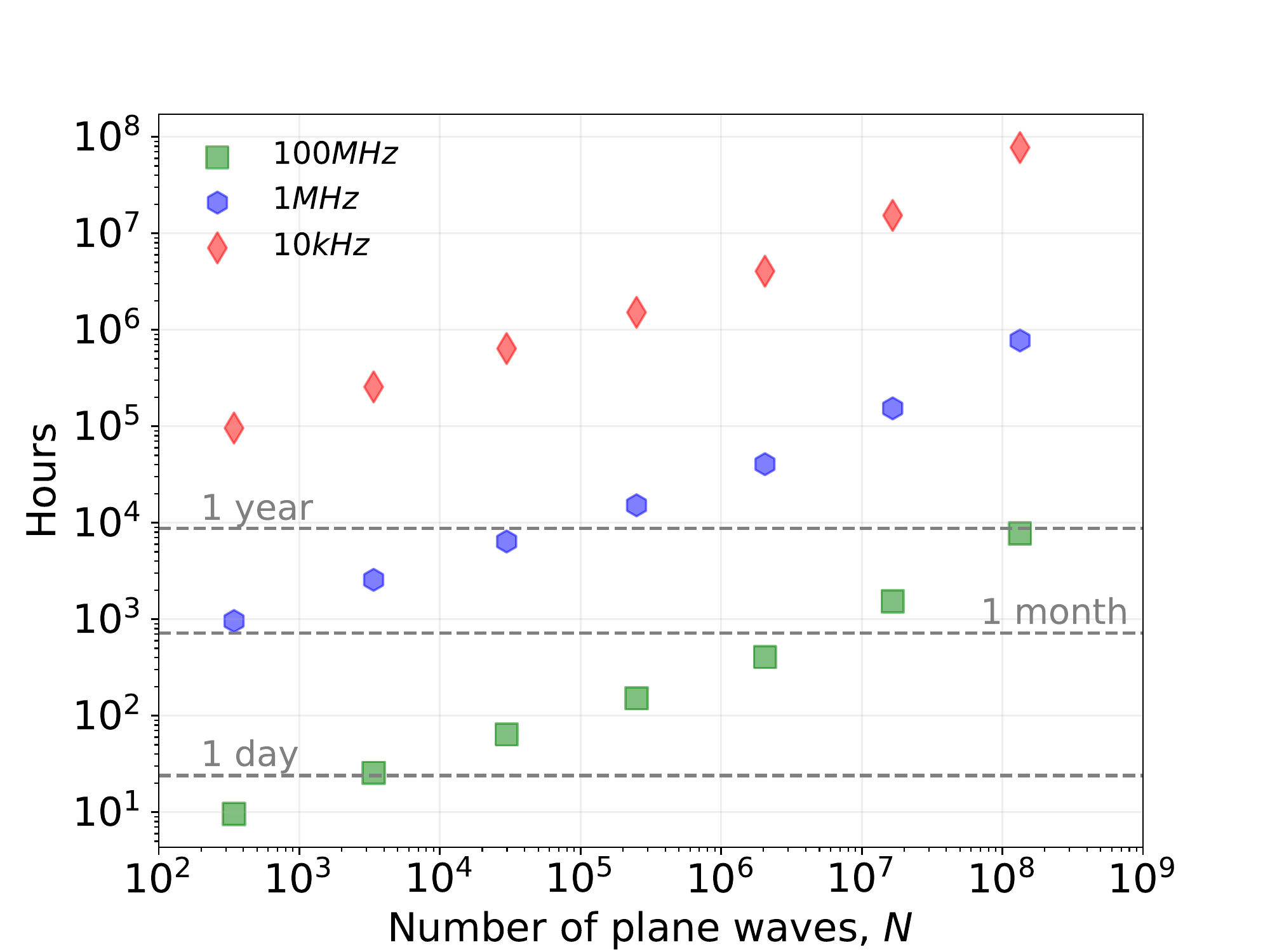}
\caption{\label{fig:time}\textbf{Execution time estimates for the qubitization quantum phase estimation algorithm}~\cite{delgado2022simulate}. Assuming chemical accuracy $\varepsilon = 0.043$ eV, this figure represents the time cost of implementing the quantum phase estimation algorithm for Li$_2$FeSiO$_4$ at varying clock rates for the synthesis of non-Clifford gates. The total number of qubits is 2,375 for $n_p=4$ and 6,652 for $n_p=9$. We compute the distillation time as the product of the number of Toffoli gates, the surface code distance $d$, and the clock frequency, all divided by a small $n_p$ factor originating from the techniques in~\cite{low2018trading} that parallelize the CSWAPs and arithmetic computations. We compute $d$ in this figure as in the moderate error case of Ref.~\cite{kim2022fault}. We emphasize that these are rough estimates whose main purpose is to provide a method to interpret the gate cost.}
\end{figure}

The main contributions of our article are three-fold: first, we slightly extend the applicability of the original algorithm from the orthonormal cubic computational cell to orthogonal, a parallelepiped. This adaptation requires minor changes in the $\Prep$ component of the algorithm, in particular of a state with amplitudes proportional to $\frac{1}{\|G_\nu\|}$, which slightly increases the cost of the algorithm.

More importantly, we explain how to perform the Givens rotations required to map the Hartree Fock from molecular orbitals to the plane wave basis in first quantization~\cite{kivlichan2018quantum} and make sure that the procedure conserves antisymmetrization. The Hartree-Fock state that we take as the initial state will only be a computational basis state in the molecular orbital basis. However, to use our Hamiltonian simulation method, we require such a state to be mapped to the plane wave basis, a procedure that has been previously explored in the literature~\cite{ortiz2001quantum,somma2002simulating,kivlichan2018quantum}. Such basis change can be written as the application of the operator
\begin{equation}
    U(u) = e^{\sum_{pq} [\log u]_{pq} a_p^\dagger a_q},
\end{equation}
according to Thouless theorem~\cite{thouless1960stability}, see~\cref{foot:Thouless}. To implement it, the Kivlichan method diagonalizes the corresponding operator matrix, decomposing it in two-orbital Givens rotations
\begin{equation}
R_{Y}(\theta_{pq}) = \begin{pmatrix}
\cos(\theta_{pq}) & -\sin(\theta_{pq}) \\
\sin(\theta_{pq}) & \cos(\theta_{pq})
\end{pmatrix},
\end{equation}
between the $\eta$ occupied orbitals and the $N-\eta$ unoccupied ones~\cite{kivlichan2018quantum}. Givens rotations are rather straightforward in second quantization, just a two-qubit rotation~\cite{arrazola2022universal}. In contrast, in first quantization, it becomes much more challenging as the information about the occupation of an orbital is delocalized among all registers. In this representation, registers track the orbital in which each electron is, not whether particular orbitals are occupied or empty. Thus, our contribution is a method to perform such rotations in first quantization.

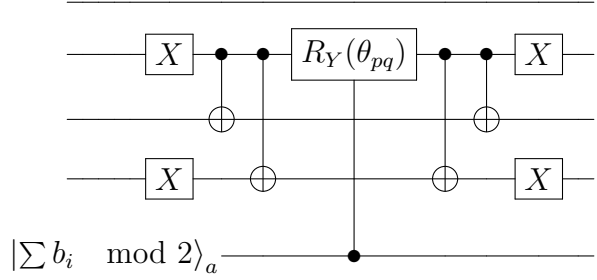
\begin{figure}[t!]
\[
\begin{array}{c}
 \Qcircuit @C=0.5em @R=0.85em { 
  &  & & & \qw& \qw& \qw & \qw & \qw & \qw & \qw & \qw & \qw & \qw & \qw &  \qw & \qw \\
  &  & & & \qw& \qw & \qw & \qw & \gate{X} & \ctrl{1} & \ctrl{2} & \gate{R_Y(\theta_{pq})} & \ctrl{2} & \ctrl{1} & \gate{X}& \qw& \qw \\ 
  &  & & &\qw& \qw & \qw & \qw & \qw & \targ & \qw & \qw & \qw & \targ & \qw & \qw & \qw  \\
  & & & & \qw& \qw& \qw & \qw & \gate{X} & \qw & \targ  & \qw & \targ & \qw & \gate{X}& \qw& \qw \\
  \\
    &   & & & & &\ket{\sum b_i \mod 2}_a   & &  &  & \qw  & \ctrl{-4} & \qw &  \qw & \qw &\qw &\qw  \\
 }
 \end{array}
\]
 \caption{\label{fig:GivensControlled}\textbf{Circuit diagram of an example controlled rotation $R_Y(\theta_{pq})$}~\cite{delgado2022simulate}. The rotation is performed on the subspace spanned by $\ket{p} = \ket{0101}$ and  $\ket{q}=\ket{0010}$. The following procedure is applied to the bits where they differ, namely the last three qubits. First, we apply $X$ gates such that $\ket{p}\rightarrow \ket{0000}$ and $\ket{q}\rightarrow\ket{0111}$. Then, CNOT gates map these states to $\ket{0000}$ and $\ket{0100}$ respectively. This allows us to perform a rotation on the second qubit controlled on the auxiliary qubit $\ket{\sum b_i \mod 2}_a$. Finally, the CNOTs and $X$ gates are uncomputed, yielding the desired controlled rotation on the subspace $\text{span}\{\ket{p} , \ket{q}\}$.}
\end{figure}

To explain how the procedure works, the key aspect to understand is that given the fermionic nature of the system, $R_Y(\theta_{pq})$ will only behave non-trivially when the occupation of $\ket{p}$ and $\ket{q}$ is different, else it acts as the identity operator. 
Consequently, we can implement the Givens rotation on $\ket{p_1,\ldots, p_\eta}$ with the following procedure~\cite{delgado2022simulate}:
\begin{enumerate}
    \item Initialize $\eta$ auxiliary qubits $\ket{b_j}_j$ in the state $\ket{0}_1 \ldots \ket{0}_{\eta}$. 
    \item For $1 \le j \le \eta$: If $p_j\in \{p,q\}$, flip the auxiliary qubit $\ket{0}_j$ to $\ket{1}_j$.
    \item For $1 \le j \le \eta-1$: Controlled on the auxiliary qubit $\ket{b_j}_j$, swap the $j$-th and $\eta$-th register.
    \item The auxiliary qubits are now in some state $\ket{b_1}_{1}\ldots\ket{b_{\eta}}_{\eta}$, where each $b_j$ indicates if $p_j\in \{p,q\}$. Controlled on the parity of $\sum_{i=1}^\eta b_i$, apply $R_Y(\theta_{pq})$ on the subspace $\text{span}\{\ket{p},\ket{q}\}$ of the $\eta$-th register. This step is illustrated for an example in Fig.~\ref{fig:GivensControlled} and can be easily generalized.
    \item Undo the controlled swaps and uncompute the auxiliary qubits by applying the same operators in steps 2 and 3.
\end{enumerate}
In the article, we show in a principled way that this procedure preserves the antisymmetry of the state, which can also be explicitly checked. 

Finally, while most elements of the algorithm presented in our article could be previously found in the literature, a third important contribution is understanding which ones to use and how to combine them into a complex algorithm capable of performing this computationally challenging task. The main conclusion of our work is that the algorithm is almost efficient enough, that provided with a quantum computer with high but achievable clock rates we could run quantum phase estimation in a reasonable amount of time for useful battery materials. However, various limitations should be addressed in future work, such as the extension to non-orthogonal cells, or going beyond Hartree-Fock as a way to prepare ground states. The latter might be especially important as the overlap between the Hartree Fock and ground states may decrease exponentially with the system size~\cite{kohn1999nobel}. Overall, this article and similar previous ones indicate that chemistry and material science might be one of the most promising applications of quantum computing.

\section{Results}

\begin{itemize}
    \item We have described Hartree-Fock, Density Functional Theory, and Coupled Cluster, and established their connections with quantum algorithms or as state preparation methods.
    \item We have described all the algorithmic choices involved in chemistry calculations, with special emphasis on Hamiltonian simulation.
    \item We have released TFermion, the library corresponding to Ref.~\cite{casares2021tfermion}, that enables researchers to easily compare quantum algorithms present in the literature, for which only complexity estimates were available. This enables, for example, the comparison of different basis functions, or Hamiltonian simulation techniques.
    \item In Ref.~\cite{delgado2022simulate}, we have described how state-of-the-art first-quantization algorithms may be applied to analyze Li-ion battery properties, such as the thermal stability or the energy capacity of the battery. First-quantization techniques are especially useful to reduce the cost of the plane wave basis. Together with qubitization, this makes for very efficient quantum algorithms. 
    \item Our results in Refs.~\cite{casares2021tfermion,delgado2022simulate} indicate that the most promising Hamiltonian simulation techniques are either:
    \begin{enumerate}
        \item Qubitization, making use of Gaussian basis functions with rank-factorization.
        \item Qubitization or interaction picture Hamiltonian simulation techniques, in plane wave basis and first quantization.
    \end{enumerate}
    Depending on the system (isolated molecules or periodic materials) one or the other might be preferable.
    \item The estimated number of logical gates for this algorithm means this is a robust quantum computing application candidate, and could be implemented in a realistic amount of time in a fault-tolerant quantum computer.
    \item In Ref.~\cite{delgado2022simulate}, we have made technical contributions to state preparations and the applicability of the first-quantization algorithm. More specifically, we have explained how to implement Givens rotations in first-quantization, without which state preparation in plane waves is not completely defined.
    \item The second technical contribution is the extension of this algorithm to orthogonal unit cells.
    \item We have understood that the main bottleneck in quantum chemistry is the state preparation problem. We have described a few techniques for that purpose.
\end{itemize}

\chapter{\label{ch:QEC}Quantum error correction}

\ifpdf
    \graphicspath{{Chapter4/Figs/Raster/}{Chapter4/Figs/PDF/}{Chapter4/Figs/}}
\else
    \graphicspath{{Chapter4/Figs/Vector/}{Chapter4/Figs/}}
\fi

\setlength{\epigraphwidth}{0.65\textwidth}
\epigraph{
    With group and eigenstate, we’ve learned to fix\newline
    Your quantum errors with our quantum tricks.
}{Daniel Gottesman, {\it \href{https://www2.perimeterinstitute.ca/personal/dgottesman/sonnet.html}{Quantum Error Correction Sonnet}.}}

\section{Objectives}

\begin{itemize}
    \item Understanding why quantum error correction plays a key enabling role in quantum chemistry, and why non-Clifford gates are often the most expensive kind of logic gate.

    \item Understanding the topological error codes and their two most important representatives: surface and color codes.

    \item Describing a procedure of how quantum fault tolerance might be achieved with error correction techniques.

    \item Analyzing how a Machine Learning decoder can be flexibly used with various topological codes.
\end{itemize}

\section{\label{sec:Error_codes}Introduction to error codes}

In the previous chapters, we have reviewed the different families of algorithms that one may wish to implement, and how they relate to each other. In particular, in \cref{ch:Chemistry} we have explored the use of these algorithms, and have explored the time and resource cost of fault-tolerantly implementing them. There, we referred to the necessity of distilling magic states to implement T gates. In this chapter, we explain the most important topological quantum error correction codes, the leading approach to ensure that quantum computing protocols can be implemented fault tolerantly. We start describing in the next section some basic concepts of quantum error correction, following the presentation in Ref.~\cite[Chapter 7]{preskill1999lecture}.

The objective of error correction might appear challenging: not only do we have to prevent decoherence of our data, but we have to ensure that we can fault tolerantly implement a continuous set of gates. Unfortunately, being continuous, even minor errors in such gates would accumulate, ultimately leading to computational failure~\cite{preskill1999lecture}. However, even if such is the case, we can decompose the possible errors in a discrete set. To see how, imagine we have an encoded qubit $\ket{\psi} = a\ket{0} + b\ket{1}$. Further, suppose that there is an arbitrary unitary operator $U$ that acts upon our qubit, and an environmental qubit is assumed to be initialized to $0$ on some basis, $\ket{0}_E$, and to which we have no access. In general, the action of $U$ is
\begin{equation}
\begin{split}
U: &\ket{0}\ket{0}_E\mapsto \ket{0}\ket{e_{00}}_E + \ket{1}\ket{e_{01}}_E\\
   &\ket{1}\ket{0}_E\mapsto \ket{0}\ket{e_{10}}_E + \ket{1}\ket{e_{11}}_E,
\end{split}
\end{equation}
where $\ket{e_{ij}}$ need not be normalized or orthogonal states. This means that $U$ transforms an arbitrary quantum state as 
\begin{equation}
\begin{split}
    U:  (a\ket{0} + b\ket{1})\ket{0}_E \mapsto &a(\ket{0}\ket{e_{00}}_E + \ket{1}\ket{e_{01}}_E) \\ + &b(\ket{0}\ket{e_{10}}_E + \ket{1}\ket{e_{11}}_E).
\end{split}
\end{equation}
We can rewrite the resulting state as
\begin{equation}
\begin{split}
    &= (a\ket{0} + b\ket{1})\otimes \frac{1}{2}(\ket{e_{00}}_E + \ket{e_{11}}_E)\\
    &+ (a\ket{0} - b\ket{1})\otimes \frac{1}{2}(\ket{e_{00}}_E - \ket{e_{11}}_E)\\
    &+ (a\ket{1} + b\ket{0})\otimes \frac{1}{2}(\ket{e_{01}}_E + \ket{e_{10}}_E)\\
    &+ (a\ket{1} - b\ket{0})\otimes \frac{1}{2}(\ket{e_{01}}_E - \ket{e_{10}}_E).
\end{split}
\end{equation}
Taking a closer look, this is equivalent to
\begin{equation}
    \bm{1}\ket{\psi}\otimes\ket{e_{\bm{1}}} + X\ket{\psi}\otimes\ket{e_{X}}+Y\ket{\psi}\otimes\ket{e_{Y}} + Z\ket{\psi}\otimes\ket{e_{Z}},
\end{equation}
where $X$, $Y$ and $Z$ represent the Pauli operators. We can always perform such an expansion because the set of Pauli operators and the identity span the state of $2\times 2$ matrix. For $n$-qubit states $\ket{\psi}$, we can similarly expand any error in the basis of tensor products of Pauli operators, $E_a\in \{\bm{1},X,Y,Z\}^{\otimes n}$ such that
\begin{equation}\label{eq:error_superoperator}
    \ket{\psi}\otimes \ket{0}_E \mapsto \sum_a E_a \ket{\psi}\otimes \ket{e_a}_E,
\end{equation}
where $\ket{e_a}_E$ need not be mutually orthogonal. If we are able to project into one of the $E_a$ possibilities and distinguish which one it was, we can implement $E_a^\dagger$ to correct our data qubits (not the environmental ones).

The errors we will aim to correct are a subset $\mathcal{E}\subset \{E_a\}$. If $E_a$ is composed of $t$ non-trivial Pauli operators, then we will say it has \textit{weight} $t$. In a binary (qubit) code, we encode $2^k$ `code words' $\ket{\bar{i}}$ in a $n$-qubit space. The code is further characterized by the \textit{code distance} $d$, the minimum weight of the operator changing the code word, e.g., 
\begin{equation}\label{eq:distance}
    d = \min_{E_a\in \mathcal{E}} t(\bm{E_a}): \braket{\bar{i}|E_a|\bar{j}} \neq C_{a}\delta_{ij}, 
\end{equation}
with $C_a$ a normalization coefficient.
Overall, quantum error corrections codes will be denoted by their properties $[[n,k,d]]$, where the double bracket identifies them in contrast to single-bracket classical codes. In general, if errors $E_a$ and $E_b$ have weight smaller than $d/2$, 
\begin{equation}
    \braket{\bar{i}|E_b^\dagger E_a|\bar{j}} = C_{ab}\delta_{ij}
\end{equation}
with $C_{ab}$ an arbitrary Hermitian matrix, is a necessary and sufficient condition to be able to recover the correct code word~\cite[Equation 7.19]{preskill1999lecture}. In other words, if the error $E_a$ has weight less than $d/2$, we will be able to correct the error using the correction procedure $E_b^{\dagger}$, which can only recover the code word $\ket{\bar{i}}$ because $E_b^{\dagger}E_a$ has weight smaller than $d$. In contrast, if one error had a weight equal to or larger than $d/2$, then it could be misidentified and corrected as a different code word.

As an introduction to quantum error correction codes, let us review one important family of classical codes, binary linear codes. In these codes, the code subspace is spanned by a set of binary vectors $\{v_i\}$, such that any message $\alpha_1,\ldots,\alpha_k$ is encoded as
\begin{equation}
    \alpha_1,\ldots,\alpha_k \mapsto \sum_i \alpha_i v_i,
\end{equation}
where summation is carried out modulo two. These generating vectors might be written as a $k\times n$ \textit{generating matrix}, $G$
\begin{equation}
    G = \begin{pmatrix}
    v_1\\
    v_2\\
    \vdots\\
    v_k
    \end{pmatrix}
\end{equation}
and encoded message as $v(\alpha) = \alpha G$. Alternatively, one can indicate $n-k$ linear constraints, which form a $(n-k)\times n$ \textit{parity check matrix} $H$, such that $Hv = 0$ for all the generating vectors. Consequently, we also have $HG^T = 0$. The parity check matrix can be used to detect errors $e$ in the code, as in that case $H(v+e) = He\neq 0$ will indicate an error syndrome.

The distance $d$ in these codes is defined as the minimum Hamming weight of vectors $\{v_i\}$, defined as the number of non-zero elements of the vectors,
\begin{equation}
    d = \min_{i} |v_i|_1.
\end{equation}
Similarly to our discussion above, the correctable errors are those of weight $t< d/2$~\cite{preskill1999lecture}. From this code $C$, we can generate its \textit{dual}, by taking the transpose of $HG^T = 0$, $GH^T=0$. In the new $n-k$ code $C^\perp$, $H^T$ is the generating matrix and $G$ is the parity check matrix.

Can we say anything more about the relation between the primal and its dual code? Since $GH^T = 0$, $\forall v\in C$ and $\forall u\in C^\perp$, we know that $v\cdot u = 0$, and consequently $(-1)^{v\cdot u} = 1$. On the other hand, if $u\notin C^\perp$ but $v = \alpha G$, it is clear that~\cite{preskill1999lecture}
\begin{equation}
    \sum_{v\in \{0,1\}^k} (-1)^{v\cdot w} = 0, \quad \forall w\neq 0\quad\Rightarrow\quad
    \sum_{v\in C}(-1)^{v\cdot u} = \sum_{\alpha \in \{0,1\}^k} (-1)^{\alpha\cdot G u} = 0.
\end{equation}
In summary,
\begin{equation}\label{eq:C_C^perp_relation}
    \sum_{v\in C}(-1)^{v\cdot u} = \begin{cases}
    2^k & u\in C^\perp,\\
    0 & u\notin C^\perp.
    \end{cases}
\end{equation}
As the eager reader might have noticed, this poses a relation between both codes via the Hadamard transformation, which we explore next.

\subsection{\label{ssec:CSS}CSS codes}
The concept of dual codes can be exploited to generate one family of quantum codes called Calderbank–Shor–Steane (or CSS) codes \cite{nielsen2002quantum}. The construction is as follows: let $C_1$ be one code defined by $(n-k_1)\times n$ parity check matrix $H_1$, and similarly $C_2$ with $(n-k_2)\times n$ parity check matrix $H_2$. We choose $C_2$ to be a subcode of $C_1$, by imposing that all constraints of $C_1$ are also obeyed by $C_2$, so $k_2<k_1$. This construction allows defining equivalence classes in $C_1$: two code words $u,v\in C_1$ are equivalent, if they are the same up to an element of $C_2$ \cite{preskill1999lecture},
\begin{equation}
     u\equiv v \iff \exists w\in C_2: u+w = v.
\end{equation}
A CSS code is a code encoding $k_1-k_2$ logical bits, where each equivalence class is one of the $2^{k_1-k_2}$ code word. The basis elements are
\begin{equation}
    \ket{\bar{w}}_F = \frac{1}{\sqrt{2^{k_2}}}\sum_{v\in C_2}\ket{v+w},
\end{equation}
where the subindex denotes that this basis will be able to protect against bit-flip errors.
As hinted on the discussion on the dual code, we can transform the code $C$ on its dual code $C^\perp$ using the bitwise Hadamard over all qubits
\begin{equation}
\begin{split}
    \ket{\bar{w}}_P = &H_2^{\otimes n}\ket{\bar{w}}_F  = \frac{1}{\sqrt{2^{n}}}\sum_u \frac{1}{\sqrt{2^{k_2}}}\left(\sum_{v\in C_2}(-1)^{u\cdot v}\right)(-1)^{u\cdot w}\ket{u}\\
    \stackrel{\eqref{eq:C_C^perp_relation}}{=}&\frac{1}{\sqrt{2^{n-k_2}}}\sum_{u\in C_2^\perp}(-1)^{u\cdot w}\ket{u}.
\end{split}
\end{equation}
Furthermore, if we shift $w$ by an element $c\in C_2$ (e.g., staying in the same equivalence class), the resulting $\ket{\bar{w}}_P$ will not change because $(-1)^{u\cdot (w+c)}= (-1)^{u\cdot w}(-1)^{u\cdot c} = (-1)^{u\cdot w}$ because $u\cdot c = 0$.

Given that the Hadamard is self-inverse, we can use it to move back and forth between the primal and dual representations, and consequently correct the phase-flip errors as bit-flip errors in the dual code. The distance for flip errors will be the minimum weight of generating vectors of $C_1$, while for phase errors it will be the minimum weight of generating vectors of $C_2^\perp$ \cite{preskill1999lecture}. The overall distance will consequently be the smallest of either.

\subsection{\label{ssec:Stabilizer}Stabilizer codes}

Another particularly important class of quantum error correction codes are so-called called stabilizer codes, which make use of the Pauli structure of errors in \eqref{eq:error_superoperator}.
Let $P = \pm \{\bm{1},X,Z,XZ =iY\}^{\otimes n}$ be the group of the $n$-fold tensor product of Pauli operators, with $iY$ in place of $Y$ so that all matrix entries are real; and let $\mathcal{S}\subset P$ be an Abelian subgroup. Then, since the elements commute and can be jointly diagonalized, we define the \textit{stabilizer code} as the $+1$ eigenspace of $\mathcal{S}$, $\mathcal{H}_\mathcal{S}$~\cite{preskill1999lecture}:
\begin{equation}
    \psi\in \mathcal{H}_\mathcal{S}\iff M\psi = \psi\quad \forall M\in \mathcal{S}.
\end{equation}
In particular, $-\bm{1}\notin \mathcal{S}$, because it has no $+1$ eigenvalue. $\mathcal{S}$ can be characterized by its generators, $\{M_i\}$. These operators can be understood as the parity check operators of the code. 

Errors, on the other hand, will map the state outside the joint $+1$ eigenstate of all generators. For each $E_a$ there will be at least one generator $M_i$ such that
\begin{equation}
    M_i E_a\ket{\psi} = -E_a\ket{\psi} = -E_a M_i\ket{\psi}.
\end{equation}
In that case, we see that the error and generator anticommute, instead of commute. This allows us to define an \textit{error syndrome} $s_{ia}$
\begin{equation}
    M_i E_a = (-1)^{s_{ia}}E_a M_i,
\end{equation}
which will be $s_{ia} = 0$ if error $E_a$ commutes with the stabilizer, and $s_{ia} = 1$ if it anticommutes.
Once identified thanks to the syndrome, we can apply a recovery procedure $E_b$ that will hopefully recover the encoded state by mapping us back to $\mathcal{H}_\mathcal{S}$.

Finally, we have to explain how logical operators will act on the code. They will be related to the normalizer of $\mathcal{S}$.
\begin{definition}[Normalizer and centralizer]
The normalizer of $\mathcal{S}$ in group $G$, $\mathcal{N}(\mathcal{S})$ is defined as
\begin{equation}
    \mathcal{N}(\mathcal{S}) = \{g\in G| g\mathcal{S} = \mathcal{S}g\} = \{g\in G| g\mathcal{S}g^{-1} = \mathcal{S}\}.
\end{equation}
In other words, the normalizer is the set of elements of the group $G$ that take components of $\mathcal{S}$ to possibly different ones of $\mathcal{S}$ under conjugation.
The centralizer is similar, but commutes element-wise with each element of $\mathcal{S}$
\begin{equation}
    \mathcal{Z}(\mathcal{S}) = \{g\in G|\forall s\in \mathcal{S}, gs = sg\} = \{g\in G| \forall s\in \mathcal{S}, gsg^{-1} = s\}.
\end{equation}
If $G$ is not clear, then we will indicate it by explicitly writing $\mathcal{Z}_G(\mathcal{S})$ and $\mathcal{N}_G(\mathcal{S})$.
\end{definition}
The group $G$ will often be the Pauli group over the physical qubits, for instance in the stabilizer surface and color codes.
Since $\mathcal{S}$ is Abelian, $\mathcal{S}\subset \mathcal{N}(\mathcal{S})$. Moreover, in the stabilizer codes we will explore next, $\mathcal{Z}(\mathcal{S})=\mathcal{N}(\mathcal{S})$.
Why is the normalizer important? It will represent all elements of $P$ that commute with but are not necessarily in $\mathcal{S}$. Since normalizer operators commute with components of the stabilizer, their syndrome will always be $+1$. On the other hand, since they do not necessarily belong to $\mathcal{S}$, they can induce logical changes in the encoded information. Thus, logical operators in stabilizer codes will be related to the $\mathcal{N}(\mathcal{S})/\mathcal{S}$ quotient group.

\section{\label{sec:Topological_error_correction}Topological error correction}

Topological codes are the most important family of both CSS and stabilizer codes~\cite{bombin2013introduction}. In these codes, the physical qubits are represented as some element of a lattice embedded in a differentiable manifold. Such elements are often given the name of $d$-cells, where $d$ represents the dimension. For example, $0$-cells are vertices, and $1$-cells are edges... Topological codes receive this name because they encode logical information in topologically non-trivial objects in the manifold. Since topological elements are robust to local deformations, these codes display great robustness and are often considered a key aspect of almost any error correction scheme. The two most important codes are the toric code (also known as surface code)~\cite{kitaev2003fault} and the color code~\cite{bombin2006topological}. In this section, we mostly follow the presentation by Ref.~\cite{bombin2013introduction} of these codes.

\subsection{\label{ssec:Homology}Homology}

Let us start defining some concepts that will, later on, be useful for understanding these error correction schemes. The first concept we will introduce are $d$-chains, subsets of $d$-cell elements $e^d_i$ that can be understood as a formal sum
\begin{equation}
    c^d = \sum_{i}c_i e_i^d, \quad c_i \in \{0,1\}.
\end{equation}
The set of $d$-chains will be called $C_d$, and has Abelian subgroup structure. In fact, using the notation $\Z_2 = \Z \mod 2$, then $C_0 \simeq \Z_2^V$ for $V$ the set of vertices. Similarly happens for $C_1$ chains, made of edges $E$, $C_2$ made of faces $F$, or $C_3$ made of cells $C$.
$d$-chains with different dimensions are connected by the boundary operators,
\begin{equation}
    \partial_d : C_d \mapsto C_{d-1},
\end{equation}
which are group homomorphisms, or in other words $\partial_d(e_i^d+ e_j^d) = \partial_d(e_i^d) + \partial_d(e_j^d)$. These operators map each $d$-chain to its boundary, for example they map an edge to its two boundary vertices.

There are also two very important classes of $d$-chains. First, we have $d$-\textit{cycles} $z^d\in Z_d\subset C_d$, defined by those $d$-chains that have no boundary: $\partial_d z^d = 0$. Second, we have $d$-\textit{boundaries} $b^d\in B_d\subset C_d$, which are themselves the boundary of a $d+1$-chain. In other words,
\begin{equation}
    Z_d = \ker \partial_d, \qquad B_d = \Ima \partial_{d+1}.
\end{equation}
Since $\partial_d \circ \partial_{d+1} = 0$, boundaries are also cycles $B_d\subset Z_d$. This relation allows defining a homology group structure,
\begin{equation}
    H_d := \frac{Z_d}{B_d} = \frac{\ker \partial_d}{\Ima \partial_{d+1}}.
\end{equation}
Interestingly, $H_d$ will only depend on the topology of the system. For example, if the manifold is an orientable surface, its Euler characteristic is
\begin{equation}
    \chi = V- E+F
\end{equation}
and the genus of the surface is\footnote{The genus is a topologically invariant of a surface. It indicates the largest number of nonintersecting topologically non-trivial closed curves that the surface can contain.}
\begin{equation}
    g = 1-\chi/2.
\end{equation}
The genus is related to $H_1$ by $H_1\simeq \Z_2^{2g}$ \cite{bombin2013introduction}.

\subsection{\label{ssec:Surface}Surface code}

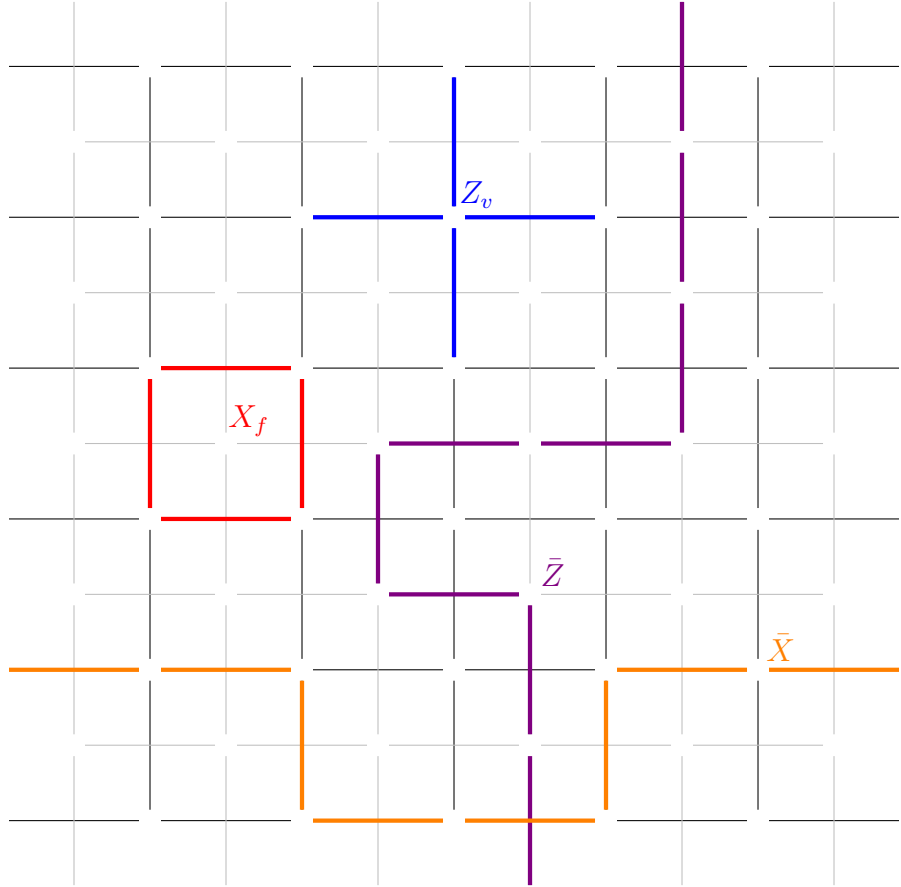
\begin{figure}[!t]
\centering
\begin{tikzpicture}
      
  \foreach \x in {0,...,6}
    \foreach \y in {0,...,5} 
       {\pgfmathtruncatemacro{\label}{\x - 5 *  \y +21}
       \node  (a\x\y) at (2*\x+2,2*\y+2) {};} 

  \foreach \x in {1,...,5}
    \foreach \y [count=\yi] in {0,...,4}  
      \draw[black, thin] (a\x\y)--(a\x\yi);
      
  \foreach \x in {0,...,5}
    \foreach \y [count=\yi] in {0,...,5}  
      \draw[black, thin] (a\y\x)--(a\yi\x);
      
  \foreach \x in {0,...,5}
    \foreach \y in {0,...,6} 
       {\pgfmathtruncatemacro{\label}{\x - 5 *  \y +21}
       \node  (b\x\y) at (2*\x+3,2*\y+1) {};}
       
  \foreach \x in {0,...,5}
    \foreach \y [count=\yi] in {0,...,5}  
      \draw[lightgray, thin] (b\x\y)--(b\x\yi);
      
  \foreach \x in {1,...,5}
    \foreach \y [count=\yi] in {0,...,4}  
      \draw[lightgray, thin] (b\y\x)--(b\yi\x);

  \draw[red, ultra thick] (a12)--(a13);
  \draw[red, ultra thick] (a12)--(a22);
  \draw[red, ultra thick] (a22)--(a23);
  \draw[red, ultra thick] (a13)--(a23);
  \node[red] at (5.3,7.3) {$X_f$};

  \draw[blue, ultra thick] (a35)--(a34);
  \draw[blue, ultra thick] (a33)--(a34);
  \draw[blue, ultra thick] (a24)--(a34);
  \draw[blue, ultra thick] (a44)--(a34);
  \node[blue] at (8.3,10.3) {$Z_v$};

  \draw[violet, ultra thick] (b46)--(b45);
  \draw[violet, ultra thick] (b45)--(b44);
  \draw[violet, ultra thick] (b44)--(b43);
  \draw[violet, ultra thick] (b33)--(b43);
  \draw[violet, ultra thick] (b23)--(b33);
  \draw[violet, ultra thick] (b22)--(b23);
  \draw[violet, ultra thick] (b32)--(b22);
  \draw[violet, ultra thick] (b32)--(b31);
  \draw[violet, ultra thick] (b31)--(b30);
  \node[violet] at (9.3,5.3) {$\bar{Z}$};
  
  \draw[orange, ultra thick] (a01)--(a11);
  \draw[orange, ultra thick] (a11)--(a21);
  \draw[orange, ultra thick] (a21)--(a20);
  \draw[orange, ultra thick] (a20)--(a30);
  \draw[orange, ultra thick] (a30)--(a40);
  \draw[orange, ultra thick] (a40)--(a41);
  \draw[orange, ultra thick] (a41)--(a51);
  \draw[orange, ultra thick] (a51)--(a61);
  \node[orange] at (12.3,4.3) {$\bar{X}$};

\end{tikzpicture}
  \caption{\textbf{Surface code} of $d=6$ encoding a single logical qubit with primal (gray) and dual (light gray) lattices. Each edge in the primal or dual lattice is a qubit. Depicted also two stabilizers $X_f, Z_v\in \mathcal{S}$, and representatives of the two logical operator classes $\bar{X},\bar{Z}\in \mathcal{N}(\mathcal{S})/\mathcal{S}$, which commute with the former because either they share an even number of qubits (eg $\bar{Z}$ and $X_f$), or because they are composed of the same Pauli operators ($\bar{Z}$ and $Z_v$). Note how $\bar{Z}$ ($\bar{X}$) starts and finishes in soft (rough) boundaries.}
  \label{fig:Surface_code}
\end{figure}

The surface code was the first topological code proposed~\cite{kitaev2003fault}, and arguably the most promising near-future quantum error-correction code, due to its high threshold~\cite{fowler2012surface}. It is also sometimes called the toric code because in its simplest version it is embedded in a torus. While it can be extended to higher dimensions, we will first explain how to describe the 2$d$ toric version. The first key component of the code is deciding where to place the qubits, which in this code will correspond to edges. For every 1-chain of qubits $c$, we can define $X$ and $Z$ Pauli operators
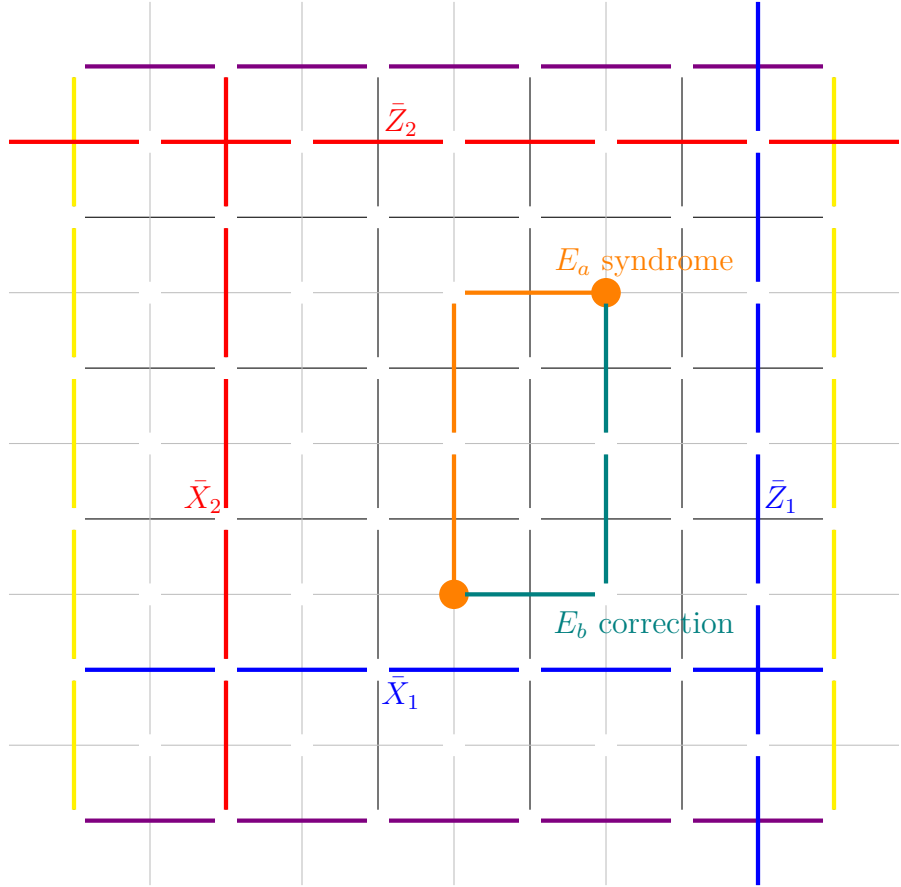
\begin{figure}[!t]
\centering
\begin{tikzpicture}
      
  \foreach \x in {0,...,5}
    \foreach \y in {0,...,5} 
       {\pgfmathtruncatemacro{\label}{\x - 5 *  \y +21}
       \node  (a\x\y) at (2*\x+2,2*\y+2) {};} 

  \foreach \x in {0,...,5}
    \foreach \y [count=\yi] in {0,...,4}  
      \draw[black, thin] (a\x\y)--(a\x\yi);
      
  \foreach \x in {0,...,5}
    \foreach \y [count=\yi] in {0,...,4}  
      \draw[black, thin] (a\y\x)--(a\yi\x);
      
  \foreach \x in {0,...,6}
    \foreach \y in {0,...,6} 
       {\pgfmathtruncatemacro{\label}{\x - 5 *  \y +21}
       \node  (b\x\y) at (2*\x+1,2*\y+1) {};}
       
  \foreach \x in {1,...,5}
    \foreach \y [count=\yi] in {0,...,5}  
      \draw[lightgray, thin] (b\x\y)--(b\x\yi);
      
  \foreach \x in {1,...,5}
    \foreach \y [count=\yi] in {0,...,5}  
      \draw[lightgray, thin] (b\y\x)--(b\yi\x);
      
  \foreach \x [count=\xi] in {0,...,4}
      \draw[violet, ultra thick] (a\x0)--(a\xi0);
  \foreach \x [count=\xi] in {0,...,4}
      \draw[violet, ultra thick] (a\x5)--(a\xi5);
      
  \foreach \x [count=\xi] in {0,...,4}
      \draw[yellow, ultra thick] (a0\x)--(a0\xi);
  \foreach \x [count=\xi] in {0,...,4}
      \draw[yellow, ultra thick] (a5\x)--(a5\xi);
      
  \foreach \x [count=\xi] in {0,...,4}
      \draw[blue, ultra thick] (a\x1)--(a\xi1);
  \node[blue] at (6.3,3.7) {$\bar{X}_1$};
  
    \foreach \x [count=\xi] in {0,...,4}
      \draw[red, ultra thick] (a1\x)--(a1\xi);
  \node[red] at (3.7,6.3) {$\bar{X}_2$};
      
  \foreach \x [count=\xi] in {0,...,5}
      \draw[red, ultra thick] (b\x5)--(b\xi5);
  \node[red] at (6.3,11.3) {$\bar{Z}_2$};
  
 \foreach \x [count=\xi] in {0,...,5}
      \draw[blue, ultra thick] (b5\x)--(b5\xi);
  \node[blue] at (11.3,6.3) {$\bar{Z}_1$};

  \draw[orange, ultra thick] (b44)--(b34);
  \draw[orange, ultra thick] (b34)--(b33);
  \draw[orange, ultra thick] (b33)--(b32);
  \node at (9,9)[circle,fill,orange]{};
  \node[orange] at (9.5,9.4) {$E_a$ syndrome};
  \node at (7,5)[circle,fill,orange]{};
  
  \draw[teal, ultra thick] (b44)--(b43);
  \draw[teal, ultra thick] (b43)--(b42);
  \draw[teal, ultra thick] (b42)--(b32);
  \node[teal] at (9.5,4.6) {$E_b$ correction};

\end{tikzpicture}
  \caption{\textbf{Toric code} with distance $d = 5$. The opposite sides of the lattice are identified. Two logical qubits can be encoded in this code, whose logical Pauli operators are depicted in blue and red. If an error happens, it will leave behind two excited stabilizers (in this case of type $X$). We can correct it by implementing $E_b$ via one of the shortest paths, such that the probability of a phase or bit flip is minimized. In the depicted case, $E_bE_a\in B_1^*$, so it does not change the logical information.  }
  \label{fig:Toric_code}
\end{figure}
\begin{equation}
    X_c = \bigotimes_{i\in c} X_i, \qquad Z_c = \bigotimes_{i\in c} Z_i, \qquad c \in C_1.
\end{equation}
From our discussion in the previous section, we define the code as states in the Homology class $H_1$. For example, if $z\in Z_1$ a closed curve without boundary, we define a logical qubit state as
\begin{equation}
    \ket{\bar{z}} = \sum_{b\in B_1}\ket{z+b} \Rightarrow \ket{\bar{z}}\in H_1.
\end{equation}
This is what gives the surface code its CSS character, identifying $B_1$ with the $C_2$ code, and $Z_1$ with the $C_1$ code, in the definition of the CSS code.
Since topological codes are also stabilizer codes, we define the stabilizer operators:
\begin{equation}
    X_f = \prod_{e\in\partial_2f}X_e, \qquad Z_v = \prod_{e|v\in\partial_1e}Z_e, \qquad \mathcal{S} = \{X_f,Z_v\}
\end{equation}
for $f$ indicating a face, and $e$ an edge, and $v$ a vertex, see \cref{fig:Surface_code}. These stabilizer generators commute with each other because they share an even number of edges, and are subject to the constraints
\begin{equation}\label{eq:Surface_constraints}
    \prod_f X_f = 1, \qquad \prod_v Z_v = 1.
\end{equation}
The number of encoded qubits will be $k = 2g$ because $|H_1| = 2^{2g}$ as we saw at the end of \cref{ssec:Homology}, or we may alternatively compute it as the number of qubits minus the number of independent stabilizers, $k = E- (V+F-2) = 2-\chi = 2g$. 

We can use the transversal Hadamard gate, as described previously, to transform the code into its dual, commonly denoted with an asterisk. Thus, we obtain dual boundary operators
\begin{equation}
    \partial_d^* : C_{d-1}^*\mapsto  C_{d}^*
\end{equation}
and dual cycles $Z_1^*$ and boundaries $B_1^*$. Conjugating by the transversal Hadamard also maps $X_f\to Z_{f^*}$ and $Z_v\to X_{v^*}$. In summary, this gate translates us to the dual code, which in this case is again the surface code.
Using both the primal and dual codes, we can explore the form of the logical operators in this code. Our objective is to relate $Z_1/B_1$ and $\mathcal{N}(\mathcal{S})/\mathcal{S}$.
One Pauli operator in the code can be written as
\begin{equation}
    A = i^\alpha X_c Z_{c^*},\qquad (c,c^*)\in C_1\times C_1^*, \qquad \alpha \in \Z\mod 4 
\end{equation}
This operator will be in $\mathcal{N}(\mathcal{S})$ if it commutes with all the operators in $\mathcal{S}$. In particular
\begin{equation}
    [X_c,Z_v] = 0 \iff v\notin \partial_1 c, \qquad [Z_{c^*},X_{f^*}] = 0 \iff f^*\notin \partial_2 c^*.
\end{equation}
Or in other words, $c$ and $c^*$ do not have boundaries
\begin{equation}
    A\in \mathcal{N}(\mathcal{S}) \iff (c,c^*)\in  Z_1 \times Z^*_1.
\end{equation}
Thus, the normalizer is composed of any closed string in the primal and dual lattices.

Now that we have characterized the normalizer, let us take a look at the stabilizer group $\mathcal{S}$. An element of the stabilizer group can be written as
\begin{equation}
    B = \prod_i X_{f_i}\prod_j Z_{v_j} = X_{\partial_2 c_2}Z_{\partial_1^* c_0^*}.
\end{equation}
for some set of faces $c_2 = \sum_i f_i$ and (dual) vertices $c_0^* = \sum_i v_i^*$. Therefore,
\begin{equation}
    B\in \mathcal{S} \iff (c,c^*)\in  B_1 \times B^*_1.
\end{equation}
Elements of the stabilizer (plaquette and vertex operators in \cref{fig:Toric_code}) will deform closed strings without modifying the encoded logical information.
Knowing this allows us to understand that the logical Pauli operators are
\begin{equation}
    \frac{\mathcal{N}(\mathcal{S})}{\mathcal{S}}\simeq \frac{Z_1}{B_1}\times \frac{Z_1^*}{B_1^*}= H_1\times H_1^* \simeq H_1^2.
\end{equation}
In a $d\times d$ torus, this means we have a $[[2d, 2, d]]$ quantum code (vertical and horizontal $\bar{X}$ and $\bar{Z}$ strings), see \cref{fig:Toric_code}. 
An alternative way of introducing a non-trivial topology is to create boundaries in the code. There are two types of boundaries, in the primal (soft) and dual (rough) lattices. However, each side of the square can only contain one kind of boundary. For this reason, the number of logical qubits is reduced to just one, and the code will be $[[2d(d-1)+1,1,d]]$, see \cref{fig:Surface_code}.

In conclusion, the surface code allows to transversally implement $\bar{X}$ and $\bar{Z}$ as the product of $X$ and $Z$ operators over strings. Using two copies of the surface code, we can also implement a transversal C-$Z$ gate. However, we cannot transversally generate the Clifford group. 
\begin{definition}[Clifford group and hierarchy~\cite{gottesman1999demonstrating}]
The Clifford group is the set of quantum gates ${C}$ such that if $P$ is a Pauli operator, then $C^{\dagger}PC$ is also a Pauli operator. In other words, it is the normalizer of the Pauli group, $\mathcal{N}(P)$. In general, the $d$-level of the Clifford hierarchy $\mathcal{L}_{d}$ is defined as those gates $C$ that map $P\in\mathcal{L}_{d-1}$ to the same level under conjugation, 
\begin{equation}
    C\in \mathcal{L}_{d}\iff \forall P \in \mathcal{L}_{d-1}, C^{\dagger}PC\in\mathcal{L}_{d-1},\qquad \mathcal{L}_d := \mathcal{N}(\mathcal{L}_{d-1})
\end{equation}
where $\mathcal{L}_{1} = \braket{i\bm{1},Z,X}$ is the Pauli group, and $\mathcal{L}_{2} = \braket{S,H,\text{C-}Z}$ is the Clifford group itself.
\end{definition}
Two well-known families of gates in the Clifford hierarchy are the single-qubit $e^{i\pi/2^d}$ rotations ($S$-family), and the $d$-qubit controlled $Z$ gates (C-$Z$ family).

What about higher dimensions? In three dimensions, for example, qubits will be attached to faces, $X$-type stabilizers to cells, and $Z$-type stabilizers to edges. Its dual-code will thus have qubits, $X$-type stabilizers, and $Z$-type stabilizers attached to edges, vertices, and faces respectively. Furthermore, in this code, it is possible to transversally implement CC-$Z$ gates~\cite{vasmer2019three}. If we go up to four dimensions, we will be able to prepare self-correcting quantum memories~\cite{alicki2010thermal,dennis2002topological,hastings2014self}. Finally, $d$ copies of the $d$-dimensional surface code will be able to transversally generate, via a mapping to color codes~\cite{kubica2015unfolding}, the $d$-qubit C-$Z$ gate, which is in the $\mathcal{L}_d$ but not in $\mathcal{L}_{d+1}$. This saturates the Bravyi-König theorem\footnote{As we shall see, due to the Eastin-Knill theorem~\cite{eastin2009restrictions}, saturating here means being able to implement one gate from $\mathcal{L}_{d}$, not any gate.}, which states that $d$-dimensional topological codes can only generate gates up to the $d^{th}$ level of the Clifford hierarchy~\cite{bravyi2013classification}. A good review on low overhead implementation of the surface code might be found in the readable Ref.~\cite{litinski2019game}.

\subsection{\label{ssec:Color}Color code}

A second important family of error correction codes is color codes. As for the surface code, we start the discussion in 2 dimensions. In such a case, their lattice fulfills two key requirements:
\begin{enumerate}
    \item Each vertex is trivalent.
    \item Each face is assigned one of three labels (red, green, and blue), so that neighbor faces have different colors.
\end{enumerate}
In the color code, each qubit is attached to the vertices, as indicated in \cref{fig:primal_dual_color_code}, and the stabilizer generators are face operators
\begin{equation}
    X_f = \prod_{v\in f}X_v; \qquad Z_f = \prod_{v\in f}Z_v
\end{equation}
\begin{figure}[!t]
    \centering
    \includegraphics[width = .6\textwidth]{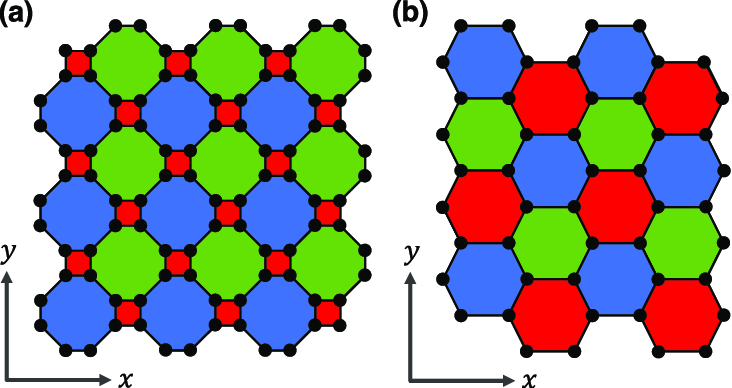}
    \caption{\textbf{Color codes.} Qubits are placed on the vertices, and stabilizers on the faces. (a) 4-8-8 color code, (b) 6-6-6 color code. Taken from Ref.~\cite{lee2022universal} under CC-BY 4.0 license.}
    \label{fig:color_codes}
\end{figure}
From this equation we can recognise that the color code is self-dual. 
Similar to \eqref{eq:Surface_constraints}, the stabilizers in the color code are subject to the following constraints
\begin{equation}\label{eq:color_constraints}
    \prod_{f\in F_r} X_f = \prod_{f\in F_g} X_f = \prod_{f\in F_b} X_f, \qquad \prod_{f\in F_r} Z_f = \prod_{f\in F_g} Z_f = \prod_{f\in F_b} Z_f.
\end{equation}
From this, the number of independent stabilizers generators is $2F-4 = 2(|F_r|+|F_b|+|F_g|)-4$.
We compute the number of encoded qubits as the number of physical qubits (vertices, $V$) minus the number of stabilizer generators ($2F-4$). Given that the lattice is trivalent, $E = \frac{3}{2}V$, and since the Euler characteristic is $\chi = 2(1-g) = V-E+F$, we have
\begin{equation}
    k = V-2F+4 = V+4 +2(V-E-2(1-g)) = 3V-2E +4 -4 +4g = 4g,
\end{equation}
doubling the amount of qubits we got in the surface code.


In the color code, there are 6 types of closed string operators, three different colors, and 2 different Pauli operators, $X$ and $Z$. Denoting by $\gamma$ a closed string, $c$ the color, and $V^\gamma_c$ the vertices bounding a $c$-colored edge in $\gamma$, these are
\begin{equation}
    X_\gamma^c = \prod_{v\in V_c^\gamma} X_v, \qquad Z_\gamma^c = \prod_{v\in V_c^\gamma} Z_v.
\end{equation}
However, not all string operators are independent, because
\begin{equation}
    X^r_\gamma X^g_\gamma X^b_\gamma = \bm{1},\qquad Z^r_\gamma Z^g_\gamma Z^b_\gamma = \bm{1},
\end{equation}
so it is sufficient to consider just two colors.
To understand the commutation rules of these string operators, the key is to check whether $X^{c'}_\gamma$ and $Z^{c}_\gamma$ coincide in an even or odd number of qubits. If they cross an even number of times $[X^{c'}_\gamma, Z^{c}_\gamma] =0$. This necessarily happens, for example, if $c = c'$~\cite{bombin2013introduction}. In contrast, if $c \neq c'$ and $Z^c_\gamma$ and $X^{c'}_\gamma$ cross an odd number of times, $\{X^{c'}_\gamma, Z^{c}_\gamma\} =0$.
\begin{figure}[!t]
    \centering
    \includegraphics[width = .8\textwidth]{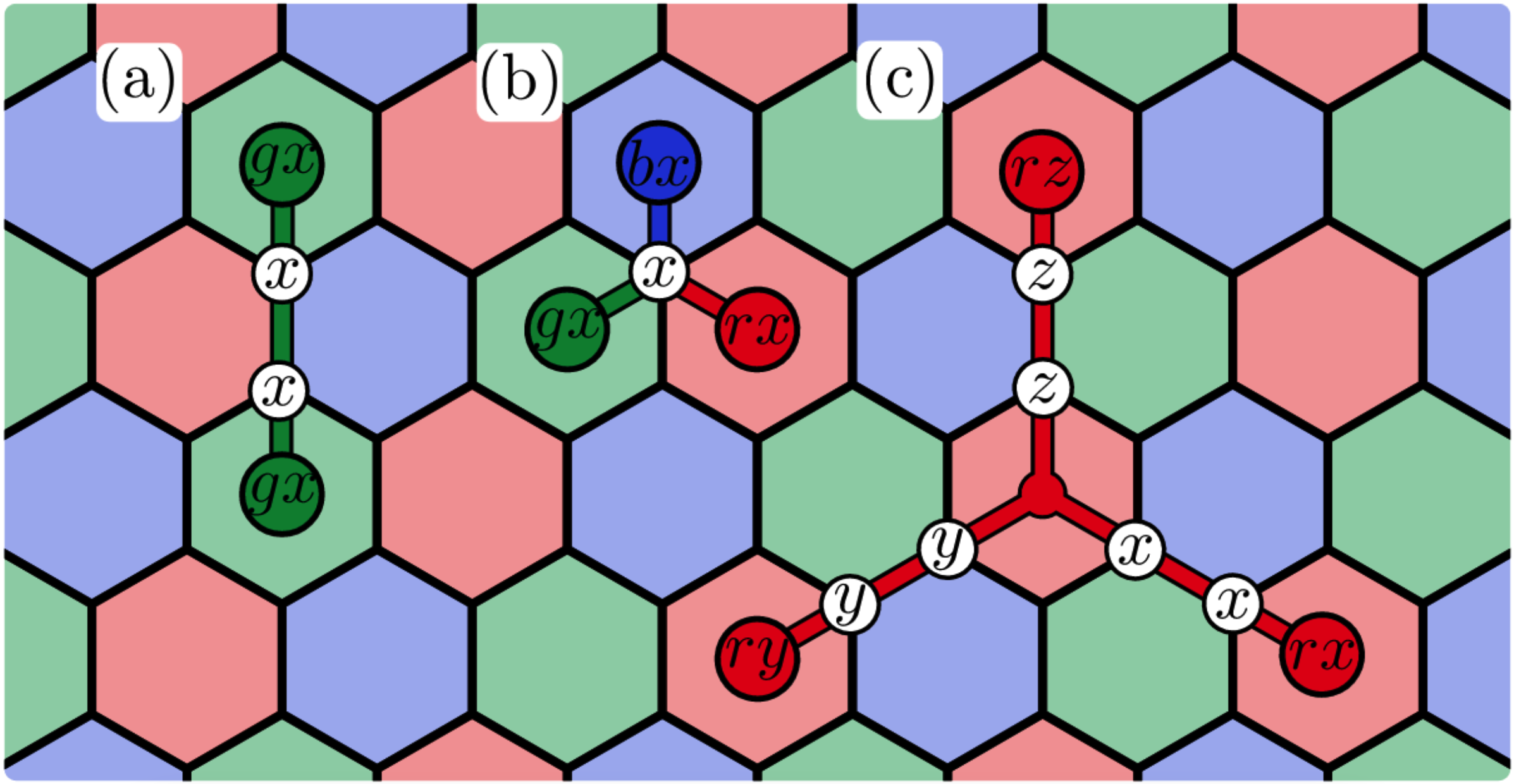}
    \caption{\textbf{Error strings in the color code}. Similar to \cref{fig:Toric_code}, the errors should be corrected by an open string of operator joining excitations, so this figure may alternatively be interpreted as fusion rules of excitations. (a) Two excitations of the same color and type $X/Z$ can be joined by a string of the same color. (b) Three excitations of the same type but different colors can be joined at the vertex they share. (c) Three excitations of types $X$, $Z$, and $XZ$, and the same color might also be joined to correct the errors.
    Figure from Ref.~\cite{kesselring2018boundaries} under CC-BY 4.0 license.}
    \label{fig:fusion_color_code}
\end{figure}

Let us now analyze the logical operators in this code. Similar to the surface code, if a closed string $\gamma$ may be contracted to a point via stabilizer generators, then it is a homologically trivial stabilizer element. As such, since only two colors are independent,
\begin{equation}
    \mathcal{S} \simeq B_1^{\text{red},X}\times B_1^{\text{red},Z}\times B_1^{\text{green},X} \times B_1^{\text{green},Z}.
\end{equation}
On the other hand, those strings that cannot be contracted but commute with the stabilizers form the normalizer, 
\begin{equation}
    \mathcal{N}(\mathcal{S}) \simeq Z_1^{\text{red},X}\times Z_1^{\text{red},Z}\times Z_1^{\text{green},X} \times Z_1^{\text{green},Z}.
\end{equation}
And finally, we obtain logical operators $\bar{X}_i$, $\bar{Z}_i$ as 
\begin{equation}
    \frac{\mathcal{N}(\mathcal{S})}{\mathcal{S}}\simeq H_1^{\text{red},X}\times H_1^{\text{red},Z}\times H_1^{\text{green},X} \times H_1^{\text{green},Z}.
\end{equation}
As such, the color code is a CSS code.
If closed strings $\gamma$ are elements of the normalizer of the stabilizer group $\mathcal{N}(\mathcal{S})$, open strings represent errors and whose boundaries will signal a $+1$ syndrome, see \cref{fig:fusion_color_code}. These errors $E_a$ can be corrected closing the string with $E_b^\dagger = E_b$, so hopefully $E_b^\dagger E_a\in \mathcal{S}$ and the logical information is preserved.

To compute the distance of a color code, we have to find the smallest non-detectable error, or in other words, the smallest weight between the logical operators. However, in this case, non-trivial elements of the normalizer do not need to be single-color strings. Rather, the string may branch into a string net.
On the other hand, similarly to the surface code, we may introduce boundaries of a given color by removing stabilizer operators of such color. Similar to the surface code, red close string operators $\gamma$ may end up in red boundaries, making them commutative with any stabilizer and therefore an element of $\mathcal{N}(\mathcal{S})$. In contrast, if a red closed string $\gamma$ only encloses red boundaries, then it is part of the stabilizer group $\mathcal{S}$, as it can be contracted to the boundary~\cite{bombin2013introduction}.

\begin{figure}[!t]
    \centering
    \includegraphics[width = .5\textwidth]{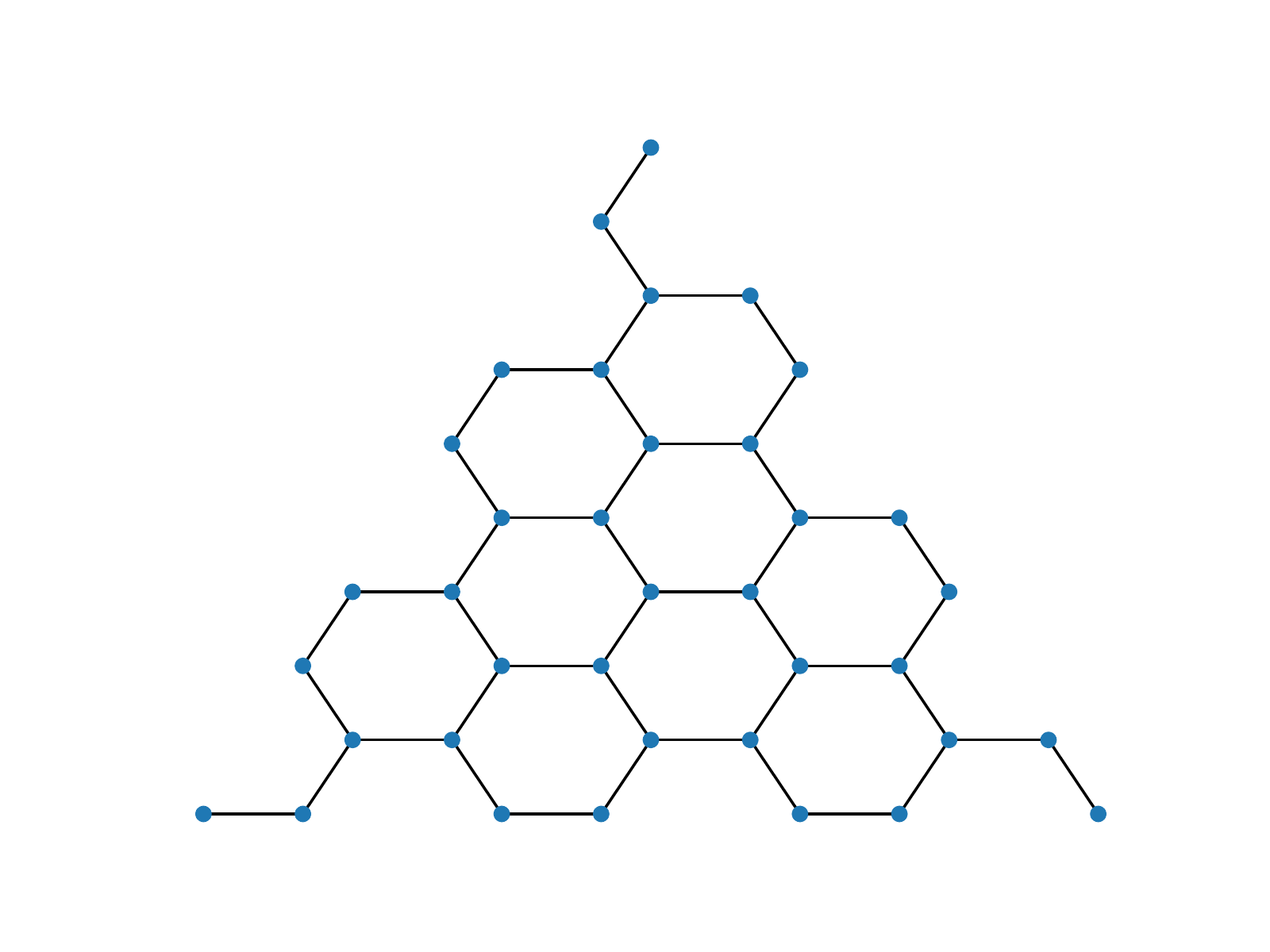}
    \includegraphics[width = .47\textwidth]{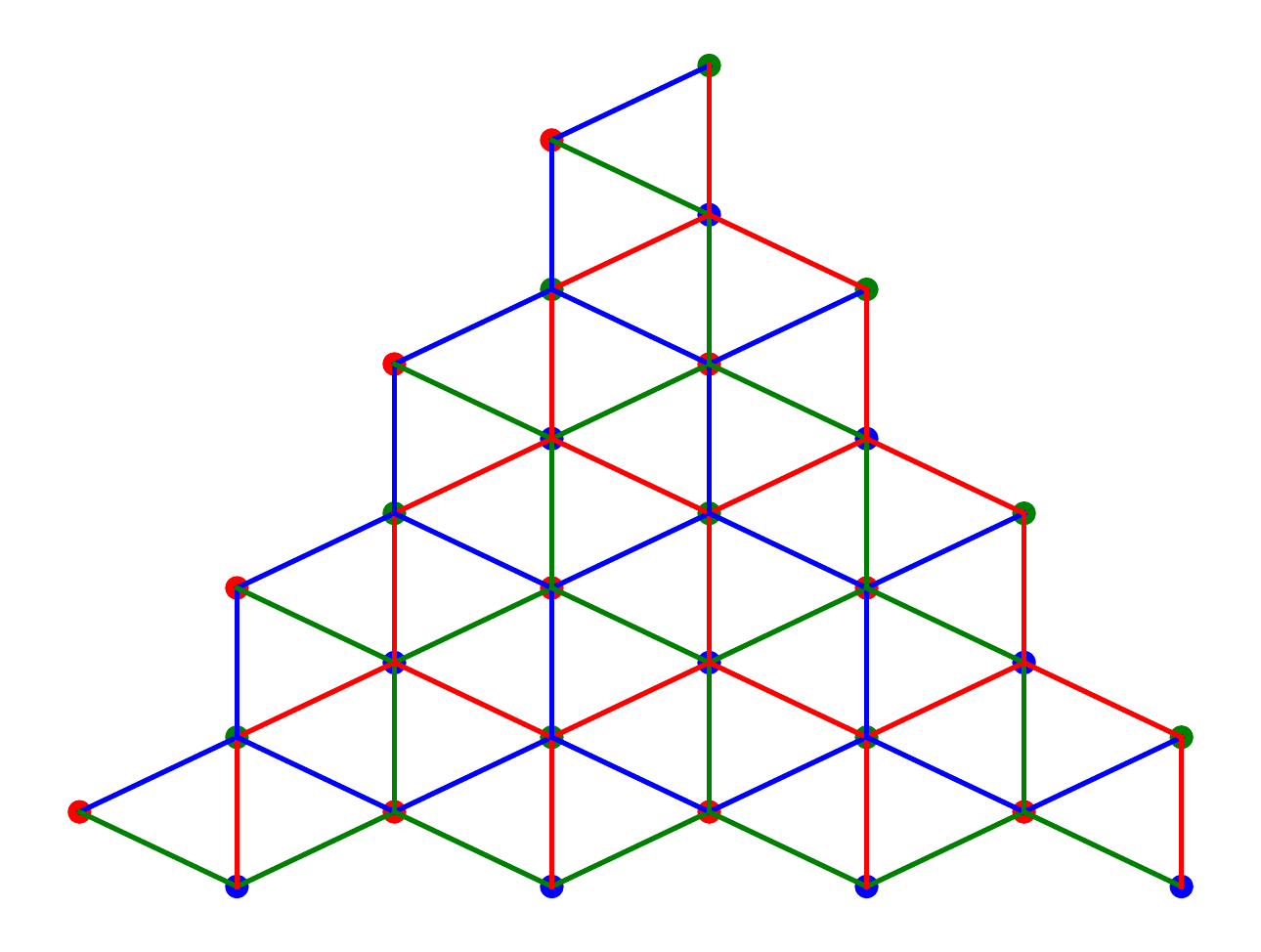}
    \caption{\textbf{6-6-6 Color code, primal and dual lattices.} Left: trivalent primal lattice, with qubits on the vertices. Right: dual lattice, with qubits on the faces. In this dual lattice, each edge can be assigned the complementary color of the vertices bounding it. To introduce boundaries allowing to encode logical qubits, we remove the stabilizers corresponding to the bottom blue vertices, upper-right green vertices, and upper-left red vertices.}
    \label{fig:primal_dual_color_code}
\end{figure}

We have already seen that the color code encodes twice the number of logical qubits than the equivalent surface code. However, the most remarkable difference between these codes is the color code capability to transversally implement the 
\begin{equation}
    S = \begin{pmatrix}
    1 & 0\\
    0 & i
    \end{pmatrix}
\end{equation}
and Hadamard gates, which together with the C-Not allow these codes to generate the Clifford group. Transversal implementation of gates is very useful because it prevents errors from propagating in the code.

\begin{figure}[!t]
\centering
\begin{tikzpicture}
\node[circle,fill=white,minimum size=2]  (1) at (0,0) {1};
\node[circle,fill=white,minimum size=2]  (2) at (3,0) {2};
\node[circle,fill=white,minimum size=2]  (3) at (6,0) {3};
\node[circle,fill=white,minimum size=2]  (4) at (3,1.63) {4};
\node[circle,fill=white,minimum size=2]  (5) at (1.5,2.598) {5};
\node[circle,fill=white,minimum size=2]  (6) at (4.5,2.598) {6};
\node[circle,fill=white,minimum size=2]  (7) at (3,5.196) {7};

\draw[white, fill = blue ] (1.center) -- (7.center) -- (2.5,5.196) -- (-0.5,0) -- cycle;
\draw[white, fill = yellow ] (3.center) -- (7.center) -- (3.5,5.196) -- (6.5,0) -- cycle;
\draw[white, fill = red ] (1.center) -- (3.center)  -- (6.25,-0.4) -- (-0.25,-0.4) -- cycle;

\draw[black, fill = yellow, thin] (1.center) -- (2.center) -- (4.center) -- (5.center) -- cycle;
\draw[black, fill = red, thin] (7.center) -- (6.center) -- (4.center) -- (5.center) -- cycle;
\draw[black, fill = blue, thin] (3.center) -- (2.center) -- (4.center) -- (6.center) -- cycle;

\foreach \x in {1,...,7}
    \filldraw[fill = white] (\x) circle (7pt) node[]{\x};
    
\draw[yellow, ultra thick] (4)--(6);
\draw[blue, ultra thick] (4)--(5);
\draw[red, ultra thick] (4)--(2);
    
\end{tikzpicture}

\caption{\textbf{Simplest color code} of distance $d=3$, with boundaries. Operator $X$ on vertices 2, 4, 5, and 6 would form a string net with the three colored edges. A weight-three logical operator would be string net red boundary -- 2-- 4 --7 -- blue and yellow boundaries.}
\label{fig:simplest_color_code}
\end{figure}
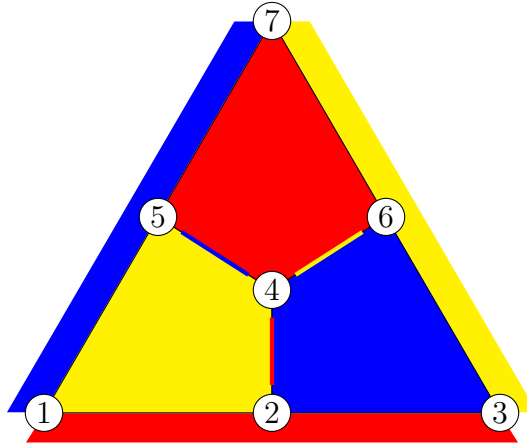

The Hadamard gate can be easily implemented on the triangular codes such as \cref{fig:primal_dual_color_code} and \cref{fig:simplest_color_code}, because both $\bar{X}$ and $\bar{Z}$ strings are applied over the same physical qubits. As such, $H^{\otimes n}\bar{X} H^{\otimes n} = \bar{Z}$, and consequently, we can take $\bar{H} = H^{\otimes n}$. As for the $\bar{S}$ gate, the desired behavior would be $\bar{S}^\dagger \bar{Z} \bar{S} = \bar{Z}$ and $\bar{S}^\dagger \bar{X} \bar{S} = \bar{Y}$. 
However, to transversally implement this gate we need to make sure that $\bar{S} = S^{\otimes n}$ does not take us outside the code space. Instead, we find that $(S^{\otimes n})^\dagger X_f S^{\otimes n} = i^{v} X_f Z_f$, where face $f$ has $v$ vertices~\cite{bombin2006topological}. Staying in the +1 subspace, so the stabilizer group is preserved, is recovered for codes where the number of qubits per face is a multiple of 4, like the 4-8-8 code in \cref{fig:color_codes}. Furthermore, if the number $n$ of qubits in the code is such that $n = 1 \mod 4$, then we obtain 
\begin{equation}
    S^{\otimes n}\bar{X}(S^\dagger)^{\otimes n} = i^{n\mod 4}\bar{X}\bar{Z} = i\bar{X}\bar{Z} = \bar{Y},\qquad  S^{\otimes n}\bar{Z}(S^\dagger)^{\otimes n} = \bar{Z}.
\end{equation}
Thus, in this code, we can implement the $\bar{S}=S^{\otimes n}$ transversally, and consequently, also the entire Clifford group.

If we have a member from the 6-6-6 color code family, \cref{fig:primal_dual_color_code} we can still find a way to transversally implement the $\bar{S}$ rotation~\cite{kubica2015universal}. The key aspect to notice is that the set of vertices can be split in two disjoint sets $S_1$ and $S_2$ under the condition that two vertices in an edge are part of different sets and $|S_1| = |S_2|+1$. Since the number of qubits is odd, but each stabilizer is applied to an even number of qubits, we can choose $\bar{S} = S^{\otimes (q\in S_1)}(S^\dagger)^{\otimes (q\in S_2)}$. Then~\cite{kubica2015universal},
\begin{equation}
    \bar{S}\bar{X}\bar{S}^\dagger = i^{|S_1|-|S_2|}\bar{X}\bar{Z} = i\bar{X}\bar{Z} = \bar{Y}, \qquad \bar{S}\bar{Z}\bar{S}^\dagger = \bar{Z}.
\end{equation}
The stabilizer subspace will similarly be preserved because in each stabilizer there is an equal number of qubits in $S_1$ and $S_2$~\cite{kubica2015universal} 
\begin{equation}
    \bar{S}X_f \bar{S}^\dagger = i^{|S_1\cap f|-|S_2\cap f|}X_f Z_f = X_f Z_f\in \mathcal{S}; \qquad \bar{S}Z_f \bar{S}^\dagger = Z_f\in \mathcal{S}.
\end{equation}
Thus, a transversal gate is also possible in the 6-6-6 codes if we take the members of the family depicted in~\cref{fig:primal_dual_color_code}.

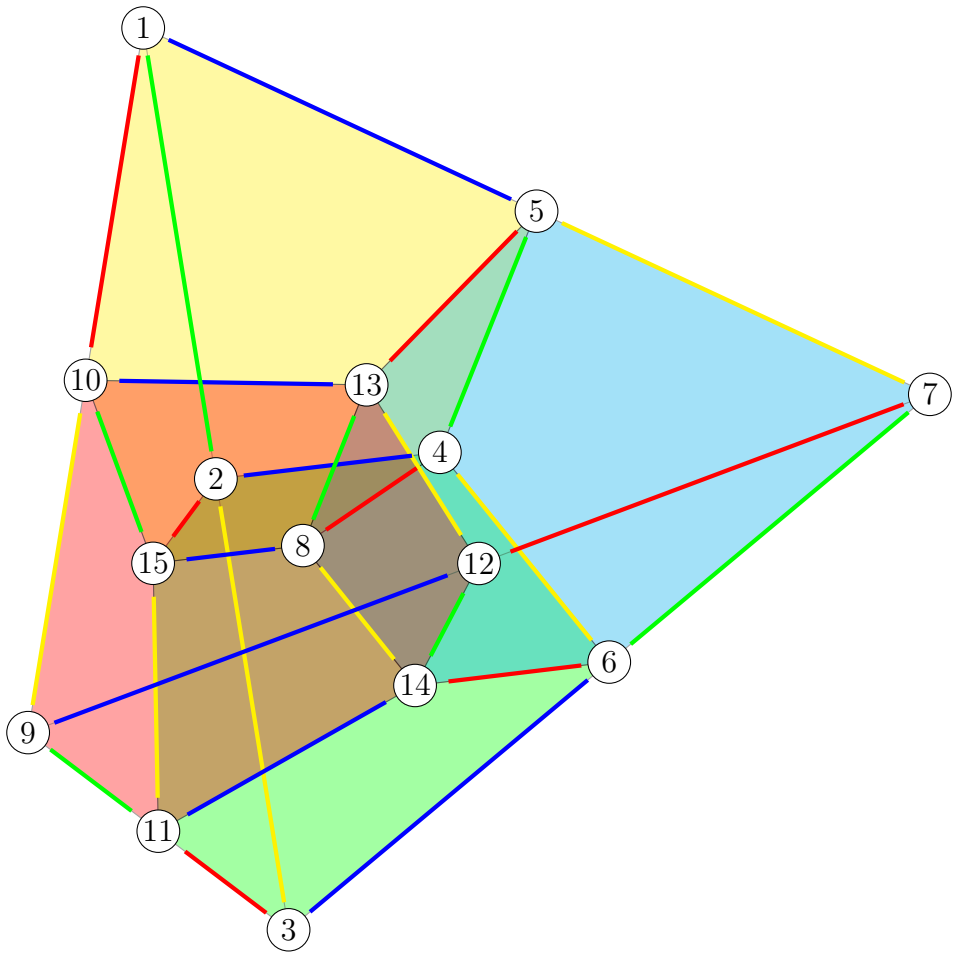
\begin{figure}[!t]
\centering
\begin{tikzpicture}[rotate around y=-10, rotate around z=280, scale=2.0]
\node[circle,fill=white,minimum size=2]  (1) at (0,0,0) {1};
\node[circle,fill=white,minimum size=2]  (2) at (3,0,0) {2};
\node[circle,fill=white,minimum size=2]  (3) at (6,0,0) {3};
\node[circle,fill=white,minimum size=2]  (4) at (3,1.63,0) {4};
\node[circle,fill=white,minimum size=2]  (5) at (1.5,2.598,0) {5};
\node[circle,fill=white,minimum size=2]  (6) at (4.5,2.598,0) {6};
\node[circle,fill=white,minimum size=2]  (7) at (3,5.196,0) {7};

\node[circle,fill=white,minimum size=2]  (8) at (3,1.63,1.63) {8}; 
\node[circle,fill=white,minimum size=2]  (9) at (3,1.63,4.90) {9}; 

\node[circle,fill=white,minimum size=2]  (10) at (1.5  , 0.815, 2.45) {10}; 
\node[circle,fill=white,minimum size=2]  (11) at (4.5  , 0.815, 2.45) {11}; 
\node[circle,fill=white,minimum size=2]  (12) at (3.   , 3.413, 2.45) {12}; 

\node[circle,fill=white,minimum size=2]  (13) at (2.        , 2.27, 1.63) {13}; 
\node[circle,fill=white,minimum size=2]  (14) at (4.        , 2.27, 1.63) {14}; 
\node[circle,fill=white,minimum size=2]  (15) at (3.        , 0.54, 1.63) {15}; 

\draw[black, fill = yellow, thin, opacity = 0.2] (1.center) -- (2.center) -- (4.center) -- (5.center) -- cycle;
\draw[black, fill = yellow, thin, opacity = 0.2] (1.center) -- (2.center) -- (15.center) -- (10.center) -- cycle;
\draw[black, fill = yellow, thin, opacity = 0.2] (1.center) -- (5.center) -- (13.center) -- (10.center) -- cycle;
\draw[black, fill = yellow, thin, opacity = 0.2] (2.center) -- (15.center) -- (8.center) -- (4.center) -- cycle;
\draw[black, fill = yellow, thin, opacity = 0.2] (8.center) -- (4.center) -- (5.center) -- (13.center) -- cycle;
\draw[black, fill = yellow, thin, opacity = 0.2] (13.center) -- (8.center) -- (15.center) -- (10.center) -- cycle;

\draw[black, fill = green, thin, opacity = 0.2] (3.center) -- (2.center) -- (4.center) -- (6.center) -- cycle;
\draw[black, fill = green, thin, opacity = 0.2] (3.center) -- (6.center) -- (14.center) -- (11.center) -- cycle;
\draw[black, fill = green, thin, opacity = 0.2] (3.center) -- (2.center) -- (15.center) -- (11.center) -- cycle;
\draw[black, fill = green, thin, opacity = 0.2] (14.center) -- (8.center) -- (4.center) -- (6.center) -- cycle;
\draw[black, fill = green, thin, opacity = 0.2] (2.center) -- (4.center) -- (8.center) -- (15.center) -- cycle;
\draw[black, fill = green, thin, opacity = 0.2] (14.center) -- (11.center) -- (15.center) -- (8.center) -- cycle;

\draw[black, fill = cyan, thin, opacity = 0.2] (7.center) -- (6.center) -- (4.center) -- (5.center) -- cycle;
\draw[black, fill = cyan, thin, opacity = 0.2] (7.center) -- (6.center) -- (14.center) -- (12.center) -- cycle;
\draw[black, fill = cyan, thin, opacity = 0.2] (7.center) -- (5.center) -- (13.center) -- (12.center) -- cycle;
\draw[black, fill = cyan, thin, opacity = 0.2] (6.center) -- (4.center) -- (8.center) -- (14.center) -- cycle;
\draw[black, fill = cyan, thin, opacity = 0.2] (4.center) -- (5.center) -- (13.center) -- (8.center) -- cycle;
\draw[black, fill = cyan, thin, opacity = 0.2] (12.center) -- (13.center) -- (8.center) -- (14.center) -- cycle;

\draw[black, fill = red, thin, opacity = 0.2] (12.center) -- (9.center) -- (10.center) -- (13.center) -- cycle;
\draw[black, fill = red, thin, opacity = 0.2] (12.center) -- (9.center) -- (11.center) -- (14.center) -- cycle;
\draw[black, fill = red, thin, opacity = 0.2] (9.center) -- (10.center) -- (15.center) -- (11.center) -- cycle;
\draw[black, fill = red, thin, opacity = 0.2] (12.center) -- (13.center) -- (8.center) -- (14.center) -- cycle;
\draw[black, fill = red, thin, opacity = 0.2] (13.center) -- (10.center) -- (15.center) -- (8.center) -- cycle;
\draw[black, fill = red, thin, opacity = 0.2] (15.center) -- (8.center) -- (14.center) -- (11.center) -- cycle;

\draw[yellow, ultra thick] (3)--(2);
\draw[yellow, ultra thick] (15)--(11);
\draw[yellow, ultra thick] (9)--(10);
\draw[yellow, ultra thick] (5)--(7);
\draw[yellow, ultra thick] (4)--(6);
\draw[yellow, ultra thick] (13)--(12);
\draw[yellow, ultra thick] (8)--(14);

\draw[red, ultra thick] (10)--(1);
\draw[red, ultra thick] (13)--(5);
\draw[red, ultra thick] (8)--(4);
\draw[red, ultra thick] (15)--(2);
\draw[red, ultra thick] (11)--(3);
\draw[red, ultra thick] (14)--(6);
\draw[red, ultra thick] (12)--(7);

\draw[blue, ultra thick] (5)--(1);
\draw[blue, ultra thick] (13)--(10);
\draw[blue, ultra thick] (2)--(4);
\draw[blue, ultra thick] (15)--(8);
\draw[blue, ultra thick] (11)--(14);
\draw[blue, ultra thick] (6)--(3);
\draw[blue, ultra thick] (12)--(9);

\draw[green, ultra thick] (11)--(9);
\draw[green, ultra thick] (15)--(10);
\draw[green, ultra thick] (2)--(1);
\draw[green, ultra thick] (4)--(5);
\draw[green, ultra thick] (6)--(7);
\draw[green, ultra thick] (14)--(12);
\draw[green, ultra thick] (8)--(13);

\foreach \x in {15,...,1}
    \filldraw[fill = white] (\x) circle (4pt) node[]{\x};

\end{tikzpicture}
\caption{\label{fig:simplest_3d_color_code}\textbf{Simplest example of the 3d color code}, similar to \cref{fig:simplest_color_code}. $X$ stabilizers are attached to cells, while $Z$ stabilizers do so in faces. Boundaries must be placed around the code, such that they are of a different color than the cells they touch.
This color code admits transversal implementation of C-Not and T gates.}
\end{figure}

Similar to the extension of surface codes we saw in the previous~\cref{ssec:Surface}, we can also generalize the surface code to a family of color codes of $d$ dimensions~\cite{bombin2007topological}. In general, the conditions will be 
\begin{enumerate}
    \item The lattice is $d$-valent: each vertex is connected to $d$ edges.
    \item $d$-cells can be $d+1$-colored such that no adjacent $d$-cells share color.
\end{enumerate}
We depict the simplest three-dimensional color code in \cref{fig:simplest_3d_color_code}. In this case, the qubits are still on the vertices, and the $Z$ stabilizers are still attached to faces, but the $X$ stabilizers are instead cell stabilizers $X_c$. The attractiveness of the 3d color code is its capability to transversally implement the gate 
\begin{equation}
    T = \sqrt{S} = \begin{pmatrix}
    1 & 0\\
    0 & i^{1/2}
    \end{pmatrix}
\end{equation}
If each stabilizer has weight multiple of 8, and $n\mod 8=:l\in \{1,3,5,7\}$ then $\bar{T} = T^{\otimes n}$ implements the logical operator
\begin{equation}
    \bar{T}\ket{\bar{0}} = \ket{\bar{0}}; \qquad \bar{T}\ket{\bar{1}} = i^{l/2}\ket{\bar{1}},
\end{equation}
and maps the code subspace to itself. 
A similar construction might be implemented for the $d$-dimensional color code, with gate
\begin{equation}\label{eq:R_d}
    R_d = \begin{pmatrix}
    1 & 0\\
    0 & 2^{\frac{i\pi}{2^d}},
    \end{pmatrix}
\end{equation}
thus saturating the Bravyi-König bound~\cite{bravyi2013classification,pastawski2015fault,watson2015qudit}.
In such color codes, the $Z$ stabilizers are attached to faces and $X$ stabilizers to $d$-cells. However, this arrangement does not allow us to transversally implement the Hadamard gate anymore, as $X$ and $Z$ stabilizers no longer have the same support. Finally, it is worth mentioning that, similarly to surface codes, in $d=4$ we similarly find a self-correcting quantum code~\cite{bombin2013self}.

\section{\label{sec:Fault tolerant quantum computing}Towards fault-tolerant quantum computing}

So far we have discussed two of the main families of error-correcting codes, the surface and color codes. Unfortunately, we have not found any code that allows us to fault-tolerantly and transversally implement a set of universal gates, $\{H,\text{C-}Z, T\}$. Is this possible? It turns out that the answer is negative: according to a theorem by Eastin and Knill, any transversal set of gates generates a finite group and therefore cannot be universal~\cite{eastin2009restrictions}.

What alternatives do we have? The most popular option is to couple a high-threshold error code, such as the surface code, with a smaller stabilizer code that is only used to generate magic states $T\ket{+}$. Each of these states will, later on, be consumed to implement one $T$ gate on the surface code via quantum teleportation~\cite{bravyi2005universal}. 
To generate these magic states, it is customary to implement a distillation procedure: starting with many noisy copies implemented via non-fault tolerant $T$ gates, one can generate a few cleaner copies. The most famous of these distillation procedures is known as the 15 to 1 protocol~\cite{bravyi2005universal}, which takes 15 noisy copies and outputs a single one with less error. In particular, assuming Pauli error $p$, the output state will be subject to error $35p^3$.

Magic state distillation is rather costly, and while there have been many advances since the early protocols, they are still often considered the bottleneck step to implementing fault-tolerant quantum algorithms. A recent review of several distillation methods may be found in Ref.~\cite{litinski2019magic}.  The alternative approach is using subsystem codes, two codes that can implement different sets of gates but are based on the same lattice and encode the same number of logical qubits. The most prominent example is gauge color codes, which we briefly review next. 

\subsection{\label{ssec:Gauge_color_codes}Gauge color codes}

\begin{figure}[!t]
    \centering
    \includegraphics[width = .45\textwidth]{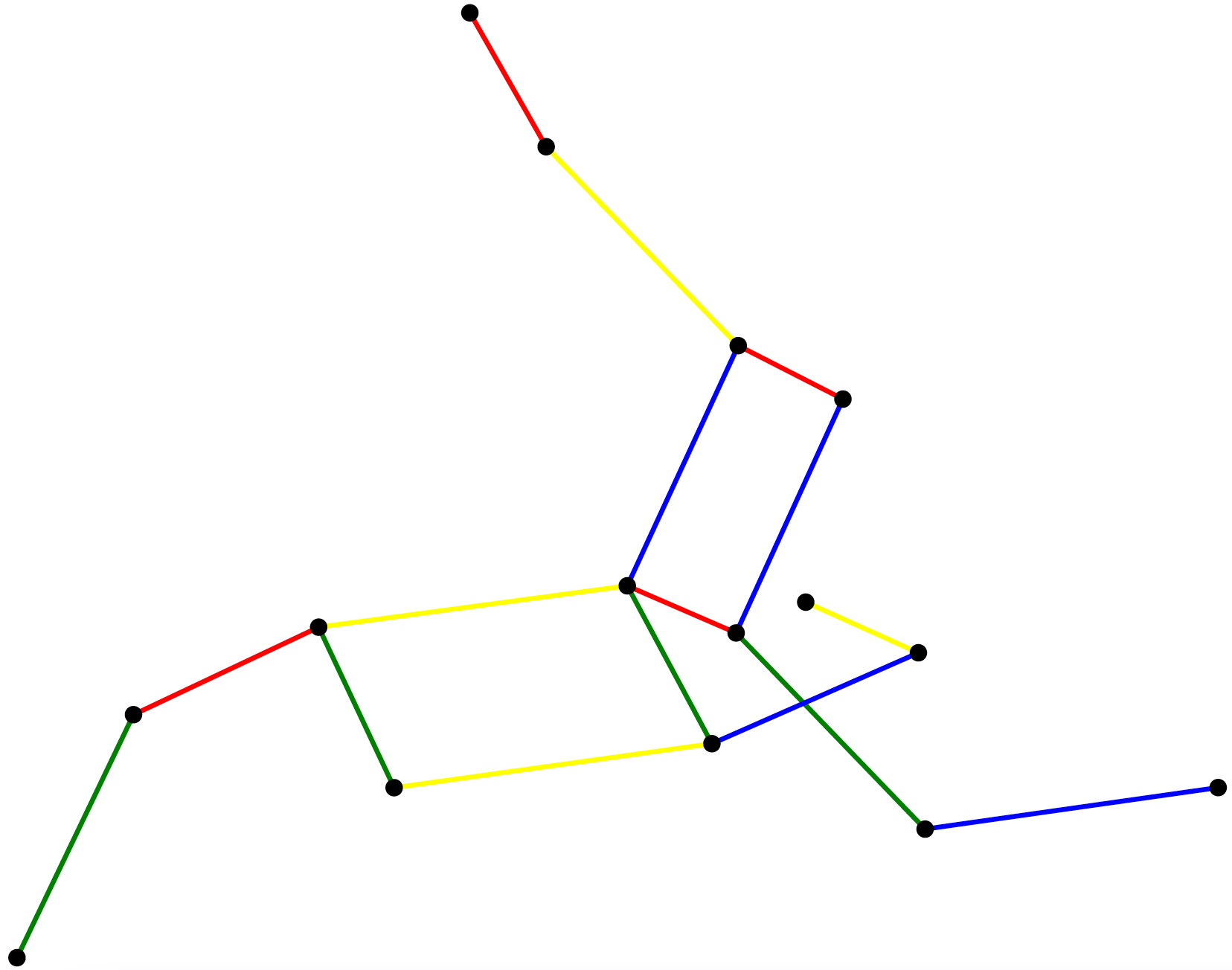}
    \includegraphics[width = .45\textwidth]{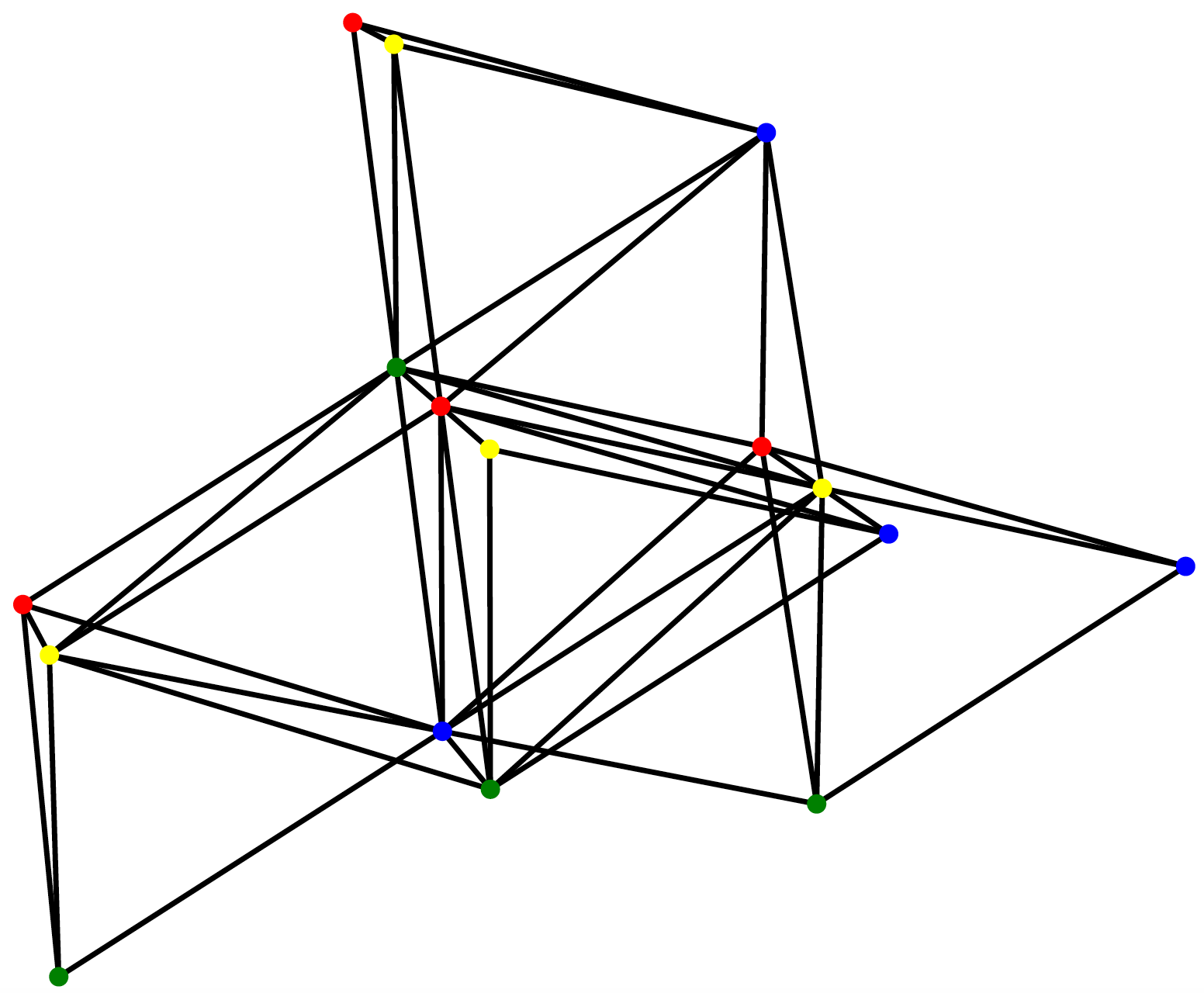}
    \caption{\textbf{Three-dimensional color code with boundaries, primal and dual lattices,} generalization of \cref{fig:primal_dual_color_code}. Left: 4-valent primal lattice, with qubits on the vertices. Edges are shown with the complementary color of the vertices in the corresponding dual face. Right: dual lattice, with qubits on the cells. Each face can be assigned the complementary color of the vertices bounding it. To introduce boundaries allowing to encode logical qubits, we remove the stabilizers corresponding to the boundary $k$-faces, see \eqref{eq:k-faces}.}
    \label{fig:3d_primal_dual_color_code}
\end{figure}

Gauge color codes are a generalization of color codes, first introduced by Héctor Bombín in Ref.~\cite{bombin2015gauge}. Here we will mostly follow their description by Ref.~\cite{kubica2015universal}, that more naturally describes them in the dual picture in \cref{fig:primal_dual_color_code} over a lattice $\mathcal{L}$. Such lattices are a collection of $d$-simplices $\delta$, defined using independent vertices $e_i$,
\begin{equation}\label{eq:k-faces}
    \delta = \left\{\sum_{i=0}^d t_i e_i\big| \sum_i t_i = 1\right\}, \quad \dim \delta := d, \quad t_i\in\mathbb{R}^+, \quad e_i = (\underbrace{0,\ldots,0}_{i-1},1,0,\ldots,0).
\end{equation}
For example, $0$-simplices are vertices, $1$-simplices are edges, $2$-simplices are faces (triangles), and $3$-simplices are cells (tetrahedra). The code is the union of a set of $d$-simplices of the same dimension $\mathcal{L} = \cup_i \delta_i$, and $\dim \mathcal{L} = \dim\delta_i$ is the dimension of the color code.

If we have a $d$-simplex $\delta$, we can also define its $k$-faces~\cite{kubica2015universal}
\begin{equation}
    \Delta_k(\delta) = \{\sigma \subset \delta |\sigma \text{ is a $k$-simplex, with }k\leq d\}.
\end{equation}
Each $d$-simplex contains $d+1 \choose k+1$ such $k$-faces. Additionally, we will denote by $\partial\mathcal{L}$ the boundary of $\mathcal{L}$, and $\Delta'_k(\mathcal{L}) := \Delta_k(\mathcal{L}\backslash \partial \mathcal{L}) $.
Color codes, as we have discussed, are also characterized by a function, called `color' that maps each vertex to $\Z_{d+1}$,
\begin{equation}
    \text{color}: \Delta_0\mathcal{L}\mapsto \Z_{d+1}.
\end{equation}
The possible values in $\Z_{d+1}$ are understood as different colors. We can also assign colors to a $d$-simplex as the union of colors of its vertices
\begin{equation}
    \col(\delta) = \bigsqcup_{v\in \Delta_0(\delta)}\col(v).
\end{equation}
Apart from $\mathcal{L}$ being ($d+1$) colorable, we need $\mathcal{L}$ to be the triangulation of a $d$-simplex. Then, we attach a qubit to each $d$-simplex in the interior of $\mathcal{L}$, such that for $\sigma\subset \mathcal{L}\backslash \partial \mathcal{L}$, we define the set of qubits connected to $\sigma$ as
\begin{equation}
    \mathcal{Q}(\sigma) = \{\delta\in \Delta_d(\mathcal{L})|\sigma\subset \delta\}.
\end{equation}
For example, in the two-dimensional color code from the dual lattice in \cref{fig:primal_dual_color_code}, $\delta$ would refer to the triangles and $\sigma$ to the edges or vertices. More concretely, to each edge, there are two triangles/qubits attached, and to each vertex, there are six, except in the border. When we say an operator is supported over a simplex $\sigma$, we mean that it is applied over the corresponding qubits, e.g., $X(\sigma) = X[\mathcal{Q}(\sigma)]$.

\begin{figure}[!t]
\begin{multicols}{2}
\centering
\noindent 
\begin{asy}
pair xyz2(real x, real y, real z)
{
return 50*(-x - z / 3+1.5, y - z / 6 - 0.1);
}

pair PrimalPoint1 = xyz2(0,0,0);
pair PrimalPoint2 = xyz2(0,0,1);
pair PrimalPoint3 = xyz2(0,1,0);
pair PrimalPoint4 = xyz2(0,1,1);
pair PrimalPoint5 = xyz2(1,0,0);
pair PrimalPoint6 = xyz2(1,0,1);
pair PrimalPoint7 = xyz2(1,1,0);
pair PrimalPoint8 = xyz2(1,1,1);

pair EvenPoint1 = (PrimalPoint1+PrimalPoint2+PrimalPoint3+PrimalPoint5) / 4;
pair EvenPoint2 = (PrimalPoint2+PrimalPoint5+PrimalPoint6+PrimalPoint8) / 4;
pair EvenPoint3 = (PrimalPoint4+PrimalPoint2+PrimalPoint3+PrimalPoint8) / 4;
pair EvenPoint4 = (PrimalPoint7+PrimalPoint8+PrimalPoint3+PrimalPoint5) / 4;
pair EvenPoint5 = (PrimalPoint5+PrimalPoint2+PrimalPoint3+PrimalPoint8) / 4;

pair OddPoint1 = (PrimalPoint1 + PrimalPoint5 + PrimalPoint6 + PrimalPoint7) / 4;
pair OddPoint2 = (PrimalPoint1 + PrimalPoint2 + PrimalPoint4 + PrimalPoint6) / 4;
pair OddPoint3 = (PrimalPoint4 + PrimalPoint8 + PrimalPoint6 + PrimalPoint7) / 4;
pair OddPoint4 = (PrimalPoint1 + PrimalPoint3 + PrimalPoint4 + PrimalPoint7) / 4;
pair OddPoint5 = (PrimalPoint1 + PrimalPoint4 + PrimalPoint6 + PrimalPoint7) / 4;

picture RedCircleCell;

filldraw(RedCircleCell, (xyz2(0,0,0)+EvenPoint3)--(xyz2(0,0,0)+EvenPoint5)--(xyz2(0,0,0)+EvenPoint2)--(xyz2(0,0,1)+OddPoint1)--(xyz2(0,0,1)+OddPoint5)--(xyz2(0,0,1)+OddPoint4)--cycle,darkred,gray);

filldraw(RedCircleCell, (xyz2(0,1,0)+OddPoint3)--(xyz2(0,1,1)+EvenPoint4)--(xyz2(1,1,1)+OddPoint4)--(xyz2(1,1,0)+EvenPoint3)--cycle,lightred,gray);

filldraw(RedCircleCell, (xyz2(0,1,1)+EvenPoint2)--(xyz2(1,1,1)+OddPoint2)--(xyz2(1,0,1)+EvenPoint3)--(xyz2(0,0,1)+OddPoint3)--cycle,red,gray);
filldraw(RedCircleCell, (xyz2(0,1,1)+EvenPoint1)--(xyz2(0,1,1)+EvenPoint5)--(xyz2(0,1,1)+EvenPoint2)--(xyz2(0,0,1)+OddPoint3)--(xyz2(0,0,1)+OddPoint5)--(xyz2(0,0,1)+OddPoint4)--cycle,heavyred,gray);

filldraw(RedCircleCell, (xyz2(0,1,1)+EvenPoint2)--(xyz2(0,1,1)+EvenPoint5)--(xyz2(0,1,1)+EvenPoint4)--(xyz2(1,1,1)+OddPoint4)--(xyz2(1,1,1)+OddPoint5)--(xyz2(1,1,1)+OddPoint2)--cycle, mediumred,gray);

filldraw(RedCircleCell, (xyz2(1,0,1)+EvenPoint3)--(xyz2(1,0,1)+EvenPoint5)--(xyz2(1,0,1)+EvenPoint4)--(xyz2(1,1,1)+OddPoint1)--(xyz2(1,1,1)+OddPoint5)--(xyz2(1,1,1)+OddPoint2)--cycle,mediumred,gray);

filldraw(RedCircleCell, (xyz2(0,1,1)+EvenPoint1)--(xyz2(0,1,1)+EvenPoint5)--(xyz2(0,1,1)+EvenPoint4)--(xyz2(0,1,0)+OddPoint3)--(xyz2(0,1,0)+OddPoint5)--(xyz2(0,1,0)+OddPoint2)--cycle,red,gray);

filldraw(RedCircleCell, (xyz2(1,0,1)+EvenPoint1)--(xyz2(1,0,1)+EvenPoint5)--(xyz2(1,0,1)+EvenPoint3)--(xyz2(0,0,1)+OddPoint3)--(xyz2(0,0,1)+OddPoint5)--(xyz2(0,0,1)+OddPoint1)--cycle,deepred,gray);

filldraw(RedCircleCell, (xyz2(0,0,0)+EvenPoint3)--(xyz2(0,0,1)+OddPoint4)--(xyz2(0,1,1)+EvenPoint1)--(xyz2(0,1,0)+OddPoint2)--cycle,heavyred,gray);

// Central cube qubits

fill(RedCircleCell,shift( (xyz2(0,0,0)+EvenPoint5) )*unitcircle, gray);
fill(RedCircleCell,shift( (xyz2(0,0,0)+EvenPoint5)  )*((0,1)..(0.84,0)..(0,-0.64)..(-0.84,0)..cycle), white);
draw(RedCircleCell,shift( (xyz2(0,0,0)+EvenPoint5)  )*unitcircle, black+0.2);

fill(RedCircleCell,shift( (xyz2(0,0,1)+EvenPoint5) )*unitcircle, gray);
fill(RedCircleCell,shift( (xyz2(0,0,1)+EvenPoint5) )*((0,1)..(0.84,0)..(0,-0.64)..(-0.84,0)..cycle), white);
draw(RedCircleCell,shift( (xyz2(0,0,1)+EvenPoint5) )*unitcircle, black+0.2);

fill(RedCircleCell,shift( (xyz2(0,1,1)+EvenPoint5) )*unitcircle, gray);
fill(RedCircleCell,shift( (xyz2(0,1,1)+EvenPoint5) )*((0,1)..(0.84,0)..(0,-0.64)..(-0.84,0)..cycle), white);
draw(RedCircleCell,shift( (xyz2(0,1,1)+EvenPoint5) )*unitcircle, black+0.2);

fill(RedCircleCell,shift( (xyz2(0,1,0)+EvenPoint5) )*unitcircle, gray);
fill(RedCircleCell,shift( (xyz2(0,1,0)+EvenPoint5) )*((0,1)..(0.84,0)..(0,-0.64)..(-0.84,0)..cycle), white);
draw(RedCircleCell,shift( (xyz2(0,1,0)+EvenPoint5) )*unitcircle, black+0.2);

fill(RedCircleCell,shift( (xyz2(1,0,1)+EvenPoint5) )*unitcircle, gray);
fill(RedCircleCell,shift( (xyz2(1,0,1)+EvenPoint5) )*((0,1)..(0.84,0)..(0,-0.64)..(-0.84,0)..cycle), white);
draw(RedCircleCell,shift( (xyz2(1,0,1)+EvenPoint5)  )*unitcircle, black+0.2);

fill(RedCircleCell,shift( (xyz2(1,1,1)+EvenPoint5) )*unitcircle, gray);
fill(RedCircleCell,shift( (xyz2(1,1,1)+EvenPoint5) )*((0,1)..(0.84,0)..(0,-0.64)..(-0.84,0)..cycle), white);
draw(RedCircleCell,shift( (xyz2(1,1,1)+EvenPoint5)  )*unitcircle, black+0.2);

// Right square qubits

fill(RedCircleCell,shift( (xyz2(1,1,1)+OddPoint1) )*unitcircle, gray);
fill(RedCircleCell,shift( (xyz2(1,1,1)+OddPoint1) )*((0,1)..(0.84,0)..(0,-0.64)..(-0.84,0)..cycle), white);
draw(RedCircleCell,shift( (xyz2(1,1,1)+OddPoint1) )*unitcircle, black+0.2);

fill(RedCircleCell,shift(  (xyz2(1,0,1)+EvenPoint4) )*unitcircle, gray);
fill(RedCircleCell,shift(  (xyz2(1,0,1)+EvenPoint4) )*((0,1)..(0.84,0)..(0,-0.64)..(-0.84,0)..cycle), white);
draw(RedCircleCell,shift(  (xyz2(1,0,1)+EvenPoint4) )*unitcircle, black+0.2);

// Bottom square qubits

fill(RedCircleCell,shift( (xyz2(0,0,1)+OddPoint1))*unitcircle, gray);
fill(RedCircleCell,shift( (xyz2(0,0,1)+OddPoint1) )*((0,1)..(0.84,0)..(0,-0.64)..(-0.84,0)..cycle), white);
draw(RedCircleCell,shift( (xyz2(0,0,1)+OddPoint1) )*unitcircle, black+0.2);

fill(RedCircleCell,shift( (xyz2(1,0,1)+EvenPoint1) )*unitcircle, gray);
fill(RedCircleCell,shift( (xyz2(1,0,1)+EvenPoint1) )*((0,1)..(0.84,0)..(0,-0.64)..(-0.84,0)..cycle), white);
draw(RedCircleCell,shift( (xyz2(1,0,1)+EvenPoint1) )*unitcircle, black+0.2);

// Side square face qubits

fill(RedCircleCell,shift( (xyz2(0,0,0)+EvenPoint3) )*unitcircle, gray);
fill(RedCircleCell,shift( (xyz2(0,0,0)+EvenPoint3) )*((0,1)..(0.84,0)..(0,-0.64)..(-0.84,0)..cycle), white);
draw(RedCircleCell,shift( (xyz2(0,0,0)+EvenPoint3) )*unitcircle, black+0.2);

fill(RedCircleCell,shift( (xyz2(0,0,1)+OddPoint4) )*unitcircle, gray);
fill(RedCircleCell,shift( (xyz2(0,0,1)+OddPoint4) )*((0,1)..(0.84,0)..(0,-0.64)..(-0.84,0)..cycle), white);
draw(RedCircleCell,shift( (xyz2(0,0,1)+OddPoint4) )*unitcircle, black+0.2);

fill(RedCircleCell,shift( (xyz2(0,1,1)+EvenPoint1) )*unitcircle, gray);
fill(RedCircleCell,shift( (xyz2(0,1,1)+EvenPoint1) )*((0,1)..(0.84,0)..(0,-0.64)..(-0.84,0)..cycle), white);
draw(RedCircleCell,shift( (xyz2(0,1,1)+EvenPoint1) )*unitcircle, black+0.2);

fill(RedCircleCell,shift( (xyz2(0,1,0)+OddPoint2) )*unitcircle, gray);
fill(RedCircleCell,shift( (xyz2(0,1,0)+OddPoint2) )*((0,1)..(0.84,0)..(0,-0.64)..(-0.84,0)..cycle), white);
draw(RedCircleCell,shift( (xyz2(0,1,0)+OddPoint2)  )*unitcircle, black+0.2);

// Top square face qubits

fill(RedCircleCell,shift( (xyz2(0,1,0)+OddPoint3) )*unitcircle, gray);
fill(RedCircleCell,shift( (xyz2(0,1,0)+OddPoint3) )*((0,1)..(0.84,0)..(0,-0.64)..(-0.84,0)..cycle), white);
draw(RedCircleCell,shift( (xyz2(0,1,0)+OddPoint3) )*unitcircle, black+0.2);

fill(RedCircleCell,shift( (xyz2(0,1,1)+EvenPoint4) )*unitcircle, gray);
fill(RedCircleCell,shift( (xyz2(0,1,1)+EvenPoint4) )*((0,1)..(0.84,0)..(0,-0.64)..(-0.84,0)..cycle), white);
draw(RedCircleCell,shift( (xyz2(0,1,1)+EvenPoint4) )*unitcircle, black+0.2);

fill(RedCircleCell,shift( (xyz2(1,1,1)+OddPoint4) )*unitcircle, gray);
fill(RedCircleCell,shift( (xyz2(1,1,1)+OddPoint4) )*((0,1)..(0.84,0)..(0,-0.64)..(-0.84,0)..cycle), white);
draw(RedCircleCell,shift( (xyz2(1,1,1)+OddPoint4)  )*unitcircle, black+0.2);

fill(RedCircleCell,shift( (xyz2(1,1,0)+EvenPoint3) )*unitcircle, gray);
fill(RedCircleCell,shift( (xyz2(1,1,0)+EvenPoint3) )*((0,1)..(0.84,0)..(0,-0.64)..(-0.84,0)..cycle), white);
draw(RedCircleCell,shift( (xyz2(1,1,0)+EvenPoint3) )*unitcircle, black+0.2);

// Front square qubits

fill(RedCircleCell,shift( (xyz2(0,0,1)+OddPoint3) )*unitcircle, gray);
fill(RedCircleCell,shift( (xyz2(0,0,1)+OddPoint3) )*((0,1)..(0.84,0)..(0,-0.64)..(-0.84,0)..cycle), white);
draw(RedCircleCell,shift( (xyz2(0,0,1)+OddPoint3)  )*unitcircle, black+0.2);

fill(RedCircleCell,shift( (xyz2(1,0,1)+EvenPoint3) )*unitcircle, gray);
fill(RedCircleCell,shift(  (xyz2(1,0,1)+EvenPoint3)  )*((0,1)..(0.84,0)..(0,-0.64)..(-0.84,0)..cycle), white);
draw(RedCircleCell,shift(  (xyz2(1,0,1)+EvenPoint3) )*unitcircle, black+0.2);

fill(RedCircleCell,shift( (xyz2(1,1,1)+OddPoint2) )*unitcircle, gray);
fill(RedCircleCell,shift( (xyz2(1,1,1)+OddPoint2) )*((0,1)..(0.84,0)..(0,-0.64)..(-0.84,0)..cycle), white);
draw(RedCircleCell,shift( (xyz2(1,1,1)+OddPoint2) )*unitcircle, black+0.2);

fill(RedCircleCell,shift( (xyz2(0,1,1)+EvenPoint2) )*unitcircle, gray);
fill(RedCircleCell,shift( (xyz2(0,1,1)+EvenPoint2) )*((0,1)..(0.84,0)..(0,-0.64)..(-0.84,0)..cycle), white);
draw(RedCircleCell,shift( (xyz2(0,1,1)+EvenPoint2)  )*unitcircle, black+0.2);

label("$ = Z_S = X_S\in\mathcal{S}$", xyz2(-3.4,0.6,0));
add(RedCircleCell, xyz2(0,0,0));

\end{asy}

\begin{asy}
pair xyz2(real x, real y, real z)
{
return 50*(-x - z / 3+1.5, y - z / 6 - 0.1);
}

pair PrimalPoint1 = xyz2(0,0,0);
pair PrimalPoint2 = xyz2(0,0,1);
pair PrimalPoint3 = xyz2(0,1,0);
pair PrimalPoint4 = xyz2(0,1,1);
pair PrimalPoint5 = xyz2(1,0,0);
pair PrimalPoint6 = xyz2(1,0,1);
pair PrimalPoint7 = xyz2(1,1,0);
pair PrimalPoint8 = xyz2(1,1,1);

pair EvenPoint1 = (PrimalPoint1+PrimalPoint2+PrimalPoint3+PrimalPoint5) / 4;
pair EvenPoint2 = (PrimalPoint2+PrimalPoint5+PrimalPoint6+PrimalPoint8) / 4;
pair EvenPoint3 = (PrimalPoint4+PrimalPoint2+PrimalPoint3+PrimalPoint8) / 4;
pair EvenPoint4 = (PrimalPoint7+PrimalPoint8+PrimalPoint3+PrimalPoint5) / 4;
pair EvenPoint5 = (PrimalPoint5+PrimalPoint2+PrimalPoint3+PrimalPoint8) / 4;

pair OddPoint1 = (PrimalPoint1 + PrimalPoint5 + PrimalPoint6 + PrimalPoint7) / 4;
pair OddPoint2 = (PrimalPoint1 + PrimalPoint2 + PrimalPoint4 + PrimalPoint6) / 4;
pair OddPoint3 = (PrimalPoint4 + PrimalPoint8 + PrimalPoint6 + PrimalPoint7) / 4;
pair OddPoint4 = (PrimalPoint1 + PrimalPoint3 + PrimalPoint4 + PrimalPoint7) / 4;
pair OddPoint5 = (PrimalPoint1 + PrimalPoint4 + PrimalPoint6 + PrimalPoint7) / 4;

picture RedCircleCell;

filldraw(RedCircleCell, (xyz2(0,0,0)+EvenPoint3)--(xyz2(0,0,0)+EvenPoint5)--(xyz2(0,0,0)+EvenPoint2)--(xyz2(0,0,1)+OddPoint1)--(xyz2(0,0,1)+OddPoint5)--(xyz2(0,0,1)+OddPoint4)--cycle,palered,gray);

filldraw(RedCircleCell, (xyz2(0,1,0)+OddPoint3)--(xyz2(0,1,1)+EvenPoint4)--(xyz2(1,1,1)+OddPoint4)--(xyz2(1,1,0)+EvenPoint3)--cycle,palered,gray);

filldraw(RedCircleCell, (xyz2(0,1,1)+EvenPoint2)--(xyz2(1,1,1)+OddPoint2)--(xyz2(1,0,1)+EvenPoint3)--(xyz2(0,0,1)+OddPoint3)--cycle,palered,gray);
filldraw(RedCircleCell, (xyz2(0,1,1)+EvenPoint1)--(xyz2(0,1,1)+EvenPoint5)--(xyz2(0,1,1)+EvenPoint2)--(xyz2(0,0,1)+OddPoint3)--(xyz2(0,0,1)+OddPoint5)--(xyz2(0,0,1)+OddPoint4)--cycle,heavyred,gray);

filldraw(RedCircleCell, (xyz2(0,1,1)+EvenPoint2)--(xyz2(0,1,1)+EvenPoint5)--(xyz2(0,1,1)+EvenPoint4)--(xyz2(1,1,1)+OddPoint4)--(xyz2(1,1,1)+OddPoint5)--(xyz2(1,1,1)+OddPoint2)--cycle, palered,gray);

filldraw(RedCircleCell, (xyz2(1,0,1)+EvenPoint3)--(xyz2(1,0,1)+EvenPoint5)--(xyz2(1,0,1)+EvenPoint4)--(xyz2(1,1,1)+OddPoint1)--(xyz2(1,1,1)+OddPoint5)--(xyz2(1,1,1)+OddPoint2)--cycle,palered,gray);

filldraw(RedCircleCell, (xyz2(0,1,1)+EvenPoint1)--(xyz2(0,1,1)+EvenPoint5)--(xyz2(0,1,1)+EvenPoint4)--(xyz2(0,1,0)+OddPoint3)--(xyz2(0,1,0)+OddPoint5)--(xyz2(0,1,0)+OddPoint2)--cycle,palered,gray);

filldraw(RedCircleCell, (xyz2(1,0,1)+EvenPoint1)--(xyz2(1,0,1)+EvenPoint5)--(xyz2(1,0,1)+EvenPoint3)--(xyz2(0,0,1)+OddPoint3)--(xyz2(0,0,1)+OddPoint5)--(xyz2(0,0,1)+OddPoint1)--cycle,palered,gray);

filldraw(RedCircleCell, (xyz2(0,0,0)+EvenPoint3)--(xyz2(0,0,1)+OddPoint4)--(xyz2(0,1,1)+EvenPoint1)--(xyz2(0,1,0)+OddPoint2)--cycle,palered,gray);

// Central cube qubits

fill(RedCircleCell,shift( (xyz2(0,0,0)+EvenPoint5) )*unitcircle, gray);
fill(RedCircleCell,shift( (xyz2(0,0,0)+EvenPoint5)  )*((0,1)..(0.84,0)..(0,-0.64)..(-0.84,0)..cycle), white);
draw(RedCircleCell,shift( (xyz2(0,0,0)+EvenPoint5)  )*unitcircle, black+0.2);

fill(RedCircleCell,shift( (xyz2(0,0,1)+EvenPoint5) )*unitcircle, gray);
fill(RedCircleCell,shift( (xyz2(0,0,1)+EvenPoint5) )*((0,1)..(0.84,0)..(0,-0.64)..(-0.84,0)..cycle), white);
draw(RedCircleCell,shift( (xyz2(0,0,1)+EvenPoint5) )*unitcircle, black+0.2);

fill(RedCircleCell,shift( (xyz2(0,1,1)+EvenPoint5) )*unitcircle, gray);
fill(RedCircleCell,shift( (xyz2(0,1,1)+EvenPoint5) )*((0,1)..(0.84,0)..(0,-0.64)..(-0.84,0)..cycle), white);
draw(RedCircleCell,shift( (xyz2(0,1,1)+EvenPoint5) )*unitcircle, black+0.2);

fill(RedCircleCell,shift( (xyz2(0,1,0)+EvenPoint5) )*unitcircle, gray);
fill(RedCircleCell,shift( (xyz2(0,1,0)+EvenPoint5) )*((0,1)..(0.84,0)..(0,-0.64)..(-0.84,0)..cycle), white);
draw(RedCircleCell,shift( (xyz2(0,1,0)+EvenPoint5) )*unitcircle, black+0.2);

fill(RedCircleCell,shift( (xyz2(1,0,1)+EvenPoint5) )*unitcircle, gray);
fill(RedCircleCell,shift( (xyz2(1,0,1)+EvenPoint5) )*((0,1)..(0.84,0)..(0,-0.64)..(-0.84,0)..cycle), white);
draw(RedCircleCell,shift( (xyz2(1,0,1)+EvenPoint5)  )*unitcircle, black+0.2);

fill(RedCircleCell,shift( (xyz2(1,1,1)+EvenPoint5) )*unitcircle, gray);
fill(RedCircleCell,shift( (xyz2(1,1,1)+EvenPoint5) )*((0,1)..(0.84,0)..(0,-0.64)..(-0.84,0)..cycle), white);
draw(RedCircleCell,shift( (xyz2(1,1,1)+EvenPoint5)  )*unitcircle, black+0.2);

// Right square qubits

fill(RedCircleCell,shift( (xyz2(1,1,1)+OddPoint1) )*unitcircle, gray);
fill(RedCircleCell,shift( (xyz2(1,1,1)+OddPoint1) )*((0,1)..(0.84,0)..(0,-0.64)..(-0.84,0)..cycle), white);
draw(RedCircleCell,shift( (xyz2(1,1,1)+OddPoint1) )*unitcircle, black+0.2);

fill(RedCircleCell,shift(  (xyz2(1,0,1)+EvenPoint4) )*unitcircle, gray);
fill(RedCircleCell,shift(  (xyz2(1,0,1)+EvenPoint4) )*((0,1)..(0.84,0)..(0,-0.64)..(-0.84,0)..cycle), white);
draw(RedCircleCell,shift(  (xyz2(1,0,1)+EvenPoint4) )*unitcircle, black+0.2);

// Bottom square qubits

fill(RedCircleCell,shift( (xyz2(0,0,1)+OddPoint1))*unitcircle, gray);
fill(RedCircleCell,shift( (xyz2(0,0,1)+OddPoint1) )*((0,1)..(0.84,0)..(0,-0.64)..(-0.84,0)..cycle), white);
draw(RedCircleCell,shift( (xyz2(0,0,1)+OddPoint1) )*unitcircle, black+0.2);

fill(RedCircleCell,shift( (xyz2(1,0,1)+EvenPoint1) )*unitcircle, gray);
fill(RedCircleCell,shift( (xyz2(1,0,1)+EvenPoint1) )*((0,1)..(0.84,0)..(0,-0.64)..(-0.84,0)..cycle), white);
draw(RedCircleCell,shift( (xyz2(1,0,1)+EvenPoint1) )*unitcircle, black+0.2);

// Side square face qubits

fill(RedCircleCell,shift( (xyz2(0,0,0)+EvenPoint3) )*unitcircle, gray);
fill(RedCircleCell,shift( (xyz2(0,0,0)+EvenPoint3) )*((0,1)..(0.84,0)..(0,-0.64)..(-0.84,0)..cycle), white);
draw(RedCircleCell,shift( (xyz2(0,0,0)+EvenPoint3) )*unitcircle, black+0.2);

fill(RedCircleCell,shift( (xyz2(0,0,1)+OddPoint4) )*unitcircle, gray);
fill(RedCircleCell,shift( (xyz2(0,0,1)+OddPoint4) )*((0,1)..(0.84,0)..(0,-0.64)..(-0.84,0)..cycle), white);
draw(RedCircleCell,shift( (xyz2(0,0,1)+OddPoint4) )*unitcircle, black+0.2);

fill(RedCircleCell,shift( (xyz2(0,1,1)+EvenPoint1) )*unitcircle, gray);
fill(RedCircleCell,shift( (xyz2(0,1,1)+EvenPoint1) )*((0,1)..(0.84,0)..(0,-0.64)..(-0.84,0)..cycle), white);
draw(RedCircleCell,shift( (xyz2(0,1,1)+EvenPoint1) )*unitcircle, black+0.2);

fill(RedCircleCell,shift( (xyz2(0,1,0)+OddPoint2) )*unitcircle, gray);
fill(RedCircleCell,shift( (xyz2(0,1,0)+OddPoint2) )*((0,1)..(0.84,0)..(0,-0.64)..(-0.84,0)..cycle), white);
draw(RedCircleCell,shift( (xyz2(0,1,0)+OddPoint2)  )*unitcircle, black+0.2);

// Top square face qubits

fill(RedCircleCell,shift( (xyz2(0,1,0)+OddPoint3) )*unitcircle, gray);
fill(RedCircleCell,shift( (xyz2(0,1,0)+OddPoint3) )*((0,1)..(0.84,0)..(0,-0.64)..(-0.84,0)..cycle), white);
draw(RedCircleCell,shift( (xyz2(0,1,0)+OddPoint3) )*unitcircle, black+0.2);

fill(RedCircleCell,shift( (xyz2(0,1,1)+EvenPoint4) )*unitcircle, gray);
fill(RedCircleCell,shift( (xyz2(0,1,1)+EvenPoint4) )*((0,1)..(0.84,0)..(0,-0.64)..(-0.84,0)..cycle), white);
draw(RedCircleCell,shift( (xyz2(0,1,1)+EvenPoint4) )*unitcircle, black+0.2);

fill(RedCircleCell,shift( (xyz2(1,1,1)+OddPoint4) )*unitcircle, gray);
fill(RedCircleCell,shift( (xyz2(1,1,1)+OddPoint4) )*((0,1)..(0.84,0)..(0,-0.64)..(-0.84,0)..cycle), white);
draw(RedCircleCell,shift( (xyz2(1,1,1)+OddPoint4)  )*unitcircle, black+0.2);

fill(RedCircleCell,shift( (xyz2(1,1,0)+EvenPoint3) )*unitcircle, gray);
fill(RedCircleCell,shift( (xyz2(1,1,0)+EvenPoint3) )*((0,1)..(0.84,0)..(0,-0.64)..(-0.84,0)..cycle), white);
draw(RedCircleCell,shift( (xyz2(1,1,0)+EvenPoint3) )*unitcircle, black+0.2);

// Front square qubits

fill(RedCircleCell,shift( (xyz2(0,0,1)+OddPoint3) )*unitcircle, gray);
fill(RedCircleCell,shift( (xyz2(0,0,1)+OddPoint3) )*((0,1)..(0.84,0)..(0,-0.64)..(-0.84,0)..cycle), white);
draw(RedCircleCell,shift( (xyz2(0,0,1)+OddPoint3)  )*unitcircle, black+0.2);

fill(RedCircleCell,shift( (xyz2(1,0,1)+EvenPoint3) )*unitcircle, gray);
fill(RedCircleCell,shift(  (xyz2(1,0,1)+EvenPoint3)  )*((0,1)..(0.84,0)..(0,-0.64)..(-0.84,0)..cycle), white);
draw(RedCircleCell,shift(  (xyz2(1,0,1)+EvenPoint3) )*unitcircle, black+0.2);

fill(RedCircleCell,shift( (xyz2(1,1,1)+OddPoint2) )*unitcircle, gray);
fill(RedCircleCell,shift( (xyz2(1,1,1)+OddPoint2) )*((0,1)..(0.84,0)..(0,-0.64)..(-0.84,0)..cycle), white);
draw(RedCircleCell,shift( (xyz2(1,1,1)+OddPoint2) )*unitcircle, black+0.2);

fill(RedCircleCell,shift( (xyz2(0,1,1)+EvenPoint2) )*unitcircle, gray);
fill(RedCircleCell,shift( (xyz2(0,1,1)+EvenPoint2) )*((0,1)..(0.84,0)..(0,-0.64)..(-0.84,0)..cycle), white);
draw(RedCircleCell,shift( (xyz2(0,1,1)+EvenPoint2)  )*unitcircle, black+0.2);

label("$ = Z_G = X_G\in\mathcal{G} $", xyz2(-3.4,0.6,0));
add(RedCircleCell, xyz2(0,0,0));

\end{asy}
\end{multicols}
\caption{\label{fig:gauge_operators} \textbf{Gauge elements in three-dimensional gauge color code} $CC_\mathcal{L}(x=0,z=0)$, depicted in the primal lattice $\mathcal{L}^*$. In the primal lattice depicted, qubits are attached to vertices, gauge operators to faces (right figure), and stabilizer operators to cells (left figure), as explained in the main text. Figure similar to those appearing in Ref.~\cite{brown2016fault}, generating code modified from the one courteously provided by Prof. Benjamin Brown.}
\end{figure}

Color codes are also CSS subsystem stabilizer codes, a generalization of stabilizer codes in which code generators do not need to commute~\cite{kubica2018ungauging}.
They are defined by their gauge group~\cite{bombin2015gauge}:
\begin{equation}
    \mathcal{G} = \braket{X(\delta),Z(\sigma)|\forall \delta \in \Delta'_{d-2-z}(\mathcal{L}), \forall\sigma \in \Delta'_{d-2-x}(\mathcal{L})},
\end{equation}
with\footnote{The $d-2$ condition is due to the minimal dimension of color codes being $d = 2$.} $x+z \leq d-2$. This induces a stabilizer group over the gauge group, $\mathcal{S} = \mathcal{Z}_{\mathcal{G}}(\mathcal{G})\subset \mathcal{G}$~\cite{bombin2015gauge},
\begin{equation}
    \mathcal{S} = \braket{X(\delta),Z(\sigma)|\forall \delta \in \Delta'_{x}(\mathcal{L}), \forall\sigma \in \Delta'_{z}(\mathcal{L})}.
\end{equation}
We refer to such color code as the $CC_{\mathcal{L}}(x,z)$. For instance, in 2 dimensions there is only one color code on $\mathcal{L}$, $CC_\mathcal{L}(0,0)$. Stabilizer generators are attached to vertices ($\Delta_0(\cal{L})$) and coincide with the gauge generators, whereas physical qubits are attached to faces ($\delta \in \Delta_{d=2}(\cal{L})$). The support of a stabilizer operator on vertex $\sigma$ are those qubits whose faces contain the vertex $\sigma$, i.e. $\cal{Q}(\sigma)$. Any color code for which $\mathcal{G}= \mathcal{S}$ is a stabilizer color code. Else, it is a subsystem color code. 

\begin{figure}[!t]
\centering
\noindent 

\begin{multicols}{2}

\begin{asy}
pair xyz2(real x, real y, real z)
{
return 24*(+x + z / 3-1.5, y - z / 4 - 0.1);
}

pair PrimalPoint1 = xyz2(0,0,0);
pair PrimalPoint2 = xyz2(0,0,1);
pair PrimalPoint3 = xyz2(0,1,0);
pair PrimalPoint4 = xyz2(0,1,1);
pair PrimalPoint5 = xyz2(1,0,0);
pair PrimalPoint6 = xyz2(1,0,1);
pair PrimalPoint7 = xyz2(1,1,0);
pair PrimalPoint8 = xyz2(1,1,1);

pair EvenPoint1 = (PrimalPoint1+PrimalPoint2+PrimalPoint3+PrimalPoint5) / 4;
pair EvenPoint2 = (PrimalPoint2+PrimalPoint5+PrimalPoint6+PrimalPoint8) / 4;
pair EvenPoint3 = (PrimalPoint4+PrimalPoint2+PrimalPoint3+PrimalPoint8) / 4;
pair EvenPoint4 = (PrimalPoint7+PrimalPoint8+PrimalPoint3+PrimalPoint5) / 4;
pair EvenPoint5 = (PrimalPoint5+PrimalPoint2+PrimalPoint3+PrimalPoint8) / 4;

pair OddPoint1 = (PrimalPoint1 + PrimalPoint5 + PrimalPoint6 + PrimalPoint7) / 4;
pair OddPoint2 = (PrimalPoint1 + PrimalPoint2 + PrimalPoint4 + PrimalPoint6) / 4;
pair OddPoint3 = (PrimalPoint4 + PrimalPoint8 + PrimalPoint6 + PrimalPoint7) / 4;
pair OddPoint4 = (PrimalPoint1 + PrimalPoint3 + PrimalPoint4 + PrimalPoint7) / 4;
pair OddPoint5 = (PrimalPoint1 + PrimalPoint4 + PrimalPoint6 + PrimalPoint7) / 4;

picture PauliX;

filldraw(PauliX, shift(xyz2(0,0,0))*scale(4)*unitcircle, white, black+1);
label(PauliX, shift(xyz2(0,0,0))*scale(3/5)*"X");

picture RedDiamondCell;

filldraw(RedDiamondCell, (xyz2(0,1,1)+OddPoint1)--(xyz2(1,1,1)+EvenPoint1)--(xyz2(1,0,1)+OddPoint4)--(xyz2(0,0,1)+EvenPoint4)--cycle, palered,gray);
filldraw(RedDiamondCell, (xyz2(0,1,1)+OddPoint1)--(xyz2(1,1,1)+EvenPoint1)--(xyz2(1,1,0)+OddPoint2)--(xyz2(0,1,0)+EvenPoint2)--cycle,palered,gray);
filldraw(RedDiamondCell, (xyz2(0,1,1)+OddPoint1)--(xyz2(0,1,0)+EvenPoint2)--(xyz2(0,0,0)+OddPoint3)--(xyz2(0,0,1)+EvenPoint4)--cycle,palered,gray);

fill(RedDiamondCell,shift( (xyz2(0,1,0)+EvenPoint2) )*unitcircle, gray);
fill(RedDiamondCell,shift( (xyz2(0,1,0)+EvenPoint2) )*((0,1)..(0.84,0)..(0,-0.64)..(-0.84,0)..cycle), white);
draw(RedDiamondCell,shift( (xyz2(0,1,0)+EvenPoint2) )*unitcircle, black+0.2);

fill(RedDiamondCell,shift( (xyz2(0,0,0)+OddPoint3) )*unitcircle, gray);
fill(RedDiamondCell,shift( (xyz2(0,0,0)+OddPoint3) )*((0,1)..(0.84,0)..(0,-0.64)..(-0.84,0)..cycle), white);
draw(RedDiamondCell,shift( (xyz2(0,0,0)+OddPoint3) )*unitcircle, black+0.2);

fill(RedDiamondCell,shift( (xyz2(1,1,0)+OddPoint2) )*unitcircle, gray);
fill(RedDiamondCell,shift( (xyz2(1,1,0)+OddPoint2) )*((0,1)..(0.84,0)..(0,-0.64)..(-0.84,0)..cycle), white);
draw(RedDiamondCell,shift( (xyz2(1,1,0)+OddPoint2) )*unitcircle, black+0.2);

fill(RedDiamondCell,shift( (xyz2(0,1,1)+OddPoint1) )*unitcircle, gray);
fill(RedDiamondCell,shift( (xyz2(0,1,1)+OddPoint1) )*((0,1)..(0.84,0)..(0,-0.64)..(-0.84,0)..cycle), white);
draw(RedDiamondCell,shift( (xyz2(0,1,1)+OddPoint1) )*unitcircle, black+0.2);

fill(RedDiamondCell,shift( (xyz2(1,1,1)+EvenPoint1) )*unitcircle, gray);
fill(RedDiamondCell,shift( (xyz2(1,1,1)+EvenPoint1) )*((0,1)..(0.84,0)..(0,-0.64)..(-0.84,0)..cycle), white);
draw(RedDiamondCell,shift( (xyz2(1,1,1)+EvenPoint1) )*unitcircle, black+0.2);

fill(RedDiamondCell,shift( (xyz2(1,0,1)+OddPoint4) )*unitcircle, gray);
fill(RedDiamondCell,shift( (xyz2(1,0,1)+OddPoint4) )*((0,1)..(0.84,0)..(0,-0.64)..(-0.84,0)..cycle), white);
draw(RedDiamondCell,shift( (xyz2(1,0,1)+OddPoint4) )*unitcircle, black+0.2);

fill(RedDiamondCell,shift((xyz2(0,0,1)+EvenPoint4) )*unitcircle, gray);
fill(RedDiamondCell,shift( (xyz2(0,0,1)+EvenPoint4)  )*((0,1)..(0.84,0)..(0,-0.64)..(-0.84,0)..cycle), white);
draw(RedDiamondCell,shift( (xyz2(0,0,1)+EvenPoint4)  )*unitcircle, black+0.2);

picture GreenDiamondCell;

filldraw(GreenDiamondCell, (xyz2(0,1,1)+OddPoint1)--(xyz2(1,1,1)+EvenPoint1)--(xyz2(1,0,1)+OddPoint4)--(xyz2(0,0,1)+EvenPoint4)--cycle, palegreen,gray);
filldraw(GreenDiamondCell, (xyz2(0,1,1)+OddPoint1)--(xyz2(1,1,1)+EvenPoint1)--(xyz2(1,1,0)+OddPoint2)--(xyz2(0,1,0)+EvenPoint2)--cycle,palegreen,gray);
filldraw(GreenDiamondCell, (xyz2(0,1,1)+OddPoint1)--(xyz2(0,1,0)+EvenPoint2)--(xyz2(0,0,0)+OddPoint3)--(xyz2(0,0,1)+EvenPoint4)--cycle, palegreen,gray);

fill(GreenDiamondCell,shift( (xyz2(0,1,0)+EvenPoint2) )*unitcircle, gray);
fill(GreenDiamondCell,shift( (xyz2(0,1,0)+EvenPoint2) )*((0,1)..(0.84,0)..(0,-0.64)..(-0.84,0)..cycle), white);
draw(GreenDiamondCell,shift( (xyz2(0,1,0)+EvenPoint2) )*unitcircle, black+0.2);

fill(GreenDiamondCell,shift( (xyz2(0,0,0)+OddPoint3) )*unitcircle, gray);
fill(GreenDiamondCell,shift( (xyz2(0,0,0)+OddPoint3) )*((0,1)..(0.84,0)..(0,-0.64)..(-0.84,0)..cycle), white);
draw(GreenDiamondCell,shift( (xyz2(0,0,0)+OddPoint3) )*unitcircle, black+0.2);

fill(GreenDiamondCell,shift( (xyz2(1,1,0)+OddPoint2) )*unitcircle, gray);
fill(GreenDiamondCell,shift( (xyz2(1,1,0)+OddPoint2) )*((0,1)..(0.84,0)..(0,-0.64)..(-0.84,0)..cycle), white);
draw(GreenDiamondCell,shift( (xyz2(1,1,0)+OddPoint2) )*unitcircle, black+0.2);

fill(GreenDiamondCell,shift( (xyz2(0,1,1)+OddPoint1) )*unitcircle, gray);
fill(GreenDiamondCell,shift( (xyz2(0,1,1)+OddPoint1) )*((0,1)..(0.84,0)..(0,-0.64)..(-0.84,0)..cycle), white);
draw(GreenDiamondCell,shift( (xyz2(0,1,1)+OddPoint1) )*unitcircle, black+0.2);

fill(GreenDiamondCell,shift( (xyz2(1,1,1)+EvenPoint1) )*unitcircle, gray);
fill(GreenDiamondCell,shift( (xyz2(1,1,1)+EvenPoint1) )*((0,1)..(0.84,0)..(0,-0.64)..(-0.84,0)..cycle), white);
draw(GreenDiamondCell,shift( (xyz2(1,1,1)+EvenPoint1) )*unitcircle, black+0.2);

fill(GreenDiamondCell,shift( (xyz2(1,0,1)+OddPoint4) )*unitcircle, gray);
fill(GreenDiamondCell,shift( (xyz2(1,0,1)+OddPoint4) )*((0,1)..(0.84,0)..(0,-0.64)..(-0.84,0)..cycle), white);
draw(GreenDiamondCell,shift( (xyz2(1,0,1)+OddPoint4) )*unitcircle, black+0.2);

fill(GreenDiamondCell,shift((xyz2(0,0,1)+EvenPoint4) )*unitcircle, gray);
fill(GreenDiamondCell,shift( (xyz2(0,0,1)+EvenPoint4)  )*((0,1)..(0.84,0)..(0,-0.64)..(-0.84,0)..cycle), white);
draw(GreenDiamondCell,shift( (xyz2(0,0,1)+EvenPoint4)  )*unitcircle, black+0.2);

picture BrightBlueDiamondCell;

filldraw(BrightBlueDiamondCell, (xyz2(0,1,1)+OddPoint1)--(xyz2(1,1,1)+EvenPoint1)--(xyz2(1,0,1)+OddPoint4)--(xyz2(0,0,1)+EvenPoint4)--cycle,mediumblue,gray);
filldraw(BrightBlueDiamondCell, (xyz2(0,1,1)+OddPoint1)--(xyz2(1,1,1)+EvenPoint1)--(xyz2(1,1,0)+OddPoint2)--(xyz2(0,1,0)+EvenPoint2)--cycle,blue,gray);
filldraw(BrightBlueDiamondCell, (xyz2(0,1,1)+OddPoint1)--(xyz2(0,1,0)+EvenPoint2)--(xyz2(0,0,0)+OddPoint3)--(xyz2(0,0,1)+EvenPoint4)--cycle,heavyblue,gray);

fill(BrightBlueDiamondCell,shift( (xyz2(0,1,0)+EvenPoint2) )*unitcircle, gray);
fill(BrightBlueDiamondCell,shift( (xyz2(0,1,0)+EvenPoint2) )*((0,1)..(0.84,0)..(0,-0.64)..(-0.84,0)..cycle), white);
draw(BrightBlueDiamondCell,shift( (xyz2(0,1,0)+EvenPoint2) )*unitcircle, black+0.2);

fill(BrightBlueDiamondCell,shift( (xyz2(0,0,0)+OddPoint3) )*unitcircle, gray);
fill(BrightBlueDiamondCell,shift( (xyz2(0,0,0)+OddPoint3) )*((0,1)..(0.84,0)..(0,-0.64)..(-0.84,0)..cycle), white);
draw(BrightBlueDiamondCell,shift( (xyz2(0,0,0)+OddPoint3) )*unitcircle, black+0.2);

fill(BrightBlueDiamondCell,shift( (xyz2(1,1,0)+OddPoint2) )*unitcircle, gray);
fill(BrightBlueDiamondCell,shift( (xyz2(1,1,0)+OddPoint2) )*((0,1)..(0.84,0)..(0,-0.64)..(-0.84,0)..cycle), white);
draw(BrightBlueDiamondCell,shift( (xyz2(1,1,0)+OddPoint2) )*unitcircle, black+0.2);

fill(BrightBlueDiamondCell,shift( (xyz2(0,1,1)+OddPoint1) )*unitcircle, gray);
fill(BrightBlueDiamondCell,shift( (xyz2(0,1,1)+OddPoint1) )*((0,1)..(0.84,0)..(0,-0.64)..(-0.84,0)..cycle), white);
draw(BrightBlueDiamondCell,shift( (xyz2(0,1,1)+OddPoint1) )*unitcircle, black+0.2);

fill(BrightBlueDiamondCell,shift( (xyz2(1,1,1)+EvenPoint1) )*unitcircle, gray);
fill(BrightBlueDiamondCell,shift( (xyz2(1,1,1)+EvenPoint1) )*((0,1)..(0.84,0)..(0,-0.64)..(-0.84,0)..cycle), white);
draw(BrightBlueDiamondCell,shift( (xyz2(1,1,1)+EvenPoint1) )*unitcircle, black+0.2);

fill(BrightBlueDiamondCell,shift( (xyz2(1,0,1)+OddPoint4) )*unitcircle, gray);
fill(BrightBlueDiamondCell,shift( (xyz2(1,0,1)+OddPoint4) )*((0,1)..(0.84,0)..(0,-0.64)..(-0.84,0)..cycle), white);
draw(BrightBlueDiamondCell,shift( (xyz2(1,0,1)+OddPoint4) )*unitcircle, black+0.2);

fill(BrightBlueDiamondCell,shift((xyz2(0,0,1)+EvenPoint4) )*unitcircle, gray);
fill(BrightBlueDiamondCell,shift( (xyz2(0,0,1)+EvenPoint4)  )*((0,1)..(0.84,0)..(0,-0.64)..(-0.84,0)..cycle), white);
draw(BrightBlueDiamondCell,shift( (xyz2(0,0,1)+EvenPoint4)  )*unitcircle, black+0.2);

picture BlueDiamondCell;

filldraw(BlueDiamondCell, (xyz2(0,1,1)+OddPoint1)--(xyz2(1,1,1)+EvenPoint1)--(xyz2(1,0,1)+OddPoint4)--(xyz2(0,0,1)+EvenPoint4)--cycle, paleblue,gray);
filldraw(BlueDiamondCell, (xyz2(0,1,1)+OddPoint1)--(xyz2(1,1,1)+EvenPoint1)--(xyz2(1,1,0)+OddPoint2)--(xyz2(0,1,0)+EvenPoint2)--cycle,paleblue,gray);
filldraw(BlueDiamondCell, (xyz2(0,1,1)+OddPoint1)--(xyz2(0,1,0)+EvenPoint2)--(xyz2(0,0,0)+OddPoint3)--(xyz2(0,0,1)+EvenPoint4)--cycle,paleblue,gray);

fill(BlueDiamondCell,shift( (xyz2(0,1,0)+EvenPoint2) )*unitcircle, gray);
fill(BlueDiamondCell,shift( (xyz2(0,1,0)+EvenPoint2) )*((0,1)..(0.84,0)..(0,-0.64)..(-0.84,0)..cycle), white);
draw(BlueDiamondCell,shift( (xyz2(0,1,0)+EvenPoint2) )*unitcircle, black+0.2);

fill(BlueDiamondCell,shift( (xyz2(0,0,0)+OddPoint3) )*unitcircle, gray);
fill(BlueDiamondCell,shift( (xyz2(0,0,0)+OddPoint3) )*((0,1)..(0.84,0)..(0,-0.64)..(-0.84,0)..cycle), white);
draw(BlueDiamondCell,shift( (xyz2(0,0,0)+OddPoint3) )*unitcircle, black+0.2);

fill(BlueDiamondCell,shift( (xyz2(1,1,0)+OddPoint2) )*unitcircle, gray);
fill(BlueDiamondCell,shift( (xyz2(1,1,0)+OddPoint2) )*((0,1)..(0.84,0)..(0,-0.64)..(-0.84,0)..cycle), white);
draw(BlueDiamondCell,shift( (xyz2(1,1,0)+OddPoint2) )*unitcircle, black+0.2);

fill(BlueDiamondCell,shift( (xyz2(0,1,1)+OddPoint1) )*unitcircle, gray);
fill(BlueDiamondCell,shift( (xyz2(0,1,1)+OddPoint1) )*((0,1)..(0.84,0)..(0,-0.64)..(-0.84,0)..cycle), white);
draw(BlueDiamondCell,shift( (xyz2(0,1,1)+OddPoint1) )*unitcircle, black+0.2);

fill(BlueDiamondCell,shift( (xyz2(1,1,1)+EvenPoint1) )*unitcircle, gray);
fill(BlueDiamondCell,shift( (xyz2(1,1,1)+EvenPoint1) )*((0,1)..(0.84,0)..(0,-0.64)..(-0.84,0)..cycle), white);
draw(BlueDiamondCell,shift( (xyz2(1,1,1)+EvenPoint1) )*unitcircle, black+0.2);

fill(BlueDiamondCell,shift( (xyz2(1,0,1)+OddPoint4) )*unitcircle, gray);
fill(BlueDiamondCell,shift( (xyz2(1,0,1)+OddPoint4) )*((0,1)..(0.84,0)..(0,-0.64)..(-0.84,0)..cycle), white);
draw(BlueDiamondCell,shift( (xyz2(1,0,1)+OddPoint4) )*unitcircle, black+0.2);

fill(BlueDiamondCell,shift((xyz2(0,0,1)+EvenPoint4) )*unitcircle, gray);
fill(BlueDiamondCell,shift( (xyz2(0,0,1)+EvenPoint4)  )*((0,1)..(0.84,0)..(0,-0.64)..(-0.84,0)..cycle), white);
draw(BlueDiamondCell,shift( (xyz2(0,0,1)+EvenPoint4)  )*unitcircle, black+0.2);

picture YellowDiamondCell;

filldraw(YellowDiamondCell, (xyz2(0,1,1)+OddPoint1)--(xyz2(1,1,1)+EvenPoint1)--(xyz2(1,0,1)+OddPoint4)--(xyz2(0,0,1)+EvenPoint4)--cycle,lightyellow,gray);
filldraw(YellowDiamondCell, (xyz2(0,1,1)+OddPoint1)--(xyz2(1,1,1)+EvenPoint1)--(xyz2(1,1,0)+OddPoint2)--(xyz2(0,1,0)+EvenPoint2)--cycle,lightyellow,gray);
filldraw(YellowDiamondCell, (xyz2(0,1,1)+OddPoint1)--(xyz2(0,1,0)+EvenPoint2)--(xyz2(0,0,0)+OddPoint3)--(xyz2(0,0,1)+EvenPoint4)--cycle,lightyellow,gray);

fill(YellowDiamondCell,shift( (xyz2(0,1,0)+EvenPoint2) )*unitcircle, gray);
fill(YellowDiamondCell,shift( (xyz2(0,1,0)+EvenPoint2) )*((0,1)..(0.84,0)..(0,-0.64)..(-0.84,0)..cycle), white);
draw(YellowDiamondCell,shift( (xyz2(0,1,0)+EvenPoint2) )*unitcircle, black+0.2);

fill(YellowDiamondCell,shift( (xyz2(0,0,0)+OddPoint3) )*unitcircle, gray);
fill(YellowDiamondCell,shift( (xyz2(0,0,0)+OddPoint3) )*((0,1)..(0.84,0)..(0,-0.64)..(-0.84,0)..cycle), white);
draw(YellowDiamondCell,shift( (xyz2(0,0,0)+OddPoint3) )*unitcircle, black+0.2);

fill(YellowDiamondCell,shift( (xyz2(1,1,0)+OddPoint2) )*unitcircle, gray);
fill(YellowDiamondCell,shift( (xyz2(1,1,0)+OddPoint2) )*((0,1)..(0.84,0)..(0,-0.64)..(-0.84,0)..cycle), white);
draw(YellowDiamondCell,shift( (xyz2(1,1,0)+OddPoint2) )*unitcircle, black+0.2);

fill(YellowDiamondCell,shift( (xyz2(0,1,1)+OddPoint1) )*unitcircle, gray);
fill(YellowDiamondCell,shift( (xyz2(0,1,1)+OddPoint1) )*((0,1)..(0.84,0)..(0,-0.64)..(-0.84,0)..cycle), white);
draw(YellowDiamondCell,shift( (xyz2(0,1,1)+OddPoint1) )*unitcircle, black+0.2);

fill(YellowDiamondCell,shift( (xyz2(1,1,1)+EvenPoint1) )*unitcircle, gray);
fill(YellowDiamondCell,shift( (xyz2(1,1,1)+EvenPoint1) )*((0,1)..(0.84,0)..(0,-0.64)..(-0.84,0)..cycle), white);
draw(YellowDiamondCell,shift( (xyz2(1,1,1)+EvenPoint1) )*unitcircle, black+0.2);

fill(YellowDiamondCell,shift( (xyz2(1,0,1)+OddPoint4) )*unitcircle, gray);
fill(YellowDiamondCell,shift( (xyz2(1,0,1)+OddPoint4) )*((0,1)..(0.84,0)..(0,-0.64)..(-0.84,0)..cycle), white);
draw(YellowDiamondCell,shift( (xyz2(1,0,1)+OddPoint4) )*unitcircle, black+0.2);

fill(YellowDiamondCell,shift((xyz2(0,0,1)+EvenPoint4) )*unitcircle, gray);
fill(YellowDiamondCell,shift( (xyz2(0,0,1)+EvenPoint4)  )*((0,1)..(0.84,0)..(0,-0.64)..(-0.84,0)..cycle), white);
draw(YellowDiamondCell,shift( (xyz2(0,0,1)+EvenPoint4)  )*unitcircle, black+0.2);

picture BrightBlueCircleCell;

filldraw(BrightBlueCircleCell, (xyz2(0,0,0)+EvenPoint3)--(xyz2(0,0,0)+EvenPoint5)--(xyz2(0,0,0)+EvenPoint2)--(xyz2(0,0,1)+OddPoint1)--(xyz2(0,0,1)+OddPoint5)--(xyz2(0,0,1)+OddPoint4)--cycle,darkblue,gray);

filldraw(BrightBlueCircleCell, (xyz2(0,1,0)+OddPoint3)--(xyz2(0,1,1)+EvenPoint4)--(xyz2(1,1,1)+OddPoint4)--(xyz2(1,1,0)+EvenPoint3)--cycle,paleblue,gray);

filldraw(BrightBlueCircleCell, (xyz2(0,1,1)+EvenPoint2)--(xyz2(1,1,1)+OddPoint2)--(xyz2(1,0,1)+EvenPoint3)--(xyz2(0,0,1)+OddPoint3)--cycle,blue,gray);
filldraw(BrightBlueCircleCell, (xyz2(0,1,1)+EvenPoint1)--(xyz2(0,1,1)+EvenPoint5)--(xyz2(0,1,1)+EvenPoint2)--(xyz2(0,0,1)+OddPoint3)--(xyz2(0,0,1)+OddPoint5)--(xyz2(0,0,1)+OddPoint4)--cycle,heavyblue,gray);

filldraw(BrightBlueCircleCell, (xyz2(0,1,1)+EvenPoint2)--(xyz2(0,1,1)+EvenPoint5)--(xyz2(0,1,1)+EvenPoint4)--(xyz2(1,1,1)+OddPoint4)--(xyz2(1,1,1)+OddPoint5)--(xyz2(1,1,1)+OddPoint2)--cycle, mediumblue,gray);

filldraw(BrightBlueCircleCell, (xyz2(1,0,1)+EvenPoint3)--(xyz2(1,0,1)+EvenPoint5)--(xyz2(1,0,1)+EvenPoint4)--(xyz2(1,1,1)+OddPoint1)--(xyz2(1,1,1)+OddPoint5)--(xyz2(1,1,1)+OddPoint2)--cycle,mediumblue,gray);

filldraw(BrightBlueCircleCell, (xyz2(0,1,1)+EvenPoint1)--(xyz2(0,1,1)+EvenPoint5)--(xyz2(0,1,1)+EvenPoint4)--(xyz2(0,1,0)+OddPoint3)--(xyz2(0,1,0)+OddPoint5)--(xyz2(0,1,0)+OddPoint2)--cycle,blue,gray);

filldraw(BrightBlueCircleCell, (xyz2(1,0,1)+EvenPoint1)--(xyz2(1,0,1)+EvenPoint5)--(xyz2(1,0,1)+EvenPoint3)--(xyz2(0,0,1)+OddPoint3)--(xyz2(0,0,1)+OddPoint5)--(xyz2(0,0,1)+OddPoint1)--cycle,deepblue,gray);

filldraw(BrightBlueCircleCell, (xyz2(0,0,0)+EvenPoint3)--(xyz2(0,0,1)+OddPoint4)--(xyz2(0,1,1)+EvenPoint1)--(xyz2(0,1,0)+OddPoint2)--cycle,heavyblue,gray);

// Central cube qubits

fill(BrightBlueCircleCell,shift( (xyz2(0,0,0)+EvenPoint5) )*unitcircle, gray);
fill(BrightBlueCircleCell,shift( (xyz2(0,0,0)+EvenPoint5)  )*((0,1)..(0.84,0)..(0,-0.64)..(-0.84,0)..cycle), white);
draw(BrightBlueCircleCell,shift( (xyz2(0,0,0)+EvenPoint5)  )*unitcircle, black+0.2);

fill(BrightBlueCircleCell,shift( (xyz2(0,0,1)+EvenPoint5) )*unitcircle, gray);
fill(BrightBlueCircleCell,shift( (xyz2(0,0,1)+EvenPoint5) )*((0,1)..(0.84,0)..(0,-0.64)..(-0.84,0)..cycle), white);
draw(BrightBlueCircleCell,shift( (xyz2(0,0,1)+EvenPoint5) )*unitcircle, black+0.2);

fill(BrightBlueCircleCell,shift( (xyz2(0,1,1)+EvenPoint5) )*unitcircle, gray);
fill(BrightBlueCircleCell,shift( (xyz2(0,1,1)+EvenPoint5) )*((0,1)..(0.84,0)..(0,-0.64)..(-0.84,0)..cycle), white);
draw(BrightBlueCircleCell,shift( (xyz2(0,1,1)+EvenPoint5) )*unitcircle, black+0.2);

fill(BrightBlueCircleCell,shift( (xyz2(0,1,0)+EvenPoint5) )*unitcircle, gray);
fill(BrightBlueCircleCell,shift( (xyz2(0,1,0)+EvenPoint5) )*((0,1)..(0.84,0)..(0,-0.64)..(-0.84,0)..cycle), white);
draw(BrightBlueCircleCell,shift( (xyz2(0,1,0)+EvenPoint5) )*unitcircle, black+0.2);

fill(BrightBlueCircleCell,shift( (xyz2(1,0,1)+EvenPoint5) )*unitcircle, gray);
fill(BrightBlueCircleCell,shift( (xyz2(1,0,1)+EvenPoint5) )*((0,1)..(0.84,0)..(0,-0.64)..(-0.84,0)..cycle), white);
draw(BrightBlueCircleCell,shift( (xyz2(1,0,1)+EvenPoint5)  )*unitcircle, black+0.2);

fill(BrightBlueCircleCell,shift( (xyz2(1,1,1)+EvenPoint5) )*unitcircle, gray);
fill(BrightBlueCircleCell,shift( (xyz2(1,1,1)+EvenPoint5) )*((0,1)..(0.84,0)..(0,-0.64)..(-0.84,0)..cycle), white);
draw(BrightBlueCircleCell,shift( (xyz2(1,1,1)+EvenPoint5)  )*unitcircle, black+0.2);

// Right square qubits

fill(BrightBlueCircleCell,shift( (xyz2(1,1,1)+OddPoint1) )*unitcircle, gray);
fill(BrightBlueCircleCell,shift( (xyz2(1,1,1)+OddPoint1) )*((0,1)..(0.84,0)..(0,-0.64)..(-0.84,0)..cycle), white);
draw(BrightBlueCircleCell,shift( (xyz2(1,1,1)+OddPoint1) )*unitcircle, black+0.2);

fill(BrightBlueCircleCell,shift(  (xyz2(1,0,1)+EvenPoint4) )*unitcircle, gray);
fill(BrightBlueCircleCell,shift(  (xyz2(1,0,1)+EvenPoint4) )*((0,1)..(0.84,0)..(0,-0.64)..(-0.84,0)..cycle), white);
draw(BrightBlueCircleCell,shift(  (xyz2(1,0,1)+EvenPoint4) )*unitcircle, black+0.2);

// Bottom square qubits

fill(BrightBlueCircleCell,shift( (xyz2(0,0,1)+OddPoint1))*unitcircle, gray);
fill(BrightBlueCircleCell,shift( (xyz2(0,0,1)+OddPoint1) )*((0,1)..(0.84,0)..(0,-0.64)..(-0.84,0)..cycle), white);
draw(BrightBlueCircleCell,shift( (xyz2(0,0,1)+OddPoint1) )*unitcircle, black+0.2);

fill(BrightBlueCircleCell,shift( (xyz2(1,0,1)+EvenPoint1) )*unitcircle, gray);
fill(BrightBlueCircleCell,shift( (xyz2(1,0,1)+EvenPoint1) )*((0,1)..(0.84,0)..(0,-0.64)..(-0.84,0)..cycle), white);
draw(BrightBlueCircleCell,shift( (xyz2(1,0,1)+EvenPoint1) )*unitcircle, black+0.2);

// Side square face qubits

fill(BrightBlueCircleCell,shift( (xyz2(0,0,0)+EvenPoint3) )*unitcircle, gray);
fill(BrightBlueCircleCell,shift( (xyz2(0,0,0)+EvenPoint3) )*((0,1)..(0.84,0)..(0,-0.64)..(-0.84,0)..cycle), white);
draw(BrightBlueCircleCell,shift( (xyz2(0,0,0)+EvenPoint3) )*unitcircle, black+0.2);

fill(BrightBlueCircleCell,shift( (xyz2(0,0,1)+OddPoint4) )*unitcircle, gray);
fill(BrightBlueCircleCell,shift( (xyz2(0,0,1)+OddPoint4) )*((0,1)..(0.84,0)..(0,-0.64)..(-0.84,0)..cycle), white);
draw(BrightBlueCircleCell,shift( (xyz2(0,0,1)+OddPoint4) )*unitcircle, black+0.2);

fill(BrightBlueCircleCell,shift( (xyz2(0,1,1)+EvenPoint1) )*unitcircle, gray);
fill(BrightBlueCircleCell,shift( (xyz2(0,1,1)+EvenPoint1) )*((0,1)..(0.84,0)..(0,-0.64)..(-0.84,0)..cycle), white);
draw(BrightBlueCircleCell,shift( (xyz2(0,1,1)+EvenPoint1) )*unitcircle, black+0.2);

fill(BrightBlueCircleCell,shift( (xyz2(0,1,0)+OddPoint2) )*unitcircle, gray);
fill(BrightBlueCircleCell,shift( (xyz2(0,1,0)+OddPoint2) )*((0,1)..(0.84,0)..(0,-0.64)..(-0.84,0)..cycle), white);
draw(BrightBlueCircleCell,shift( (xyz2(0,1,0)+OddPoint2)  )*unitcircle, black+0.2);

// Top square face qubits

fill(BrightBlueCircleCell,shift( (xyz2(0,1,0)+OddPoint3) )*unitcircle, gray);
fill(BrightBlueCircleCell,shift( (xyz2(0,1,0)+OddPoint3) )*((0,1)..(0.84,0)..(0,-0.64)..(-0.84,0)..cycle), white);
draw(BrightBlueCircleCell,shift( (xyz2(0,1,0)+OddPoint3) )*unitcircle, black+0.2);

fill(BrightBlueCircleCell,shift( (xyz2(0,1,1)+EvenPoint4) )*unitcircle, gray);
fill(BrightBlueCircleCell,shift( (xyz2(0,1,1)+EvenPoint4) )*((0,1)..(0.84,0)..(0,-0.64)..(-0.84,0)..cycle), white);
draw(BrightBlueCircleCell,shift( (xyz2(0,1,1)+EvenPoint4) )*unitcircle, black+0.2);

fill(BrightBlueCircleCell,shift( (xyz2(1,1,1)+OddPoint4) )*unitcircle, gray);
fill(BrightBlueCircleCell,shift( (xyz2(1,1,1)+OddPoint4) )*((0,1)..(0.84,0)..(0,-0.64)..(-0.84,0)..cycle), white);
draw(BrightBlueCircleCell,shift( (xyz2(1,1,1)+OddPoint4)  )*unitcircle, black+0.2);

fill(BrightBlueCircleCell,shift( (xyz2(1,1,0)+EvenPoint3) )*unitcircle, gray);
fill(BrightBlueCircleCell,shift( (xyz2(1,1,0)+EvenPoint3) )*((0,1)..(0.84,0)..(0,-0.64)..(-0.84,0)..cycle), white);
draw(BrightBlueCircleCell,shift( (xyz2(1,1,0)+EvenPoint3) )*unitcircle, black+0.2);

// Front square qubits

fill(BrightBlueCircleCell,shift( (xyz2(0,0,1)+OddPoint3) )*unitcircle, gray);
fill(BrightBlueCircleCell,shift( (xyz2(0,0,1)+OddPoint3) )*((0,1)..(0.84,0)..(0,-0.64)..(-0.84,0)..cycle), white);
draw(BrightBlueCircleCell,shift( (xyz2(0,0,1)+OddPoint3)  )*unitcircle, black+0.2);

fill(BrightBlueCircleCell,shift( (xyz2(1,0,1)+EvenPoint3) )*unitcircle, gray);
fill(BrightBlueCircleCell,shift(  (xyz2(1,0,1)+EvenPoint3)  )*((0,1)..(0.84,0)..(0,-0.64)..(-0.84,0)..cycle), white);
draw(BrightBlueCircleCell,shift(  (xyz2(1,0,1)+EvenPoint3) )*unitcircle, black+0.2);

fill(BrightBlueCircleCell,shift( (xyz2(1,1,1)+OddPoint2) )*unitcircle, gray);
fill(BrightBlueCircleCell,shift( (xyz2(1,1,1)+OddPoint2) )*((0,1)..(0.84,0)..(0,-0.64)..(-0.84,0)..cycle), white);
draw(BrightBlueCircleCell,shift( (xyz2(1,1,1)+OddPoint2) )*unitcircle, black+0.2);

fill(BrightBlueCircleCell,shift( (xyz2(0,1,1)+EvenPoint2) )*unitcircle, gray);
fill(BrightBlueCircleCell,shift( (xyz2(0,1,1)+EvenPoint2) )*((0,1)..(0.84,0)..(0,-0.64)..(-0.84,0)..cycle), white);
draw(BrightBlueCircleCell,shift( (xyz2(0,1,1)+EvenPoint2)  )*unitcircle, black+0.2);

picture BlueCircleCell;

filldraw(BlueCircleCell, (xyz2(0,0,0)+EvenPoint3)--(xyz2(0,0,0)+EvenPoint5)--(xyz2(0,0,0)+EvenPoint2)--(xyz2(0,0,1)+OddPoint1)--(xyz2(0,0,1)+OddPoint5)--(xyz2(0,0,1)+OddPoint4)--cycle,paleblue,gray);

filldraw(BlueCircleCell, (xyz2(0,1,0)+OddPoint3)--(xyz2(0,1,1)+EvenPoint4)--(xyz2(1,1,1)+OddPoint4)--(xyz2(1,1,0)+EvenPoint3)--cycle,paleblue,gray);

filldraw(BlueCircleCell, (xyz2(0,1,1)+EvenPoint2)--(xyz2(1,1,1)+OddPoint2)--(xyz2(1,0,1)+EvenPoint3)--(xyz2(0,0,1)+OddPoint3)--cycle,paleblue,gray);
filldraw(BlueCircleCell, (xyz2(0,1,1)+EvenPoint1)--(xyz2(0,1,1)+EvenPoint5)--(xyz2(0,1,1)+EvenPoint2)--(xyz2(0,0,1)+OddPoint3)--(xyz2(0,0,1)+OddPoint5)--(xyz2(0,0,1)+OddPoint4)--cycle,paleblue,gray);

filldraw(BlueCircleCell, (xyz2(0,1,1)+EvenPoint2)--(xyz2(0,1,1)+EvenPoint5)--(xyz2(0,1,1)+EvenPoint4)--(xyz2(1,1,1)+OddPoint4)--(xyz2(1,1,1)+OddPoint5)--(xyz2(1,1,1)+OddPoint2)--cycle, paleblue,gray);

filldraw(BlueCircleCell, (xyz2(1,0,1)+EvenPoint3)--(xyz2(1,0,1)+EvenPoint5)--(xyz2(1,0,1)+EvenPoint4)--(xyz2(1,1,1)+OddPoint1)--(xyz2(1,1,1)+OddPoint5)--(xyz2(1,1,1)+OddPoint2)--cycle,paleblue,gray);

filldraw(BlueCircleCell, (xyz2(0,1,1)+EvenPoint1)--(xyz2(0,1,1)+EvenPoint5)--(xyz2(0,1,1)+EvenPoint4)--(xyz2(0,1,0)+OddPoint3)--(xyz2(0,1,0)+OddPoint5)--(xyz2(0,1,0)+OddPoint2)--cycle,paleblue,gray);

filldraw(BlueCircleCell, (xyz2(1,0,1)+EvenPoint1)--(xyz2(1,0,1)+EvenPoint5)--(xyz2(1,0,1)+EvenPoint3)--(xyz2(0,0,1)+OddPoint3)--(xyz2(0,0,1)+OddPoint5)--(xyz2(0,0,1)+OddPoint1)--cycle, paleblue,gray);

filldraw(BlueCircleCell, (xyz2(0,0,0)+EvenPoint3)--(xyz2(0,0,1)+OddPoint4)--(xyz2(0,1,1)+EvenPoint1)--(xyz2(0,1,0)+OddPoint2)--cycle,lightblue,gray);

// Central cube qubits

fill(BlueCircleCell,shift( (xyz2(0,0,0)+EvenPoint5) )*unitcircle, gray);
fill(BlueCircleCell,shift( (xyz2(0,0,0)+EvenPoint5)  )*((0,1)..(0.84,0)..(0,-0.64)..(-0.84,0)..cycle), white);
draw(BlueCircleCell,shift( (xyz2(0,0,0)+EvenPoint5)  )*unitcircle, black+0.2);

fill(BlueCircleCell,shift( (xyz2(0,0,1)+EvenPoint5) )*unitcircle, gray);
fill(BlueCircleCell,shift( (xyz2(0,0,1)+EvenPoint5) )*((0,1)..(0.84,0)..(0,-0.64)..(-0.84,0)..cycle), white);
draw(BlueCircleCell,shift( (xyz2(0,0,1)+EvenPoint5) )*unitcircle, black+0.2);

fill(BlueCircleCell,shift( (xyz2(0,1,1)+EvenPoint5) )*unitcircle, gray);
fill(BlueCircleCell,shift( (xyz2(0,1,1)+EvenPoint5) )*((0,1)..(0.84,0)..(0,-0.64)..(-0.84,0)..cycle), white);
draw(BlueCircleCell,shift( (xyz2(0,1,1)+EvenPoint5) )*unitcircle, black+0.2);

fill(BlueCircleCell,shift( (xyz2(0,1,0)+EvenPoint5) )*unitcircle, gray);
fill(BlueCircleCell,shift( (xyz2(0,1,0)+EvenPoint5) )*((0,1)..(0.84,0)..(0,-0.64)..(-0.84,0)..cycle), white);
draw(BlueCircleCell,shift( (xyz2(0,1,0)+EvenPoint5) )*unitcircle, black+0.2);

fill(BlueCircleCell,shift( (xyz2(1,0,1)+EvenPoint5) )*unitcircle, gray);
fill(BlueCircleCell,shift( (xyz2(1,0,1)+EvenPoint5) )*((0,1)..(0.84,0)..(0,-0.64)..(-0.84,0)..cycle), white);
draw(BlueCircleCell,shift( (xyz2(1,0,1)+EvenPoint5)  )*unitcircle, black+0.2);

fill(BlueCircleCell,shift( (xyz2(1,1,1)+EvenPoint5) )*unitcircle, gray);
fill(BlueCircleCell,shift( (xyz2(1,1,1)+EvenPoint5) )*((0,1)..(0.84,0)..(0,-0.64)..(-0.84,0)..cycle), white);
draw(BlueCircleCell,shift( (xyz2(1,1,1)+EvenPoint5)  )*unitcircle, black+0.2);

// Right square qubits

fill(BlueCircleCell,shift( (xyz2(1,1,1)+OddPoint1) )*unitcircle, gray);
fill(BlueCircleCell,shift( (xyz2(1,1,1)+OddPoint1) )*((0,1)..(0.84,0)..(0,-0.64)..(-0.84,0)..cycle), white);
draw(BlueCircleCell,shift( (xyz2(1,1,1)+OddPoint1) )*unitcircle, black+0.2);

fill(BlueCircleCell,shift(  (xyz2(1,0,1)+EvenPoint4) )*unitcircle, gray);
fill(BlueCircleCell,shift(  (xyz2(1,0,1)+EvenPoint4) )*((0,1)..(0.84,0)..(0,-0.64)..(-0.84,0)..cycle), white);
draw(BlueCircleCell,shift(  (xyz2(1,0,1)+EvenPoint4) )*unitcircle, black+0.2);

// Bottom square qubits

fill(BlueCircleCell,shift( (xyz2(0,0,1)+OddPoint1))*unitcircle, gray);
fill(BlueCircleCell,shift( (xyz2(0,0,1)+OddPoint1) )*((0,1)..(0.84,0)..(0,-0.64)..(-0.84,0)..cycle), white);
draw(BlueCircleCell,shift( (xyz2(0,0,1)+OddPoint1) )*unitcircle, black+0.2);

fill(BlueCircleCell,shift( (xyz2(1,0,1)+EvenPoint1) )*unitcircle, gray);
fill(BlueCircleCell,shift( (xyz2(1,0,1)+EvenPoint1) )*((0,1)..(0.84,0)..(0,-0.64)..(-0.84,0)..cycle), white);
draw(BlueCircleCell,shift( (xyz2(1,0,1)+EvenPoint1) )*unitcircle, black+0.2);

// Side square face qubits

fill(BlueCircleCell,shift( (xyz2(0,0,0)+EvenPoint3) )*unitcircle, gray);
fill(BlueCircleCell,shift( (xyz2(0,0,0)+EvenPoint3) )*((0,1)..(0.84,0)..(0,-0.64)..(-0.84,0)..cycle), white);
draw(BlueCircleCell,shift( (xyz2(0,0,0)+EvenPoint3) )*unitcircle, black+0.2);

fill(BlueCircleCell,shift( (xyz2(0,0,1)+OddPoint4) )*unitcircle, gray);
fill(BlueCircleCell,shift( (xyz2(0,0,1)+OddPoint4) )*((0,1)..(0.84,0)..(0,-0.64)..(-0.84,0)..cycle), white);
draw(BlueCircleCell,shift( (xyz2(0,0,1)+OddPoint4) )*unitcircle, black+0.2);

fill(BlueCircleCell,shift( (xyz2(0,1,1)+EvenPoint1) )*unitcircle, gray);
fill(BlueCircleCell,shift( (xyz2(0,1,1)+EvenPoint1) )*((0,1)..(0.84,0)..(0,-0.64)..(-0.84,0)..cycle), white);
draw(BlueCircleCell,shift( (xyz2(0,1,1)+EvenPoint1) )*unitcircle, black+0.2);

fill(BlueCircleCell,shift( (xyz2(0,1,0)+OddPoint2) )*unitcircle, gray);
fill(BlueCircleCell,shift( (xyz2(0,1,0)+OddPoint2) )*((0,1)..(0.84,0)..(0,-0.64)..(-0.84,0)..cycle), white);
draw(BlueCircleCell,shift( (xyz2(0,1,0)+OddPoint2)  )*unitcircle, black+0.2);

// Top square face qubits

fill(BlueCircleCell,shift( (xyz2(0,1,0)+OddPoint3) )*unitcircle, gray);
fill(BlueCircleCell,shift( (xyz2(0,1,0)+OddPoint3) )*((0,1)..(0.84,0)..(0,-0.64)..(-0.84,0)..cycle), white);
draw(BlueCircleCell,shift( (xyz2(0,1,0)+OddPoint3) )*unitcircle, black+0.2);

fill(BlueCircleCell,shift( (xyz2(0,1,1)+EvenPoint4) )*unitcircle, gray);
fill(BlueCircleCell,shift( (xyz2(0,1,1)+EvenPoint4) )*((0,1)..(0.84,0)..(0,-0.64)..(-0.84,0)..cycle), white);
draw(BlueCircleCell,shift( (xyz2(0,1,1)+EvenPoint4) )*unitcircle, black+0.2);

fill(BlueCircleCell,shift( (xyz2(1,1,1)+OddPoint4) )*unitcircle, gray);
fill(BlueCircleCell,shift( (xyz2(1,1,1)+OddPoint4) )*((0,1)..(0.84,0)..(0,-0.64)..(-0.84,0)..cycle), white);
draw(BlueCircleCell,shift( (xyz2(1,1,1)+OddPoint4)  )*unitcircle, black+0.2);

fill(BlueCircleCell,shift( (xyz2(1,1,0)+EvenPoint3) )*unitcircle, gray);
fill(BlueCircleCell,shift( (xyz2(1,1,0)+EvenPoint3) )*((0,1)..(0.84,0)..(0,-0.64)..(-0.84,0)..cycle), white);
draw(BlueCircleCell,shift( (xyz2(1,1,0)+EvenPoint3) )*unitcircle, black+0.2);

// Front square qubits

fill(BlueCircleCell,shift( (xyz2(0,0,1)+OddPoint3) )*unitcircle, gray);
fill(BlueCircleCell,shift( (xyz2(0,0,1)+OddPoint3) )*((0,1)..(0.84,0)..(0,-0.64)..(-0.84,0)..cycle), white);
draw(BlueCircleCell,shift( (xyz2(0,0,1)+OddPoint3)  )*unitcircle, black+0.2);

fill(BlueCircleCell,shift( (xyz2(1,0,1)+EvenPoint3) )*unitcircle, gray);
fill(BlueCircleCell,shift(  (xyz2(1,0,1)+EvenPoint3)  )*((0,1)..(0.84,0)..(0,-0.64)..(-0.84,0)..cycle), white);
draw(BlueCircleCell,shift(  (xyz2(1,0,1)+EvenPoint3) )*unitcircle, black+0.2);

fill(BlueCircleCell,shift( (xyz2(1,1,1)+OddPoint2) )*unitcircle, gray);
fill(BlueCircleCell,shift( (xyz2(1,1,1)+OddPoint2) )*((0,1)..(0.84,0)..(0,-0.64)..(-0.84,0)..cycle), white);
draw(BlueCircleCell,shift( (xyz2(1,1,1)+OddPoint2) )*unitcircle, black+0.2);

fill(BlueCircleCell,shift( (xyz2(0,1,1)+EvenPoint2) )*unitcircle, gray);
fill(BlueCircleCell,shift( (xyz2(0,1,1)+EvenPoint2) )*((0,1)..(0.84,0)..(0,-0.64)..(-0.84,0)..cycle), white);
draw(BlueCircleCell,shift( (xyz2(0,1,1)+EvenPoint2)  )*unitcircle, black+0.2);

picture YellowCircleCell;

// Right edges with no qubits shown on the shape

filldraw(YellowCircleCell, (xyz2(0,0,0)+EvenPoint3)--(xyz2(0,0,0)+EvenPoint5)--(xyz2(0,0,0)+EvenPoint2)--(xyz2(0,0,1)+OddPoint1)--(xyz2(0,0,1)+OddPoint5)--(xyz2(0,0,1)+OddPoint4)--cycle, lightyellow,gray);

filldraw(YellowCircleCell, (xyz2(0,1,0)+OddPoint3)--(xyz2(0,1,1)+EvenPoint4)--(xyz2(1,1,1)+OddPoint4)--(xyz2(1,1,0)+EvenPoint3)--cycle,lightyellow,gray);

filldraw(YellowCircleCell, (xyz2(0,1,1)+EvenPoint2)--(xyz2(1,1,1)+OddPoint2)--(xyz2(1,0,1)+EvenPoint3)--(xyz2(0,0,1)+OddPoint3)--cycle,lightyellow,gray);
filldraw(YellowCircleCell, (xyz2(0,1,1)+EvenPoint1)--(xyz2(0,1,1)+EvenPoint5)--(xyz2(0,1,1)+EvenPoint2)--(xyz2(0,0,1)+OddPoint3)--(xyz2(0,0,1)+OddPoint5)--(xyz2(0,0,1)+OddPoint4)--cycle,lightyellow,gray);

filldraw(YellowCircleCell, (xyz2(0,1,1)+EvenPoint2)--(xyz2(0,1,1)+EvenPoint5)--(xyz2(0,1,1)+EvenPoint4)--(xyz2(1,1,1)+OddPoint4)--(xyz2(1,1,1)+OddPoint5)--(xyz2(1,1,1)+OddPoint2)--cycle, lightyellow,gray);

filldraw(YellowCircleCell, (xyz2(1,0,1)+EvenPoint3)--(xyz2(1,0,1)+EvenPoint5)--(xyz2(1,0,1)+EvenPoint4)--(xyz2(1,1,1)+OddPoint1)--(xyz2(1,1,1)+OddPoint5)--(xyz2(1,1,1)+OddPoint2)--cycle, lightyellow,gray);

filldraw(YellowCircleCell, (xyz2(0,1,1)+EvenPoint1)--(xyz2(0,1,1)+EvenPoint5)--(xyz2(0,1,1)+EvenPoint4)--(xyz2(0,1,0)+OddPoint3)--(xyz2(0,1,0)+OddPoint5)--(xyz2(0,1,0)+OddPoint2)--cycle,lightyellow,gray);

filldraw(YellowCircleCell, (xyz2(1,0,1)+EvenPoint1)--(xyz2(1,0,1)+EvenPoint5)--(xyz2(1,0,1)+EvenPoint3)--(xyz2(0,0,1)+OddPoint3)--(xyz2(0,0,1)+OddPoint5)--(xyz2(0,0,1)+OddPoint1)--cycle,lightyellow,gray);

filldraw(YellowCircleCell, (xyz2(0,0,0)+EvenPoint3)--(xyz2(0,0,1)+OddPoint4)--(xyz2(0,1,1)+EvenPoint1)--(xyz2(0,1,0)+OddPoint2)--cycle,lightyellow,gray);

// Central cube qubits

fill(YellowCircleCell,shift( (xyz2(0,0,0)+EvenPoint5) )*unitcircle, gray);
fill(YellowCircleCell,shift( (xyz2(0,0,0)+EvenPoint5)  )*((0,1)..(0.84,0)..(0,-0.64)..(-0.84,0)..cycle), white);
draw(YellowCircleCell,shift( (xyz2(0,0,0)+EvenPoint5)  )*unitcircle, black+0.2);

fill(YellowCircleCell,shift( (xyz2(0,0,1)+EvenPoint5) )*unitcircle, gray);
fill(YellowCircleCell,shift( (xyz2(0,0,1)+EvenPoint5) )*((0,1)..(0.84,0)..(0,-0.64)..(-0.84,0)..cycle), white);
draw(YellowCircleCell,shift( (xyz2(0,0,1)+EvenPoint5) )*unitcircle, black+0.2);

fill(YellowCircleCell,shift( (xyz2(0,1,1)+EvenPoint5) )*unitcircle, gray);
fill(YellowCircleCell,shift( (xyz2(0,1,1)+EvenPoint5) )*((0,1)..(0.84,0)..(0,-0.64)..(-0.84,0)..cycle), white);
draw(YellowCircleCell,shift( (xyz2(0,1,1)+EvenPoint5) )*unitcircle, black+0.2);

fill(YellowCircleCell,shift( (xyz2(0,1,0)+EvenPoint5) )*unitcircle, gray);
fill(YellowCircleCell,shift( (xyz2(0,1,0)+EvenPoint5) )*((0,1)..(0.84,0)..(0,-0.64)..(-0.84,0)..cycle), white);
draw(YellowCircleCell,shift( (xyz2(0,1,0)+EvenPoint5) )*unitcircle, black+0.2);

fill(YellowCircleCell,shift( (xyz2(1,0,1)+EvenPoint5) )*unitcircle, gray);
fill(YellowCircleCell,shift( (xyz2(1,0,1)+EvenPoint5) )*((0,1)..(0.84,0)..(0,-0.64)..(-0.84,0)..cycle), white);
draw(YellowCircleCell,shift( (xyz2(1,0,1)+EvenPoint5)  )*unitcircle, black+0.2);

fill(YellowCircleCell,shift( (xyz2(1,1,1)+EvenPoint5) )*unitcircle, gray);
fill(YellowCircleCell,shift( (xyz2(1,1,1)+EvenPoint5) )*((0,1)..(0.84,0)..(0,-0.64)..(-0.84,0)..cycle), white);
draw(YellowCircleCell,shift( (xyz2(1,1,1)+EvenPoint5)  )*unitcircle, black+0.2);

// Right square qubits

fill(YellowCircleCell,shift( (xyz2(1,1,1)+OddPoint1) )*unitcircle, gray);
fill(YellowCircleCell,shift( (xyz2(1,1,1)+OddPoint1) )*((0,1)..(0.84,0)..(0,-0.64)..(-0.84,0)..cycle), white);
draw(YellowCircleCell,shift( (xyz2(1,1,1)+OddPoint1) )*unitcircle, black+0.2);

fill(YellowCircleCell,shift(  (xyz2(1,0,1)+EvenPoint4) )*unitcircle, gray);
fill(YellowCircleCell,shift(  (xyz2(1,0,1)+EvenPoint4) )*((0,1)..(0.84,0)..(0,-0.64)..(-0.84,0)..cycle), white);
draw(YellowCircleCell,shift(  (xyz2(1,0,1)+EvenPoint4) )*unitcircle, black+0.2);

// Bottom square qubits

fill(YellowCircleCell,shift( (xyz2(0,0,1)+OddPoint1))*unitcircle, gray);
fill(YellowCircleCell,shift( (xyz2(0,0,1)+OddPoint1) )*((0,1)..(0.84,0)..(0,-0.64)..(-0.84,0)..cycle), white);
draw(YellowCircleCell,shift( (xyz2(0,0,1)+OddPoint1) )*unitcircle, black+0.2);

fill(YellowCircleCell,shift( (xyz2(1,0,1)+EvenPoint1) )*unitcircle, gray);
fill(YellowCircleCell,shift( (xyz2(1,0,1)+EvenPoint1) )*((0,1)..(0.84,0)..(0,-0.64)..(-0.84,0)..cycle), white);
draw(YellowCircleCell,shift( (xyz2(1,0,1)+EvenPoint1) )*unitcircle, black+0.2);

// Side square face qubits

fill(YellowCircleCell,shift( (xyz2(0,0,0)+EvenPoint3) )*unitcircle, gray);
fill(YellowCircleCell,shift( (xyz2(0,0,0)+EvenPoint3) )*((0,1)..(0.84,0)..(0,-0.64)..(-0.84,0)..cycle), white);
draw(YellowCircleCell,shift( (xyz2(0,0,0)+EvenPoint3) )*unitcircle, black+0.2);

fill(YellowCircleCell,shift( (xyz2(0,0,1)+OddPoint4) )*unitcircle, gray);
fill(YellowCircleCell,shift( (xyz2(0,0,1)+OddPoint4) )*((0,1)..(0.84,0)..(0,-0.64)..(-0.84,0)..cycle), white);
draw(YellowCircleCell,shift( (xyz2(0,0,1)+OddPoint4) )*unitcircle, black+0.2);

fill(YellowCircleCell,shift( (xyz2(0,1,1)+EvenPoint1) )*unitcircle, gray);
fill(YellowCircleCell,shift( (xyz2(0,1,1)+EvenPoint1) )*((0,1)..(0.84,0)..(0,-0.64)..(-0.84,0)..cycle), white);
draw(YellowCircleCell,shift( (xyz2(0,1,1)+EvenPoint1) )*unitcircle, black+0.2);

fill(YellowCircleCell,shift( (xyz2(0,1,0)+OddPoint2) )*unitcircle, gray);
fill(YellowCircleCell,shift( (xyz2(0,1,0)+OddPoint2) )*((0,1)..(0.84,0)..(0,-0.64)..(-0.84,0)..cycle), white);
draw(YellowCircleCell,shift( (xyz2(0,1,0)+OddPoint2)  )*unitcircle, black+0.2);

// Top square face qubits

fill(YellowCircleCell,shift( (xyz2(0,1,0)+OddPoint3) )*unitcircle, gray);
fill(YellowCircleCell,shift( (xyz2(0,1,0)+OddPoint3) )*((0,1)..(0.84,0)..(0,-0.64)..(-0.84,0)..cycle), white);
draw(YellowCircleCell,shift( (xyz2(0,1,0)+OddPoint3) )*unitcircle, black+0.2);

fill(YellowCircleCell,shift( (xyz2(0,1,1)+EvenPoint4) )*unitcircle, gray);
fill(YellowCircleCell,shift( (xyz2(0,1,1)+EvenPoint4) )*((0,1)..(0.84,0)..(0,-0.64)..(-0.84,0)..cycle), white);
draw(YellowCircleCell,shift( (xyz2(0,1,1)+EvenPoint4) )*unitcircle, black+0.2);

fill(YellowCircleCell,shift( (xyz2(1,1,1)+OddPoint4) )*unitcircle, gray);
fill(YellowCircleCell,shift( (xyz2(1,1,1)+OddPoint4) )*((0,1)..(0.84,0)..(0,-0.64)..(-0.84,0)..cycle), white);
draw(YellowCircleCell,shift( (xyz2(1,1,1)+OddPoint4)  )*unitcircle, black+0.2);

fill(YellowCircleCell,shift( (xyz2(1,1,0)+EvenPoint3) )*unitcircle, gray);
fill(YellowCircleCell,shift( (xyz2(1,1,0)+EvenPoint3) )*((0,1)..(0.84,0)..(0,-0.64)..(-0.84,0)..cycle), white);
draw(YellowCircleCell,shift( (xyz2(1,1,0)+EvenPoint3) )*unitcircle, black+0.2);

// Front square qubits

fill(YellowCircleCell,shift( (xyz2(0,0,1)+OddPoint3) )*unitcircle, gray);
fill(YellowCircleCell,shift( (xyz2(0,0,1)+OddPoint3) )*((0,1)..(0.84,0)..(0,-0.64)..(-0.84,0)..cycle), white);
draw(YellowCircleCell,shift( (xyz2(0,0,1)+OddPoint3)  )*unitcircle, black+0.2);

fill(YellowCircleCell,shift( (xyz2(1,0,1)+EvenPoint3) )*unitcircle, gray);
fill(YellowCircleCell,shift(  (xyz2(1,0,1)+EvenPoint3)  )*((0,1)..(0.84,0)..(0,-0.64)..(-0.84,0)..cycle), white);
draw(YellowCircleCell,shift(  (xyz2(1,0,1)+EvenPoint3) )*unitcircle, black+0.2);

fill(YellowCircleCell,shift( (xyz2(1,1,1)+OddPoint2) )*unitcircle, gray);
fill(YellowCircleCell,shift( (xyz2(1,1,1)+OddPoint2) )*((0,1)..(0.84,0)..(0,-0.64)..(-0.84,0)..cycle), white);
draw(YellowCircleCell,shift( (xyz2(1,1,1)+OddPoint2) )*unitcircle, black+0.2);

fill(YellowCircleCell,shift( (xyz2(0,1,1)+EvenPoint2) )*unitcircle, gray);
fill(YellowCircleCell,shift( (xyz2(0,1,1)+EvenPoint2) )*((0,1)..(0.84,0)..(0,-0.64)..(-0.84,0)..cycle), white);
draw(YellowCircleCell,shift( (xyz2(0,1,1)+EvenPoint2)  )*unitcircle, black+0.2);

picture RedCircleCell;

// Right edges with no qubits shown on the shape

filldraw(RedCircleCell, (xyz2(0,0,0)+EvenPoint3)--(xyz2(0,0,0)+EvenPoint5)--(xyz2(0,0,0)+EvenPoint2)--(xyz2(0,0,1)+OddPoint1)--(xyz2(0,0,1)+OddPoint5)--(xyz2(0,0,1)+OddPoint4)--cycle,palered,gray);

filldraw(RedCircleCell, (xyz2(0,1,0)+OddPoint3)--(xyz2(0,1,1)+EvenPoint4)--(xyz2(1,1,1)+OddPoint4)--(xyz2(1,1,0)+EvenPoint3)--cycle,palered,gray);

filldraw(RedCircleCell, (xyz2(0,1,1)+EvenPoint2)--(xyz2(1,1,1)+OddPoint2)--(xyz2(1,0,1)+EvenPoint3)--(xyz2(0,0,1)+OddPoint3)--cycle,palered,gray);
filldraw(RedCircleCell, (xyz2(0,1,1)+EvenPoint1)--(xyz2(0,1,1)+EvenPoint5)--(xyz2(0,1,1)+EvenPoint2)--(xyz2(0,0,1)+OddPoint3)--(xyz2(0,0,1)+OddPoint5)--(xyz2(0,0,1)+OddPoint4)--cycle,palered,gray);

filldraw(RedCircleCell, (xyz2(0,1,1)+EvenPoint2)--(xyz2(0,1,1)+EvenPoint5)--(xyz2(0,1,1)+EvenPoint4)--(xyz2(1,1,1)+OddPoint4)--(xyz2(1,1,1)+OddPoint5)--(xyz2(1,1,1)+OddPoint2)--cycle, palered,gray);

filldraw(RedCircleCell, (xyz2(1,0,1)+EvenPoint3)--(xyz2(1,0,1)+EvenPoint5)--(xyz2(1,0,1)+EvenPoint4)--(xyz2(1,1,1)+OddPoint1)--(xyz2(1,1,1)+OddPoint5)--(xyz2(1,1,1)+OddPoint2)--cycle, palered,gray);

filldraw(RedCircleCell, (xyz2(0,1,1)+EvenPoint1)--(xyz2(0,1,1)+EvenPoint5)--(xyz2(0,1,1)+EvenPoint4)--(xyz2(0,1,0)+OddPoint3)--(xyz2(0,1,0)+OddPoint5)--(xyz2(0,1,0)+OddPoint2)--cycle,palered,gray);

filldraw(RedCircleCell, (xyz2(1,0,1)+EvenPoint1)--(xyz2(1,0,1)+EvenPoint5)--(xyz2(1,0,1)+EvenPoint3)--(xyz2(0,0,1)+OddPoint3)--(xyz2(0,0,1)+OddPoint5)--(xyz2(0,0,1)+OddPoint1)--cycle, palered,gray);

filldraw(RedCircleCell, (xyz2(0,0,0)+EvenPoint3)--(xyz2(0,0,1)+OddPoint4)--(xyz2(0,1,1)+EvenPoint1)--(xyz2(0,1,0)+OddPoint2)--cycle, palered,gray);

// Central cube qubits

fill(RedCircleCell,shift( (xyz2(0,0,0)+EvenPoint5) )*unitcircle, gray);
fill(RedCircleCell,shift( (xyz2(0,0,0)+EvenPoint5)  )*((0,1)..(0.84,0)..(0,-0.64)..(-0.84,0)..cycle), white);
draw(RedCircleCell,shift( (xyz2(0,0,0)+EvenPoint5)  )*unitcircle, black+0.2);

fill(RedCircleCell,shift( (xyz2(0,0,1)+EvenPoint5) )*unitcircle, gray);
fill(RedCircleCell,shift( (xyz2(0,0,1)+EvenPoint5) )*((0,1)..(0.84,0)..(0,-0.64)..(-0.84,0)..cycle), white);
draw(RedCircleCell,shift( (xyz2(0,0,1)+EvenPoint5) )*unitcircle, black+0.2);

fill(RedCircleCell,shift( (xyz2(0,1,1)+EvenPoint5) )*unitcircle, gray);
fill(RedCircleCell,shift( (xyz2(0,1,1)+EvenPoint5) )*((0,1)..(0.84,0)..(0,-0.64)..(-0.84,0)..cycle), white);
draw(RedCircleCell,shift( (xyz2(0,1,1)+EvenPoint5) )*unitcircle, black+0.2);

fill(RedCircleCell,shift( (xyz2(0,1,0)+EvenPoint5) )*unitcircle, gray);
fill(RedCircleCell,shift( (xyz2(0,1,0)+EvenPoint5) )*((0,1)..(0.84,0)..(0,-0.64)..(-0.84,0)..cycle), white);
draw(RedCircleCell,shift( (xyz2(0,1,0)+EvenPoint5) )*unitcircle, black+0.2);

fill(RedCircleCell,shift( (xyz2(1,0,1)+EvenPoint5) )*unitcircle, gray);
fill(RedCircleCell,shift( (xyz2(1,0,1)+EvenPoint5) )*((0,1)..(0.84,0)..(0,-0.64)..(-0.84,0)..cycle), white);
draw(RedCircleCell,shift( (xyz2(1,0,1)+EvenPoint5)  )*unitcircle, black+0.2);

fill(RedCircleCell,shift( (xyz2(1,1,1)+EvenPoint5) )*unitcircle, gray);
fill(RedCircleCell,shift( (xyz2(1,1,1)+EvenPoint5) )*((0,1)..(0.84,0)..(0,-0.64)..(-0.84,0)..cycle), white);
draw(RedCircleCell,shift( (xyz2(1,1,1)+EvenPoint5)  )*unitcircle, black+0.2);

// Right square qubits

fill(RedCircleCell,shift( (xyz2(1,1,1)+OddPoint1) )*unitcircle, gray);
fill(RedCircleCell,shift( (xyz2(1,1,1)+OddPoint1) )*((0,1)..(0.84,0)..(0,-0.64)..(-0.84,0)..cycle), white);
draw(RedCircleCell,shift( (xyz2(1,1,1)+OddPoint1) )*unitcircle, black+0.2);

fill(RedCircleCell,shift(  (xyz2(1,0,1)+EvenPoint4) )*unitcircle, gray);
fill(RedCircleCell,shift(  (xyz2(1,0,1)+EvenPoint4) )*((0,1)..(0.84,0)..(0,-0.64)..(-0.84,0)..cycle), white);
draw(RedCircleCell,shift(  (xyz2(1,0,1)+EvenPoint4) )*unitcircle, black+0.2);

// Bottom square qubits

fill(RedCircleCell,shift( (xyz2(0,0,1)+OddPoint1))*unitcircle, gray);
fill(RedCircleCell,shift( (xyz2(0,0,1)+OddPoint1) )*((0,1)..(0.84,0)..(0,-0.64)..(-0.84,0)..cycle), white);
draw(RedCircleCell,shift( (xyz2(0,0,1)+OddPoint1) )*unitcircle, black+0.2);

fill(RedCircleCell,shift( (xyz2(1,0,1)+EvenPoint1) )*unitcircle, gray);
fill(RedCircleCell,shift( (xyz2(1,0,1)+EvenPoint1) )*((0,1)..(0.84,0)..(0,-0.64)..(-0.84,0)..cycle), white);
draw(RedCircleCell,shift( (xyz2(1,0,1)+EvenPoint1) )*unitcircle, black+0.2);

// Side square face qubits

fill(RedCircleCell,shift( (xyz2(0,0,0)+EvenPoint3) )*unitcircle, gray);
fill(RedCircleCell,shift( (xyz2(0,0,0)+EvenPoint3) )*((0,1)..(0.84,0)..(0,-0.64)..(-0.84,0)..cycle), white);
draw(RedCircleCell,shift( (xyz2(0,0,0)+EvenPoint3) )*unitcircle, black+0.2);

fill(RedCircleCell,shift( (xyz2(0,0,1)+OddPoint4) )*unitcircle, gray);
fill(RedCircleCell,shift( (xyz2(0,0,1)+OddPoint4) )*((0,1)..(0.84,0)..(0,-0.64)..(-0.84,0)..cycle), white);
draw(RedCircleCell,shift( (xyz2(0,0,1)+OddPoint4) )*unitcircle, black+0.2);

fill(RedCircleCell,shift( (xyz2(0,1,1)+EvenPoint1) )*unitcircle, gray);
fill(RedCircleCell,shift( (xyz2(0,1,1)+EvenPoint1) )*((0,1)..(0.84,0)..(0,-0.64)..(-0.84,0)..cycle), white);
draw(RedCircleCell,shift( (xyz2(0,1,1)+EvenPoint1) )*unitcircle, black+0.2);

fill(RedCircleCell,shift( (xyz2(0,1,0)+OddPoint2) )*unitcircle, gray);
fill(RedCircleCell,shift( (xyz2(0,1,0)+OddPoint2) )*((0,1)..(0.84,0)..(0,-0.64)..(-0.84,0)..cycle), white);
draw(RedCircleCell,shift( (xyz2(0,1,0)+OddPoint2)  )*unitcircle, black+0.2);

// Top square face qubits

fill(RedCircleCell,shift( (xyz2(0,1,0)+OddPoint3) )*unitcircle, gray);
fill(RedCircleCell,shift( (xyz2(0,1,0)+OddPoint3) )*((0,1)..(0.84,0)..(0,-0.64)..(-0.84,0)..cycle), white);
draw(RedCircleCell,shift( (xyz2(0,1,0)+OddPoint3) )*unitcircle, black+0.2);

fill(RedCircleCell,shift( (xyz2(0,1,1)+EvenPoint4) )*unitcircle, gray);
fill(RedCircleCell,shift( (xyz2(0,1,1)+EvenPoint4) )*((0,1)..(0.84,0)..(0,-0.64)..(-0.84,0)..cycle), white);
draw(RedCircleCell,shift( (xyz2(0,1,1)+EvenPoint4) )*unitcircle, black+0.2);

fill(RedCircleCell,shift( (xyz2(1,1,1)+OddPoint4) )*unitcircle, gray);
fill(RedCircleCell,shift( (xyz2(1,1,1)+OddPoint4) )*((0,1)..(0.84,0)..(0,-0.64)..(-0.84,0)..cycle), white);
draw(RedCircleCell,shift( (xyz2(1,1,1)+OddPoint4)  )*unitcircle, black+0.2);

fill(RedCircleCell,shift( (xyz2(1,1,0)+EvenPoint3) )*unitcircle, gray);
fill(RedCircleCell,shift( (xyz2(1,1,0)+EvenPoint3) )*((0,1)..(0.84,0)..(0,-0.64)..(-0.84,0)..cycle), white);
draw(RedCircleCell,shift( (xyz2(1,1,0)+EvenPoint3) )*unitcircle, black+0.2);

// Front square qubits

fill(RedCircleCell,shift( (xyz2(0,0,1)+OddPoint3) )*unitcircle, gray);
fill(RedCircleCell,shift( (xyz2(0,0,1)+OddPoint3) )*((0,1)..(0.84,0)..(0,-0.64)..(-0.84,0)..cycle), white);
draw(RedCircleCell,shift( (xyz2(0,0,1)+OddPoint3)  )*unitcircle, black+0.2);

fill(RedCircleCell,shift( (xyz2(1,0,1)+EvenPoint3) )*unitcircle, gray);
fill(RedCircleCell,shift(  (xyz2(1,0,1)+EvenPoint3)  )*((0,1)..(0.84,0)..(0,-0.64)..(-0.84,0)..cycle), white);
draw(RedCircleCell,shift(  (xyz2(1,0,1)+EvenPoint3) )*unitcircle, black+0.2);

fill(RedCircleCell,shift( (xyz2(1,1,1)+OddPoint2) )*unitcircle, gray);
fill(RedCircleCell,shift( (xyz2(1,1,1)+OddPoint2) )*((0,1)..(0.84,0)..(0,-0.64)..(-0.84,0)..cycle), white);
draw(RedCircleCell,shift( (xyz2(1,1,1)+OddPoint2) )*unitcircle, black+0.2);

fill(RedCircleCell,shift( (xyz2(0,1,1)+EvenPoint2) )*unitcircle, gray);
fill(RedCircleCell,shift( (xyz2(0,1,1)+EvenPoint2) )*((0,1)..(0.84,0)..(0,-0.64)..(-0.84,0)..cycle), white);
draw(RedCircleCell,shift( (xyz2(0,1,1)+EvenPoint2)  )*unitcircle, black+0.2);

picture GreenCircleCell;

// Right edges with no qubits shown on the shape

filldraw(GreenCircleCell, (xyz2(0,0,0)+EvenPoint3)--(xyz2(0,0,0)+EvenPoint5)--(xyz2(0,0,0)+EvenPoint2)--(xyz2(0,0,1)+OddPoint1)--(xyz2(0,0,1)+OddPoint5)--(xyz2(0,0,1)+OddPoint4)--cycle, palegreen,gray);

filldraw(GreenCircleCell, (xyz2(0,1,0)+OddPoint3)--(xyz2(0,1,1)+EvenPoint4)--(xyz2(1,1,1)+OddPoint4)--(xyz2(1,1,0)+EvenPoint3)--cycle, palegreen,gray);

filldraw(GreenCircleCell, (xyz2(0,1,1)+EvenPoint2)--(xyz2(1,1,1)+OddPoint2)--(xyz2(1,0,1)+EvenPoint3)--(xyz2(0,0,1)+OddPoint3)--cycle,palegreen,gray);
filldraw(GreenCircleCell, (xyz2(0,1,1)+EvenPoint1)--(xyz2(0,1,1)+EvenPoint5)--(xyz2(0,1,1)+EvenPoint2)--(xyz2(0,0,1)+OddPoint3)--(xyz2(0,0,1)+OddPoint5)--(xyz2(0,0,1)+OddPoint4)--cycle, palegreen,gray);

filldraw(GreenCircleCell, (xyz2(0,1,1)+EvenPoint2)--(xyz2(0,1,1)+EvenPoint5)--(xyz2(0,1,1)+EvenPoint4)--(xyz2(1,1,1)+OddPoint4)--(xyz2(1,1,1)+OddPoint5)--(xyz2(1,1,1)+OddPoint2)--cycle, palegreen,gray);

filldraw(GreenCircleCell, (xyz2(1,0,1)+EvenPoint3)--(xyz2(1,0,1)+EvenPoint5)--(xyz2(1,0,1)+EvenPoint4)--(xyz2(1,1,1)+OddPoint1)--(xyz2(1,1,1)+OddPoint5)--(xyz2(1,1,1)+OddPoint2)--cycle, palegreen,gray);

filldraw(GreenCircleCell, (xyz2(0,1,1)+EvenPoint1)--(xyz2(0,1,1)+EvenPoint5)--(xyz2(0,1,1)+EvenPoint4)--(xyz2(0,1,0)+OddPoint3)--(xyz2(0,1,0)+OddPoint5)--(xyz2(0,1,0)+OddPoint2)--cycle, palegreen,gray);

filldraw(GreenCircleCell, (xyz2(1,0,1)+EvenPoint1)--(xyz2(1,0,1)+EvenPoint5)--(xyz2(1,0,1)+EvenPoint3)--(xyz2(0,0,1)+OddPoint3)--(xyz2(0,0,1)+OddPoint5)--(xyz2(0,0,1)+OddPoint1)--cycle, palegreen,gray);

filldraw(GreenCircleCell, (xyz2(0,0,0)+EvenPoint3)--(xyz2(0,0,1)+OddPoint4)--(xyz2(0,1,1)+EvenPoint1)--(xyz2(0,1,0)+OddPoint2)--cycle, palegreen,gray);

// Central cube qubits

fill(GreenCircleCell,shift( (xyz2(0,0,0)+EvenPoint5) )*unitcircle, gray);
fill(GreenCircleCell,shift( (xyz2(0,0,0)+EvenPoint5)  )*((0,1)..(0.84,0)..(0,-0.64)..(-0.84,0)..cycle), white);
draw(GreenCircleCell,shift( (xyz2(0,0,0)+EvenPoint5)  )*unitcircle, black+0.2);

fill(GreenCircleCell,shift( (xyz2(0,0,1)+EvenPoint5) )*unitcircle, gray);
fill(GreenCircleCell,shift( (xyz2(0,0,1)+EvenPoint5) )*((0,1)..(0.84,0)..(0,-0.64)..(-0.84,0)..cycle), white);
draw(GreenCircleCell,shift( (xyz2(0,0,1)+EvenPoint5) )*unitcircle, black+0.2);

fill(GreenCircleCell,shift( (xyz2(0,1,1)+EvenPoint5) )*unitcircle, gray);
fill(GreenCircleCell,shift( (xyz2(0,1,1)+EvenPoint5) )*((0,1)..(0.84,0)..(0,-0.64)..(-0.84,0)..cycle), white);
draw(GreenCircleCell,shift( (xyz2(0,1,1)+EvenPoint5) )*unitcircle, black+0.2);

fill(GreenCircleCell,shift( (xyz2(0,1,0)+EvenPoint5) )*unitcircle, gray);
fill(GreenCircleCell,shift( (xyz2(0,1,0)+EvenPoint5) )*((0,1)..(0.84,0)..(0,-0.64)..(-0.84,0)..cycle), white);
draw(GreenCircleCell,shift( (xyz2(0,1,0)+EvenPoint5) )*unitcircle, black+0.2);

fill(GreenCircleCell,shift( (xyz2(1,0,1)+EvenPoint5) )*unitcircle, gray);
fill(GreenCircleCell,shift( (xyz2(1,0,1)+EvenPoint5) )*((0,1)..(0.84,0)..(0,-0.64)..(-0.84,0)..cycle), white);
draw(GreenCircleCell,shift( (xyz2(1,0,1)+EvenPoint5)  )*unitcircle, black+0.2);

fill(GreenCircleCell,shift( (xyz2(1,1,1)+EvenPoint5) )*unitcircle, gray);
fill(GreenCircleCell,shift( (xyz2(1,1,1)+EvenPoint5) )*((0,1)..(0.84,0)..(0,-0.64)..(-0.84,0)..cycle), white);
draw(GreenCircleCell,shift( (xyz2(1,1,1)+EvenPoint5)  )*unitcircle, black+0.2);

// Right square qubits

fill(GreenCircleCell,shift( (xyz2(1,1,1)+OddPoint1) )*unitcircle, gray);
fill(GreenCircleCell,shift( (xyz2(1,1,1)+OddPoint1) )*((0,1)..(0.84,0)..(0,-0.64)..(-0.84,0)..cycle), white);
draw(GreenCircleCell,shift( (xyz2(1,1,1)+OddPoint1) )*unitcircle, black+0.2);

fill(GreenCircleCell,shift(  (xyz2(1,0,1)+EvenPoint4) )*unitcircle, gray);
fill(GreenCircleCell,shift(  (xyz2(1,0,1)+EvenPoint4) )*((0,1)..(0.84,0)..(0,-0.64)..(-0.84,0)..cycle), white);
draw(GreenCircleCell,shift(  (xyz2(1,0,1)+EvenPoint4) )*unitcircle, black+0.2);

// Bottom square qubits

fill(GreenCircleCell,shift( (xyz2(0,0,1)+OddPoint1))*unitcircle, gray);
fill(GreenCircleCell,shift( (xyz2(0,0,1)+OddPoint1) )*((0,1)..(0.84,0)..(0,-0.64)..(-0.84,0)..cycle), white);
draw(GreenCircleCell,shift( (xyz2(0,0,1)+OddPoint1) )*unitcircle, black+0.2);

fill(GreenCircleCell,shift( (xyz2(1,0,1)+EvenPoint1) )*unitcircle, gray);
fill(GreenCircleCell,shift( (xyz2(1,0,1)+EvenPoint1) )*((0,1)..(0.84,0)..(0,-0.64)..(-0.84,0)..cycle), white);
draw(GreenCircleCell,shift( (xyz2(1,0,1)+EvenPoint1) )*unitcircle, black+0.2);

// Side square face qubits

fill(GreenCircleCell,shift( (xyz2(0,0,0)+EvenPoint3) )*unitcircle, gray);
fill(GreenCircleCell,shift( (xyz2(0,0,0)+EvenPoint3) )*((0,1)..(0.84,0)..(0,-0.64)..(-0.84,0)..cycle), white);
draw(GreenCircleCell,shift( (xyz2(0,0,0)+EvenPoint3) )*unitcircle, black+0.2);

fill(GreenCircleCell,shift( (xyz2(0,0,1)+OddPoint4) )*unitcircle, gray);
fill(GreenCircleCell,shift( (xyz2(0,0,1)+OddPoint4) )*((0,1)..(0.84,0)..(0,-0.64)..(-0.84,0)..cycle), white);
draw(GreenCircleCell,shift( (xyz2(0,0,1)+OddPoint4) )*unitcircle, black+0.2);

fill(GreenCircleCell,shift( (xyz2(0,1,1)+EvenPoint1) )*unitcircle, gray);
fill(GreenCircleCell,shift( (xyz2(0,1,1)+EvenPoint1) )*((0,1)..(0.84,0)..(0,-0.64)..(-0.84,0)..cycle), white);
draw(GreenCircleCell,shift( (xyz2(0,1,1)+EvenPoint1) )*unitcircle, black+0.2);

fill(GreenCircleCell,shift( (xyz2(0,1,0)+OddPoint2) )*unitcircle, gray);
fill(GreenCircleCell,shift( (xyz2(0,1,0)+OddPoint2) )*((0,1)..(0.84,0)..(0,-0.64)..(-0.84,0)..cycle), white);
draw(GreenCircleCell,shift( (xyz2(0,1,0)+OddPoint2)  )*unitcircle, black+0.2);

// Top square face qubits

fill(GreenCircleCell,shift( (xyz2(0,1,0)+OddPoint3) )*unitcircle, gray);
fill(GreenCircleCell,shift( (xyz2(0,1,0)+OddPoint3) )*((0,1)..(0.84,0)..(0,-0.64)..(-0.84,0)..cycle), white);
draw(GreenCircleCell,shift( (xyz2(0,1,0)+OddPoint3) )*unitcircle, black+0.2);

fill(GreenCircleCell,shift( (xyz2(0,1,1)+EvenPoint4) )*unitcircle, gray);
fill(GreenCircleCell,shift( (xyz2(0,1,1)+EvenPoint4) )*((0,1)..(0.84,0)..(0,-0.64)..(-0.84,0)..cycle), white);
draw(GreenCircleCell,shift( (xyz2(0,1,1)+EvenPoint4) )*unitcircle, black+0.2);

fill(GreenCircleCell,shift( (xyz2(1,1,1)+OddPoint4) )*unitcircle, gray);
fill(GreenCircleCell,shift( (xyz2(1,1,1)+OddPoint4) )*((0,1)..(0.84,0)..(0,-0.64)..(-0.84,0)..cycle), white);
draw(GreenCircleCell,shift( (xyz2(1,1,1)+OddPoint4)  )*unitcircle, black+0.2);

fill(GreenCircleCell,shift( (xyz2(1,1,0)+EvenPoint3) )*unitcircle, gray);
fill(GreenCircleCell,shift( (xyz2(1,1,0)+EvenPoint3) )*((0,1)..(0.84,0)..(0,-0.64)..(-0.84,0)..cycle), white);
draw(GreenCircleCell,shift( (xyz2(1,1,0)+EvenPoint3) )*unitcircle, black+0.2);

// Front square qubits

fill(GreenCircleCell,shift( (xyz2(0,0,1)+OddPoint3) )*unitcircle, gray);
fill(GreenCircleCell,shift( (xyz2(0,0,1)+OddPoint3) )*((0,1)..(0.84,0)..(0,-0.64)..(-0.84,0)..cycle), white);
draw(GreenCircleCell,shift( (xyz2(0,0,1)+OddPoint3)  )*unitcircle, black+0.2);

fill(GreenCircleCell,shift( (xyz2(1,0,1)+EvenPoint3) )*unitcircle, gray);
fill(GreenCircleCell,shift(  (xyz2(1,0,1)+EvenPoint3)  )*((0,1)..(0.84,0)..(0,-0.64)..(-0.84,0)..cycle), white);
draw(GreenCircleCell,shift(  (xyz2(1,0,1)+EvenPoint3) )*unitcircle, black+0.2);

fill(GreenCircleCell,shift( (xyz2(1,1,1)+OddPoint2) )*unitcircle, gray);
fill(GreenCircleCell,shift( (xyz2(1,1,1)+OddPoint2) )*((0,1)..(0.84,0)..(0,-0.64)..(-0.84,0)..cycle), white);
draw(GreenCircleCell,shift( (xyz2(1,1,1)+OddPoint2) )*unitcircle, black+0.2);

fill(GreenCircleCell,shift( (xyz2(0,1,1)+EvenPoint2) )*unitcircle, gray);
fill(GreenCircleCell,shift( (xyz2(0,1,1)+EvenPoint2) )*((0,1)..(0.84,0)..(0,-0.64)..(-0.84,0)..cycle), white);
draw(GreenCircleCell,shift( (xyz2(0,1,1)+EvenPoint2)  )*unitcircle, black+0.2);

add(YellowCircleCell, xyz2(7,0,0));

add(BlueDiamondCell, xyz2(6,0,0));

add(YellowCircleCell, xyz2(7,2,0));
add(GreenCircleCell, xyz2(6,1,0));
add(YellowCircleCell, xyz2(5,0,0));

add(BlueDiamondCell, xyz2(4,0,0));
add(RedDiamondCell, xyz2(5,1,0));
add(BlueDiamondCell, xyz2(6,2,0));

add(GreenCircleCell, xyz2(6,3,0));
add(YellowCircleCell, xyz2(5,2,0));
add(GreenCircleCell, xyz2(4,1,0));
add(YellowCircleCell, xyz2(3,0,0));

add(BlueDiamondCell, xyz2(6,4,0));
add(RedDiamondCell, xyz2(5,3,0));
add(BlueDiamondCell, xyz2(4,2,0));
add(RedDiamondCell, xyz2(3,1,0));
add(BlueDiamondCell, xyz2(2,0,0));

add(YellowCircleCell, xyz2(5,4,0));
add(GreenCircleCell, xyz2(4,3,0));
add(YellowCircleCell, xyz2(3,2,0));
add(GreenCircleCell, xyz2(2,1,0));

add(BlueDiamondCell, xyz2(4,4,0));
add(RedDiamondCell, xyz2(3,3,0));
add(BlueDiamondCell, xyz2(2,2,0));

add(GreenDiamondCell, xyz2(7,0,1));
add(BlueCircleCell, xyz2(7,1,1));
add(RedCircleCell, xyz2(6,0,1));
add(GreenDiamondCell, xyz2(5,0,1));
add(YellowDiamondCell, xyz2(6,1,1));
add(RedCircleCell, xyz2(6,2,1));
add(BlueCircleCell, xyz2(5,1,1));
add(RedCircleCell, xyz2(4,0,1));
add(GreenDiamondCell, xyz2(3,0,1));
add(RedCircleCell, xyz2(2,0,1));

add(YellowDiamondCell, xyz2(4,1,1));
add(GreenDiamondCell, xyz2(5,2,1));
add(RedDiamondCell, xyz2(7,1,2));

\end{asy}

\begin{asy}
pair xyz(real x, real y, real z)
{
return 16*(x + z / 2.4, -y + z / 3.2);
}

pair xyz2(real x, real y, real z)
{
return 30*(-x - z / 3+1.5, -y + z / 6);
}

pen Outlines = black+0.2;

//label(scale(2/3)*"(a)", (-124,75));
//label(scale(2/3)*"(b)", (-50,75));
//label(scale(2/3)*"(c)", (50,75));

real ysep = 120;
real xsep = 160;

pair Point1 = xyz(0,0,0);
pair Point2 = xyz(0,0,1);
pair Point3 = xyz(0,1,0);
pair Point4 = xyz(0,1,1);
pair Point5 = xyz(1,0,0);
pair Point6 = xyz(1,0,1);
pair Point7 = xyz(1,1,0);
pair Point8 = xyz(1,1,1);

pair PrimalPoint1 = xyz2(0,0,0);
pair PrimalPoint2 = xyz2(0,0,1);
pair PrimalPoint3 = xyz2(0,1,0);
pair PrimalPoint4 = xyz2(0,1,1);
pair PrimalPoint5 = xyz2(1,0,0);
pair PrimalPoint6 = xyz2(1,0,1);
pair PrimalPoint7 = xyz2(1,1,0);
pair PrimalPoint8 = xyz2(1,1,1);

pair EvenPoint1 = (PrimalPoint1+PrimalPoint2+PrimalPoint3+PrimalPoint5) / 4;
pair EvenPoint2 = (PrimalPoint2+PrimalPoint5+PrimalPoint6+PrimalPoint8) / 4;
pair EvenPoint3 = (PrimalPoint4+PrimalPoint2+PrimalPoint3+PrimalPoint8) / 4;
pair EvenPoint4 = (PrimalPoint7+PrimalPoint8+PrimalPoint3+PrimalPoint5) / 4;
pair EvenPoint5 = (PrimalPoint5+PrimalPoint2+PrimalPoint3+PrimalPoint8) / 4;

pair OddPoint1 = (PrimalPoint1 + PrimalPoint5 + PrimalPoint6 + PrimalPoint7) / 4;
pair OddPoint2 = (PrimalPoint1 + PrimalPoint2 + PrimalPoint4 + PrimalPoint6) / 4;
pair OddPoint3 = (PrimalPoint4 + PrimalPoint8 + PrimalPoint6 + PrimalPoint7) / 4;
pair OddPoint4 = (PrimalPoint1 + PrimalPoint3 + PrimalPoint4 + PrimalPoint7) / 4;
pair OddPoint5 = (PrimalPoint1 + PrimalPoint4 + PrimalPoint6 + PrimalPoint7) / 4;

picture RedLeftFace;
fill(RedLeftFace, (xyz2(0,0,0)+OddPoint3)--(xyz2(0,0,1)+EvenPoint4)--(xyz2(0,1,1)+OddPoint1)--(xyz2(0,1,0)+EvenPoint2)--cycle, mediumblue);

picture RedRightFace;
fill(RedRightFace, (xyz2(1,0,0)+EvenPoint3)--(xyz2(1,0,1)+OddPoint4)--(xyz2(1,1,1)+EvenPoint1)--(xyz2(1,1,0)+OddPoint2)--cycle, 
mediumblue);

picture GreenTopFace;
filldraw(GreenTopFace, (xyz2(0,1,0)+EvenPoint2)--(xyz2(0,1,1)+OddPoint1)--(xyz2(1,1,1)+EvenPoint1)--(xyz2(1,1,0)+OddPoint2)--cycle, mediumblue);

picture GreenBottomFace;
fill(GreenBottomFace, (xyz2(0,0,0)+OddPoint3)--(xyz2(0,0,1)+EvenPoint4)--(xyz2(1,0,1)+OddPoint4)--(xyz2(1,0,0)+EvenPoint3)--cycle, mediumblue);

picture BlueTopFace;
filldraw(BlueTopFace, (xyz2(0,1,0)+EvenPoint2)--(xyz2(0,1,1)+OddPoint1)--(xyz2(1,1,1)+EvenPoint1)--(xyz2(1,1,0)+OddPoint2)--cycle, lightolive);

picture BlueBottomFace;
fill(BlueBottomFace, (xyz2(0,0,0)+OddPoint3)--(xyz2(0,0,1)+EvenPoint4)--(xyz2(1,0,1)+OddPoint4)--(xyz2(1,0,0)+EvenPoint3)--cycle, lightolive);

picture GreenBackFace;
fill(GreenBackFace, (xyz2(0,0,0)+OddPoint3)--(xyz2(1,0,0)+EvenPoint3)--(xyz2(1,1,0)+OddPoint2)--(xyz2(0,1,0)+EvenPoint2)--cycle, deepgreen);

picture GreenFrontFace;
filldraw(GreenFrontFace, (xyz2(0,0,1)+EvenPoint4)--(xyz2(1,0,1)+OddPoint4)--(xyz2(1,1,1)+EvenPoint1)--(xyz2(0,1,1)+OddPoint1)--cycle, heavygreen);

picture RightBlueFace;

fill(RightBlueFace, (xyz2(0,0,0)+EvenPoint3)--(xyz2(0,0,0)+EvenPoint5)--(xyz2(0,0,0)+EvenPoint2)--(xyz2(0,0,1)+OddPoint1)--(xyz2(0,0,1)+OddPoint5)--(xyz2(0,0,1)+OddPoint4)--cycle,heavygreen);

picture TopHexagonFace;

filldraw(TopHexagonFace, (xyz2(0,1,1)+EvenPoint2)--(xyz2(0,1,1)+EvenPoint5)--(xyz2(0,1,1)+EvenPoint4)--(xyz2(1,1,1)+OddPoint4)--(xyz2(1,1,1)+OddPoint5)--(xyz2(1,1,1)+OddPoint2)--cycle, lightblue);

picture RedDiamondCell;

filldraw(RedDiamondCell, (xyz2(0,1,1)+OddPoint1)--(xyz2(1,1,1)+EvenPoint1)--(xyz2(1,0,1)+OddPoint4)--(xyz2(0,0,1)+EvenPoint4)--cycle,mediumblue,Outlines);
filldraw(RedDiamondCell, (xyz2(0,1,1)+OddPoint1)--(xyz2(1,1,1)+EvenPoint1)--(xyz2(1,1,0)+OddPoint2)--(xyz2(0,1,0)+EvenPoint2)--cycle, heavygreen,Outlines);
filldraw(RedDiamondCell, (xyz2(0,1,1)+OddPoint1)--(xyz2(0,1,0)+EvenPoint2)--(xyz2(0,0,0)+OddPoint3)--(xyz2(0,0,1)+EvenPoint4)--cycle,heavyred,Outlines);

pen LineColor = black;
pen LineColor2 = heavyred+0.8;
pen LineColor3 = heavygreen+0.8;

picture GreenDiamondCell;

draw(GreenDiamondCell, (xyz2(0,1,1)+OddPoint1)--(xyz2(1,1,1)+EvenPoint1)--(xyz2(1,0,1)+OddPoint4)--(xyz2(0,0,1)+EvenPoint4)--cycle, LineColor);

draw(GreenDiamondCell, (xyz2(0,1,1)+OddPoint1)--(xyz2(1,1,1)+EvenPoint1)--(xyz2(1,1,0)+OddPoint2)--(xyz2(0,1,0)+EvenPoint2)--cycle, LineColor);
draw(GreenDiamondCell, (xyz2(0,1,1)+OddPoint1)--(xyz2(0,1,0)+EvenPoint2)--(xyz2(0,0,0)+OddPoint3)--(xyz2(0,0,1)+EvenPoint4)--cycle, LineColor);

draw(GreenDiamondCell, (xyz2(1,0,1)+OddPoint4)--(xyz2(1,0,0)+EvenPoint3)--(xyz2(1,1,0)+OddPoint2)--(xyz2(1,1,1)+EvenPoint1)--cycle, LineColor);

draw(GreenDiamondCell, (xyz2(0,0,0)+OddPoint3)--(xyz2(1,0,0)+EvenPoint3)--(xyz2(1,0,1)+OddPoint4)--(xyz2(0,0,1)+EvenPoint4)--cycle, LineColor);

/*
filldraw(GreenDiamondCell,shift( (xyz2(0,1,0)+EvenPoint2) )*unitcircle,white,black);
filldraw(GreenDiamondCell,shift( (xyz2(0,0,0)+OddPoint3) )*unitcircle,white,black);
filldraw(GreenDiamondCell,shift( (xyz2(1,1,0)+OddPoint2) )*unitcircle,white,black);

filldraw(GreenDiamondCell,shift( (xyz2(0,1,1)+OddPoint1) )*unitcircle,white,black);
filldraw(GreenDiamondCell,shift( (xyz2(1,1,1)+EvenPoint1) )*unitcircle,white,black);
filldraw(GreenDiamondCell,shift( (xyz2(1,0,1)+OddPoint4) )*unitcircle,white,black);
filldraw(GreenDiamondCell,shift( (xyz2(0,0,1)+EvenPoint4) )*unitcircle,white,black);
*/

picture BlueDiamondCell;

filldraw(BlueDiamondCell, (xyz2(0,1,1)+OddPoint1)--(xyz2(1,1,1)+EvenPoint1)--(xyz2(1,0,1)+OddPoint4)--(xyz2(0,0,1)+EvenPoint4)--cycle,mediumblue,gray);
filldraw(BlueDiamondCell, (xyz2(0,1,1)+OddPoint1)--(xyz2(1,1,1)+EvenPoint1)--(xyz2(1,1,0)+OddPoint2)--(xyz2(0,1,0)+EvenPoint2)--cycle,blue,gray);
filldraw(BlueDiamondCell, (xyz2(0,1,1)+OddPoint1)--(xyz2(0,1,0)+EvenPoint2)--(xyz2(0,0,0)+OddPoint3)--(xyz2(0,0,1)+EvenPoint4)--cycle,heavyblue,gray);

filldraw(BlueDiamondCell,shift( (xyz2(0,1,0)+EvenPoint2) )*unitcircle,white,black);
filldraw(BlueDiamondCell,shift( (xyz2(0,0,0)+OddPoint3) )*unitcircle,white,black);
filldraw(BlueDiamondCell,shift( (xyz2(1,1,0)+OddPoint2) )*unitcircle,white,black);

filldraw(BlueDiamondCell,shift( (xyz2(0,1,1)+OddPoint1) )*unitcircle,white,black);
filldraw(BlueDiamondCell,shift( (xyz2(1,1,1)+EvenPoint1) )*unitcircle,white,black);
filldraw(BlueDiamondCell,shift( (xyz2(1,0,1)+OddPoint4) )*unitcircle,white,black);
filldraw(BlueDiamondCell,shift( (xyz2(0,0,1)+EvenPoint4) )*unitcircle,white,black);

picture YellowDiamondCell;

filldraw(YellowDiamondCell, (xyz2(0,1,1)+OddPoint1)--(xyz2(1,1,1)+EvenPoint1)--(xyz2(1,0,1)+OddPoint4)--(xyz2(0,0,1)+EvenPoint4)--cycle,yellow,gray);
filldraw(YellowDiamondCell, (xyz2(0,1,1)+OddPoint1)--(xyz2(1,1,1)+EvenPoint1)--(xyz2(1,1,0)+OddPoint2)--(xyz2(0,1,0)+EvenPoint2)--cycle,lightyellow,gray);
filldraw(YellowDiamondCell, (xyz2(0,1,1)+OddPoint1)--(xyz2(0,1,0)+EvenPoint2)--(xyz2(0,0,0)+OddPoint3)--(xyz2(0,0,1)+EvenPoint4)--cycle,olive,gray);

filldraw(YellowDiamondCell,shift( (xyz2(0,1,0)+EvenPoint2) )*unitcircle,white,black);
filldraw(YellowDiamondCell,shift( (xyz2(0,0,0)+OddPoint3) )*unitcircle,white,black);
filldraw(YellowDiamondCell,shift( (xyz2(1,1,0)+OddPoint2) )*unitcircle,white,black);

filldraw(YellowDiamondCell,shift( (xyz2(0,1,1)+OddPoint1) )*unitcircle,white,black);
filldraw(YellowDiamondCell,shift( (xyz2(1,1,1)+EvenPoint1) )*unitcircle,white,black);
filldraw(YellowDiamondCell,shift( (xyz2(1,0,1)+OddPoint4) )*unitcircle,white,black);
filldraw(YellowDiamondCell,shift( (xyz2(0,0,1)+EvenPoint4) )*unitcircle,white,black);

picture BlueCircleCell;

// Right edges with no qubits shown on the shape

draw(BlueCircleCell, (xyz2(0,0,0)+EvenPoint3)--(xyz2(0,0,0)+EvenPoint5)--(xyz2(0,0,0)+EvenPoint2)--(xyz2(0,0,1)+OddPoint1)--(xyz2(0,0,1)+OddPoint5)--(xyz2(0,0,1)+OddPoint4)--cycle, LineColor2);

draw(BlueCircleCell, (xyz2(1,0,0)+OddPoint3)--(xyz2(1,0,0)+OddPoint5)--(xyz2(1,0,0)+OddPoint2)--(xyz2(1,0,1)+EvenPoint1)--(xyz2(1,0,1)+EvenPoint5)--(xyz2(1,0,1)+EvenPoint4)--cycle, LineColor2);

draw(BlueCircleCell, (xyz2(0,1,0)+OddPoint3)--(xyz2(0,1,1)+EvenPoint4)--(xyz2(1,1,1)+OddPoint4)--(xyz2(1,1,0)+EvenPoint3)--cycle, LineColor2);

draw(BlueCircleCell, (xyz2(0,0,0)+EvenPoint2)--(xyz2(0,0,1)+OddPoint1)--(xyz2(1,0,1)+EvenPoint1)--(xyz2(1,0,0)+OddPoint2)--cycle, LineColor2);

draw(BlueCircleCell, (xyz2(0,1,1)+EvenPoint2)--(xyz2(1,1,1)+OddPoint2)--(xyz2(1,0,1)+EvenPoint3)--(xyz2(0,0,1)+OddPoint3)--cycle, LineColor2);

draw(BlueCircleCell, (xyz2(0,1,1)+EvenPoint1)--(xyz2(0,1,1)+EvenPoint5)--(xyz2(0,1,1)+EvenPoint2)--(xyz2(0,0,1)+OddPoint3)--(xyz2(0,0,1)+OddPoint5)--(xyz2(0,0,1)+OddPoint4)--cycle, LineColor2);

draw(BlueCircleCell, (xyz2(0,1,1)+EvenPoint2)--(xyz2(0,1,1)+EvenPoint5)--(xyz2(0,1,1)+EvenPoint4)--(xyz2(1,1,1)+OddPoint4)--(xyz2(1,1,1)+OddPoint5)--(xyz2(1,1,1)+OddPoint2)--cycle, LineColor2);

draw(BlueCircleCell, (xyz2(1,0,1)+EvenPoint3)--(xyz2(1,0,1)+EvenPoint5)--(xyz2(1,0,1)+EvenPoint4)--(xyz2(1,1,1)+OddPoint1)--(xyz2(1,1,1)+OddPoint5)--(xyz2(1,1,1)+OddPoint2)--cycle, LineColor2);

draw(BlueCircleCell, (xyz2(1,1,1)+OddPoint1)--(xyz2(1,1,1)+OddPoint5)--(xyz2(1,1,1)+OddPoint4)--(xyz2(1,1,0)+EvenPoint3)--(xyz2(1,1,0)+EvenPoint5)--(xyz2(1,1,0)+EvenPoint2)--cycle, LineColor2);

draw(BlueCircleCell, (xyz2(0,1,1)+EvenPoint1)--(xyz2(0,1,1)+EvenPoint5)--(xyz2(0,1,1)+EvenPoint4)--(xyz2(0,1,0)+OddPoint3)--(xyz2(0,1,0)+OddPoint5)--(xyz2(0,1,0)+OddPoint2)--cycle, LineColor2);

draw(BlueCircleCell, (xyz2(1,0,1)+EvenPoint1)--(xyz2(1,0,1)+EvenPoint5)--(xyz2(1,0,1)+EvenPoint3)--(xyz2(0,0,1)+OddPoint3)--(xyz2(0,0,1)+OddPoint5)--(xyz2(0,0,1)+OddPoint1)--cycle, LineColor2);

draw(BlueCircleCell, (xyz2(0,0,0)+EvenPoint3)--(xyz2(0,0,1)+OddPoint4)--(xyz2(0,1,1)+EvenPoint1)--(xyz2(0,1,0)+OddPoint2)--cycle, LineColor2);

/*
// Central cube qubits

filldraw(BlueCircleCell,shift(  (xyz2(0,0,0)+EvenPoint5))*unitcircle,white,black);
filldraw(BlueCircleCell,shift(  (xyz2(0,0,1)+EvenPoint5))*unitcircle,white,black);
filldraw(BlueCircleCell,shift(  (xyz2(0,1,1)+EvenPoint5))*unitcircle,white,black);
filldraw(BlueCircleCell,shift(  (xyz2(0,1,0)+EvenPoint5))*unitcircle,white,black);
filldraw(BlueCircleCell,shift(  (xyz2(1,0,1)+EvenPoint5))*unitcircle,white,black);
filldraw(BlueCircleCell,shift(  (xyz2(1,1,1)+EvenPoint5))*unitcircle,white,black);

// Right square qubits

filldraw(BlueCircleCell,shift(  (xyz2(1,1,1)+OddPoint1))*unitcircle,white,black);
filldraw(BlueCircleCell,shift(  (xyz2(1,0,1)+EvenPoint4))*unitcircle,white,black);

// Bottom square qubits

filldraw(BlueCircleCell,shift(  (xyz2(0,0,1)+OddPoint1))*unitcircle,white,black);
filldraw(BlueCircleCell,shift(  (xyz2(1,0,1)+EvenPoint1))*unitcircle,white,black);

// Side square face qubits

filldraw(BlueCircleCell,shift(  (xyz2(0,0,0)+EvenPoint3))*unitcircle,white,black);
filldraw(BlueCircleCell,shift( (xyz2(0,0,1)+OddPoint4) )*unitcircle,white,black);
filldraw(BlueCircleCell,shift( (xyz2(0,1,1)+EvenPoint1) )*unitcircle,white,black);
filldraw(BlueCircleCell,shift( (xyz2(0,1,0)+OddPoint2) )*unitcircle,white,black);

// Top square face qubits

filldraw(BlueCircleCell,shift(  (xyz2(0,1,0)+OddPoint3))*unitcircle,white,black);
filldraw(BlueCircleCell,shift( (xyz2(0,1,1)+EvenPoint4) )*unitcircle,white,black);
filldraw(BlueCircleCell,shift( (xyz2(1,1,1)+OddPoint4))*unitcircle,white,black);
filldraw(BlueCircleCell,shift((xyz2(1,1,0)+EvenPoint3) )*unitcircle,white,black);

// Front square qubits

filldraw(BlueCircleCell,shift( (xyz2(0,0,1)+OddPoint3) )*unitcircle,white,black);
filldraw(BlueCircleCell,shift( (xyz2(1,0,1)+EvenPoint3) )*unitcircle,white,black);
filldraw(BlueCircleCell,shift( (xyz2(1,1,1)+OddPoint2) )*unitcircle,white,black);
filldraw(BlueCircleCell,shift( (xyz2(0,1,1)+EvenPoint2) )*unitcircle,white,black);
*/

picture YellowCircleCell;

// Right edges with no qubits shown on the shape

draw(YellowCircleCell, (xyz2(0,0,0)+EvenPoint3)--(xyz2(0,0,0)+EvenPoint5)--(xyz2(0,0,0)+EvenPoint2)--(xyz2(0,0,1)+OddPoint1)--(xyz2(0,0,1)+OddPoint5)--(xyz2(0,0,1)+OddPoint4)--cycle, LineColor3);

draw(YellowCircleCell, (xyz2(1,0,0)+OddPoint3)--(xyz2(1,0,0)+OddPoint5)--(xyz2(1,0,0)+OddPoint2)--(xyz2(1,0,1)+EvenPoint1)--(xyz2(1,0,1)+EvenPoint5)--(xyz2(1,0,1)+EvenPoint4)--cycle, LineColor3);

draw(YellowCircleCell, (xyz2(0,1,0)+OddPoint3)--(xyz2(0,1,1)+EvenPoint4)--(xyz2(1,1,1)+OddPoint4)--(xyz2(1,1,0)+EvenPoint3)--cycle, LineColor3);

draw(YellowCircleCell, (xyz2(0,0,0)+EvenPoint2)--(xyz2(0,0,1)+OddPoint1)--(xyz2(1,0,1)+EvenPoint1)--(xyz2(1,0,0)+OddPoint2)--cycle, LineColor3);

draw(YellowCircleCell, (xyz2(0,1,1)+EvenPoint2)--(xyz2(1,1,1)+OddPoint2)--(xyz2(1,0,1)+EvenPoint3)--(xyz2(0,0,1)+OddPoint3)--cycle, LineColor3);

draw(YellowCircleCell, (xyz2(0,1,1)+EvenPoint1)--(xyz2(0,1,1)+EvenPoint5)--(xyz2(0,1,1)+EvenPoint2)--(xyz2(0,0,1)+OddPoint3)--(xyz2(0,0,1)+OddPoint5)--(xyz2(0,0,1)+OddPoint4)--cycle, LineColor3);

draw(YellowCircleCell, (xyz2(0,1,1)+EvenPoint2)--(xyz2(0,1,1)+EvenPoint5)--(xyz2(0,1,1)+EvenPoint4)--(xyz2(1,1,1)+OddPoint4)--(xyz2(1,1,1)+OddPoint5)--(xyz2(1,1,1)+OddPoint2)--cycle, LineColor3);

draw(YellowCircleCell, (xyz2(1,0,1)+EvenPoint3)--(xyz2(1,0,1)+EvenPoint5)--(xyz2(1,0,1)+EvenPoint4)--(xyz2(1,1,1)+OddPoint1)--(xyz2(1,1,1)+OddPoint5)--(xyz2(1,1,1)+OddPoint2)--cycle, LineColor3);

draw(YellowCircleCell, (xyz2(0,1,1)+EvenPoint1)--(xyz2(0,1,1)+EvenPoint5)--(xyz2(0,1,1)+EvenPoint4)--(xyz2(0,1,0)+OddPoint3)--(xyz2(0,1,0)+OddPoint5)--(xyz2(0,1,0)+OddPoint2)--cycle, LineColor3);

draw(YellowCircleCell, (xyz2(1,1,1)+OddPoint1)--(xyz2(1,1,1)+OddPoint5)--(xyz2(1,1,1)+OddPoint4)--(xyz2(1,1,0)+EvenPoint3)--(xyz2(1,1,0)+EvenPoint5)--(xyz2(1,1,0)+EvenPoint2)--cycle, LineColor3);

draw(YellowCircleCell, (xyz2(1,0,1)+EvenPoint1)--(xyz2(1,0,1)+EvenPoint5)--(xyz2(1,0,1)+EvenPoint3)--(xyz2(0,0,1)+OddPoint3)--(xyz2(0,0,1)+OddPoint5)--(xyz2(0,0,1)+OddPoint1)--cycle, LineColor3);

draw(YellowCircleCell, (xyz2(0,0,0)+EvenPoint3)--(xyz2(0,0,1)+OddPoint4)--(xyz2(0,1,1)+EvenPoint1)--(xyz2(0,1,0)+OddPoint2)--cycle, LineColor3);

draw(YellowCircleCell, (xyz2(1,0,0)+OddPoint3)--(xyz2(1,0,1)+EvenPoint4)--(xyz2(1,1,1)+OddPoint1)--(xyz2(1,1,0)+EvenPoint2)--cycle, LineColor3);

/*
// Central cube qubits

filldraw(YellowCircleCell,shift(  (xyz2(0,0,0)+EvenPoint5))*unitcircle,white,black);
filldraw(YellowCircleCell,shift(  (xyz2(0,0,1)+EvenPoint5))*unitcircle,white,black);
filldraw(YellowCircleCell,shift(  (xyz2(0,1,1)+EvenPoint5))*unitcircle,white,black);
filldraw(YellowCircleCell,shift(  (xyz2(0,1,0)+EvenPoint5))*unitcircle,white,black);
filldraw(YellowCircleCell,shift(  (xyz2(1,0,1)+EvenPoint5))*unitcircle,white,black);
filldraw(YellowCircleCell,shift(  (xyz2(1,1,1)+EvenPoint5))*unitcircle,white,black);

// Right square qubits

filldraw(YellowCircleCell,shift(  (xyz2(1,1,1)+OddPoint1))*unitcircle,white,black);
filldraw(YellowCircleCell,shift(  (xyz2(1,0,1)+EvenPoint4))*unitcircle,white,black);

// Bottom square qubits

filldraw(YellowCircleCell,shift(  (xyz2(0,0,1)+OddPoint1))*unitcircle,white,black);
filldraw(YellowCircleCell,shift(  (xyz2(1,0,1)+EvenPoint1))*unitcircle,white,black);

// Side square face qubits

filldraw(YellowCircleCell,shift(  (xyz2(0,0,0)+EvenPoint3))*unitcircle,white,black);
filldraw(YellowCircleCell,shift( (xyz2(0,0,1)+OddPoint4) )*unitcircle,white,black);
filldraw(YellowCircleCell,shift( (xyz2(0,1,1)+EvenPoint1) )*unitcircle,white,black);
filldraw(YellowCircleCell,shift( (xyz2(0,1,0)+OddPoint2) )*unitcircle,white,black);

// Top square face qubits

filldraw(YellowCircleCell,shift(  (xyz2(0,1,0)+OddPoint3))*unitcircle,white,black);
filldraw(YellowCircleCell,shift( (xyz2(0,1,1)+EvenPoint4) )*unitcircle,white,black);
filldraw(YellowCircleCell,shift( (xyz2(1,1,1)+OddPoint4))*unitcircle,white,black);
filldraw(YellowCircleCell,shift((xyz2(1,1,0)+EvenPoint3) )*unitcircle,white,black);

// Front square qubits

filldraw(YellowCircleCell,shift( (xyz2(0,0,1)+OddPoint3) )*unitcircle,white,black);
filldraw(YellowCircleCell,shift( (xyz2(1,0,1)+EvenPoint3) )*unitcircle,white,black);
filldraw(YellowCircleCell,shift( (xyz2(1,1,1)+OddPoint2) )*unitcircle,white,black);
filldraw(YellowCircleCell,shift( (xyz2(0,1,1)+EvenPoint2) )*unitcircle,white,black);
*/

picture RedCircleCell;

// Right edges with no qubits shown on the shape

filldraw(RedCircleCell, (xyz2(0,0,0)+EvenPoint3)--(xyz2(0,0,0)+EvenPoint5)--(xyz2(0,0,0)+EvenPoint2)--(xyz2(0,0,1)+OddPoint1)--(xyz2(0,0,1)+OddPoint5)--(xyz2(0,0,1)+OddPoint4)--cycle,deepgreen, Outlines);

filldraw(RedCircleCell, (xyz2(0,1,0)+OddPoint3)--(xyz2(0,1,1)+EvenPoint4)--(xyz2(1,1,1)+OddPoint4)--(xyz2(1,1,0)+EvenPoint3)--cycle,blue, Outlines);

filldraw(RedCircleCell, (xyz2(0,1,1)+EvenPoint2)--(xyz2(1,1,1)+OddPoint2)--(xyz2(1,0,1)+EvenPoint3)--(xyz2(0,0,1)+OddPoint3)--cycle,green, Outlines);

filldraw(RedCircleCell, (xyz2(0,1,1)+EvenPoint1)--(xyz2(0,1,1)+EvenPoint5)--(xyz2(0,1,1)+EvenPoint2)--(xyz2(0,0,1)+OddPoint3)--(xyz2(0,0,1)+OddPoint5)--(xyz2(0,0,1)+OddPoint4)--cycle,heavyblue, Outlines);

filldraw(RedCircleCell, (xyz2(0,1,1)+EvenPoint2)--(xyz2(0,1,1)+EvenPoint5)--(xyz2(0,1,1)+EvenPoint4)--(xyz2(1,1,1)+OddPoint4)--(xyz2(1,1,1)+OddPoint5)--(xyz2(1,1,1)+OddPoint2)--cycle, mediumred, Outlines);

filldraw(RedCircleCell, (xyz2(1,0,1)+EvenPoint3)--(xyz2(1,0,1)+EvenPoint5)--(xyz2(1,0,1)+EvenPoint4)--(xyz2(1,1,1)+OddPoint1)--(xyz2(1,1,1)+OddPoint5)--(xyz2(1,1,1)+OddPoint2)--cycle,mediumblue, Outlines);

filldraw(RedCircleCell, (xyz2(0,1,1)+EvenPoint1)--(xyz2(0,1,1)+EvenPoint5)--(xyz2(0,1,1)+EvenPoint4)--(xyz2(0,1,0)+OddPoint3)--(xyz2(0,1,0)+OddPoint5)--(xyz2(0,1,0)+OddPoint2)--cycle,heavygreen, Outlines);

filldraw(RedCircleCell, (xyz2(1,0,1)+EvenPoint1)--(xyz2(1,0,1)+EvenPoint5)--(xyz2(1,0,1)+EvenPoint3)--(xyz2(0,0,1)+OddPoint3)--(xyz2(0,0,1)+OddPoint5)--(xyz2(0,0,1)+OddPoint1)--cycle,brown, Outlines);

filldraw(RedCircleCell, (xyz2(0,0,0)+EvenPoint3)--(xyz2(0,0,1)+OddPoint4)--(xyz2(0,1,1)+EvenPoint1)--(xyz2(0,1,0)+OddPoint2)--cycle,heavyred, Outlines);

picture GreenCircleCell;

// Right edges with no qubits shown on the shape

draw(GreenCircleCell, (xyz2(0,0,0)+EvenPoint3)--(xyz2(0,0,0)+EvenPoint5)--(xyz2(0,0,0)+EvenPoint2)--(xyz2(0,0,1)+OddPoint1)--(xyz2(0,0,1)+OddPoint5)--(xyz2(0,0,1)+OddPoint4)--cycle, LineColor);

draw(GreenCircleCell, (xyz2(0,1,0)+OddPoint3)--(xyz2(0,1,1)+EvenPoint4)--(xyz2(1,1,1)+OddPoint4)--(xyz2(1,1,0)+EvenPoint3)--cycle, LineColor);

draw(GreenCircleCell, (xyz2(0,1,1)+EvenPoint2)--(xyz2(1,1,1)+OddPoint2)--(xyz2(1,0,1)+EvenPoint3)--(xyz2(0,0,1)+OddPoint3)--cycle, LineColor);
draw(GreenCircleCell, (xyz2(0,1,1)+EvenPoint1)--(xyz2(0,1,1)+EvenPoint5)--(xyz2(0,1,1)+EvenPoint2)--(xyz2(0,0,1)+OddPoint3)--(xyz2(0,0,1)+OddPoint5)--(xyz2(0,0,1)+OddPoint4)--cycle, LineColor);

draw(GreenCircleCell, (xyz2(0,1,1)+EvenPoint2)--(xyz2(0,1,1)+EvenPoint5)--(xyz2(0,1,1)+EvenPoint4)--(xyz2(1,1,1)+OddPoint4)--(xyz2(1,1,1)+OddPoint5)--(xyz2(1,1,1)+OddPoint2)--cycle, LineColor);

draw(GreenCircleCell, (xyz2(1,0,1)+EvenPoint3)--(xyz2(1,0,1)+EvenPoint5)--(xyz2(1,0,1)+EvenPoint4)--(xyz2(1,1,1)+OddPoint1)--(xyz2(1,1,1)+OddPoint5)--(xyz2(1,1,1)+OddPoint2)--cycle, LineColor);

draw(GreenCircleCell, (xyz2(0,1,1)+EvenPoint1)--(xyz2(0,1,1)+EvenPoint5)--(xyz2(0,1,1)+EvenPoint4)--(xyz2(0,1,0)+OddPoint3)--(xyz2(0,1,0)+OddPoint5)--(xyz2(0,1,0)+OddPoint2)--cycle, LineColor);

draw(GreenCircleCell, (xyz2(1,0,1)+EvenPoint1)--(xyz2(1,0,1)+EvenPoint5)--(xyz2(1,0,1)+EvenPoint3)--(xyz2(0,0,1)+OddPoint3)--(xyz2(0,0,1)+OddPoint5)--(xyz2(0,0,1)+OddPoint1)--cycle, LineColor);

draw(GreenCircleCell, (xyz2(0,0,0)+EvenPoint3)--(xyz2(0,0,1)+OddPoint4)--(xyz2(0,1,1)+EvenPoint1)--(xyz2(0,1,0)+OddPoint2)--cycle, LineColor);

/*
// Central cube qubits

filldraw(GreenCircleCell,shift(  (xyz2(0,0,0)+EvenPoint5))*unitcircle,white,black);
filldraw(GreenCircleCell,shift(  (xyz2(0,0,1)+EvenPoint5))*unitcircle,white,black);
filldraw(GreenCircleCell,shift(  (xyz2(0,1,1)+EvenPoint5))*unitcircle,white,black);
filldraw(GreenCircleCell,shift(  (xyz2(0,1,0)+EvenPoint5))*unitcircle,white,black);
filldraw(GreenCircleCell,shift(  (xyz2(1,0,1)+EvenPoint5))*unitcircle,white,black);
filldraw(GreenCircleCell,shift(  (xyz2(1,1,1)+EvenPoint5))*unitcircle,white,black);

// Right square qubits

filldraw(GreenCircleCell,shift(  (xyz2(1,1,1)+OddPoint1))*unitcircle,white,black);
filldraw(GreenCircleCell,shift(  (xyz2(1,0,1)+EvenPoint4))*unitcircle,white,black);

// Bottom square qubits

filldraw(GreenCircleCell,shift(  (xyz2(0,0,1)+OddPoint1))*unitcircle,white,black);
filldraw(GreenCircleCell,shift(  (xyz2(1,0,1)+EvenPoint1))*unitcircle,white,black);

// Side square face qubits

filldraw(GreenCircleCell,shift(  (xyz2(0,0,0)+EvenPoint3))*unitcircle,white,black);
filldraw(GreenCircleCell,shift( (xyz2(0,0,1)+OddPoint4) )*unitcircle,white,black);
filldraw(GreenCircleCell,shift( (xyz2(0,1,1)+EvenPoint1) )*unitcircle,white,black);
filldraw(GreenCircleCell,shift( (xyz2(0,1,0)+OddPoint2) )*unitcircle,white,black);

// Top square face qubits

filldraw(GreenCircleCell,shift(  (xyz2(0,1,0)+OddPoint3))*unitcircle,white,black);
filldraw(GreenCircleCell,shift( (xyz2(0,1,1)+EvenPoint4) )*unitcircle,white,black);
filldraw(GreenCircleCell,shift( (xyz2(1,1,1)+OddPoint4))*unitcircle,white,black);
filldraw(GreenCircleCell,shift((xyz2(1,1,0)+EvenPoint3) )*unitcircle,white,black);

// Front square qubits

filldraw(GreenCircleCell,shift( (xyz2(0,0,1)+OddPoint3) )*unitcircle,white,black);
filldraw(GreenCircleCell,shift( (xyz2(1,0,1)+EvenPoint3) )*unitcircle,white,black);
filldraw(GreenCircleCell,shift( (xyz2(1,1,1)+OddPoint2) )*unitcircle,white,black);
filldraw(GreenCircleCell,shift( (xyz2(0,1,1)+EvenPoint2) )*unitcircle,white,black);
*/

add(GreenCircleCell, xyz2(4, 1.14, 0));
add(TopHexagonFace, xyz2(4,1.14,0));

add(RedRightFace, xyz2(3,1.14,0));
add(GreenDiamondCell, xyz2(3, 1.14, 0));
add(RedLeftFace,xyz2(3,1.14,0));

add(BlueCircleCell, xyz2(2, 1.14, 0));
add(RightBlueFace, xyz2(2,1.14,0));
add(BlueBottomFace, xyz2(2,2.14,0));

add(GreenDiamondCell, xyz2(2, 2.14, 0));
add(BlueTopFace, xyz2(2,2.14,0));

add(GreenCircleCell, xyz2(1, 0.14, 0));

add(GreenBackFace, xyz2(1,0.14,1));
add(GreenDiamondCell, xyz2(1, 0.14, 1));
add(GreenFrontFace, xyz2(1,0.14,1));

add(YellowCircleCell, xyz2(6.4, -0.36, 0));

add(GreenBottomFace, xyz2(6.4,0.64,0));
add(GreenDiamondCell, xyz2(6.4, 0.64, 0));
add(GreenTopFace, xyz2(6.4,0.64,0));

add(YellowCircleCell, xyz2(6.4, 1.64, 0));

\end{asy}

\end{multicols}

\caption{\label{fig:Gauge_color_code} \textbf{Gauge color code}. Left: structure of the primal lattice $\mathcal{L}^*$. The lattice is composed of truncated octahedrons and cubes, colored with four possible values. Qubits lie in the vertices. Right: Errors in the color code form either a single string, of color blue-green as in the first figure; or a stringnet such as the one shown in the second figure of colors red-blue, red-yellow and red-green. Strings and stringnets may end in a boundary. Each colored face indicates a $-1$ gauge face measurement. Furthermore, at each cell, the product of the gauge measurements of each color gives us repeated syndrome measurements that enable single-shot error correction~\cite{bombin2015single}.
Figures similar to those appearing in Ref.~\cite{brown2016fault}, generating code modified from the one courteously provided by Prof. Benjamin Brown.}
\end{figure}

Let us give two more examples: the three-dimensional color code with $X$ stabilizers attached to vertices and $Z$ stabilizers attached to edges is the stabilizer color code $CC_\mathcal{L}(0,1)$. On the other hand, the $CC_\mathcal{L}(0,0)$ subsystem color code has, in the primal (dual) lattice $\mathcal{L}^*$ ($\mathcal{L}$), gauge generators attached to faces (edges) and stabilizer generators attached to cells (vertices), see \cref{fig:gauge_operators}. Since two faces (edges) may share an odd number of qubits, gauge stabilizers do not need to commute. In three dimensions, there are four edges arriving at each vertex, so the corresponding faces do not need to share an edge.

Code words in CSS codes are those that lie in the +1 eigenspace of the element of the stabilizer. Furthermore, two bit-strings are equivalent if they are connected by an element of $\mathcal{G}\backslash\mathcal{S}$~\cite{kubica2015universal}. As a consequence, one may formally decompose the space as a tensor product of logical and gauge qubits: $\ket{\psi}\ket{g}$. This allows to distinguish two types of logic gates. \textit{Bare} logical operators $\bar{U}_{\text{bare}}: \ket{\psi}\ket{g}\mapsto (\bar{U}\ket{\psi})\ket{g}$ do not affect the gauge qubits. We denote by $\bar{X} = X^{\otimes n}$ and $\bar{Z} = Z^{\otimes n}$ the bare logical Pauli operators $X$ and $Z$~\cite{kubica2015universal}.  In contrast, \textit{dressed} logical may also affect them, $\bar{U}_{\text{dressed}}: \ket{\psi}\ket{g}\mapsto (\bar{U}\ket{\psi})\ket{g'}$. We then define the logical states as
\begin{equation}\label{eq:gx_logical_qubit}
    \ket{\bar{0}}\ket{g_x} := \sum_{X_G\in\mathcal{G}}X_G\ket{0}^{\otimes n}, \qquad  \ket{\bar{1}}\ket{g_x} := \bar{X}\ket{\bar{0}}\ket{g_x},
\end{equation}
where we have used that elements of the gauge group do not change the logical information, and $\ket{\bar{0}}\ket{g_x}$ and $\ket{\bar{1}}\ket{g_x}$ are $+1$ eigenstates of the $X_G\in\mathcal{G}$ generators~\cite{kubica2015universal}.
Alternatively, we can encode the logical qubits using
\begin{equation}\label{eq:gz_logical_qubit}
    \ket{\bar{0}}\ket{g_z} := \sum_{X_S\in\mathcal{S}}X_S\ket{0}^{\otimes n}, \qquad  \ket{\bar{1}}\ket{g_z} := \bar{X}\ket{\bar{0}}\ket{g_z},
\end{equation}
with $\ket{\bar{0}}\ket{g_z}$ and $\ket{\bar{1}}\ket{g_z}$ are $+1$ eigenstates of the $Z_G\in\mathcal{G}$ generators, as $Z_G X_S \ket{0}^{\otimes n} =  X_S Z_G \ket{0}^{\otimes n} = X_S \ket{0}^{\otimes n}$~\cite{kubica2015universal}.

Apart from Pauli $\bar{X}$ and $\bar{Z}$, we want to know what gates we can implement transversally. C-$Z$ is transversal because it is a CSS code and can therefore be implemented physical-qubit-wise between two copies of the code. Similarly, if the code is self-dual, $CC_\mathcal{L}(x,x)$, then $\bar{H}= H^{\otimes n}$ transforms the $\bar{X}$ into $\bar{Z}$ as strings have support over the same physical qubits. However, it is also important to notice that this operator is a dressed logical operator, as it exchanges $X_G\leftrightarrow Z_G$~\cite{kubica2015universal}.

Finally, to complete a universal set of gates, we want to explore the possibility of applying $\bar{R}_\ell$, defined in \eqref{eq:R_d}. As we did for the 6-6-6 two-dimensional color code, we divide the physical qubits in two disjoint sets $S_1\cup S_2 = \mathcal{Q}(\mathcal{L})$ such that the logical gate is defined as $\bar{R}_\ell = (R_\ell^k)^{\otimes (q\in S_1)} (R_\ell^{-k})^{\otimes (q\in S_2)}$, with $k\in\{1,\ldots,2^{n}-1\}$ ensuring
\begin{equation}
    k(|S_1|-|S_2|) = 1\mod 2^\ell.
\end{equation}
Finding such a $k$ is always possible, \cite[Page 7]{kubica2015universal} and it will implement the desired behavior. On the other hand, though, we want to make sure that $\bar{R}_\ell$ commutes with gauge operators, so we need~\cite{kubica2015universal}
\begin{equation}\label{eq:equal_support_R_d}
   \forall X_G\in\mathcal{G},\quad |S_1\cap \mathcal{Q}(X_G)|=|S_2\cap \mathcal{Q}(X_G)|\mod 2^\ell.
\end{equation}
In the two-dimensional color code, this condition meant that stabilizer/gauge operators had equal support over $S_1$ and $S_2$ elements. By imposing this constraint, we are making sure that $\bar{R}_\ell$ will be a bare logical operator. One way of enforcing \eqref{eq:equal_support_R_d} is to satisfy that, for any subset of gauge operators $\{X_{G_1},\ldots,X_{G_m}\}$, \cite[Equation 3]{bombin2015gauge} and \cite[Lemma 6]{kubica2015universal}
\begin{equation} \label{eq:R_d_transversal_sufficiency_condition}
    \left|S_1\bigcap_{i=1}^m \mathcal{Q}(X_{G_i})\right| =  \left|S_2\bigcap_{i=1}^m \mathcal{Q}(X_{G_i})\right| \mod 2^{\ell-m+1}.
\end{equation}
Since gauge operators $X_{G_i}$ are attached to $d-2-z$ simplices $\delta_i$, we can analyze the qubits to which they are attached. Let
\begin{equation}
    \bigcap_{i=1}^m \mathcal{Q}(\delta_i) = \mathcal{Q}(\tau)\text{ or }\emptyset.
\end{equation}
If the intersection is not an empty set, then $\cup_{i}\delta_i\subset \tau\in \mathcal{L}\backslash\partial \mathcal{L}$. Now, if $\dim\tau = m < d$, we have~\cite[Lemma 7]{kubica2015universal}
\begin{equation}
   \left|S_1\cap\mathcal{Q}(\tau)\right|  = \left|S_2\cap\mathcal{Q}(\tau)\right|,
\end{equation}
and we can use this fact to prove \eqref{eq:R_d_transversal_sufficiency_condition}:
\begin{equation}
    \left|S_1\bigcap_{i=1}^m \mathcal{Q}(X_{G_i})\right| -  \left|S_2\bigcap_{i=1}^m \mathcal{Q}(X_{G_i})\right| = \left|S_1\cap\mathcal{Q}(\tau)\right| - \left|S_2\cap\mathcal{Q}(\tau)\right| = 0.
\end{equation}
In summary, to transversally implement $\bar{R}_\ell$ one possibility is to ensure that the union of up to $\ell$ gauge operators is contained in a simplex $\tau$ of dimension smaller than $d$. Let us see what this entails with respect to the coloring. We know that $d\geq \col (\tau) \geq \col (\cup_i \delta_i)\geq m(d-1-z)$, because each $c$-simplex is colored with $c+1$ colors. Taking $m = \ell$, we find that
\begin{equation}
    \ell \leq \frac{d}{d-1-z}
\end{equation}
Since $x+z\leq d-2$, $\ell$ is maximized for $z=d-2$, which makes self-duality impossible but ensures that $\ell = \frac{d}{d-1-z} = d$. Therefore, we have seen that self-dual gauge color codes allow transversal implementation of $\bar{H}$, and $CC_\mathcal{L}(0,d-2)$ that of $\bar{R}_d$, but both conditions are incompatible for $d>2$, as would be expected from Eastin-Knill theorem~\cite{eastin2009restrictions}. We can also arithmetically check that we fail to simultaneously obtain self-duality and transversal implementation of $\bar{R}_3$ in any dimension, as we would need
\begin{equation}
    x=z\geq\frac{2d-3}{3} \Rightarrow x+z \geq \frac{4}{3}d-2>d-2 \Rightarrow\!\Leftarrow x+z < d-2.
\end{equation}
There is a way out, however, called \textit{gauge fixing}, which helps connect the self-dual codes $CC_\mathcal{L}(z,z)$, where we can transversally apply $\bar{H}$, and $CC_\mathcal{L}(0,d-2)$, where we can transversally apply $\bar{R}_d$.

To understand how it works, consider the three-dimensional case $CC_\mathcal{L}(0,0)$ and $CC_\mathcal{L}(0,1)$. In the former, stabilizer operators are applied over cells in $\mathcal{L}^*$, while in the latter, $Z_S\in\mathcal{S}$ stabilizers are applied over faces. This means imposing additional constraints over the $Z$ measurement of such faces, over the subsystems code $CC_\mathcal{L}(0,0)$. Thus, the gauge fixing procedure checks such face stabilizers and corrects them, so they are in the $+1$ subspace of the $Z$ face stabilizers of $CC_\mathcal{L}(0,1)$. 
In other words, if the gauge qubits are in the state $\ket{g_z}$, we can implement $\bar{R}_3$ transversally. However, when we apply $\bar{H} = H^{\otimes n}$ we transform:
\begin{equation}
    \ket{\psi}\ket{g_z}\mapsto (\bar{H}\ket{\psi})\ket{g_x}.
\end{equation}
Gauge fixing immediately after (measuring the face stabilizers and correcting them) ensures we can recover the $\ket{g_z}$ state necessary to implement any further $\bar{R}_3$ gates. The entire procedure can be applied transversally:
\begin{equation}
    \ket{\psi}\ket{g_z}\xmapsto{H^{\otimes n}} (\bar{H}\ket{\psi})\ket{g_x}\xmapsto{\text{gauge fix}} (\bar{H}\ket{\psi})\ket{g_z}.
\end{equation}

It is possible to use gauge fixing between $CC_\mathcal{L}(x,z)$ and $CC_\mathcal{L}(x',z')$ as long as they encode the same number of logical qubits, and $x\geq x'$ and $z\geq z'$. In that case $\mathcal{S}'\subseteq \mathcal{S}$ and $\mathcal{G}\subseteq \mathcal{G}'$~\cite{kubica2015universal}. Gauge color codes are also interesting because they allow for so-called `single-shot error correction': instead of measuring high-weight stabilizer operators in 3 dimensions, we can measure the gauge operators, from which the stabilizer operators' eigenvalues can be deduced. Furthermore, since each cell contains faces of different colors, we automatically recover several parity checks that allow us to identify measurement errors~\cite{bombin2015single}. Fascinatingly, one may also create gauge subsystem codes and achieve single-shot error corrections from the toric code~\cite{kubica2021single}. Finally, (gauge) color codes allow switching back and forth between 2 and 3 dimensions, instead of between two three-dimensional color codes~\cite{bombin2016dimensional}. This procedure is called \textit{code-switching} and similarly allows for a universal and transversal set of gates. Though beautiful, in the high physical error regime magic state distillation has still been found to offer an advantage over this procedure~\cite{beverland2021cost}.

\section{Concluding thoughts}

In this chapter, we have reviewed the basics of quantum error correction, with special attention to the most popular topological stabilizer codes: the surface and color codes. Not only that, but we have built our way up and explained how to achieve fault tolerance via gauge fixing in the latter. However, there are many topics we have not discussed. For example, there are many other codes~\cite{dauphinais2019quantum}, some with the same geometry as some superconducting quantum chips~\cite{gidney2021fault}, others specifically designed for photons~\cite{gottesman2001encoding,noh2022low,tzitrin2020progress} which can be embedded in full fault-tolerant quantum computing architectures~\cite{bartolucci2021fusion,bourassa2021blueprint}. Photonic architectures in particular may additionally benefit from the use of hyperbolic surface~\cite{breuckmann2016constructions,breuckmann2017hyperbolic,lavasani2018low} and color codes~\cite{ soares2018hyperbolic}, which display a rate $\lim_{n\rightarrow \infty }k/n \geq c_1$ at the expense of a slower code distance growth $d\geq c_2 \log n$ and non-local interactions~\cite{breuckmann2016constructions}, for $c_1, c_2$ constants. This follows a general tradeoff bound $kd^2 = k\leq c(\log k)^2n $~\cite{delfosse2013tradeoffs} with $c$ a constant. Alternatively, there have also been quantum computing proposals that make use of the topological properties of Majorana fermions~\cite{litinski2018quantum}. 

However, what is most exciting is that just a few years after Google's quantum supremacy experiment~\cite{arute2019quantum,zhong2020quantum,madsen2022quantum} we are now approaching the physical error required to start implementing quantum error correction. There have been several experiments approaching the error-correction threshold~\cite{krinner2022realizing,ryan2021realization,postler2022demonstration,zhao2021realizing,acharya2022suppressing}, which open the door to the implementation of many fault-tolerant quantum algorithms discussed in this thesis. In conclusion, we can only say that we are living in an exciting time for quantum computing.

\section{Results}

\begin{itemize}
    \item We have explained the basics of classical and quantum error correction, with a focus on Stabilizer and CSS codes.

    \item We have presented the family of Topological error-correcting codes, and their two well-known members: the surface and color codes. We have indicated how both encode logical qubits and how to implement logic gates.

    \item We have also described how Clifford gates can be efficiently implemented in a color code, but also that non-Clifford gates require either magic state distillation or gauge fixing techniques. Further, we have explained that the theorem of Easting and Knill forbids the existence of a topological code with a set of universal and transversal logic gates, even if do not restrict ourselves to three dimensions~\cite{eastin2009restrictions}.

    \item We have described the gauge color codes as a way to achieve fault tolerance, via the procedure of gauge fixing. This method allows the implementation of non-Clifford gates in the 3d color code, while the rest of the operations are implemented in its two-dimensional version.

    \item We are working on a Machine Learning decoder, following work previously carried out in our group.
\end{itemize}


\begin{spacing}{0.9}


\bibliographystyle{apalike}
\cleardoublepage
\bibliography{References/references} 

\begin{thebibliography}{}

\bibitem[Aaronson and Ambainis, 2009]{aaronson2009need}
Aaronson, S. and Ambainis, A. (2009).
\newblock The need for structure in quantum speedups.
\newblock {\em arXiv preprint arXiv:0911.0996}.

\bibitem[Acharya et~al., 2022]{acharya2022suppressing}
Acharya, R., Aleiner, I., Allen, R., Andersen, T.~I., Ansmann, M., Arute, F.,
  Arya, K., Asfaw, A., Atalaya, J., Babbush, R., et~al. (2022).
\newblock Suppressing quantum errors by scaling a surface code logical qubit.
\newblock {\em arXiv preprint arXiv:2207.06431}.

\bibitem[Alcaide, 2019]{ericalcaide2019minifold}
Alcaide, E. (2019).
\newblock Minifold: a deeplearning-based mini protein folding engine.
\newblock \url{https://github.com/EricAlcaide/MiniFold/}.

\bibitem[Alicki et~al., 2010]{alicki2010thermal}
Alicki, R., Horodecki, M., Horodecki, P., and Horodecki, R. (2010).
\newblock On thermal stability of topological qubit in kitaev's 4d model.
\newblock {\em Open Systems \& Information Dynamics}, 17(01):1--20.

\bibitem[Ambainis, 2004]{ambainis2004quantum}
Ambainis, A. (2004).
\newblock Quantum walk algorithm for element distinctness.
\newblock In {\em Proceedings of the 45th Annual IEEE Symposium on Foundations
  of Computer Science}, FOCS '04, page 22–31, USA. IEEE Computer Society.

\bibitem[Ambainis, 2010]{ambainis2010variable}
Ambainis, A. (2010).
\newblock Variable time amplitude amplification and a faster quantum algorithm
  for solving systems of linear equations.
\newblock {\em arXiv preprint arXiv:1010.4458}.

\bibitem[Ambainis et~al., 2020]{ambainis2020quadratic}
Ambainis, A., Gily{\'e}n, A., Jeffery, S., and Kokainis, M. (2020).
\newblock Quadratic speedup for finding marked vertices by quantum walks.
\newblock In {\em Proceedings of the 52nd Annual ACM SIGACT Symposium on Theory
  of Computing}, pages 412--424.

\bibitem[Anand et~al., 2022]{anand2022quantum}
Anand, A., Schleich, P., Alperin-Lea, S., Jensen, P.~W., Sim, S.,
  D{\'\i}az-Tinoco, M., Kottmann, J.~S., Degroote, M., Izmaylov, A.~F., and
  Aspuru-Guzik, A. (2022).
\newblock A quantum computing view on unitary coupled cluster theory.
\newblock {\em Chemical Society Reviews}.

\bibitem[Anfinsen, 1973]{anfinsen1973principles}
Anfinsen, C.~B. (1973).
\newblock Principles that govern the folding of protein chains.
\newblock {\em Science}, 181(4096):223--230.

\bibitem[Anfinsen et~al., 1961]{anfinsen1961kinetics}
Anfinsen, C.~B., Haber, E., Sela, M., and White~Jr, F. (1961).
\newblock The kinetics of formation of native ribonuclease during oxidation of
  the reduced polypeptide chain.
\newblock {\em Proceedings of the National Academy of Sciences of the United
  States of America}, 47(9):1309.

\bibitem[Apers and Sarlette, 2019]{apers2019quantum}
Apers, S. and Sarlette, A. (2019).
\newblock Quantum fast-forwarding: Markov chains and graph property testing.
\newblock {\em Quantum Information \& Computation}, 19(3–4):181–213.

\bibitem[Arrazola et~al., 2022]{arrazola2022universal}
Arrazola, J.~M., Di~Matteo, O., Quesada, N., Jahangiri, S., Delgado, A., and
  Killoran, N. (2022).
\newblock Universal quantum circuits for quantum chemistry.
\newblock {\em Quantum}, 6:742.

\bibitem[Arute et~al., 2019]{arute2019quantum}
Arute, F., Arya, K., Babbush, R., Bacon, D., Bardin, J.~C., Barends, R.,
  Biswas, R., Boixo, S., Brandao, F.~G., Buell, D.~A., et~al. (2019).
\newblock Quantum supremacy using a programmable superconducting processor.
\newblock {\em Nature}, 574(7779):505--510.

\bibitem[Augustino et~al., 2021]{augustino2021infeasible}
Augustino, B., Nannicini, G., Terlaky, T., and Zuluaga, L.~F. (2021).
\newblock An infeasible-inexact quantum interior point method for convex
  quadratic symmetric cone optimization.

\bibitem[Babai, 2016]{babai2016graph}
Babai, L. (2016).
\newblock Graph isomorphism in quasipolynomial time.
\newblock In {\em Proceedings of the forty-eighth annual ACM Symposium on
  Theory of Computing}, pages 684--697.

\bibitem[Babbush et~al., 2019]{babbush2019quantum}
Babbush, R., Berry, D.~W., McClean, J.~R., and Neven, H. (2019).
\newblock Quantum simulation of chemistry with sublinear scaling in basis size.
\newblock {\em npj Quantum Information}, 5(1):1--7.

\bibitem[Babbush et~al., 2017]{babbush2017exponentially}
Babbush, R., Berry, D.~W., Sanders, Y.~R., Kivlichan, I.~D., Scherer, A., Wei,
  A.~Y., Love, P.~J., and Aspuru-Guzik, A. (2017).
\newblock Exponentially more precise quantum simulation of fermions in the
  configuration interaction representation.
\newblock {\em Quantum Science and Technology}, 3(1):015006.

\bibitem[Babbush et~al., 2018a]{babbush2018encoding}
Babbush, R., Gidney, C., Berry, D.~W., Wiebe, N., McClean, J.~R., Paler, A.,
  Fowler, A., and Neven, H. (2018a).
\newblock Encoding electronic spectra in quantum circuits with linear t
  complexity.
\newblock {\em Physical Review X}, 8(4):041015.

\bibitem[Babbush et~al., 2023]{babbush2023quantum}
Babbush, R., Huggins, W.~J., Berry, D.~W., Ung, S.~F., Zhao, A., Reichman,
  D.~R., Neven, H., Baczewski, A.~D., and Lee, J. (2023).
\newblock Quantum simulation of exact electron dynamics can be more efficient
  than classical mean-field methods.
\newblock {\em arXiv preprint arXiv:2301.01203}.

\bibitem[Babbush et~al., 2021]{babbush2021focus}
Babbush, R., McClean, J.~R., Newman, M., Gidney, C., Boixo, S., and Neven, H.
  (2021).
\newblock Focus beyond quadratic speedups for error-corrected quantum
  advantage.
\newblock {\em PRX Quantum}, 2(1):010103.

\bibitem[Babbush et~al., 2012]{babbush2012construction}
Babbush, R., Perdomo-Ortiz, A., O'Gorman, B., Macready, W., and Aspuru-Guzik,
  A. (2012).
\newblock Construction of energy functions for lattice heteropolymer models: a
  case study in constraint satisfaction programming and adiabatic quantum
  optimization.
\newblock {\em arXiv preprint arXiv:1211.3422}.

\bibitem[Babbush et~al., 2018b]{babbush2018low}
Babbush, R., Wiebe, N., McClean, J.~R., McClain, J., Neven, H., and Chan, G.
  K.-L. (2018b).
\newblock Low-depth quantum simulation of materials.
\newblock {\em Physical Review X}, 8(1):011044.

\bibitem[Babej et~al., 2018]{babej2018coarse}
Babej, T., Fingerhuth, M., et~al. (2018).
\newblock Coarse-grained lattice protein folding on a quantum annealer.
\newblock {\em arXiv preprint arXiv:1811.00713}.

\bibitem[Banchi et~al., 2020]{banchi2020molecular}
Banchi, L., Fingerhuth, M., Babej, T., Ing, C., and Arrazola, J.~M. (2020).
\newblock Molecular docking with gaussian boson sampling.
\newblock {\em Science Advances}, 6(23):eaax1950.

\bibitem[Bank and Scott, 1989]{bank1989conditioning}
Bank, R.~E. and Scott, L.~R. (1989).
\newblock On the conditioning of finite element equations with highly refined
  meshes.
\newblock {\em SIAM Journal on Numerical Analysis}, 26(6):1383--1394.

\bibitem[Bartolucci et~al., 2021]{bartolucci2021fusion}
Bartolucci, S., Birchall, P., Bombin, H., Cable, H., Dawson, C.,
  Gimeno-Segovia, M., Johnston, E., Kieling, K., Nickerson, N.~H., Pant, M.,
  et~al. (2021).
\newblock Fusion-based quantum computation.
\newblock {\em arXiv preprint arXiv:2101.09310}.

\bibitem[Bateman, 1953]{bateman1953higher}
Bateman, H. (1953).
\newblock {\em Higher transcendental functions, volume II}, volume~1.
\newblock McGraw-Hill Book Company.

\bibitem[Becke, 1988]{becke1988density}
Becke, A.~D. (1988).
\newblock Density-functional exchange-energy approximation with correct
  asymptotic behavior.
\newblock {\em Physical Review A}, 38(6):3098.

\bibitem[Becke, 1993a]{becke1993density}
Becke, A.~D. (1993a).
\newblock Density‐functional thermochemistry. iii. the role of exact
  exchange.
\newblock {\em The Journal of Chemical Physics}, 98(7):5648--5652.

\bibitem[Becke, 1993b]{becke1993new}
Becke, A.~D. (1993b).
\newblock A new mixing of hartree--fock and local density-functional theories.
\newblock {\em The Journal of Chemical Physics}, 98(2):1372--1377.

\bibitem[Bennett et~al., 1997]{bennett1997strengths}
Bennett, C.~H., Bernstein, E., Brassard, G., and Vazirani, U. (1997).
\newblock Strengths and weaknesses of quantum computing.
\newblock {\em SIAM journal on Computing}, 26(5):1510--1523.

\bibitem[Berger and Leighton, 1998]{berger1998protein}
Berger, B. and Leighton, T. (1998).
\newblock Protein folding in the hydrophobic-hydrophilic (hp) is np-complete.
\newblock In {\em Proceedings of the second annual International Conference on
  Computational Molecular Biology}, pages 30--39.

\bibitem[Bergholm et~al., 2018]{bergholm2018pennylane}
Bergholm, V., Izaac, J., Schuld, M., Gogolin, C., Alam, M.~S., Ahmed, S.,
  Arrazola, J.~M., Blank, C., Delgado, A., Jahangiri, S., et~al. (2018).
\newblock Pennylane: Automatic differentiation of hybrid quantum-classical
  computations.
\newblock {\em arXiv preprint arXiv:1811.04968}.

\bibitem[Berry et~al., 2007]{berry2007efficient}
Berry, D.~W., Ahokas, G., Cleve, R., and Sanders, B.~C. (2007).
\newblock Efficient quantum algorithms for simulating sparse hamiltonians.
\newblock {\em Communications in Mathematical Physics}, 270(2):359--371.

\bibitem[Berry et~al., 2015]{berry2015simulating}
Berry, D.~W., Childs, A.~M., Cleve, R., Kothari, R., and Somma, R.~D. (2015).
\newblock Simulating hamiltonian dynamics with a truncated taylor series.
\newblock {\em Physical Review Letters}, 114(9):090502.

\bibitem[Berry et~al., 2019]{berry2019qubitization}
Berry, D.~W., Gidney, C., Motta, M., McClean, J.~R., and Babbush, R. (2019).
\newblock Qubitization of arbitrary basis quantum chemistry leveraging sparsity
  and low rank factorization.
\newblock {\em Quantum}, 3:208.

\bibitem[Berry et~al., 2018]{berry2018improved}
Berry, D.~W., Kieferov{\'a}, M., Scherer, A., Sanders, Y.~R., Low, G.~H.,
  Wiebe, N., Gidney, C., and Babbush, R. (2018).
\newblock Improved techniques for preparing eigenstates of fermionic
  hamiltonians.
\newblock {\em npj Quantum Information}, 4(1):1--7.

\bibitem[Bespalova and Kyriienko, 2021]{bespalova2021hamiltonian}
Bespalova, T.~A. and Kyriienko, O. (2021).
\newblock Hamiltonian operator approximation for energy measurement and
  ground-state preparation.
\newblock {\em PRX Quantum}, 2(3):030318.

\bibitem[Beverland et~al., 2021]{beverland2021cost}
Beverland, M.~E., Kubica, A., and Svore, K.~M. (2021).
\newblock Cost of universality: A comparative study of the overhead of state
  distillation and code switching with color codes.
\newblock {\em PRX Quantum}, 2(2):020341.

\bibitem[Bharti et~al., 2022]{bharti2021noisy}
Bharti, K., Cervera-Lierta, A., Kyaw, T.~H., Haug, T., Alperin-Lea, S., Anand,
  A., Degroote, M., Heimonen, H., Kottmann, J.~S., Menke, T., et~al. (2022).
\newblock Noisy intermediate-scale quantum algorithms.
\newblock {\em Reviews of Modern Physics}, 94(1):015004.

\bibitem[Bloch, 1929]{bloch1929quantenmechanik}
Bloch, F. (1929).
\newblock {\"U}ber die quantenmechanik der elektronen in kristallgittern.
\newblock {\em Zeitschrift f{\"u}r physik}, 52(7):555--600.

\bibitem[Bomb{\'\i}n, 2013]{bombin2013introduction}
Bomb{\'\i}n, H. (2013).
\newblock An introduction to topological quantum codes.
\newblock {\em arXiv preprint arXiv:1311.0277}.

\bibitem[Bomb{\'\i}n, 2015a]{bombin2015gauge}
Bomb{\'\i}n, H. (2015a).
\newblock Gauge color codes: optimal transversal gates and gauge fixing in
  topological stabilizer codes.
\newblock {\em New Journal of Physics}, 17(8):083002.

\bibitem[Bomb{\'\i}n, 2015b]{bombin2015single}
Bomb{\'\i}n, H. (2015b).
\newblock Single-shot fault-tolerant quantum error correction.
\newblock {\em Physical Review X}, 5(3):031043.

\bibitem[Bomb{\'\i}n, 2016]{bombin2016dimensional}
Bomb{\'\i}n, H. (2016).
\newblock Dimensional jump in quantum error correction.
\newblock {\em New Journal of Physics}, 18(4):043038.

\bibitem[Bomb{\'\i}n et~al., 2013]{bombin2013self}
Bomb{\'\i}n, H., Chhajlany, R.~W., Horodecki, M., and Martin-Delgado, M.-A.
  (2013).
\newblock Self-correcting quantum computers.
\newblock {\em New Journal of Physics}, 15(5):055023.

\bibitem[Bomb{\'\i}n and Martin-Delgado, 2006]{bombin2006topological}
Bomb{\'\i}n, H. and Martin-Delgado, M.~A. (2006).
\newblock Topological quantum distillation.
\newblock {\em Physical Review Letters}, 97(18):180501.

\bibitem[Bomb{\'\i}n and Martin-Delgado, 2007]{bombin2007topological}
Bomb{\'\i}n, H. and Martin-Delgado, M.~A. (2007).
\newblock Topological computation without braiding.
\newblock {\em Physical Review Letters}, 98(16):160502.

\bibitem[Born and Oppenheimer, 1927]{born1927quantentheorie}
Born, M. and Oppenheimer, R. (1927).
\newblock Zur quantentheorie der molekeln.
\newblock {\em Annalen der Physik}, 389(20):457--484.

\bibitem[Bourassa et~al., 2021]{bourassa2021blueprint}
Bourassa, J.~E., Alexander, R.~N., Vasmer, M., Patil, A., Tzitrin, I.,
  Matsuura, T., Su, D., Baragiola, B.~Q., Guha, S., Dauphinais, G., et~al.
  (2021).
\newblock Blueprint for a scalable photonic fault-tolerant quantum computer.
\newblock {\em Quantum}, 5:392.

\bibitem[Boyer et~al., 1998]{boyer1998tight}
Boyer, M., Brassard, G., H{\o}yer, P., and Tapp, A. (1998).
\newblock Tight bounds on quantum searching.
\newblock {\em Fortschritte der Physik: Progress of Physics}, 46(4-5):493--505.

\bibitem[Brassard et~al., 2002]{brassard2002quantum}
Brassard, G., Hoyer, P., Mosca, M., and Tapp, A. (2002).
\newblock Quantum amplitude amplification and estimation.
\newblock {\em Contemporary Mathematics}, 305:53--74.

\bibitem[Bravyi and Kitaev, 2002]{bravyi2002fermionic}
Bravyi, S. and Kitaev, A.~Y. (2002).
\newblock Fermionic quantum computation.
\newblock {\em Annals of Physics}, 298(1):210--226.

\bibitem[Bravyi and Kitaev, 2005]{bravyi2005universal}
Bravyi, S. and Kitaev, A.~Y. (2005).
\newblock Universal quantum computation with ideal clifford gates and noisy
  ancillas.
\newblock {\em Physical Review A}, 71(2):022316.

\bibitem[Bravyi and K{\"o}nig, 2013]{bravyi2013classification}
Bravyi, S. and K{\"o}nig, R. (2013).
\newblock Classification of topologically protected gates for local stabilizer
  codes.
\newblock {\em Physical Review Letters}, 110(17):170503.

\bibitem[Brenner et~al., 2008]{brenner2008mathematical}
Brenner, S.~C., Scott, L.~R., and Scott, L.~R. (2008).
\newblock {\em The mathematical theory of finite element methods}, volume~3.
\newblock Springer.

\bibitem[Breuckmann and Terhal, 2016]{breuckmann2016constructions}
Breuckmann, N.~P. and Terhal, B.~M. (2016).
\newblock Constructions and noise threshold of hyperbolic surface codes.
\newblock {\em IEEE transactions on Information Theory}, 62(6):3731--3744.

\bibitem[Breuckmann et~al., 2017]{breuckmann2017hyperbolic}
Breuckmann, N.~P., Vuillot, C., Campbell, E., Krishna, A., and Terhal, B.~M.
  (2017).
\newblock Hyperbolic and semi-hyperbolic surface codes for quantum storage.
\newblock {\em Quantum Science and Technology}, 2(3):035007.

\bibitem[Brown et~al., 2016]{brown2016fault}
Brown, B.~J., Nickerson, N.~H., and Browne, D.~E. (2016).
\newblock Fault-tolerant error correction with the gauge color code.
\newblock {\em Nature Communications}, 7(1):1--8.

\bibitem[Buhrman et~al., 2001]{buhrman2001quantum}
Buhrman, H., Cleve, R., Watrous, J., and De~Wolf, R. (2001).
\newblock Quantum fingerprinting.
\newblock {\em Physical Review Letters}, 87(16):167902.

\bibitem[Campbell, 2019]{campbell2019random}
Campbell, E.~T. (2019).
\newblock Random compiler for fast hamiltonian simulation.
\newblock {\em Physical Review Letters}, 123(7):070503.

\bibitem[Campbell, 2021]{campbell2021early}
Campbell, E.~T. (2021).
\newblock Early fault-tolerant simulations of the hubbard model.
\newblock {\em Quantum Science and Technology}, 7(1):015007.

\bibitem[Campos et~al., 2022]{campos2022quantum}
Campos, R., Casares, P. A.~M., and Martin-Delgado, M.~A. (2022).
\newblock Quantum metropolis solver: A quantum walks approach to optimization
  problems.
\newblock {\em arXiv preprint arXiv:2207.06462}.

\bibitem[Cao et~al., 2019]{cao2019quantum}
Cao, Y., Romero, J., Olson, J.~P., Degroote, M., Johnson, P.~D., Kieferov{\'a},
  M., Kivlichan, I.~D., Menke, T., Peropadre, B., Sawaya, N.~P., et~al. (2019).
\newblock Quantum chemistry in the age of quantum computing.
\newblock {\em Chemical Reviews}, 119(19):10856--10915.

\bibitem[Casares et~al., 2022a]{casares2022qfold}
Casares, P. A.~M., Campos, R., and Martin-Delgado, M.~A. (2022a).
\newblock Qfold: quantum walks and deep learning to solve protein folding.
\newblock {\em Quantum Science and Technology}.

\bibitem[Casares et~al., 2022b]{casares2021tfermion}
Casares, P. A.~M., Campos, R., and Martin-Delgado, M.~A. (2022b).
\newblock {TF}ermion: {A} non-{C}lifford gate cost assessment library of
  quantum phase estimation algorithms for quantum chemistry.
\newblock {\em {Quantum}}, 6:768.

\bibitem[Casares and Martin-Delgado, 2020a]{casares2020active}
Casares, P. A.~M. and Martin-Delgado, M.~A. (2020a).
\newblock A quantum active learning algorithm for sampling against adversarial
  attacks.
\newblock {\em New Journal of Physics}, 22(7):073026.

\bibitem[Casares and Martin-Delgado, 2020b]{casares2020IP}
Casares, P. A.~M. and Martin-Delgado, M.~A. (2020b).
\newblock A quantum interior-point predictor--corrector algorithm for linear
  programming.
\newblock {\em Journal of Physics A: Mathematical and Theoretical},
  53(44):445305.

\bibitem[Casares et~al., 2022c]{casares2022how}
Casares, P. A.~M., Sheng~Loe, B., Burden, J., hEigeartaigh, S., and
  Hern{\'a}ndez-Orallo, J. (2022c).
\newblock How general-purpose is a language model? usefulness and safety with
  human prompters in the wild.
\newblock {\em Proceedings of the 36th AAAI Conference on Artificial
  Intelligence}, 36(5):5295--5303.

\bibitem[Cerezo et~al., 2021]{cerezo2021variational}
Cerezo, M., Arrasmith, A., Babbush, R., Benjamin, S.~C., Endo, S., Fujii, K.,
  McClean, J.~R., Mitarai, K., Yuan, X., Cincio, L., et~al. (2021).
\newblock Variational quantum algorithms.
\newblock {\em Nature Reviews Physics}, 3(9):625--644.

\bibitem[Chakraborty et~al., 2020a]{chakraborty2020analog}
Chakraborty, S., Luh, K., and Roland, J. (2020a).
\newblock Analog quantum algorithms for the mixing of markov chains.
\newblock {\em Physical Review A}, 102(2):022423.

\bibitem[Chakraborty et~al., 2020b]{chakraborty2020fast}
Chakraborty, S., Luh, K., and Roland, J. (2020b).
\newblock How fast do quantum walks mix?
\newblock {\em Physical Review Letters}, 124(5):050501.

\bibitem[Chao et~al., 2020]{chao2020finding}
Chao, R., Ding, D., Gilyen, A., Huang, C., and Szegedy, M. (2020).
\newblock Finding angles for quantum signal processing with machine precision.
\newblock {\em arXiv preprint arXiv:2003.02831}.

\bibitem[Childs et~al., 2017]{childs2017quantum}
Childs, A.~M., Kothari, R., and Somma, R.~D. (2017).
\newblock Quantum algorithm for systems of linear equations with exponentially
  improved dependence on precision.
\newblock {\em SIAM Journal on Computing}, 46(6):1920--1950.

\bibitem[Childs et~al., 2019]{childs2019faster}
Childs, A.~M., Ostrander, A., and Su, Y. (2019).
\newblock Faster quantum simulation by randomization.
\newblock {\em Quantum}, 3:182.

\bibitem[Chow, 2000]{chow2000priori}
Chow, E. (2000).
\newblock A priori sparsity patterns for parallel sparse approximate inverse
  preconditioners.
\newblock {\em SIAM Journal on Scientific Computing}, 21(5):1804--1822.

\bibitem[Clader et~al., 2013]{clader2013preconditioned}
Clader, B.~D., Jacobs, B.~C., and Sprouse, C.~R. (2013).
\newblock Preconditioned quantum linear system algorithm.
\newblock {\em Physical Review Letters}, 110(25):250504.

\bibitem[Cleve et~al., 1998]{cleve1998quantum}
Cleve, R., Ekert, A., Macchiavello, C., and Mosca, M. (1998).
\newblock Quantum algorithms revisited.
\newblock {\em Proceedings of the Royal Society of London. Series A:
  Mathematical, Physical and Engineering Sciences}, 454(1969):339--354.

\bibitem[Clinton et~al., 2022]{clinton2022towards}
Clinton, L., Cubitt, T., Flynn, B., Gambetta, F.~M., Klassen, J., Montanaro,
  A., Piddock, S., Santos, R.~A., and Sheridan, E. (2022).
\newblock Towards near-term quantum simulation of materials.
\newblock {\em arXiv preprint arXiv:2205.15256}.

\bibitem[Cooley and Tukey, 1965]{cooley1965algorithm}
Cooley, J.~W. and Tukey, J.~W. (1965).
\newblock An algorithm for the machine calculation of complex fourier series.
\newblock {\em Mathematics of Computation}, 19(90):297--301.

\bibitem[Coudron and Menda, 2020]{coudron2020computations}
Coudron, M. and Menda, S. (2020).
\newblock Computations with greater quantum depth are strictly more powerful
  (relative to an oracle).
\newblock In {\em Proceedings of the 52nd Annual ACM SIGACT Symposium on Theory
  of Computing}, pages 889--901.

\bibitem[Dantzig, 1951]{dantzig1951application}
Dantzig, G.~B. (1951).
\newblock Application of the simplex method to a transportation problem.
\newblock {\em Activity Analysis of Production and Allocation}.

\bibitem[Das and Baker, 2008]{Rosetta}
Das, R. and Baker, D. (2008).
\newblock Macromolecular modeling with rosetta.
\newblock {\em Annual Review of Biochemistry}, 77:363--382.

\bibitem[Das et~al., 2007]{das2007rosetta@home}
Das, R., Qian, B., Raman, S., Vernon, R., Thompson, J., Bradley, P., Khare, S.,
  Tyka, M.~D., Bhat, D., Chivian, D., et~al. (2007).
\newblock Structure prediction for casp7 targets using extensive all-atom
  refinement with rosetta@ home.
\newblock {\em Proteins: Structure, Function, and Bioinformatics},
  69(S8):118--128.

\bibitem[Dauphinais et~al., 2019]{dauphinais2019quantum}
Dauphinais, G., Ortiz, L., Varona, S., and Martin-Delgado, M.~A. (2019).
\newblock Quantum error correction with the semion code.
\newblock {\em New Journal of Physics}, 21(5):053035.

\bibitem[De~Wolf, 2019]{de2019quantum}
De~Wolf, R. (2019).
\newblock Quantum computing: Lecture notes.
\newblock {\em arXiv preprint arXiv:1907.09415}.

\bibitem[Delfosse, 2013]{delfosse2013tradeoffs}
Delfosse, N. (2013).
\newblock Tradeoffs for reliable quantum information storage in surface codes
  and color codes.
\newblock In {\em 2013 IEEE International Symposium on Information Theory},
  pages 917--921. IEEE.

\bibitem[Delgado et~al., 2022]{delgado2022simulate}
Delgado, A., Casares, P. A.~M., dos Reis, R., Zini, M.~S., Campos, R.,
  Cruz-Hern\'andez, N., Voigt, A.-C., Lowe, A., Jahangiri, S., Martin-Delgado,
  M.~A., Mueller, J.~E., and Arrazola, J.~M. (2022).
\newblock Simulating key properties of lithium-ion batteries with a
  fault-tolerant quantum computer.
\newblock {\em Phys. Rev. A}, 106:032428.

\bibitem[Dennis et~al., 2002]{dennis2002topological}
Dennis, E., Kitaev, A.~Y., Landahl, A., and Preskill, J. (2002).
\newblock Topological quantum memory.
\newblock {\em Journal of Mathematical Physics}, 43(9):4452--4505.

\bibitem[Derby and Klassen, 2021]{derby2021acompact}
Derby, C. and Klassen, J. (2021).
\newblock A compact fermion to qubit mapping part 2: Alternative lattice
  geometries.
\newblock {\em arXiv preprint arXiv:2101.10735}.

\bibitem[Derby et~al., 2021]{derby2021compact}
Derby, C., Klassen, J., Bausch, J., and Cubitt, T. (2021).
\newblock Compact fermion to qubit mappings.
\newblock {\em Physical Review B}, 104(3):035118.

\bibitem[Dirac, 1930]{dirac1930note}
Dirac, P. A.~M. (1930).
\newblock Note on exchange phenomena in the thomas atom.
\newblock {\em Mathematical Proceedings of the Cambridge Philosophical
  Society}, 26(3):376–385.

\bibitem[Eastin and Knill, 2009]{eastin2009restrictions}
Eastin, B. and Knill, E. (2009).
\newblock Restrictions on transversal encoded quantum gate sets.
\newblock {\em Physical Review Letters}, 102(11):110502.

\bibitem[Efron, 1992]{efron1992bootstrap}
Efron, B. (1992).
\newblock Bootstrap methods: another look at the jackknife.
\newblock In {\em Breakthroughs in Statistics}, pages 569--593. Springer.

\bibitem[Ettinger et~al., 2004]{ettinger2004quantum}
Ettinger, M., H{\o}yer, P., and Knill, E. (2004).
\newblock The quantum query complexity of the hidden subgroup problem is
  polynomial.
\newblock {\em Information Processing Letters}, 91(1):43--48.

\bibitem[Farhi et~al., 2014]{farhi2014quantum}
Farhi, E., Goldstone, J., and Gutmann, S. (2014).
\newblock A quantum approximate optimization algorithm.
\newblock {\em arXiv preprint arXiv:1411.4028}.

\bibitem[Ferris, 2014]{ferris2014fourier}
Ferris, A.~J. (2014).
\newblock Fourier transform for fermionic systems and the spectral tensor
  network.
\newblock {\em Physical Review Letters}, 113(1):010401.

\bibitem[Fingerhuth et~al., 2018]{fingerhuth2018quantum}
Fingerhuth, M., Babej, T., et~al. (2018).
\newblock A quantum alternating operator ansatz with hard and soft constraints
  for lattice protein folding.
\newblock {\em arXiv preprint arXiv:1810.13411}.

\bibitem[Fowler et~al., 2012]{fowler2012surface}
Fowler, A.~G., Mariantoni, M., Martinis, J.~M., and Cleland, A.~N. (2012).
\newblock Surface codes: Towards practical large-scale quantum computation.
\newblock {\em Physical Review A}, 86(3):032324.

\bibitem[Ge et~al., 2019]{ge2019faster}
Ge, Y., Tura, J., and Cirac, J.~I. (2019).
\newblock Faster ground state preparation and high-precision ground energy
  estimation with fewer qubits.
\newblock {\em Journal of Mathematical Physics}, 60(2):022202.

\bibitem[Geman and Geman, 1984]{geman1984stochastic}
Geman, S. and Geman, D. (1984).
\newblock Stochastic relaxation, gibbs distributions, and the bayesian
  restoration of images.
\newblock {\em IEEE Transactions on Pattern Analysis and Machine Intelligence},
  PAMI-6(6):721--741.

\bibitem[Gidney and Fowler, 2019]{gidney2019efficient}
Gidney, C. and Fowler, A.~G. (2019).
\newblock Efficient magic state factories with a catalyzed $\ket{CCZ}$ to 2
  $\ket{T}$ transformation.
\newblock {\em Quantum}, 3:135.

\bibitem[Gidney et~al., 2021]{gidney2021fault}
Gidney, C., Newman, M., Fowler, A., and Broughton, M. (2021).
\newblock A fault-tolerant honeycomb memory.
\newblock {\em Quantum}, 5:605.

\bibitem[Gingrich et~al., 2000]{gingrich2000generalized}
Gingrich, R.~M., Williams, C.~P., and Cerf, N.~J. (2000).
\newblock Generalized quantum search with parallelism.
\newblock {\em Physical Review A}, 61(5):052313.

\bibitem[Giovannetti et~al., 2008]{giovannetti2008quantum}
Giovannetti, V., Lloyd, S., and Maccone, L. (2008).
\newblock Quantum random access memory.
\newblock {\em Physical Review Letters}, 100(16):160501.

\bibitem[Goings et~al., 2022]{goings2022reliably}
Goings, J.~J., White, A., Lee, J., Tautermann, C.~S., Degroote, M., Gidney, C.,
  Shiozaki, T., Babbush, R., and Rubin, N.~C. (2022).
\newblock Reliably assessing the electronic structure of cytochrome p450 on
  today's classical computers and tomorrow's quantum computers.
\newblock {\em Proceedings of the National Academy of Sciences},
  119(38):e2203533119.

\bibitem[Goodfellow et~al., 2014]{goodfellow2014explaining}
Goodfellow, I.~J., Shlens, J., and Szegedy, C. (2014).
\newblock Explaining and harnessing adversarial examples.
\newblock {\em arXiv preprint arXiv:1412.6572}.

\bibitem[Gottesman and Chuang, 1999]{gottesman1999demonstrating}
Gottesman, D. and Chuang, I.~L. (1999).
\newblock Demonstrating the viability of universal quantum computation using
  teleportation and single-qubit operations.
\newblock {\em Nature}, 402(6760):390--393.

\bibitem[Gottesman et~al., 2001]{gottesman2001encoding}
Gottesman, D., Kitaev, A.~Y., and Preskill, J. (2001).
\newblock Encoding a qubit in an oscillator.
\newblock {\em Physical Review A}, 64(1):012310.

\bibitem[Greengard and Rokhlin, 1987]{greengard1987fast}
Greengard, L. and Rokhlin, V. (1987).
\newblock A fast algorithm for particle simulations.
\newblock {\em Journal of Computational Physics}, 73(2):325--348.

\bibitem[Griffiths and Schroeter, 2018]{griffiths2018introduction}
Griffiths, D.~J. and Schroeter, D.~F. (2018).
\newblock {\em Introduction to quantum mechanics}.
\newblock Cambridge university press.

\bibitem[Grimsley et~al., 2019]{grimsley2019adaptive}
Grimsley, H.~R., Economou, S.~E., Barnes, E., and Mayhall, N.~J. (2019).
\newblock An adaptive variational algorithm for exact molecular simulations on
  a quantum computer.
\newblock {\em Nature Communications}, 10(1):1--9.

\bibitem[Grote and Huckle, 1997]{grote1997parallel}
Grote, M.~J. and Huckle, T. (1997).
\newblock Parallel preconditioning with sparse approximate inverses.
\newblock {\em SIAM Journal on Scientific Computing}, 18(3):838--853.

\bibitem[Grover, 1998]{grover1998quantum}
Grover, L.~K. (1998).
\newblock Quantum computers can search rapidly by using almost any
  transformation.
\newblock {\em Physical Review Letters}, 80(19):4329.

\bibitem[Grover, 2005]{grover2005fixed}
Grover, L.~K. (2005).
\newblock Fixed-point quantum search.
\newblock {\em Physical Review Letters}, 95(15):150501.

\bibitem[Grover et~al., 2006]{grover2006quantum}
Grover, L.~K., Patel, A., and Tulsi, T. (2006).
\newblock Quantum algorithms with fixed points: The case of database search.
\newblock {\em arXiv preprint quant-ph/0603132}.

\bibitem[Haah, 2019]{haah2019product}
Haah, J. (2019).
\newblock Product decomposition of periodic functions in quantum signal
  processing.
\newblock {\em Quantum}, 3:190.

\bibitem[Hadfield et~al., 2019]{hadfield2019quantum}
Hadfield, S., Wang, Z., O’gorman, B., Rieffel, E.~G., Venturelli, D., and
  Biswas, R. (2019).
\newblock From the quantum approximate optimization algorithm to a quantum
  alternating operator ansatz.
\newblock {\em Algorithms}, 12(2):34.

\bibitem[Hallgren et~al., 2003]{hallgren2003hidden}
Hallgren, S., Russell, A., and Ta-Shma, A. (2003).
\newblock The hidden subgroup problem and quantum computation using group
  representations.
\newblock {\em SIAM Journal on Computing}, 32(4):916--934.

\bibitem[Harrow et~al., 2009]{harrow2009quantum}
Harrow, A.~W., Hassidim, A., and Lloyd, S. (2009).
\newblock Quantum algorithm for linear systems of equations.
\newblock {\em Physical Review Letters}, 103(15):150502.

\bibitem[Hart and Istrail, 1997]{hart1997robust}
Hart, W.~E. and Istrail, S. (1997).
\newblock Robust proofs of np-hardness for protein folding: general lattices
  and energy potentials.
\newblock {\em Journal of Computational Biology}, 4(1):1--22.

\bibitem[Hastings et~al., 2014]{hastings2014self}
Hastings, M.~B., Watson, G.~H., and Melko, R.~G. (2014).
\newblock Self-correcting quantum memories beyond the percolation threshold.
\newblock {\em Physical Review Letters}, 112(7):070501.

\bibitem[Hastings, 1970]{hastings1970monte}
Hastings, W.~K. (1970).
\newblock Monte carlo sampling methods using markov chains and their
  applications.
\newblock {\em Biometrika}, 57(1).

\bibitem[Havl{\'\i}{\v{c}}ek et~al., 2017]{havlivcek2017operator}
Havl{\'\i}{\v{c}}ek, V., Troyer, M., and Whitfield, J.~D. (2017).
\newblock Operator locality in the quantum simulation of fermionic models.
\newblock {\em Physical Review A}, 95(3):032332.

\bibitem[Hohenberg and Kohn, 1964]{hohenberg1964inhomogeneous}
Hohenberg, P. and Kohn, W. (1964).
\newblock Inhomogeneous electron gas.
\newblock {\em Physical Review}, 136(3B):B864.

\bibitem[Huang et~al., 2021]{huang2021provably}
Huang, H.-Y., Kueng, R., Torlai, G., Albert, V.~V., and Preskill, J. (2021).
\newblock Provably efficient machine learning for quantum many-body problems.
\newblock {\em arXiv preprint arXiv:2106.12627}.

\bibitem[Ivanyos et~al., 2008]{ivanyos2008efficient}
Ivanyos, G., Sanselme, L., and Santha, M. (2008).
\newblock An efficient quantum algorithm for the hidden subgroup problem in
  nil-2 groups.
\newblock In {\em Latin American Symposium on Theoretical Informatics}, pages
  759--771. Springer.

\bibitem[Jensen, 2013]{jensen2013atomic}
Jensen, F. (2013).
\newblock Atomic orbital basis sets.
\newblock {\em Wiley Interdisciplinary Reviews: Computational Molecular
  Science}, 3(3):273--295.

\bibitem[Jin, 2015]{jin2015finite}
Jin, J.-M. (2015).
\newblock {\em The finite element method in electromagnetics}.
\newblock John Wiley \& Sons.

\bibitem[Jumper et~al., 2021]{AlphaFoldv2}
Jumper, J., Evans, R., Pritzel, A., Green, T., Figurnov, M., Ronneberger, O.,
  Tunyasuvunakool, K., Bates, R., {\v{Z}}{\'\i}dek, A., Potapenko, A., et~al.
  (2021).
\newblock Highly accurate protein structure prediction with alphafold.
\newblock {\em Nature}, 596(7873):583--589.

\bibitem[Karmarkar, 1984]{karmarkar1984new}
Karmarkar, N. (1984).
\newblock A new polynomial-time algorithm for linear programming.
\newblock In {\em Proceedings of the sixteenth annual ACM Symposium on Theory
  of Computing}, pages 302--311.

\bibitem[Kassal et~al., 2008]{kassal2008polynomial}
Kassal, I., Jordan, S.~P., Love, P.~J., Mohseni, M., and Aspuru-Guzik, A.
  (2008).
\newblock Polynomial-time quantum algorithm for the simulation of chemical
  dynamics.
\newblock {\em Proceedings of the National Academy of Sciences},
  105(48):18681--18686.

\bibitem[Keen et~al., 2021]{keen2021quantum}
Keen, T., Dumitrescu, E., and Wang, Y. (2021).
\newblock Quantum algorithms for ground-state preparation and green's function
  calculation.
\newblock {\em arXiv preprint arXiv:2112.05731}.

\bibitem[Kempe et~al., 2006]{kempe2006complexity}
Kempe, J., Kitaev, A.~Y., and Regev, O. (2006).
\newblock The complexity of the local hamiltonian problem.
\newblock {\em SIAM Journal of Computing}, 35(5):1070--1097.

\bibitem[Kerenidis and Prakash, 2016]{kerenidis2016quantum}
Kerenidis, I. and Prakash, A. (2016).
\newblock Quantum recommendation systems.
\newblock {\em arXiv preprint arXiv:1603.08675}.

\bibitem[Kesselring et~al., 2018]{kesselring2018boundaries}
Kesselring, M.~S., Pastawski, F., Eisert, J., and Brown, B.~J. (2018).
\newblock The boundaries and twist defects of the color code and their
  applications to topological quantum computation.
\newblock {\em Quantum}, 2:101.

\bibitem[Khachiyan, 1979]{khachiyan1979polynomial}
Khachiyan, L.~G. (1979).
\newblock A polynomial algorithm in linear programming.
\newblock In {\em Doklady Akademii Nauk}, volume 244, pages 1093--1096. Russian
  Academy of Sciences.

\bibitem[Khoury and Hadfield-Menell, 2018]{khoury2018geometry}
Khoury, M. and Hadfield-Menell, D. (2018).
\newblock On the geometry of adversarial examples.
\newblock {\em arXiv preprint arXiv:1811.00525}.

\bibitem[Kieferov{\'a} et~al., 2019]{kieferova2019simulating}
Kieferov{\'a}, M., Scherer, A., and Berry, D.~W. (2019).
\newblock Simulating the dynamics of time-dependent hamiltonians with a
  truncated dyson series.
\newblock {\em Physical Review A}, 99(4):042314.

\bibitem[Kim et~al., 2022]{kim2022fault}
Kim, I.~H., Liu, Y.-H., Pallister, S., Pol, W., Roberts, S., and Lee, E.
  (2022).
\newblock Fault-tolerant resource estimate for quantum chemical simulations:
  Case study on li-ion battery electrolyte molecules.
\newblock {\em Physical Review Research}, 4(2):023019.

\bibitem[Kirby et~al., 2022]{kirby2022second}
Kirby, W., Fuller, B., Hadfield, C., and Mezzacapo, A. (2022).
\newblock Second-quantized fermionic operators with polylogarithmic qubit and
  gate complexity.
\newblock {\em PRX Quantum}, 3:020351.

\bibitem[Kirkpatrick et~al., 2021]{kirkpatrick2021pushing}
Kirkpatrick, J., McMorrow, B., Turban, D.~H., Gaunt, A.~L., Spencer, J.~S.,
  Matthews, A.~G., Obika, A., Thiry, L., Fortunato, M., Pfau, D., et~al.
  (2021).
\newblock Pushing the frontiers of density functionals by solving the
  fractional electron problem.
\newblock {\em Science}, 374(6573):1385--1389.

\bibitem[Kirkpatrick et~al., 1983]{kirkpatrick1983optimization}
Kirkpatrick, S., Gelatt~Jr, C.~D., and Vecchi, M.~P. (1983).
\newblock Optimization by simulated annealing.
\newblock {\em Science}, 220(4598):671--680.

\bibitem[Kitaev, 1995]{kitaev1995quantum}
Kitaev, A.~Y. (1995).
\newblock Quantum measurements and the abelian stabilizer problem.
\newblock {\em arXiv preprint quant-ph/9511026}.

\bibitem[Kitaev, 2003]{kitaev2003fault}
Kitaev, A.~Y. (2003).
\newblock Fault-tolerant quantum computation by anyons.
\newblock {\em Annals of Physics}, 303(1):2--30.

\bibitem[Kivlichan et~al., 2020]{kivlichan2020improved}
Kivlichan, I.~D., Gidney, C., Berry, D.~W., Wiebe, N., McClean, J.~R., Sun, W.,
  Jiang, Z., Rubin, N., Fowler, A., Aspuru-Guzik, A., et~al. (2020).
\newblock Improved fault-tolerant quantum simulation of condensed-phase
  correlated electrons via trotterization.
\newblock {\em Quantum}, 4:296.

\bibitem[Kivlichan et~al., 2018]{kivlichan2018quantum}
Kivlichan, I.~D., McClean, J.~R., Wiebe, N., Gidney, C., Aspuru-Guzik, A.,
  Chan, G. K.-L., and Babbush, R. (2018).
\newblock Quantum simulation of electronic structure with linear depth and
  connectivity.
\newblock {\em Physical Review Letters}, 120(11):110501.

\bibitem[Kohn, 1985]{kohn1985highlights}
Kohn, W. (1985).
\newblock Highlights of condensed matter theory.
\newblock In {\em International School of Physics” Enrico Fermi}, pages
  1--15.

\bibitem[Kohn, 1999]{kohn1999nobel}
Kohn, W. (1999).
\newblock Nobel lecture: Electronic structure of matter—wave functions and
  density functionals.
\newblock {\em Reviews of Modern Physics}, 71(5):1253.

\bibitem[Kohn and Sham, 1965]{kohn1965self}
Kohn, W. and Sham, L.~J. (1965).
\newblock Self-consistent equations including exchange and correlation effects.
\newblock {\em Physical Review}, 140(4A):A1133.

\bibitem[Koralov and Sinai, 2007]{koralov2007theory}
Koralov, L. and Sinai, Y.~G. (2007).
\newblock {\em Theory of probability and random processes}.
\newblock Springer Science \& Business Media.

\bibitem[Krinner et~al., 2022]{krinner2022realizing}
Krinner, S., Lacroix, N., Remm, A., Di~Paolo, A., Genois, E., Leroux, C.,
  Hellings, C., Lazar, S., Swiadek, F., Herrmann, J., et~al. (2022).
\newblock Realizing repeated quantum error correction in a distance-three
  surface code.
\newblock {\em Nature}, 605(7911):669--674.

\bibitem[Krovi et~al., 2016]{krovi2016quantum}
Krovi, H., Magniez, F., Ozols, M., and Roland, J. (2016).
\newblock Quantum walks can find a marked element on any graph.
\newblock {\em Algorithmica}, 74(2):851--907.

\bibitem[Kubica and Beverland, 2015]{kubica2015universal}
Kubica, A. and Beverland, M.~E. (2015).
\newblock Universal transversal gates with color codes: A simplified approach.
\newblock {\em Physical Review A}, 91(3):032330.

\bibitem[Kubica and Vasmer, 2021]{kubica2021single}
Kubica, A. and Vasmer, M. (2021).
\newblock Single-shot quantum error correction with the three-dimensional
  subsystem toric code.
\newblock {\em arXiv preprint arXiv:2106.02621}.

\bibitem[Kubica and Yoshida, 2018]{kubica2018ungauging}
Kubica, A. and Yoshida, B. (2018).
\newblock Ungauging quantum error-correcting codes.
\newblock {\em arXiv preprint arXiv:1805.01836}.

\bibitem[Kubica et~al., 2015]{kubica2015unfolding}
Kubica, A., Yoshida, B., and Pastawski, F. (2015).
\newblock Unfolding the color code.
\newblock {\em New Journal of Physics}, 17(8):083026.

\bibitem[Kutzelnigg and Morgan~III, 1992]{kutzelnigg1992rates}
Kutzelnigg, W. and Morgan~III, J.~D. (1992).
\newblock Rates of convergence of the partial-wave expansions of atomic
  correlation energies.
\newblock {\em The Journal of Chemical Physics}, 96(6):4484--4508.

\bibitem[Kyriienko, 2020]{kyriienko2020quantum}
Kyriienko, O. (2020).
\newblock Quantum inverse iteration algorithm for programmable quantum
  simulators.
\newblock {\em npj Quantum Information}, 6(1):1--8.

\bibitem[Lavasani and Barkeshli, 2018]{lavasani2018low}
Lavasani, A. and Barkeshli, M. (2018).
\newblock Low overhead clifford gates from joint measurements in surface,
  color, and hyperbolic codes.
\newblock {\em Physical Review A}, 98(5):052319.

\bibitem[Lee et~al., 1988]{lee1988development}
Lee, C., Yang, W., and Parr, R.~G. (1988).
\newblock Development of the colle-salvetti correlation-energy formula into a
  functional of the electron density.
\newblock {\em Physical Review B}, 37(2):785.

\bibitem[Lee et~al., 2021]{lee2020even}
Lee, J., Berry, D.~W., Gidney, C., Huggins, W.~J., McClean, J.~R., Wiebe, N.,
  and Babbush, R. (2021).
\newblock Even more efficient quantum computations of chemistry through tensor
  hypercontraction.
\newblock {\em PRX Quantum}, 2(3):030305.

\bibitem[Lee et~al., 2022]{lee2022there}
Lee, S., Lee, J., Zhai, H., Tong, Y., Dalzell, A.~M., Kumar, A., Helms, P.,
  Gray, J., Cui, Z.-H., Liu, W., et~al. (2022).
\newblock Is there evidence for exponential quantum advantage in quantum
  chemistry?
\newblock {\em arXiv preprint arXiv:2208.02199}.

\bibitem[Lee and Jeong, 2022]{lee2022universal}
Lee, S.-H. and Jeong, H. (2022).
\newblock Universal hardware-efficient topological measurement-based quantum
  computation via color-code-based cluster states.
\newblock {\em Physical Review Research}, 4(1):013010.

\bibitem[Lemieux et~al., 2021]{lemieux2021resource}
Lemieux, J., Duclos-Cianci, G., S{\'e}n{\'e}chal, D., and Poulin, D. (2021).
\newblock Resource estimate for quantum many-body ground-state preparation on a
  quantum computer.
\newblock {\em Physical Review A}, 103(5):052408.

\bibitem[Lemieux et~al., 2020]{lemieux2019efficient}
Lemieux, J., Heim, B., Poulin, D., Svore, K., and Troyer, M. (2020).
\newblock Efficient {Q}uantum {W}alk {C}ircuits for {M}etropolis-{H}astings
  {A}lgorithm.
\newblock {\em {Quantum}}, 4:287.

\bibitem[Lin and Tong, 2020]{lin2020near}
Lin, L. and Tong, Y. (2020).
\newblock Near-optimal ground state preparation.
\newblock {\em Quantum}, 4:372.

\bibitem[Litinski, 2019a]{litinski2019game}
Litinski, D. (2019a).
\newblock A game of surface codes: Large-scale quantum computing with lattice
  surgery.
\newblock {\em Quantum}, 3:128.

\bibitem[Litinski, 2019b]{litinski2019magic}
Litinski, D. (2019b).
\newblock Magic state distillation: Not as costly as you think.
\newblock {\em Quantum}, 3:205.

\bibitem[Litinski and von Oppen, 2018]{litinski2018quantum}
Litinski, D. and von Oppen, F. (2018).
\newblock Quantum computing with majorana fermion codes.
\newblock {\em Physical Review B}, 97(20):205404.

\bibitem[Loaiza et~al., 2022]{loaiza2022reducing}
Loaiza, I., Khah, A.~M., Wiebe, N., and Izmaylov, A.~F. (2022).
\newblock Reducing molecular electronic hamiltonian simulation cost for linear
  combination of unitaries approaches.
\newblock {\em arXiv preprint arXiv:2208.08272}.

\bibitem[Low and Chuang, 2017]{low2017optimal}
Low, G.~H. and Chuang, I.~L. (2017).
\newblock Optimal hamiltonian simulation by quantum signal processing.
\newblock {\em Physical Review Letters}, 118(1):010501.

\bibitem[Low and Chuang, 2019]{low2019qubitization}
Low, G.~H. and Chuang, I.~L. (2019).
\newblock Hamiltonian simulation by qubitization.
\newblock {\em Quantum}, 3:163.

\bibitem[Low et~al., 2018]{low2018trading}
Low, G.~H., Kliuchnikov, V., and Schaeffer, L. (2018).
\newblock Trading t-gates for dirty qubits in state preparation and unitary
  synthesis.
\newblock {\em arXiv preprint arXiv:1812.00954}.

\bibitem[Low and Wiebe, 2018]{low2018hamiltonian}
Low, G.~H. and Wiebe, N. (2018).
\newblock Hamiltonian simulation in the interaction picture.
\newblock {\em arXiv preprint arXiv:1805.00675}.

\bibitem[Madsen et~al., 2022]{madsen2022quantum}
Madsen, L.~S., Laudenbach, F., Askarani, M.~F., Rortais, F., Vincent, T.,
  Bulmer, J.~F., Miatto, F.~M., Neuhaus, L., Helt, L.~G., Collins, M.~J.,
  et~al. (2022).
\newblock Quantum computational advantage with a programmable photonic
  processor.
\newblock {\em Nature}, 606(7912):75--81.

\bibitem[Magniez et~al., 2011]{magniez2011search}
Magniez, F., Nayak, A., Roland, J., and Santha, M. (2011).
\newblock Search via quantum walk.
\newblock {\em SIAM Journal on Computing}, 40(1):142--164.

\bibitem[Manthiram, 2020]{manthiram2020reflection}
Manthiram, A. (2020).
\newblock A reflection on lithium-ion battery cathode chemistry.
\newblock {\em Nature Communications}, 11(1):1--9.

\bibitem[Martin, 2020]{martin2020electronic}
Martin, R.~M. (2020).
\newblock {\em Electronic structure: basic theory and practical methods}.
\newblock Cambridge university press.

\bibitem[Martyn et~al., 2021]{martyn2021grand}
Martyn, J.~M., Rossi, Z.~M., Tan, A.~K., and Chuang, I.~L. (2021).
\newblock Grand unification of quantum algorithms.
\newblock {\em PRX Quantum}, 2(4):040203.

\bibitem[Mas et~al., 2023]{escrig2022parameter}
Mas, G.~E., Campos, R., Casares, P. A.~M., and Martin-Delgado, M.~A. (2023).
\newblock Parameter estimation of gravitational waves with a quantum metropolis
  algorithm.
\newblock {\em Classical and Quantum Gravity}.

\bibitem[Matousek and G{\"a}rtner, 2006]{matousek2006understanding}
Matousek, J. and G{\"a}rtner, B. (2006).
\newblock {\em Understanding and using linear programming}.
\newblock Springer Science \& Business Media.

\bibitem[McArdle et~al., 2022]{mcardle2022exploiting}
McArdle, S., Campbell, E.~T., and Su, Y. (2022).
\newblock Exploiting fermion number in factorized decompositions of the
  electronic structure hamiltonian.
\newblock {\em Physical Review A}, 105(1):012403.

\bibitem[McClain et~al., 2017]{mcclain2017gaussian}
McClain, J., Sun, Q., Chan, G. K.-L., and Berkelbach, T.~C. (2017).
\newblock Gaussian-based coupled-cluster theory for the ground-state and band
  structure of solids.
\newblock {\em Journal of Chemical Theory and Computation}, 13(3):1209--1218.

\bibitem[McClean et~al., 2018]{mcclean2018barren}
McClean, J.~R., Boixo, S., Smelyanskiy, V.~N., Babbush, R., and Neven, H.
  (2018).
\newblock Barren plateaus in quantum neural network training landscapes.
\newblock {\em Nature Communications}, 9(1):1--6.

\bibitem[McClean et~al., 2021]{mcclean2021foundations}
McClean, J.~R., Rubin, N.~C., Lee, J., Harrigan, M.~P., O’Brien, T.~E.,
  Babbush, R., Huggins, W.~J., and Huang, H.-Y. (2021).
\newblock What the foundations of quantum computer science teach us about
  chemistry.
\newblock {\em The Journal of Chemical Physics}, 155(15):150901.

\bibitem[Metropolis et~al., 1953]{metropolis1953equation}
Metropolis, N., Rosenbluth, A.~W., Rosenbluth, M.~N., Teller, A.~H., and
  Teller, E. (1953).
\newblock Equation of state calculations by fast computing machines.
\newblock {\em The Journal of Chemical Physics}, 21(6):1087--1092.

\bibitem[Mohammadisiahroudi et~al., 2022]{mohammadisiahroudi2022efficient}
Mohammadisiahroudi, M., Fakhimi, R., and Terlaky, T. (2022).
\newblock Efficient use of quantum linear system algorithms in interior point
  methods for linear optimization.
\newblock {\em arXiv preprint arXiv:2205.01220}.

\bibitem[M{\o}ller and Plesset, 1934]{moller1934note}
M{\o}ller, C. and Plesset, M.~S. (1934).
\newblock Note on an approximation treatment for many-electron systems.
\newblock {\em Physical Review}, 46(7):618.

\bibitem[Moore et~al., 2008]{moore2008symmetric}
Moore, C., Russell, A., and Schulman, L.~J. (2008).
\newblock The symmetric group defies strong fourier sampling.
\newblock {\em SIAM Journal on Computing}, 37(6):1842--1864.

\bibitem[Moore, 1920]{moore1920reciprocal}
Moore, E.~H. (1920).
\newblock On the reciprocal of the general algebraic matrix.
\newblock {\em Bulletin of the American Mathematical Society}, 26:394--395.

\bibitem[Mosca and Ekert, 1998]{mosca1998hidden}
Mosca, M. and Ekert, A. (1998).
\newblock The hidden subgroup problem and eigenvalue estimation on a quantum
  computer.
\newblock In {\em NASA International Conference on Quantum Computing and
  Quantum Communications}, pages 174--188. Springer.

\bibitem[Motta et~al., 2020]{motta2020determining}
Motta, M., Sun, C., Tan, A.~T., O’Rourke, M.~J., Ye, E., Minnich, A.~J.,
  Brandao, F.~G., and Chan, G.~K. (2020).
\newblock Determining eigenstates and thermal states on a quantum computer
  using quantum imaginary time evolution.
\newblock {\em Nature Physics}, 16(2):205--210.

\bibitem[Moult et~al., 1995]{CASP}
Moult, J., Pedersen, J.~T., Judson, R., and Fidelis, K. (1995).
\newblock A large-scale experiment to assess protein structure prediction
  methods.
\newblock {\em Proteins: Structure, Function, and Bioinformatics},
  23(3):ii--iv.

\bibitem[Mulligan et~al., 2020]{mulligan2020designing}
Mulligan, V.~K., Melo, H., Merritt, H.~I., Slocum, S., Weitzner, B.~D.,
  Watkins, A.~M., Renfrew, P.~D., Pelissier, C., Arora, P.~S., and Bonneau, R.
  (2020).
\newblock Designing peptides on a quantum computer.
\newblock {\em bioRxiv}, page 752485.

\bibitem[Murty, 1983]{murty1983linear}
Murty, K.~G. (1983).
\newblock {\em Linear programming}.
\newblock Springer.

\bibitem[Nagaj et~al., 2009]{nagaj2009fast}
Nagaj, D., Wocjan, P., and Zhang, Y. (2009).
\newblock Fast amplification of qma.
\newblock {\em Quantum Information \& Computation}, 9(11):1053–1068.

\bibitem[Nielsen and Chuang, 2002]{nielsen2002quantum}
Nielsen, M.~A. and Chuang, I.~L. (2002).
\newblock {\em Quantum Computation and Quantum Information}.
\newblock American Association of Physics Teachers.

\bibitem[Noh et~al., 2022]{noh2022low}
Noh, K., Chamberland, C., and Brand{\~a}o, F.~G. (2022).
\newblock Low-overhead fault-tolerant quantum error correction with the
  surface-gkp code.
\newblock {\em PRX Quantum}, 3(1):010315.

\bibitem[Ortiz et~al., 2001]{ortiz2001quantum}
Ortiz, G., Gubernatis, J.~E., Knill, E., and Laflamme, R. (2001).
\newblock Quantum algorithms for fermionic simulations.
\newblock {\em Physical Review A}, 64(2):022319.

\bibitem[Oumarou et~al., 2022]{oumarou2022accelerating}
Oumarou, O., Scheurer, M., Parrish, R.~M., Hohenstein, E.~G., and Gogolin, C.
  (2022).
\newblock Accelerating quantum computations of chemistry through regularized
  compressed double factorization.
\newblock {\em arXiv preprint arXiv:2212.07957}.

\bibitem[Outeiral et~al., 2021]{outeiral2020investigating}
Outeiral, C., Morris, G.~M., Shi, J., Strahm, M., Benjamin, S.~C., and Deane,
  C.~M. (2021).
\newblock Investigating the potential for a limited quantum speedup on protein
  lattice problems.
\newblock {\em New Journal of Physics}, 23(10):103030.

\bibitem[Pastawski and Yoshida, 2015]{pastawski2015fault}
Pastawski, F. and Yoshida, B. (2015).
\newblock Fault-tolerant logical gates in quantum error-correcting codes.
\newblock {\em Physical Review A}, 91(1):012305.

\bibitem[Perdomo-Ortiz et~al., 2012]{perdomo2012finding}
Perdomo-Ortiz, A., Dickson, N., Drew-Brook, M., Rose, G., and Aspuru-Guzik, A.
  (2012).
\newblock Finding low-energy conformations of lattice protein models by quantum
  annealing.
\newblock {\em Scientific Reports}, 2:571.

\bibitem[Perdomo-Ortiz et~al., 2008]{perdomo2008construction}
Perdomo-Ortiz, A., Truncik, C., Tubert-Brohman, I., Rose, G., and Aspuru-Guzik,
  A. (2008).
\newblock Construction of model hamiltonians for adiabatic quantum computation
  and its application to finding low-energy conformations of lattice protein
  models.
\newblock {\em Physical Review A}, 78(1):012320.

\bibitem[Peruzzo et~al., 2014]{peruzzo2014variational}
Peruzzo, A., McClean, J.~R., Shadbolt, P., Yung, M.-H., Zhou, X.-Q., Love,
  P.~J., Aspuru-Guzik, A., and O’brien, J.~L. (2014).
\newblock A variational eigenvalue solver on a photonic quantum processor.
\newblock {\em Nature Communications}, 5(1):1--7.

\bibitem[Portugal, 2013]{portugal2013quantum}
Portugal, R. (2013).
\newblock {\em Quantum walks and search algorithms}, volume~19.
\newblock Springer.

\bibitem[Postler et~al., 2022]{postler2022demonstration}
Postler, L., Heu$\beta$en, S., Pogorelov, I., Rispler, M., Feldker, T., Meth,
  M., Marciniak, C.~D., Stricker, R., Ringbauer, M., Blatt, R., et~al. (2022).
\newblock Demonstration of fault-tolerant universal quantum gate operations.
\newblock {\em Nature}, 605(7911):675--680.

\bibitem[Potra and Wright, 2000]{potra2000interior}
Potra, F.~A. and Wright, S.~J. (2000).
\newblock Interior-point methods.
\newblock {\em Journal of Computational and Applied Mathematics},
  124(1-2):281--302.

\bibitem[Preskill, 2022]{preskill1999lecture}
Preskill, J. (2022).
\newblock Lecture notes for physics 219: Quantum computation.
\newblock {\em Caltech Lecture Notes}.

\bibitem[Rebentrost et~al., 2014]{rebentrost2014quantum}
Rebentrost, P., Mohseni, M., and Lloyd, S. (2014).
\newblock Quantum support vector machine for big data classification.
\newblock {\em Physical Review Letters}, 113(13):130503.

\bibitem[Reiher et~al., 2017]{reiher2017elucidating}
Reiher, M., Wiebe, N., Svore, K.~M., Wecker, D., and Troyer, M. (2017).
\newblock Elucidating reaction mechanisms on quantum computers.
\newblock {\em Proceedings of the National Academy of Sciences},
  114(29):7555--7560.

\bibitem[Remler and Madden, 1990]{remler1990molecular}
Remler, D.~K. and Madden, P.~A. (1990).
\newblock Molecular dynamics without effective potentials via the
  car-parrinello approach.
\newblock {\em Molecular Physics}, 70(6):921--966.

\bibitem[Robert et~al., 2021]{robert2021resource}
Robert, A., Barkoutsos, P.~K., Woerner, S., and Tavernelli, I. (2021).
\newblock Resource-efficient quantum algorithm for protein folding.
\newblock {\em npj Quantum Information}, 7(1):1--5.

\bibitem[Romero et~al., 2018]{romero2018strategies}
Romero, J., Babbush, R., McClean, J.~R., Hempel, C., Love, P.~J., and
  Aspuru-Guzik, A. (2018).
\newblock Strategies for quantum computing molecular energies using the unitary
  coupled cluster ansatz.
\newblock {\em Quantum Science and Technology}, 4(1):014008.

\bibitem[Roothaan, 1951]{roothaan1951new}
Roothaan, C. C.~J. (1951).
\newblock New developments in molecular orbital theory.
\newblock {\em Reviews of Modern Physics}, 23(2):69.

\bibitem[Ryabinkin et~al., 2018]{ryabinkin2018qubit}
Ryabinkin, I.~G., Yen, T.-C., Genin, S.~N., and Izmaylov, A.~F. (2018).
\newblock Qubit coupled cluster method: a systematic approach to quantum
  chemistry on a quantum computer.
\newblock {\em Journal of Chemical Theory and Computation}, 14(12):6317--6326.

\bibitem[Ryan-Anderson et~al., 2021]{ryan2021realization}
Ryan-Anderson, C., Bohnet, J., Lee, K., Gresh, D., Hankin, A., Gaebler, J.,
  Francois, D., Chernoguzov, A., Lucchetti, D., Brown, N., et~al. (2021).
\newblock Realization of real-time fault-tolerant quantum error correction.
\newblock {\em Physical Review X}, 11(4):041058.

\bibitem[Schuch and Verstraete, 2009]{schuch2009computational}
Schuch, N. and Verstraete, F. (2009).
\newblock Computational complexity of interacting electrons and fundamental
  limitations of density functional theory.
\newblock {\em Nature Physics}, 5(10):732--735.

\bibitem[Schuld et~al., 2019]{schuld2019evaluating}
Schuld, M., Bergholm, V., Gogolin, C., Izaac, J., and Killoran, N. (2019).
\newblock Evaluating analytic gradients on quantum hardware.
\newblock {\em Physical Review A}, 99(3):032331.

\bibitem[Seeley et~al., 2012]{seeley2012bravyi}
Seeley, J.~T., Richard, M.~J., and Love, P.~J. (2012).
\newblock The bravyi-kitaev transformation for quantum computation of
  electronic structure.
\newblock {\em The Journal of Chemical Physics}, 137(22):224109.

\bibitem[Senior et~al., 2020]{AlphaFold}
Senior, A., Evans, R., Jumper, J., Kirkpatrick, J., Sifre, L., Green, T., Qin,
  C., Zidek, A., Nelson, A., Bridgland, A., et~al. (2020).
\newblock Improved protein structure prediction using potentials from deep
  learning.
\newblock {\em Nature}.

\bibitem[Shavitt and Bartlett, 2009]{shavitt2009many}
Shavitt, I. and Bartlett, R.~J. (2009).
\newblock {\em Many-body methods in chemistry and physics: MBPT and
  coupled-cluster theory}.
\newblock Cambridge university press.

\bibitem[Shepherd et~al., 2012]{shepherd2012convergence}
Shepherd, J.~J., Gr{\"u}neis, A., Booth, G.~H., Kresse, G., and Alavi, A.
  (2012).
\newblock Convergence of many-body wave-function expansions using a plane-wave
  basis: From homogeneous electron gas to solid state systems.
\newblock {\em Physical Review B}, 86(3):035111.

\bibitem[Shewchuk et~al., 1994]{shewchuk1994introduction}
Shewchuk, J.~R. et~al. (1994).
\newblock An introduction to the conjugate gradient method without the
  agonizing pain.

\bibitem[Shor, 1999]{shor1999polynomial}
Shor, P.~W. (1999).
\newblock Polynomial-time algorithms for prime factorization and discrete
  logarithms on a quantum computer.
\newblock {\em SIAM review}, 41(2):303--332.

\bibitem[Simon, 1997]{simon1997power}
Simon, D.~R. (1997).
\newblock On the power of quantum computation.
\newblock {\em SIAM Journal on Computing}, 26(5):1474--1483.

\bibitem[Soares~Jr and Da~Silva, 2018]{soares2018hyperbolic}
Soares~Jr, W.~S. and Da~Silva, E.~B. (2018).
\newblock Hyperbolic quantum color codes.
\newblock {\em Quantum Information and Computation}, 18(3 and 4).

\bibitem[Somma et~al., 2007]{somma2007quantum}
Somma, R.~D., Boixo, S., and Barnum, H. (2007).
\newblock Quantum simulated annealing.
\newblock {\em arXiv preprint arXiv:0712.1008}.

\bibitem[Somma et~al., 2008]{somma2008quantum}
Somma, R.~D., Boixo, S., Barnum, H., and Knill, E. (2008).
\newblock Quantum simulations of classical annealing processes.
\newblock {\em Physical Review Letters}, 101(13):130504.

\bibitem[Somma et~al., 2002]{somma2002simulating}
Somma, R.~D., Ortiz, G., Gubernatis, J.~E., Knill, E., and Laflamme, R. (2002).
\newblock Simulating physical phenomena by quantum networks.
\newblock {\em Physical Review A}, 65(4):042323.

\bibitem[Stair and Evangelista, 2021]{stair2021simulating}
Stair, N.~H. and Evangelista, F.~A. (2021).
\newblock Simulating many-body systems with a projective quantum eigensolver.
\newblock {\em PRX Quantum}, 2(3):030301.

\bibitem[Stair et~al., 2020]{stair2020multireference}
Stair, N.~H., Huang, R., and Evangelista, F.~A. (2020).
\newblock A multireference quantum krylov algorithm for strongly correlated
  electrons.
\newblock {\em Journal of Chemical Theory and Computation}, 16(4):2236--2245.

\bibitem[Stephens et~al., 1994]{stephens1994ab}
Stephens, P.~J., Devlin, F.~J., Chabalowski, C.~F., and Frisch, M.~J. (1994).
\newblock Ab initio calculation of vibrational absorption and circular
  dichroism spectra using density functional force fields.
\newblock {\em The Journal of Physical Chemistry}, 98(45):11623--11627.

\bibitem[Steudtner, 2019]{steudtner2019methods}
Steudtner, M. (2019).
\newblock {\em Methods to simulate fermions on quantum computers with hardware
  limitations}.
\newblock PhD thesis, Leiden.

\bibitem[Steudtner and Wehner, 2018]{steudtner2018fermion}
Steudtner, M. and Wehner, S. (2018).
\newblock Fermion-to-qubit mappings with varying resource requirements for
  quantum simulation.
\newblock {\em New Journal of Physics}, 20(6):063010.

\bibitem[Student, 1908]{student1908ttest}
Student (1908).
\newblock The probable error of a mean.
\newblock {\em Biometrika}, pages 1--25.

\bibitem[Su et~al., 2021a]{su2021fault}
Su, Y., Berry, D.~W., Wiebe, N., Rubin, N., and Babbush, R. (2021a).
\newblock Fault-tolerant quantum simulations of chemistry in first
  quantization.
\newblock {\em PRX Quantum}, 2(4):040332.

\bibitem[Su et~al., 2021b]{su2021nearly}
Su, Y., Huang, H.-Y., and Campbell, E. (2021b).
\newblock Nearly tight trotterization of interacting electrons.
\newblock {\em Quantum}, 5:495.

\bibitem[Suzuki, 1991]{suzuki1991general}
Suzuki, M. (1991).
\newblock General theory of fractal path integrals with applications to
  many-body theories and statistical physics.
\newblock {\em Journal of Mathematical Physics}, 32(2):400--407.

\bibitem[Szabo and Ostlund, 2012]{szabo2012modern}
Szabo, A. and Ostlund, N.~S. (2012).
\newblock {\em Modern quantum chemistry: introduction to advanced electronic
  structure theory}.
\newblock Courier Corporation.

\bibitem[Szegedy et~al., 2013]{szegedy2013intriguing}
Szegedy, C., Zaremba, W., Sutskever, I., Bruna, J., Erhan, D., Goodfellow, I.,
  and Fergus, R. (2013).
\newblock Intriguing properties of neural networks.
\newblock {\em arXiv preprint arXiv:1312.6199}.

\bibitem[Szegedy, 2004]{szegedy2004quantum}
Szegedy, M. (2004).
\newblock Quantum speed-up of markov chain based algorithms.
\newblock In {\em 45th Annual IEEE Symposium on Foundations of Computer
  Science}, pages 32--41. IEEE.

\bibitem[Tang, 2019]{tang2019quantum}
Tang, E. (2019).
\newblock A quantum-inspired classical algorithm for recommendation systems.
\newblock In {\em Proceedings of the 51st Annual ACM SIGACT Symposium on Theory
  of Computing}, pages 217--228.

\bibitem[Tang, 2021]{tang2018pca}
Tang, E. (2021).
\newblock Quantum principal component analysis only achieves an exponential
  speedup because of its state preparation assumptions.
\newblock {\em Physical Review Letters}, 127(6):060503.

\bibitem[Thouless, 1960]{thouless1960stability}
Thouless, D.~J. (1960).
\newblock Stability conditions and nuclear rotations in the hartree-fock
  theory.
\newblock {\em Nuclear Physics}, 21:225--232.

\bibitem[Tubman et~al., 2018]{tubman2018postponing}
Tubman, N.~M., Mejuto-Zaera, C., Epstein, J.~M., Hait, D., Levine, D.~S.,
  Huggins, W., Jiang, Z., McClean, J.~R., Babbush, R., Head-Gordon, M., et~al.
  (2018).
\newblock Postponing the orthogonality catastrophe: efficient state preparation
  for electronic structure simulations on quantum devices.
\newblock {\em arXiv preprint arXiv:1809.05523}.

\bibitem[Tzitrin et~al., 2020]{tzitrin2020progress}
Tzitrin, I., Bourassa, J.~E., Menicucci, N.~C., and Sabapathy, K.~K. (2020).
\newblock Progress towards practical qubit computation using approximate
  gottesman-kitaev-preskill codes.
\newblock {\em Physical Review A}, 101(3):032315.

\bibitem[{University of Washington}, 2021]{Rosetta@home}
{University of Washington} (2021).
\newblock Rosetta@home.
\newblock \url{boinc.bakerlab.org}.

\bibitem[Van~Laarhoven and Aarts, 1987]{van1987simulated}
Van~Laarhoven, P.~J. and Aarts, E.~H. (1987).
\newblock Simulated annealing.
\newblock In {\em Simulated Annealing: Theory and Applications}, pages 7--15.
  Springer.

\bibitem[Varadi et~al., 2022]{AlphaFold_database}
Varadi, M., Anyango, S., Deshpande, M., Nair, S., Natassia, C., Yordanova, G.,
  Yuan, D., Stroe, O., Wood, G., Laydon, A., et~al. (2022).
\newblock Alphafold protein structure database: Massively expanding the
  structural coverage of protein-sequence space with high-accuracy models.
\newblock {\em Nucleic Acids Research}, 50(D1):D439--D444.

\bibitem[Vasmer and Browne, 2019]{vasmer2019three}
Vasmer, M. and Browne, D.~E. (2019).
\newblock Three-dimensional surface codes: Transversal gates and fault-tolerant
  architectures.
\newblock {\em Physical Review A}, 100(1):012312.

\bibitem[Verstraete and Cirac, 2005]{verstraete2005mapping}
Verstraete, F. and Cirac, J.~I. (2005).
\newblock Mapping local hamiltonians of fermions to local hamiltonians of
  spins.
\newblock {\em Journal of Statistical Mechanics: Theory and Experiment},
  2005(09):P09012.

\bibitem[von Burg et~al., 2021]{von2021quantum}
von Burg, V., Low, G.~H., H{\"a}ner, T., Steiger, D.~S., Reiher, M., Roetteler,
  M., and Troyer, M. (2021).
\newblock Quantum computing enhanced computational catalysis.
\newblock {\em Physical Review Research}, 3(3):033055.

\bibitem[Vosko et~al., 1980]{vosko1980accurate}
Vosko, S.~H., Wilk, L., and Nusair, M. (1980).
\newblock Accurate spin-dependent electron liquid correlation energies for
  local spin density calculations: a critical analysis.
\newblock {\em Canadian Journal of Physics}, 58(8):1200--1211.

\bibitem[Wan et~al., 2022]{wan2021randomized}
Wan, K., Berta, M., and Campbell, E.~T. (2022).
\newblock Randomized quantum algorithm for statistical phase estimation.
\newblock {\em Physical Review Letters}, 129(3):030503.

\bibitem[Wang et~al., 2022]{wang2022quantum}
Wang, G., Stilck-Fran{\c{c}}a, D., Zhang, R., Zhu, S., and Johnson, P.~D.
  (2022).
\newblock Quantum algorithm for ground state energy estimation using circuit
  depth with exponentially improved dependence on precision.
\newblock {\em arXiv preprint arXiv:2209.06811}.

\bibitem[Wannier, 1937]{wannier1937structure}
Wannier, G.~H. (1937).
\newblock The structure of electronic excitation levels in insulating crystals.
\newblock {\em Physical Review}, 52(3):191.

\bibitem[Watrous, 2001]{watrous2001quantum}
Watrous, J. (2001).
\newblock Quantum algorithms for solvable groups.
\newblock In {\em Proceedings of the thirty-third annual ACM Symposium on
  Theory of Computing}, pages 60--67.

\bibitem[Watson et~al., 2015]{watson2015qudit}
Watson, F.~H., Campbell, E.~T., Anwar, H., and Browne, D.~E. (2015).
\newblock Qudit color codes and gauge color codes in all spatial dimensions.
\newblock {\em Physical Review A}, 92(2):022312.

\bibitem[Werner et~al., 2003]{werner2003fast}
Werner, H.-J., Manby, F.~R., and Knowles, P.~J. (2003).
\newblock Fast linear scaling second-order m{\o}ller-plesset perturbation
  theory (mp2) using local and density fitting approximations.
\newblock {\em The Journal of Chemical Physics}, 118(18):8149--8160.

\bibitem[Whitfield et~al., 2016]{whitfield2016local}
Whitfield, J.~D., Havl{\'\i}{\v{c}}ek, V., and Troyer, M. (2016).
\newblock Local spin operators for fermion simulations.
\newblock {\em Physical Review A}, 94(3):030301.

\bibitem[Whitfield et~al., 2014]{whitfield2014computational}
Whitfield, J.~D., Yung, M., Tempel, D.~G., Boixo, S., and Aspuru-Guzik, A.
  (2014).
\newblock Computational complexity of time-dependent density functional theory.
\newblock {\em New Journal of Physics}, 16(8):083035.

\bibitem[Wiebe and Granade, 2016]{wiebe2016efficient}
Wiebe, N. and Granade, C. (2016).
\newblock Efficient bayesian phase estimation.
\newblock {\em Physical Review Letters}, 117(1):010503.

\bibitem[Wierichs et~al., 2022]{wierichs2022general}
Wierichs, D., Izaac, J., Wang, C., and Lin, C. Y.-Y. (2022).
\newblock General parameter-shift rules for quantum gradients.
\newblock {\em Quantum}, 6:677.

\bibitem[Wigner and Jordan, 1928]{wigner1928paulische}
Wigner, E. and Jordan, P. (1928).
\newblock {\"U}ber das paulische {\"a}quivalenzverbot.
\newblock {\em Zeitschrift für Physik}, 47:631.

\bibitem[Wocjan and Abeyesinghe, 2008]{wocjan2008speedup}
Wocjan, P. and Abeyesinghe, A. (2008).
\newblock Speedup via quantum sampling.
\newblock {\em Physical Review A}, 78(4):042336.

\bibitem[Wocjan and Zhang, 2006]{wocjan2006several}
Wocjan, P. and Zhang, S. (2006).
\newblock Several natural bqp-complete problems.
\newblock {\em arXiv preprint quant-ph/0606179}.

\bibitem[Wong and Chang, 2021]{wong2021quantum}
Wong, R. and Chang, W.-L. (2021).
\newblock Quantum speedup for protein structure prediction.
\newblock {\em IEEE Transactions on NanoBioscience}.

\bibitem[Wossnig et~al., 2018]{wossnig2018quantum}
Wossnig, L., Zhao, Z., and Prakash, A. (2018).
\newblock Quantum linear system algorithm for dense matrices.
\newblock {\em Physical Review Letters}, 120(5):050502.

\bibitem[Ye et~al., 1994]{ye1994nl}
Ye, Y., Todd, M.~J., and Mizuno, S. (1994).
\newblock An {$O(\sqrt{n}L)$}-iteration homogeneous and self-dual linear
  programming algorithm.
\newblock {\em Mathematics of Operations Research}, 19(1):53--67.

\bibitem[Yoder et~al., 2014]{yoder2014fixed}
Yoder, T.~J., Low, G.~H., and Chuang, I.~L. (2014).
\newblock Fixed-point quantum search with an optimal number of queries.
\newblock {\em Physical Review Letters}, 113(21):210501.

\bibitem[Yung and Aspuru-Guzik, 2012]{yung2012quantum}
Yung, M.-H. and Aspuru-Guzik, A. (2012).
\newblock A quantum--quantum metropolis algorithm.
\newblock {\em Proceedings of the National Academy of Sciences},
  109(3):754--759.

\bibitem[Zhang et~al., 2022]{zhang2022computing}
Zhang, R., Wang, G., and Johnson, P. (2022).
\newblock Computing ground state properties with early fault-tolerant quantum
  computers.
\newblock {\em Quantum}, 6:761.

\bibitem[Zhao et~al., 2022]{zhao2021realizing}
Zhao, Y., Ye, Y., Huang, H.-L., Zhang, Y., Wu, D., Guan, H., Zhu, Q., Wei, Z.,
  He, T., Cao, S., et~al. (2022).
\newblock Realization of an error-correcting surface code with superconducting
  qubits.
\newblock {\em Physical Review Letters}, 129(3):030501.

\bibitem[Zhong et~al., 2020]{zhong2020quantum}
Zhong, H.-S., Wang, H., Deng, Y.-H., Chen, M.-C., Peng, L.-C., Luo, Y.-H., Qin,
  J., Wu, D., Ding, X., Hu, Y., et~al. (2020).
\newblock Quantum computational advantage using photons.
\newblock {\em Science}, 370(6523):1460--1463.

\end{thebibliography}




\end{spacing}


\begin{appendices} 

\end{appendices}

\printthesisindex 

\end{document}